\documentclass{article}
\usepackage[margin=1in]{geometry}
\usepackage{amsthm,amsmath,amssymb}
\usepackage{graphicx}
\usepackage[hidelinks,breaklinks]{hyperref}
\newtheorem{theorem}{Theorem}[]

\newtheorem{remark1}[theorem]{Remark}

\newlength{\vertsep}
\newlength{\imsize}
\newlength{\imsized}
\title{Controlling for multiple covariates}
\author{Mark Tygert\\{\normalsize Facebook Artificial Intelligence Research}\\
{\normalsize 1 Facebook Way, Menlo Park, CA 94025}\\
{\normalsize Main e-mail address:\ \ {\tt mark@tygert.com}}}

\begin{document}

\maketitle

\begin{abstract}
A fundamental problem in statistics is to compare
the outcomes attained by members of subpopulations,
whether comparing a subpopulation to the full population
or comparing two distinct subpopulations.
This problem arises in the analysis of randomized controlled trials,
in the analysis of A/B tests, and in the assessment of fairness and bias
in the treatment of sensitive subpopulations,
and is becoming especially important in measuring the effects
of algorithms and machine-learned systems.
Very often the comparison makes the most sense when performed
separately for individuals who are similar according to certain characteristics
given by the values of covariates of interest;
the separate comparisons can also be aggregated in various ways
to compare across all values of the covariates.
Separating, segmenting, or stratifying into those with similar values
of the covariates is also known as ``conditioning on'' or ``controlling for''
those covariates.
For instance, controlling for age or annual income is common.

Two standard methods of controlling for covariates are (1) binning
and (2) regression modeling.
Binning requires making fairly arbitrary, yet frequently highly influential
choices, and is unsatisfactorily temperamental in multiple dimensions,
with multiple covariates.
Regression analysis works wonderfully when there is good reason to believe
in a particular regression model or classifier (such as logistic regression).
Regression typically assumes a parametric or semi-parametric model,
possibly validated with non-parametric goodness-of-fit tests.
Goodness-of-fit only very indirectly characterizes differences
of a subpopulation from the full population
or between two distinct subpopulations.
Thus, there appears to be no extant canonical fully non-parametric regression
for the comparison of a subpopulation to the full population
or for the comparison of two distinct subpopulations,
not while conditioning on multiple specified covariates.
Existing methods rely on analysts to make choices,
and those choices can be debatable;
analysts can deceive others or even themselves,
whether purposefully or unintentionally.
The present paper aims to provide
an essentially unique fully non-parametric method
for such comparisons, combining two ingredients:
(1) recently developed methodologies for such comparisons
that already exist when conditioning on a single scalar covariate and
(2) the Hilbert space-filling curve that maps continuously
from one dimension to multiple dimensions.
\end{abstract}

\section{Introduction}
\label{intro}

Controlling for specified covariates during analysis of differences
between two subpopulations is essential in comparisons
throughout biomedicine, the fairness of algorithms,
and the social sciences.
Formally, controlling for specified covariates refers to conditioning
on those covariates (that is, on the ``independent variables'') when analyzing
the differences in responses (that is, in ``dependent variables'');
the point is to compare responses between individuals
who have similar values for the covariates.
Many studies in biomedicine, the fairness of algorithms,
and the social sciences aim to compare individuals
whose socioeconomic status is similar, for instance.
After comparing the responses from individuals whose covariates are similar,
the obtained differences are easy to summarize across all values of covariates
via standard methods for aggregation (for example, simply sum across all values
of the covariates the differences for every set of values for the covariates).

Canonical examples of controlling for specified covariates
occur with randomized controlled trials, A/B tests, and assessing fairness
in the treatment of a sensitive subpopulation (where ``sensitive''
can refer to protected classes, such as those defined by race, color, religion,
gender, national origin, age, disability, veteran status, genetic information,
or political affiliation). In biomedicine, the subpopulation of interest
often is diseased, infected, treated, or recovered (while the control
or full population typically has no such special characteristic).
Alternatively, the covariates on which the analysis gets conditioned
can involve attributes including those just mentioned (such as age),
in addition to other defining characteristics (such as annual income).

Binning together similar values of the covariates and analyzing in each bin
the differences between the subpopulations' responses is perhaps
the most direct way to condition on the covariates.
Unfortunately, binning requires making rather arbitrary,
yet often highly influential decisions about the number of bins
and where to set the bins' boundaries,
even with only a single scalar covariate;
worse, the potential pitfalls multiply as the number of covariates grows.
Closely related to binning are smoothing methods based
on kernel density estimation, as detailed and compared to binning
by~\cite{srihera-stute}, Chapter~8 of~\cite{wilks}, and others;
all such methods require making choices similar to those required when binning.
Fortunately, the methods of~\cite{tygert_full} and~\cite{tygert_two}
specifically avoid binning and smoothing kernels,
and \cite{tygert_full} and~\cite{tygert_two} provide
copious examples of problems when binning or smoothing; unfortunately,
the methods of~\cite{tygert_full} and~\cite{tygert_two} on their own
are limited to a single scalar covariate.

Another way to control for specified covariates
is to perform regression analysis with an assumed regression model,
and such regression can work very well even with multiple covariates.
When the responses are discrete,
standard models such as logistic regression or Poisson regression
are often appropriate, and classifiers based on neural networks
are very popular, as reviewed, for example,
by~\cite{hastie-tibshirani-friedman}.
Such methods are great and often highly informative.
However, regression modeling relies on the validity
of a parametric (or semi-parametric) statistical model.
Goodness-of-fit tests can partially assess
the validity of the parametric model, but do not directly quantify
the differences between subpopulations being analyzed.

The present paper proposes a fully non-parametric method
for analyzing the differences between two subpopulations' discrete responses
(or between the responses from a subpopulation and from the full population)
while conditioning on specified covariates.
The method is canonical --- essentially unique --- there are no parameters
to manipulate, aside from the ordering of the covariates
(and numerical experiments indicate that the results of such analyses
are relatively insensitive to the choice of the ordering of the covariates).
Thus, the non-parametric methodology of the present paper
enables adjustment for selected covariates in comparisons
between subpopulations or between a subpopulation and the full population,
and the adjustment is almost completely canonical,
with essentially no knobs to turn or parameters to set;
analysts cannot deceive anyone (including themselves)
when using the methods proposed in the present paper.
This may make the methods especially appealing for regulatory compliance.
Unlike in regression analysis, there is no need to choose and believe a model.
(In cases for which there is good reason to believe in a particular model,
though, the analyst may wish to take advantage of the additional power afforded
by parametric modeling.)

For detailed investigations, the present paper provides a graphical method.
For investigators' convenience, the present paper also provides
statistics that summarize the graph into a single scalar (namely, an analogue
of the Kolmogorov-Smirnov or Kuiper metrics familiar from comparisons
of probability distributions) and that gauge the statistical significance
of deviations displayed across the full range of the graph.
The graph and scalar summary statistics
come from~\cite{tygert_full} and~\cite{tygert_two}.

The methods of~\cite{tygert_full} and~\cite{tygert_two}
are already fully non-parametric and canonical when conditioning
on a single real-valued covariate.
The present paper simply extends those methods to controlling
for multiple covariates, by imposing a total order
on the values of the covariates via a Hilbert space-filling curve.
The Hilbert curve is unique aside from the ordering of the covariates
(that is, aside from the ordering of the dimensions in which the curve
is embedded), and is optimal according to many criteria, such as those
discussed by~\cite{moon-jagadish-faloutsos-saltz}.
The extent to which the Hilbert curve is unique is what makes the scheme
of the present paper essentially unique and canonical, with no parameters
to set except for the ordering of the dimensions in which the Hilbert curve
is embedded.
Furthermore, the numerical examples of Section~\ref{results} below indicate
that the scheme yields similar results irrespective of the ordering
of the dimensions.

The Hilbert curve is ideal for clustering multi-dimensional data,
as long observed empirically and proven rigorously
by~\cite{moon-jagadish-faloutsos-saltz}.
Indexing and sharding in databases have often imposed a total ordering
on the data via a space-filling curve and then fed the results
into one-dimensional schemes; this reduction of the problem
from multi-dimensional to one-dimensional is the primary application
discussed by~\cite{moon-jagadish-faloutsos-saltz}.
A prominent recent deployment of this using the Hilbert curve
is Google's S2 Geometry Library, which drives Google Maps, Foursquare, MongoDB,
et al.\footnote{Google's S2 Geometry Library is described
at \url{https://s2geometry.io/devguide/s2cell_hierarchy.html}}
Visualization of multi-dimensional data is another popular application,
as illustrated and reviewed by~\cite{castro-burns}.
Other applications related to the use in the present paper
include numerical integration in high-dimensional spaces,
as coupled with Markov-chain Monte-Carlo methods by~\cite{skilling2}.
The simplest, most efficient algorithms for the Hilbert curve
are due to~\cite{skilling}. The numerical experiments reported
in Section~\ref{results} below leverage the Python package
{\tt hilbertcurve},\footnote{The Python module {\tt hilbertcurve}
is available at \url{https://github.com/galtay/hilbertcurve}
under the permissive MIT copyright license.}
which implements the algorithms of~\cite{skilling}.

The remainder of the paper has the following structure:
Section~\ref{methods} first reviews the methods
of~\cite{hilbert} (the Hilbert space-filling curve),
of~\cite{tygert_full} (comparing a subpopulation to the full population
while controlling for a single scalar covariate),
and of~\cite{tygert_two} (comparing two subpopulations
while conditioning on a single scalar covariate);
Section~\ref{methods} also combines these methods
in order to obtain the main methodology proposed here
for conditioning on several covariates.
Section~\ref{results} then presents the results
of several numerical experiments, some using synthetic data
and others using public surveys.\footnote{Permissively licensed
open-source software implementing these methods in Python
--- software that also reproduces all figures and statistics reported below
--- is available at \url{https://github.com/facebookresearch/metamulti}}
Finally, Section~\ref{conclusion} recapitulates the main findings to conclude.

\section{Methods}
\label{methods}

This section presents the methodology of the present paper,
beginning with three subsections reviewing the required ingredients.
The first, Subsection~\ref{Hilbert}, collects together well-known facts
about Hilbert space-filling curves.
Then, Subsection~\ref{subpop} summarizes methods of~\cite{tygert_full}
for analyzing deviation of a subpopulation from the full population.
Similarly, Subsection~\ref{subpops} summarizes methods of~\cite{tygert_two}
for analyzing deviations between different subpopulations.
Finally, Subsection~\ref{combo} chains together
first Subsections~\ref{Hilbert} and~\ref{subpop}
and then Subsections~\ref{Hilbert} and~\ref{subpops},
yielding the main solution to the problem posed in the introduction.

\subsection{Preliminaries on Hilbert curves}
\label{Hilbert}

This subsection reviews the construction and properties of Hilbert curves.

Hilbert curves are so-called ``space-filling'' curves similar
to the original Peano curves introduced by~\cite{peano}.
\cite{hilbert} originally introduced these curves
for the two-dimensional plane, but they generalize straightforwardly
to any finite-dimensional Euclidean space.
A detailed presentation of what the present subsection summarizes
is available from~\cite{moon-jagadish-faloutsos-saltz}.

A canonical space-filling curve in $p$ dimensions is a continuous mapping $h$
from the unit interval $(0, 1)$ onto the unit hypercube $(0, 1)^p$,
where ``onto'' means that the curve covers every point in the hypercube
--- the mapping is surjective.
A mapping $g$ from the unit hypercube $(0, 1)^p$
to the unit interval $(0, 1)$ that is a right inverse for $h$
accompanies the space-filling curve: $h(g(x)) = x$
for any point $x$ in the unit hypercube $(0, 1)^p$.
However, the ``inverse'' mapping $g$ cannot be continuous when $p > 1$:
the existence of a left inverse function ($h$) for $g$
implies that $g$ is injective (that is, $g$ is one-to-one),
so $g$ cannot be continuous without violating the topological invariance
of dimension (or the Brouwer invariance of domain).\footnote{See, for example,
Corollary~3 and its proof on Terence Tao's blog at
\url{https://terrytao.wordpress.com/2011/06/13/brouwers-fixed-point-and-invariance-of-domain-theorems-and-hilberts-fifth-problem}}

The continuity of the mapping $h$ from the unit interval $(0, 1)$
onto the unit hypercube $(0, 1)^p$ ensures that,
for any points $t$ and $u$ from the unit interval $(0, 1)$,
if $t$ and $u$ are close, then so are $h(t)$ and $h(u)$.
Therefore, for any real-valued function $f$ on the unit hypercube $(0, 1)^p$,
local averages of $f \circ h$ will also be local averages of $f$
(with ``local'' defined by the usual Euclidean metric),
where $f \circ h$ is the composition of $f$ and $h$, that is,
$(f \circ h)(t) = f(h(t))$ for any point $t$ in the unit interval $(0, 1)$.
The constructions below impose a total order on the unit hypercube $(0, 1)^p$
via the inverse mapping $g$: if $x$ and $y$ are two points
in the unit hypercube $(0, 1)^p$, then $x < y$ means that $g(x) < g(y)$.
A local average of $f$ under this ordering (with ``local'' defined
by the perhaps counterintuitive total order ``$<$'' imposed via $g$
as just mentioned, rather than via the usual Euclidean metric)
is a local average of $f \circ h$ and thus is also a local average of $f$
under the usual Euclidean metric (where this latter consequence
is simply restating the second sentence of this paragraph).
If $g$ could be continuous, then any local average of $f$
under the usual Euclidean metric would also be a local average of $f$
under the total ordering \dots\,but $g$ cannot be continuous,
as discussed in the previous paragraph.
The constructions below consider all the local averages of $f$
under the total ordering, without needing to consider
every single other possible local average of $f$
under the usual Euclidean metric.

A Hilbert curve $h$ is an instance from a specific class
of space-filling curves in which the points are ordered according
to a depth-first traversal of the canonical $2^p$-ary (dyadic) tree
on the unit hypercube $(0, 1)^p$
--- this is the canonical binary tree for $p = 1$,
the canonical quad-tree for $p = 2$, the canonical oct-tree for $p = 3$,
and so on. A Hilbert curve minimizes the Lipschitz constant of order $1/p$
beyond that for the original Peano curve of~\cite{peano}
(the Lipschitz constant is also known as the constant of H\"older continuity
to order $1/p$); indeed, the Hilbert curve possesses many favorable properties,
as discussed by~\cite{moon-jagadish-faloutsos-saltz}.

\begin{figure}
\begin{centering}
\hfil\parbox{0.71\textwidth}
{\includegraphics[width=0.71\textwidth]{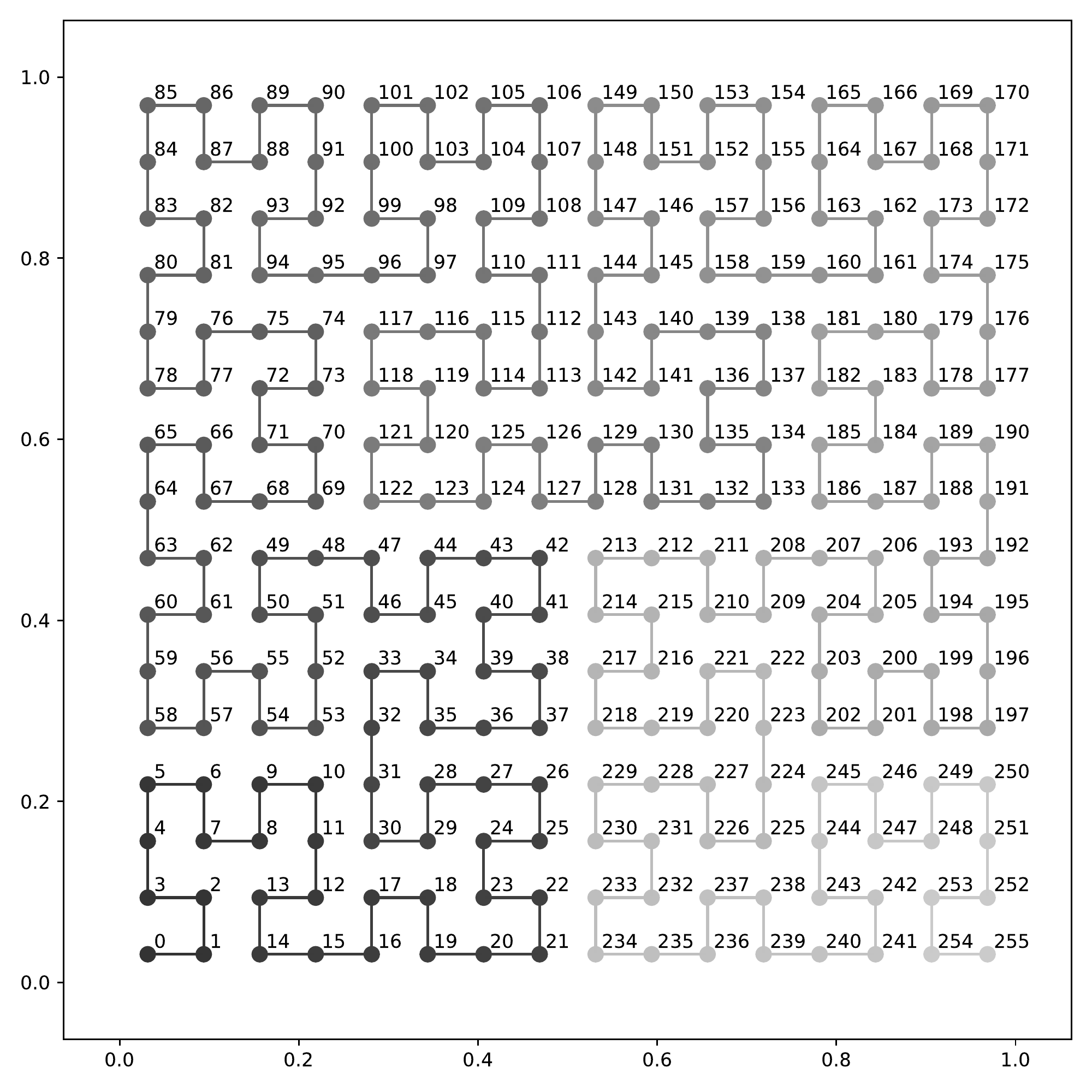}}
\end{centering}
\caption{An approximation to the Hilbert curve in $p = 2$ dimensions
with $255$ line segments --- the numbers of the points on the curve
specify the total ordering.}
\label{hilbert}
\end{figure}

Figure~\ref{hilbert} displays an approximation to the image of a Hilbert curve.
The traditional construction of a Hilbert curve leverages approximations
with more and more line segments, where every line segment
in the same approximation has the same length as every other line segment
in that same approximation. The mapping from one dimension to $p$ dimensions 
for each approximation is continuous, and the approximations converge uniformly
as the numbers of line segments increase, so the limit exists
and is continuous. The limit is the Hilbert curve.
For calculations with 64-bit variables (unsigned 8-byte integers),
we use an approximation with $2^{64} - 1$ line segments.

\subsection{Preliminaries on cumulative differences of a subpopulation
from the full population}
\label{subpop}

This subsection reviews cumulative methods developed by~\cite{tygert_full}
for assessing deviation of a subpopulation from the full population.

Suppose that $S_1$,~$S_2$, \dots, $S_m$ are real numbers (known as ``scores''),
$R_1$,~$R_2$, \dots, $R_m$ are real numbers (known as ``responses,''
``results,'' ``regressands,'' or ``outcomes''),
and $W_1$,~$W_2$, \dots, $W_m$ are positive real numbers (known as ``weights'')
such that $S_1 < S_2 < \dots < S_m$ and $R_1$,~$R_2$, \dots, $R_m$
are random variables.
Suppose also that $i_1$,~$i_2$, \dots, $i_n$ are integers (with $n < m$)
such that $1 \le i_1 < i_2 < \dots < i_n \le m$; these indices specify
a subpopulation of the full population.
The probability distributions of $R_1$,~$R_2$, \dots, $R_m$ will be discrete
in the examples considered below, and the scores $S_1$,~$S_2$, \dots, $S_m$,
weights $W_1$,~$W_2$, \dots, $W_n$, and indices $i_1$,~$i_2$, \dots, $i_n$
will be deterministic, not viewed as random.
The following sets up notation in order to define a graph.

The cumulative weighted response for the subpopulation is
\begin{equation}
F_k = \frac{\sum_{j=1}^k W_{i_j} R_{i_j}}{\sum_{j=1}^n W_{i_j}}
\end{equation}
for $k = 1$, $2$, \dots, $n$.
The weighted average response for the full population in a narrow bin
around a score $S_{i_k}$ from the subpopulation is
\begin{equation}
\tilde{R}_{i_k} = \frac{\sum_{i : B_{k-1} < S_i \le B_k} W_i R_i}
                       {\sum_{i : B_{k-1} < S_i \le B_k} W_i}
\end{equation}
for $k = 1$,~$2$, \dots, $n$, where the thresholds for the bins are
\begin{equation}
B_k = \frac{S_{i_k} + S_{i_{k+1}}}{2}
\end{equation}
for $k = 0$,~$1$,~$2$, \dots, $n$, using the notation that
$S_{i_0} = -\infty$ and $S_{i_{n+1}} = \infty$
(and so $B_0 = -\infty$ and $B_n = \infty$).
Then, the cumulative weighted response for the full population
averaged to the scores from the subpopulation is
\begin{equation}
\tilde{F}_k = \frac{\sum_{j=1}^k W_{i_j} \tilde{R}_{i_j}}{\sum_{j=1}^n W_{i_j}}
\end{equation}
for $k = 1$, $2$, \dots, $n$.
The cumulative weight is the aggregate
\begin{equation}
A_k = \frac{\sum_{j=1}^k W_{i_j}}{\sum_{j=1}^n W_{i_j}}
\end{equation}
for $k = 1$, $2$, \dots, $n$.

In a plot where $A_k$ is the abscissa (that is, the horizontal coordinate)
and $F_k - \tilde{F}_k$ is the ordinate (that is, the vertical coordinate),
the expected slope of a secant line connecting two points on the graph is equal
to the weighted average difference in responses between the subpopulation
and the full population over the range of scores between the two points.
Thus, deviation between the subpopulation and the full population
is simply the slope. A long range of steep slopes in the graph
indicates significant weighted average deviation over that range.
If the weights are all identical ---
$W_1 = W_2 = \dots = W_m$ --- then $A_k = k/n$, so the graph
of $F_k - \tilde{F}_k$ versus $A_k$ is then the same as the usual graph
of $F_k - \tilde{F}_k$ (with equispaced abscissae).

If the differences are all small, then the graph should be fairly flat
and hence not deviate much overall.
Two standard metrics which summarize the overall deviation across all scores
are the statistic of Kolmogorov and Smirnov, the maximal absolute deviation
\begin{equation}
\label{Kolmogorov-Smirnov}
G = \max_{1 \le k \le n} |F_k - \tilde{F}_k|,
\end{equation}
and the statistic of Kuiper, the size of the range of deviations
\begin{equation}
\label{Kuiper}
H = \max_{0 \le k \le n} (F_k - \tilde{F}_k)
  - \min_{0 \le k \le n} (F_k - \tilde{F}_k),
\end{equation}
with $F_0 = \tilde{F}_0 = 0$.
As detailed by~\cite{tygert_full},
the expected value of $G$ is roughly $1.25\sigma$
under the null hypothesis of no deviation between the subpopulation
and the full population in their responses'
underlying probability distributions (and similarly for $H$),
where estimation of $\sigma$ is discussed at length by~\cite{tygert_full};
normalizing $G$ and $H$ by $\sigma$ therefore characterizes
statistical significance (as both $G$ and $H$ are sub-Gaussian
under the null hypothesis).
Values of $G/\sigma$ much greater than 1.25 indicate
highly statistically significant deviation,
while values of $G/\sigma$ near 0 give little indication
of statistically significant deviation.

\subsection{Preliminaries on cumulative differences
between distinct subpopulations}
\label{subpops}

This subsection reviews cumulative methods developed by~\cite{tygert_two}
for assessing deviations between different subpopulations.

Similar to the set-up from the previous subsection,
the present subsection considers a strictly increasing sequence
of distinct real numbers known as ``scores,'' where each score
is associated with another real number known as a ``response,'' ``result,''
``regressand,'' or ``outcome,'' as well as with a third, positive real number
known as a ``weight.''
Each score will be designated as belonging either to subpopulation 0
or to subpopulation 1 (always one or the other, but never both).
As in the previous subsection, the probability distributions
of the responses will be discrete in the examples considered below,
and the scores, weights, and subpopulation designations will be deterministic,
not viewed as random.
The following sets up notation in order to define a graph.

\begin{figure}
\begin{centering}
\hfil\parbox{0.65\textwidth}
{\includegraphics[width=0.65\textwidth]{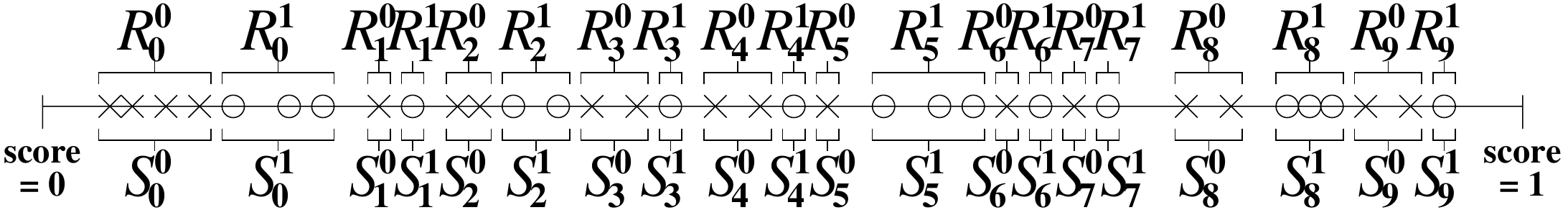}}
\end{centering}
\caption{The crosses (``x'') indicate the scores for subpopulation 0
while the circles (``o'') indicate the scores for subpopulation 1.
The weighted averages of the scores for subpopulation 0
for the indicated blocks of observed scores are
$S^0_0$, $S^0_1$, \dots, $S^0_9$,
while the weighted averages of the scores for subpopulation 1 are
$S^1_0$, $S^1_1$, \dots, $S^1_9$.
The weighted averages of the responses for subpopulation 0 corresponding
to the indicated blocks of observed scores are
$R^0_0$, $R^0_1$, \dots, $R^0_9$, while the weighted averages of the responses
for subpopulation 1 are $R^1_0$, $R^1_1$, \dots, $R^1_9$.
}
\label{partition}
\end{figure}

Figure~\ref{partition} illustrates the first stage of processing for this data:
in each contiguous block of scores from one of the subpopulations,
the weighted average of responses gets labeled $R^0_k$ or $R^1_k$,
the weighted average of scores gets labeled $S^0_k$ or $S^1_k$,
and the sum total of the weights gets labeled $T^0_k$ or $T^1_k$
(with the superscript chosen to match the designation as either subpopulation 0
or subpopulation 1 associated with the block, and with the subscript $k$
chosen such that
$S^0_0 < S^1_0 < S^0_1 < S^1_1 < S^0_2 < S^1_2 < S^0_3 < S^1_3 < \dots$).

\begin{figure}
\vspace{.2in}
\begin{centering}
\hfil
(a) \parbox{0.111\textwidth}
{\includegraphics[width=0.111\textwidth]{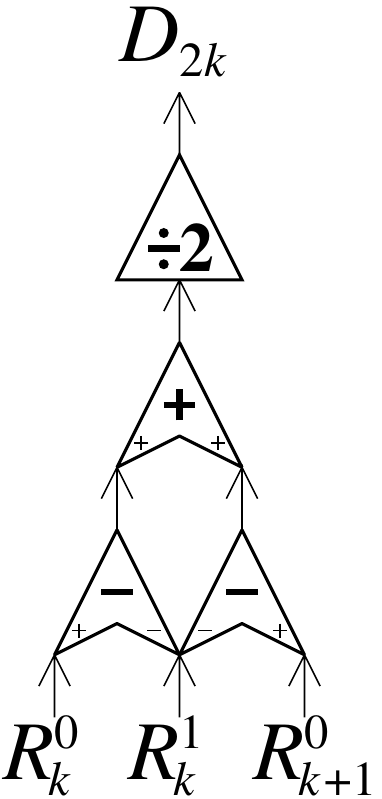}}
\hfil
(b) \parbox{0.111\textwidth}
{\includegraphics[width=0.111\textwidth]{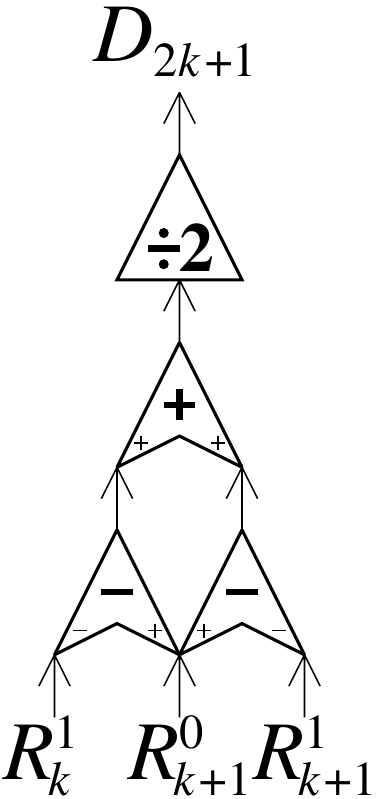}}
\end{centering}
\caption{In each of these subfigures, the operation indicated by ``$+$'' sums
its two inputs and the operations indicated by ``$-$'' subtract their inputs,
with one of these ``$-$'' operations subtracting its rightmost input
from its leftmost input, while the other subtracts its leftmost input
from its rightmost input.
In all cases, the operations indicated by ``$-$'' subtract
subpopulation 1 from subpopulation 0, in that order.
The operation indicated by ``$\div 2$'' divides its input by 2.
}
\label{diffs}
\end{figure}

\begin{figure}
\vspace{.2in}
\begin{centering}
\hfil
(a) \parbox{0.111\textwidth}
{\includegraphics[width=0.111\textwidth]{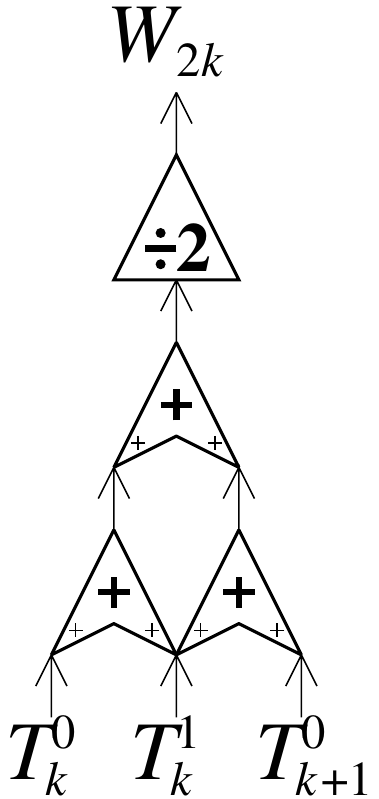}}
\hfil
(b) \parbox{0.111\textwidth}
{\includegraphics[width=0.111\textwidth]{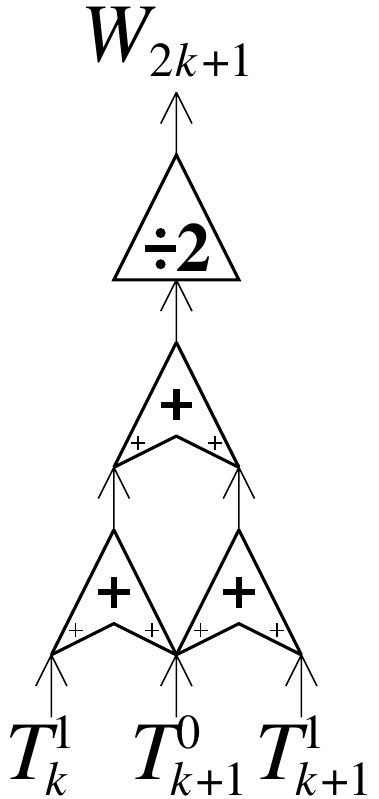}}
\end{centering}
\caption{In each of these subfigures, the operation indicated by ``$+$'' sums
its two inputs. The operation indicated by ``$\div 2$'' divides its input by 2.
}
\label{sums}
\end{figure}

Figures~\ref{diffs} and~\ref{sums} illustrate the second stage
of processing for this data: construct the differences
with even-indexed entries
\begin{equation}
\label{diff_even}
D_{2k} = \frac{(R^0_k - R^1_k) + (R^0_{k+1} - R^1_k)}{2}
       = \frac{R^0_k + R^0_{k+1} - 2R^1_k}{2}
\end{equation}
and odd-indexed entries
\begin{equation}
\label{diff_odd}
D_{2k+1} = \frac{(R^0_{k+1} - R^1_k) + (R^0_{k+1} - R^1_{k+1})}{2}
         = \frac{2R^0_{k+1} - R^1_k - R^1_{k+1}}{2},
\end{equation}
as well as the sums with even-indexed entries
\begin{equation}
\label{sum_even}
W_{2k} = \frac{T^0_k + T^0_{k+1} + 2T^1_k}{2}
\end{equation}
and odd-indexed entries
\begin{equation}
\label{sum_odd}
W_{2k+1} = \frac{2T^0_{k+1} + T^1_k + T^1_{k+1}}{2}.
\end{equation}

The abscissae (that is, the horizontal coordinates)
for a graph consist of the normalized aggregated weights
\begin{equation}
\label{abscissae}
A_j = \frac{\sum_{k=0}^{j-1} W_k}{\sum_{k=0}^{n-1} W_k}
\end{equation}
for $j = 1$, $2$, \dots, $n$.
The ordinates (that is, the vertical coordinates)
for the graph are the cumulative differences
\begin{equation}
\label{cumulativew}
C_j = \frac{\sum_{k=0}^{j-1} W_k D_k}{\sum_{k=0}^{n-1} W_k}
\end{equation}
for $j = 1$, $2$, \dots, $n$.

In a plot where $A_k$ is the abscissa (that is, the horizontal coordinate)
and $C_k$ is the ordinate (that is, the vertical coordinate),
the expected slope of a secant line connecting two points on the graph is equal
to the weighted average difference in responses
between the two subpopulations over the range of scores between the two points.
Thus, deviation between the subpopulations is simply the slope.
A long range of steep slopes in the graph indicates significant
weighted average deviation between the subpopulations over that range.
If the weights are all identical ---
$W_0 = W_1 = \dots = W_{n-1}$ --- then $A_k = k/n$, so the graph
of $C_k$ versus $A_k$ is then the same as the usual graph
of $C_k$ (with equispaced abscissae).

If the differences are all small, then the graph should be fairly flat
and hence not deviate much overall.
Two standard metrics which summarize the overall deviation across all scores
are the statistic of Kolmogorov and Smirnov, the maximal absolute deviation
\begin{equation}
\label{Kolmogorov-Smirnov2}
G = \max_{1 \le k \le n} |C_k|,
\end{equation}
and the statistic of Kuiper, the size of the range of deviations
\begin{equation}
\label{Kuiper2}
H = \max_{0 \le k \le n} C_k - \min_{0 \le k \le n} C_k,
\end{equation}
with $C_0 = 0$.
As in the last paragraph of Subsection~\ref{subpop},
normalizing $G$ and $H$ by an estimate $\sigma$ characterizes
statistical significance under the null hypothesis
of no difference between the probability distributions
underlying the responses of the two subpopulations,
where the estimation of $\sigma$ is detailed by~\cite{tygert_two}.

\subsection{Combination}
\label{combo}

This subsection combines the methods summarized
in Subsections~\ref{Hilbert}, \ref{subpop}, and~\ref{subpops},
yielding the main methodology of the present paper.

The cumulative graphs and scalar summary statistics
of Subsections~\ref{subpop} and~\ref{subpops}
require as inputs both scores and responses
(together with weights when the sampling is weighted).
Each example considered in the present paper
will detail a specific choice of scalar, real-valued responses (and weights,
when using non-uniform weights), chosen similarly to what~\cite{tygert_full}
and~\cite{tygert_two} set as precedents.
The novelty of the combined approach that the present subsection proposes
lies solely in the choice of scores:

For the scores, we use the mapping $g$ from Subsection~\ref{Hilbert},
first applying $g$ to the vectors of covariates
for the members of the populations and then normalizing the results
to range from 0 to 1 (the normalization is affine --- linear plus a constant
--- simply subtracting the minimum of all values
and then dividing by the original maximum minus the original minimum).
These scores directly account for the geometry of the data
as embedded in the ambient space of covariates, since the Hilbert curve snakes
through the ambient space in accord with the geometry of the ambient space.
These scores also account for the intrinsic geometry of the data, or
at least for the density of data points in various parts of the ambient space
of covariates. Indeed, while the Hilbert curve is the same for every data set
with the same number of covariates, the positioning of the points
along the curve is specific to each data set.

\section{Results and discussion}
\label{results}

This section illustrates the methods of the previous section
via several numerical examples.\footnote{Permissively licensed
open-source Python scripts that reproduce all figures and statistics
reported here is available at
\url{https://github.com/facebookresearch/metamulti}}
Subsection~\ref{synthetic} considers synthetic, toy examples
whose construction is especially easy to understand.
Subsection~\ref{kddcup98} analyzes A/B tests from a classic marketing campaign.
Subsection~\ref{acs} analyzes the latest (2019) American Community Survey
from the U.S.\ Census Bureau, whose data arises from weighted sampling.
Subsection~\ref{outlook} discusses the results and proposes avenues
for further development of the methods.

The implementation combines version 2 of the Python package
{\tt hilbertcurve}\footnote{The Python package {\tt hilbertcurve}
is available at \url{https://github.com/galtay/hilbertcurve}
under the permissive MIT license.} with the Python modules
{\tt fbcdgraph}\footnote{The Python module {\tt fbcdgraph}
is available at \url{https://github.com/facebookresearch/fbcdgraph}
under the MIT copyright license.} and
{\tt fbcddisgraph}.\footnote{The Python module {\tt fbcddisgraph}
is available at \url{https://github.com/facebookresearch/fbcddisgraph}
copyrighted under the MIT license.}

In every plot for a subpopulation
versus the full population and whose title begins,
``subpop.\ deviation is the slope as a function of\dots,''
the slope of the secant line connecting two points on the displayed graph
converges to the average deviation of the subpopulation's responses
from the full population's responses,
with the average taken over the range of index $k$ between the two points
on the graph, as the range becomes large.
Similarly, in every plot for one subpopulation
versus another distinct subpopulation and whose title begins,
``subpop.\ deviation is the slope as a function of\dots,''
the slope of the secant line connecting two points on the displayed graph
converges to the average deviation of the first subpopulation's responses
from the other subpopulation's responses,
with the average taken over the range of index $k$ between the two points
on the graph, as the range becomes large.
If a slope is positive, then the deviation is positive;
if a slope is negative, then the deviation is negative.
Flat horizontal lines correspond to lack of deviation.
The height of the triangle displayed at the origin of each plot indicates
the size of deviations over the full range of scores
which would be statistically significant at about the 95\% confidence level,
accounting for random fluctuations expected in the responses
(recall that the responses are random variables).

The values of $G$ and $H$ reported beneath the corresponding plots
refer to the Kolmogorov-Smirnov and Kuiper statistics
defined in~(\ref{Kolmogorov-Smirnov}) and~(\ref{Kuiper}), respectively,
when comparing a subpopulation to the full population.
The values of $G$ and $H$ refer to the Kolmogorov-Smirnov and Kuiper metrics
defined in~(\ref{Kolmogorov-Smirnov2}) and~(\ref{Kuiper2}), respectively,
when comparing two different subpopulations directly.
As discussed at the ends of Subsections~\ref{subpop} and~\ref{subpops},
the normalized statistics $G/\sigma$ and $H/\sigma$ characterize
the statistical significance of the deviations;
for example, values of $G/\sigma$ much greater than 1.25 indicate
highly statistically significant deviation,
while values of $G/\sigma$ near 0 give little indication
of statistically significant deviation.

\subsection{Synthetic}
\label{synthetic}

\begin{figure}
\begin{centering}

(2) \parbox{\imsize}{\includegraphics[width=\imsize,
                                      trim={0pt 0pt 0pt 3pt}, clip]
{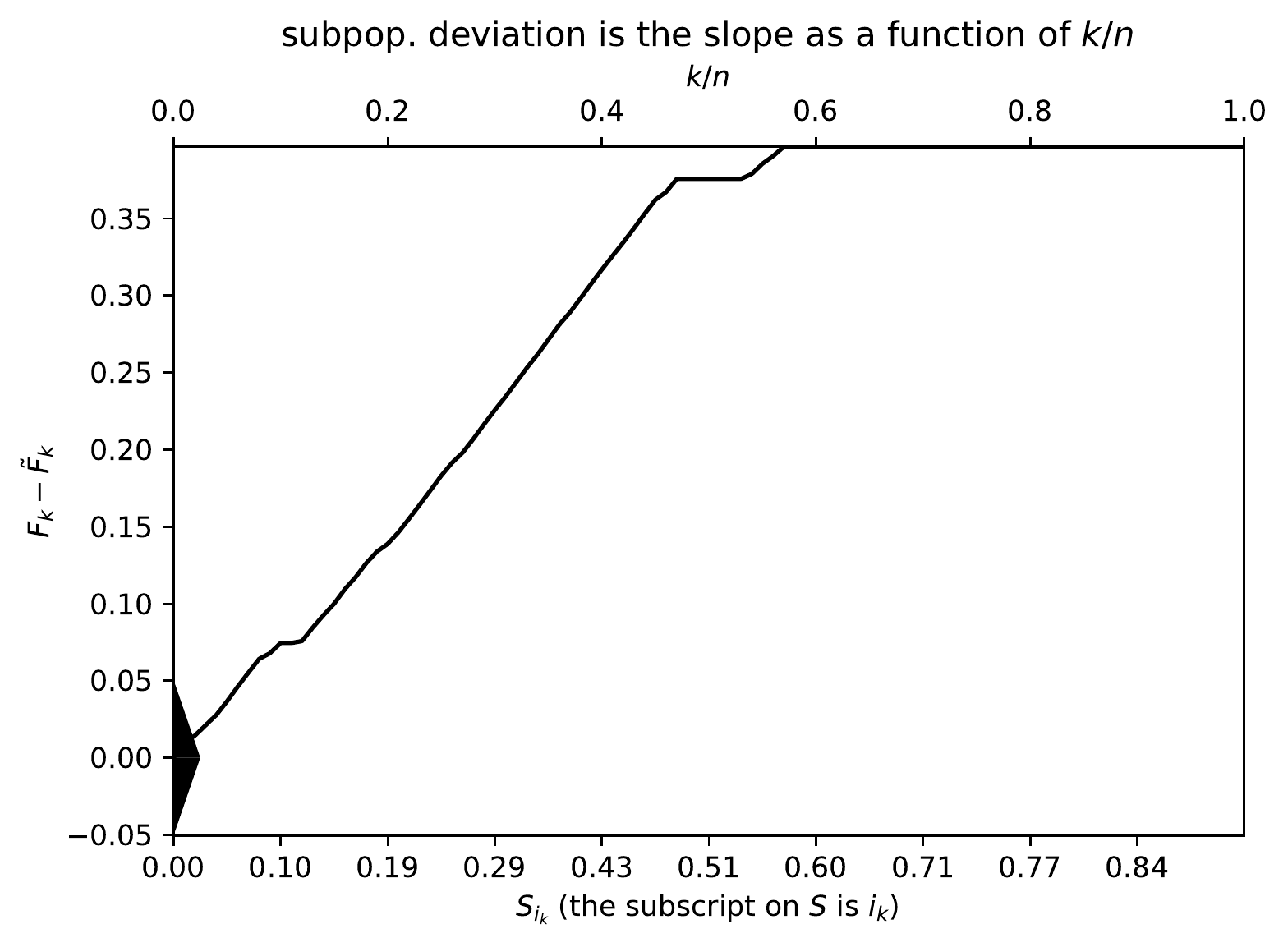}}
\quad\quad
(32) \parbox{\imsize}{\includegraphics[width=\imsize,
                                       trim={0pt 0pt 0pt 3pt}, clip]
{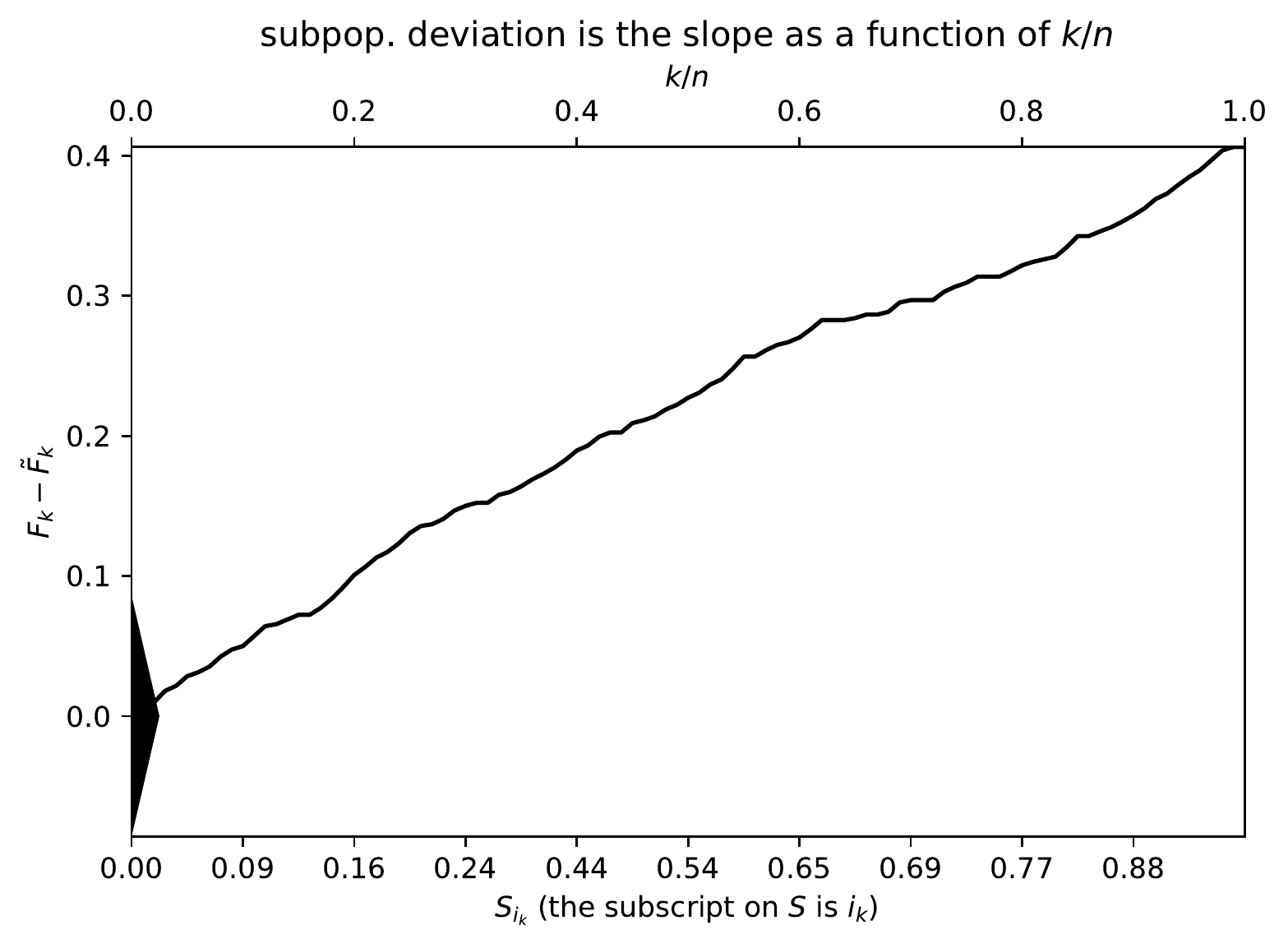}}

\parbox{\imsized}{\hfil \footnotesize $G$ = 0.3965; $H$ = 0.3965;
$G/\sigma$ = 15.66; $H/\sigma$ = 15.66}
\parbox{\imsized}{\hfil \footnotesize $G$ = 0.4063; $H$ = 0.4063;
$G/\sigma$ = 9.450; $H/\sigma$ = 9.450}

\vspace{\vertsep}

(4) \parbox{\imsize}{\includegraphics[width=\imsize,
                                      trim={0pt 0pt 0pt 2pt}, clip]
{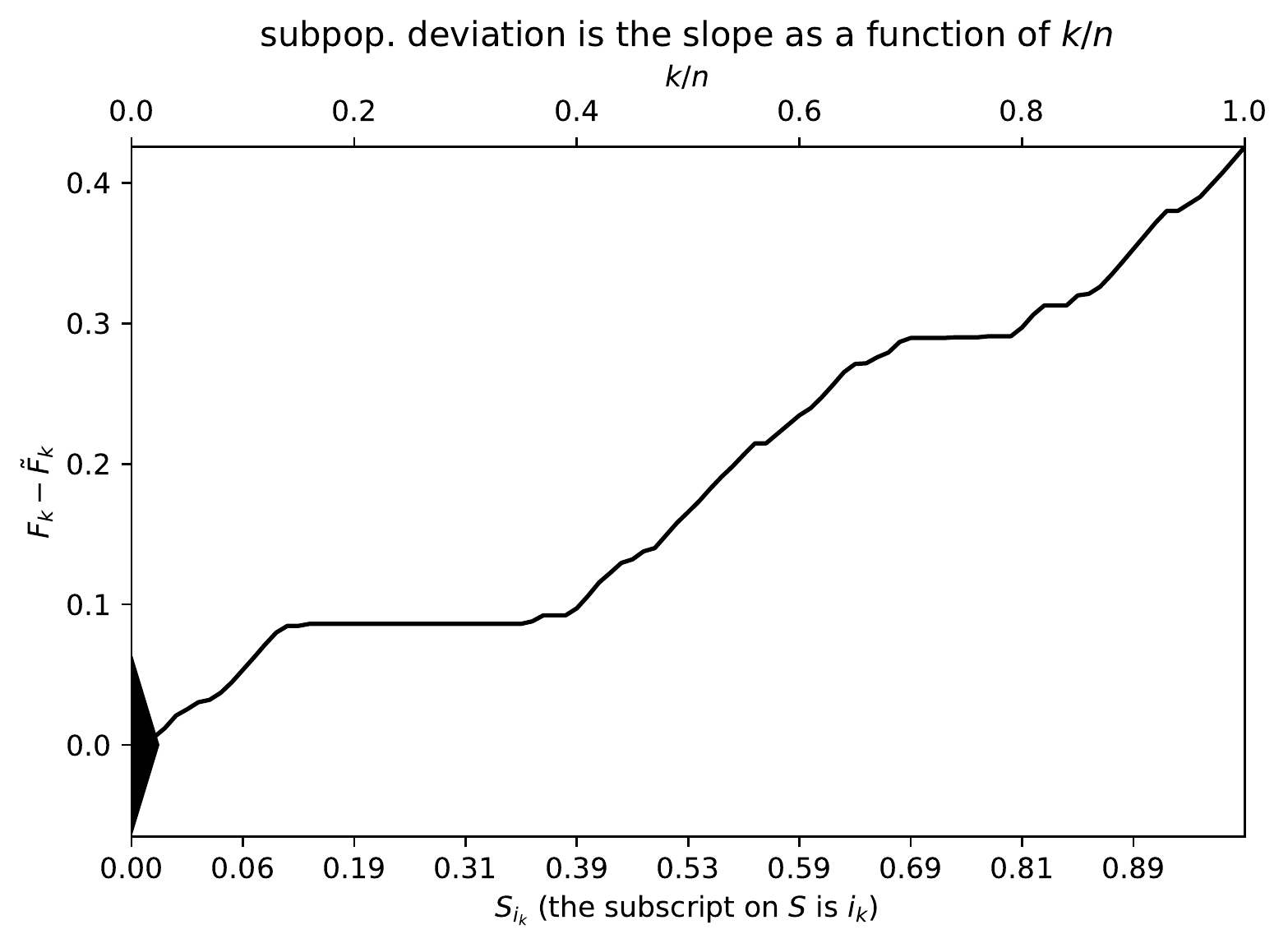}}
\quad\quad
(64) \parbox{\imsize}{\includegraphics[width=\imsize,
                                       trim={0pt 0pt 0pt 2pt}, clip]
{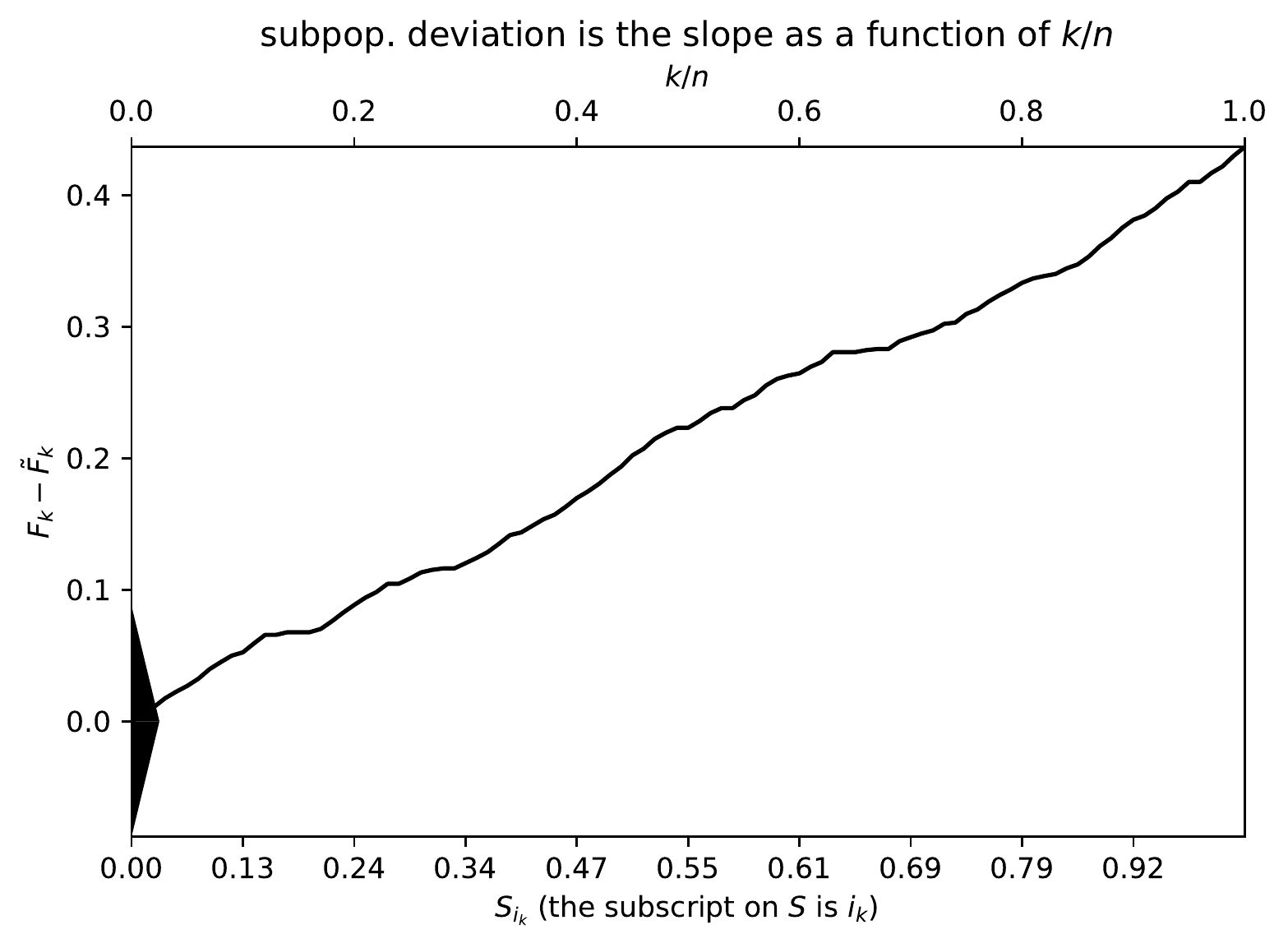}}

\parbox{\imsized}{\hfil \footnotesize $G$ = 0.4256; $H$ = 0.4256;
$G/\sigma$ = 13.03; $H/\sigma$ = 13.03}
\parbox{\imsized}{\hfil \footnotesize $G$ = 0.4368; $H$ = 0.4368;
$G/\sigma$ = 9.967; $H/\sigma$ = 9.967}

\vspace{\vertsep}

(8) \parbox{\imsize}{\includegraphics[width=\imsize,
                                      trim={0pt 0pt 0pt 2pt}, clip]
{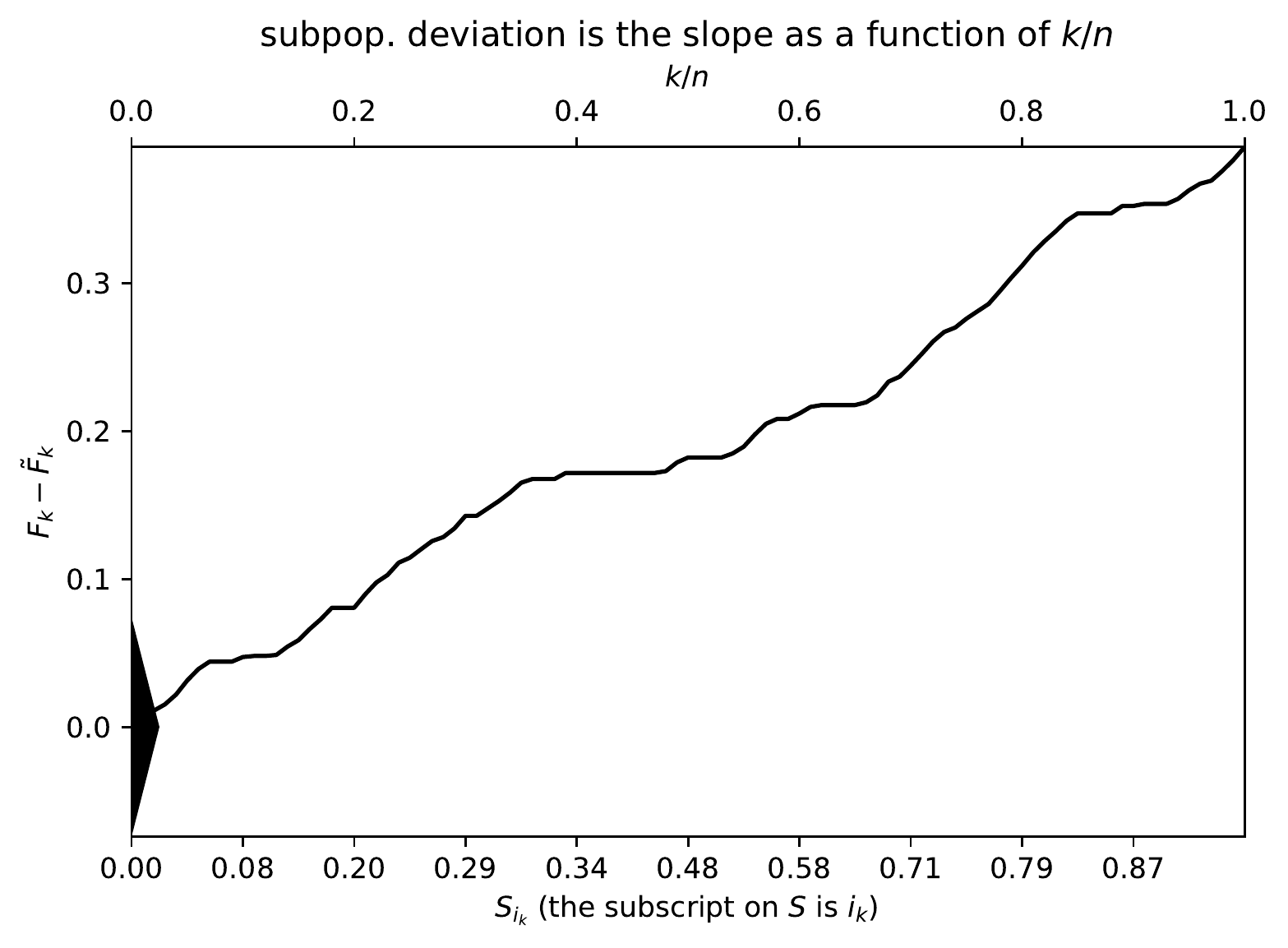}}
\quad\quad
(128) \parbox{\imsize}{\includegraphics[width=\imsize,
                                        trim={0pt 0pt 0pt 2pt}, clip]
{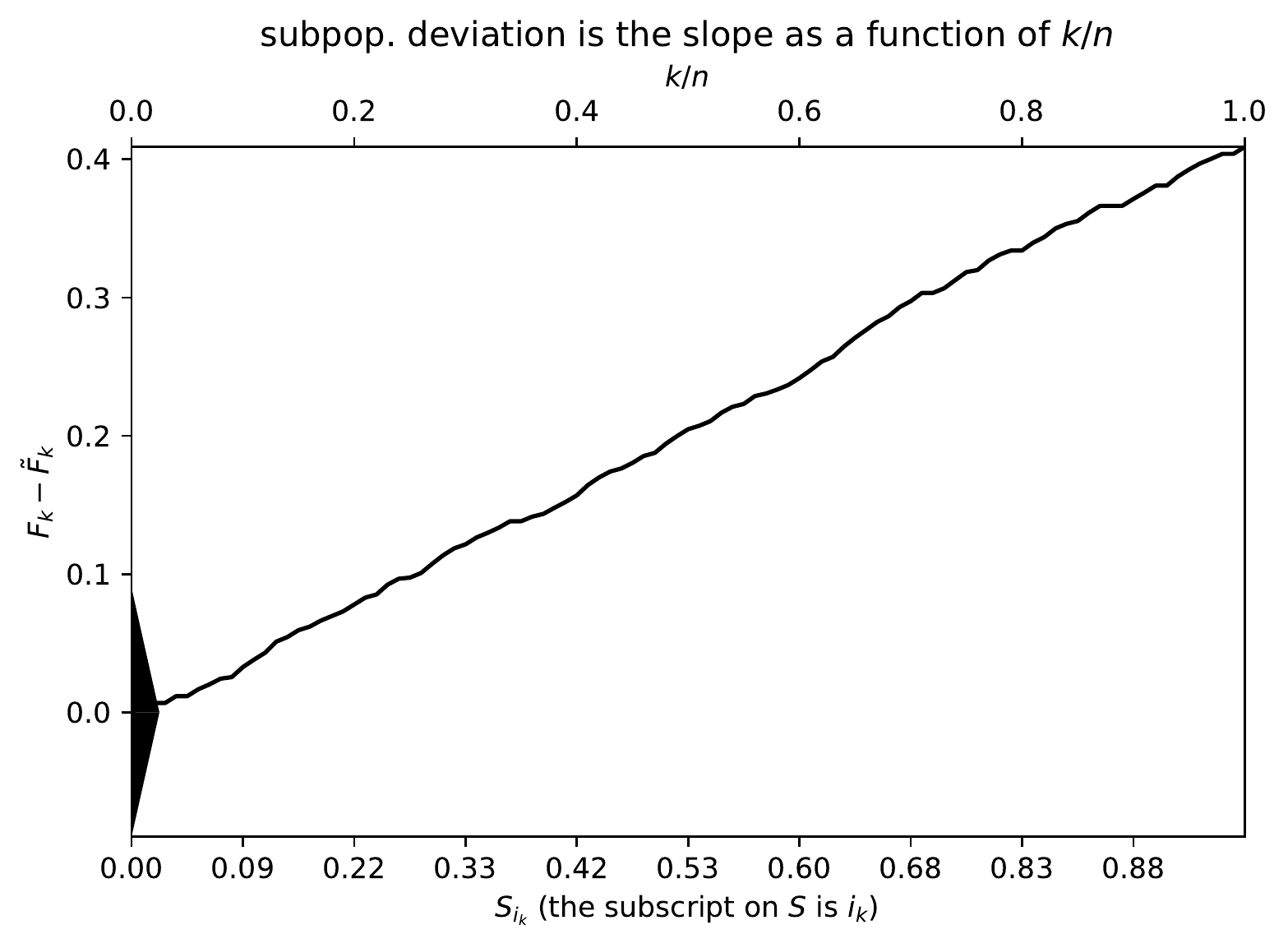}}

\parbox{\imsized}{\hfil \footnotesize $G$ = 0.3924; $H$ = 0.3924;
$G/\sigma$ = 10.58; $H/\sigma$ = 10.58}
\parbox{\imsized}{\hfil \footnotesize $G$ = 0.4090; $H$ = 0.4090;
$G/\sigma$ = 9.099; $H/\sigma$ = 9.099}

\vspace{\vertsep}

(16) \parbox{\imsize}{\includegraphics[width=\imsize,
                                       trim={0pt 0pt 0pt 2pt}, clip]
{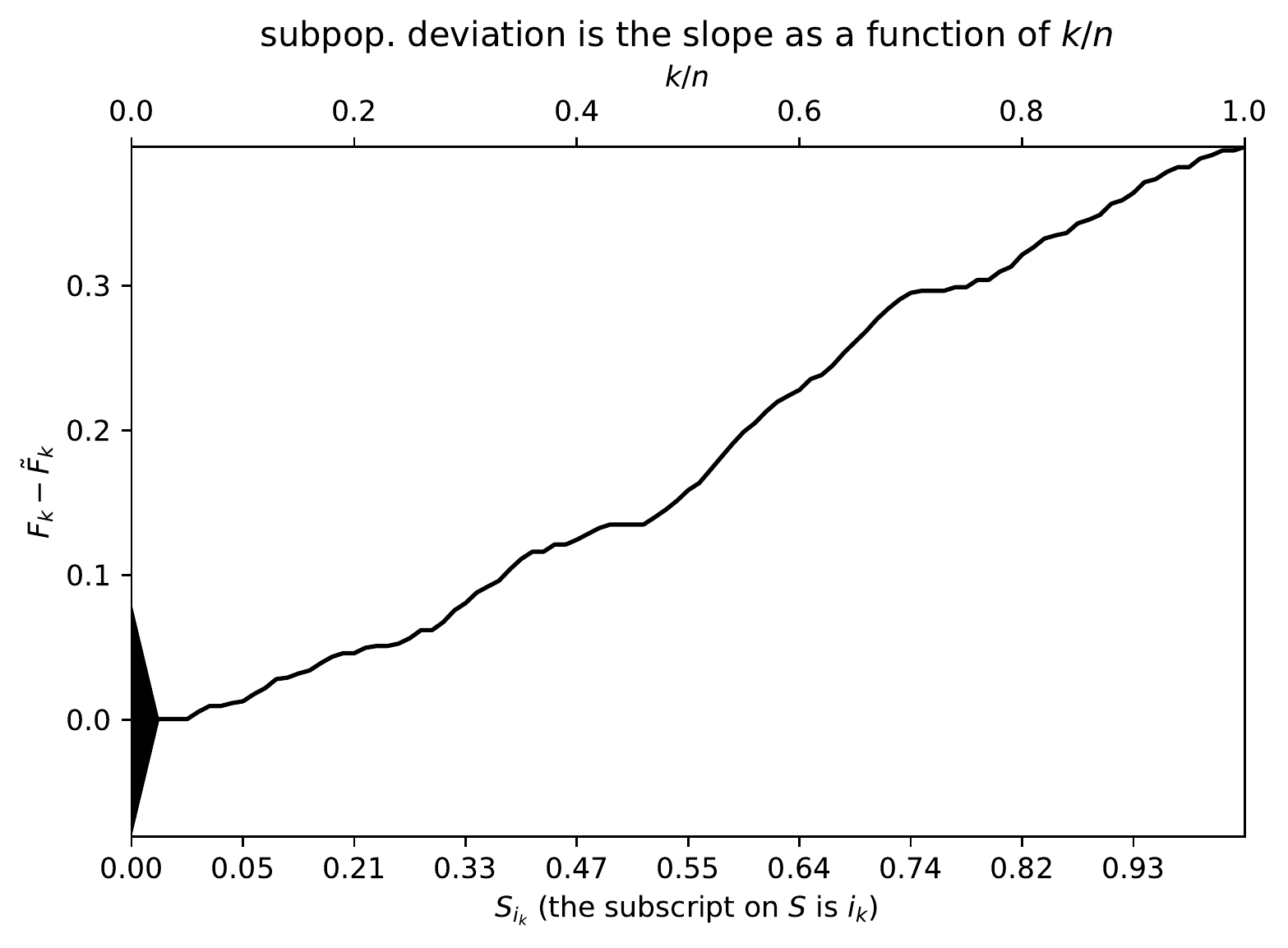}}
\quad\quad
(256) \parbox{\imsize}{\includegraphics[width=\imsize,
                                        trim={0pt 0pt 0pt 2pt}, clip]
{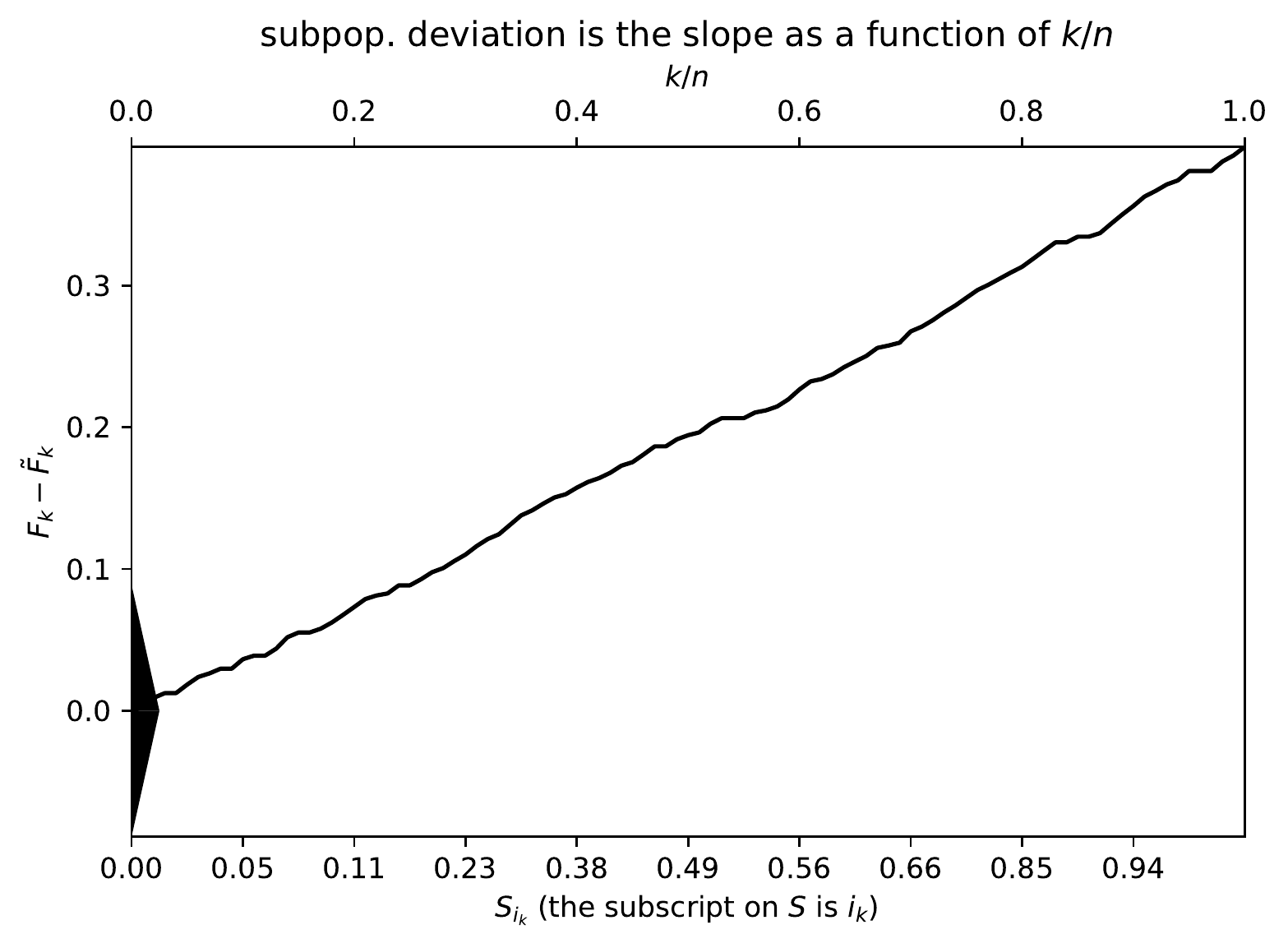}}

\parbox{\imsized}{\hfil \footnotesize $G$ = 0.3960; $H$ = 0.3960;
$G/\sigma$ = 9.833; $H/\sigma$ = 9.833}
\parbox{\imsized}{\hfil \footnotesize $G$ = 0.3978; $H$ = 0.3978;
$G/\sigma$ = 8.962; $H/\sigma$ = 8.962}

\end{centering}
\caption{Synthetic examples with a varying number of covariates
($m =$ 1,000; $n =$ 100)}
\label{synth}
\end{figure}

\begin{figure}
\begin{centering}
(512) \parbox{\imsize}{\includegraphics[width=\imsize]
{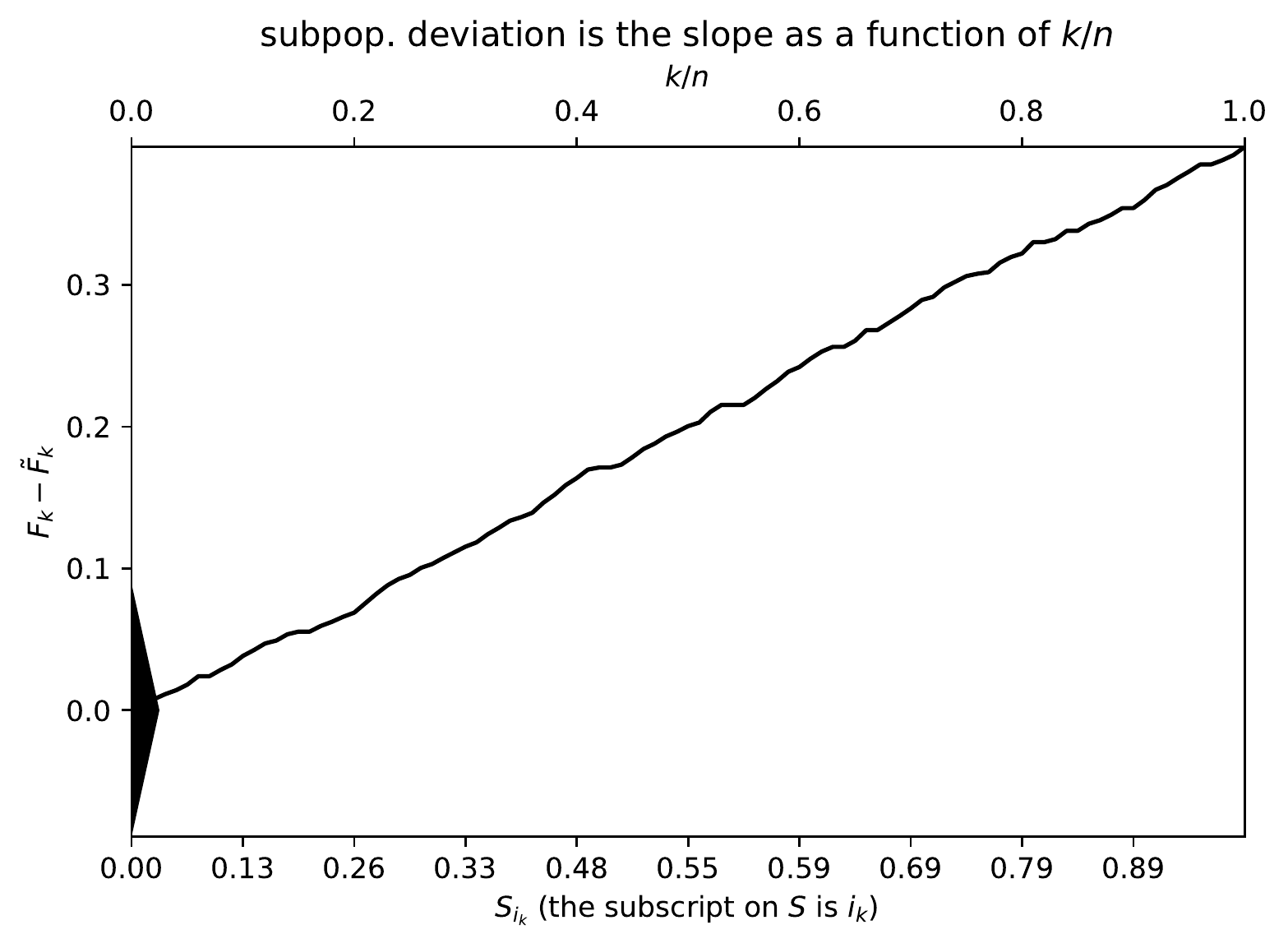}}
\quad
(2048) \parbox{\imsize}{\includegraphics[width=\imsize]
{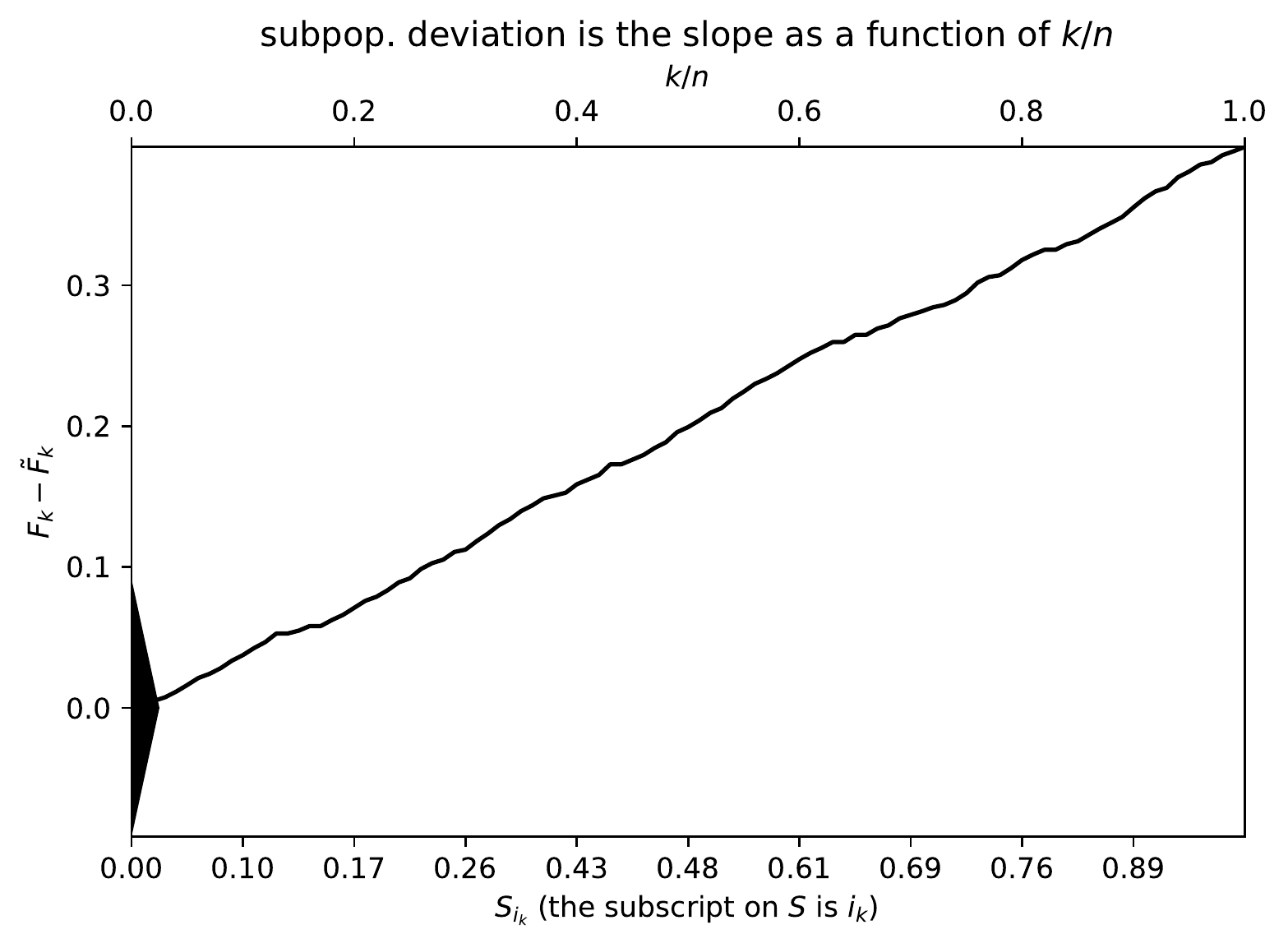}}

\parbox{\imsized}{\hfil \footnotesize $G$ = 0.3975; $H$ = 0.3975;
$G/\sigma$ = 8.916; $H/\sigma$ = 8.916}
\parbox{\imsized}{\hfil \footnotesize $G$ = 0.3983; $H$ = 0.3983;
$G/\sigma$ = 8.716; $H/\sigma$ = 8.716}

\vspace{\vertsep}

(1024) \parbox{\imsize}{\includegraphics[width=\imsize]
{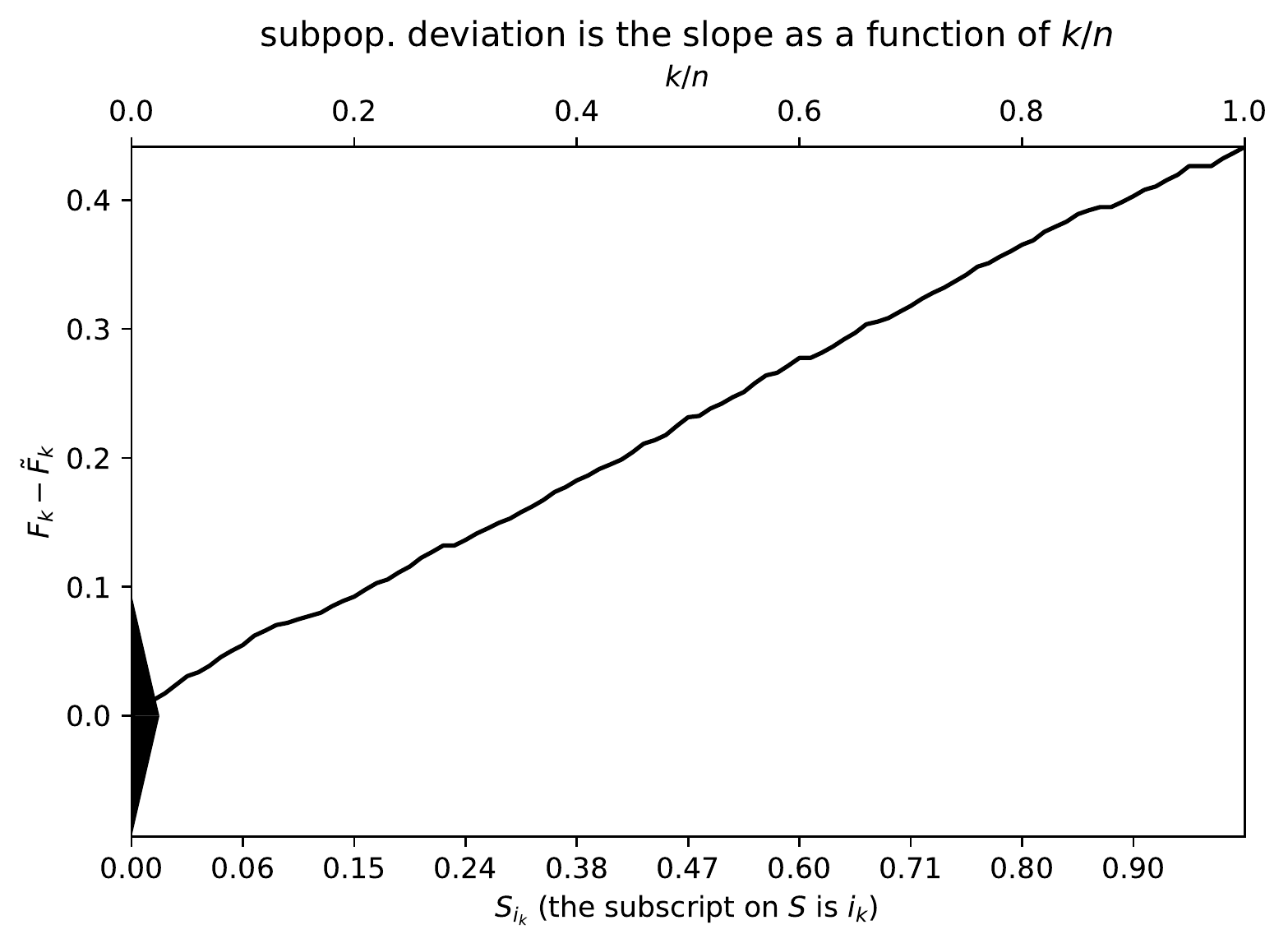}}
\quad
(4096) \parbox{\imsize}{\includegraphics[width=\imsize]
{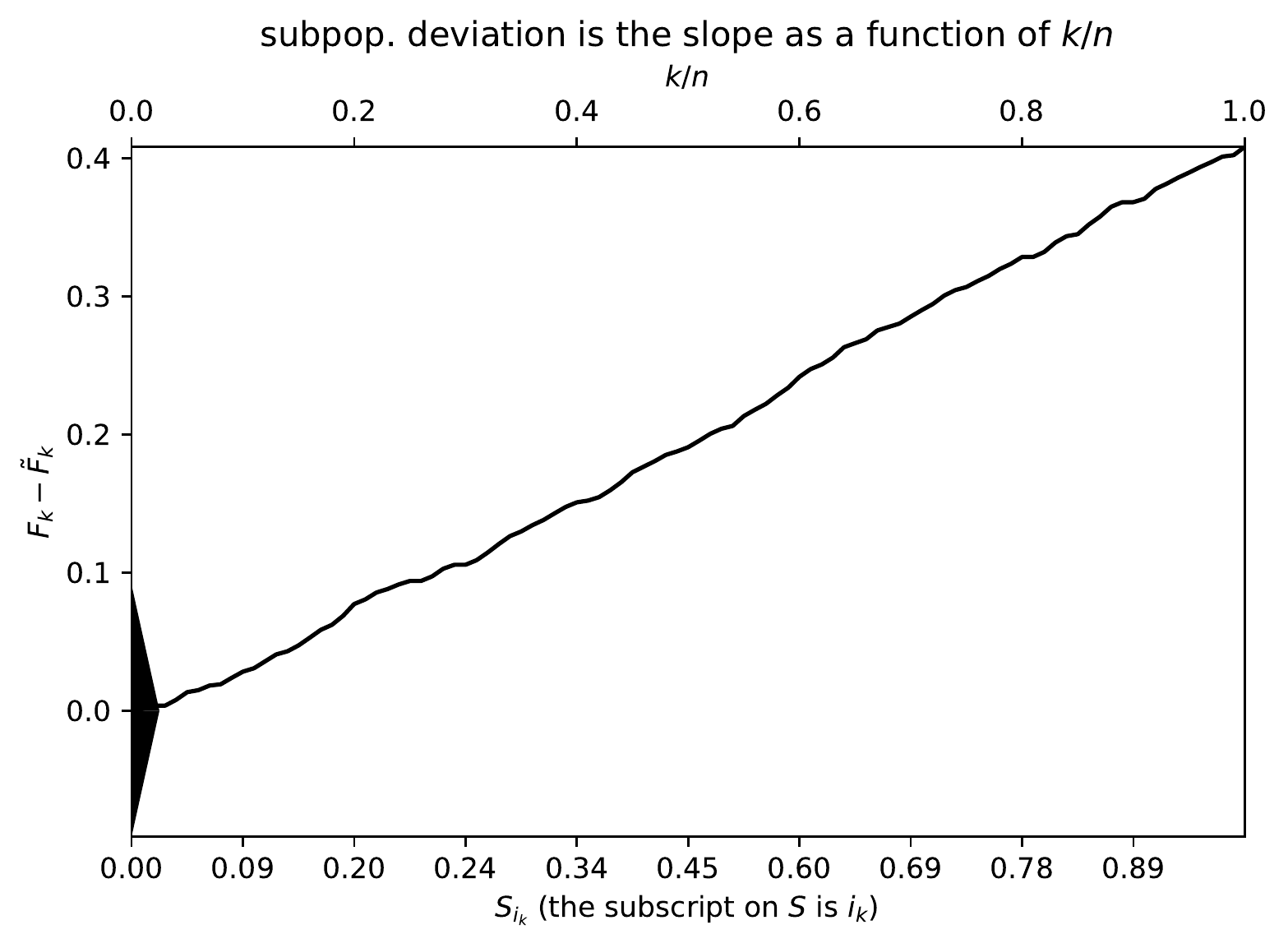}}

\parbox{\imsized}{\hfil \footnotesize $G$ = 0.4411; $H$ = 0.4411;
$G/\sigma$ = 9.428; $H/\sigma$ = 9.428}
\parbox{\imsized}{\hfil \footnotesize $G$ = 0.4082; $H$ = 0.4082;
$G/\sigma$ = 8.962; $H/\sigma$ = 8.962}
\end{centering}
\caption{Synthetic examples with even more covariates ($m =$ 1,000; $n =$ 100)}
\label{synths}
\end{figure}

\begin{figure}
\begin{centering}

(2) \parbox{\imsize}{\includegraphics[width=\imsize,
                                      trim={0pt 0pt 0pt 3pt}, clip]
{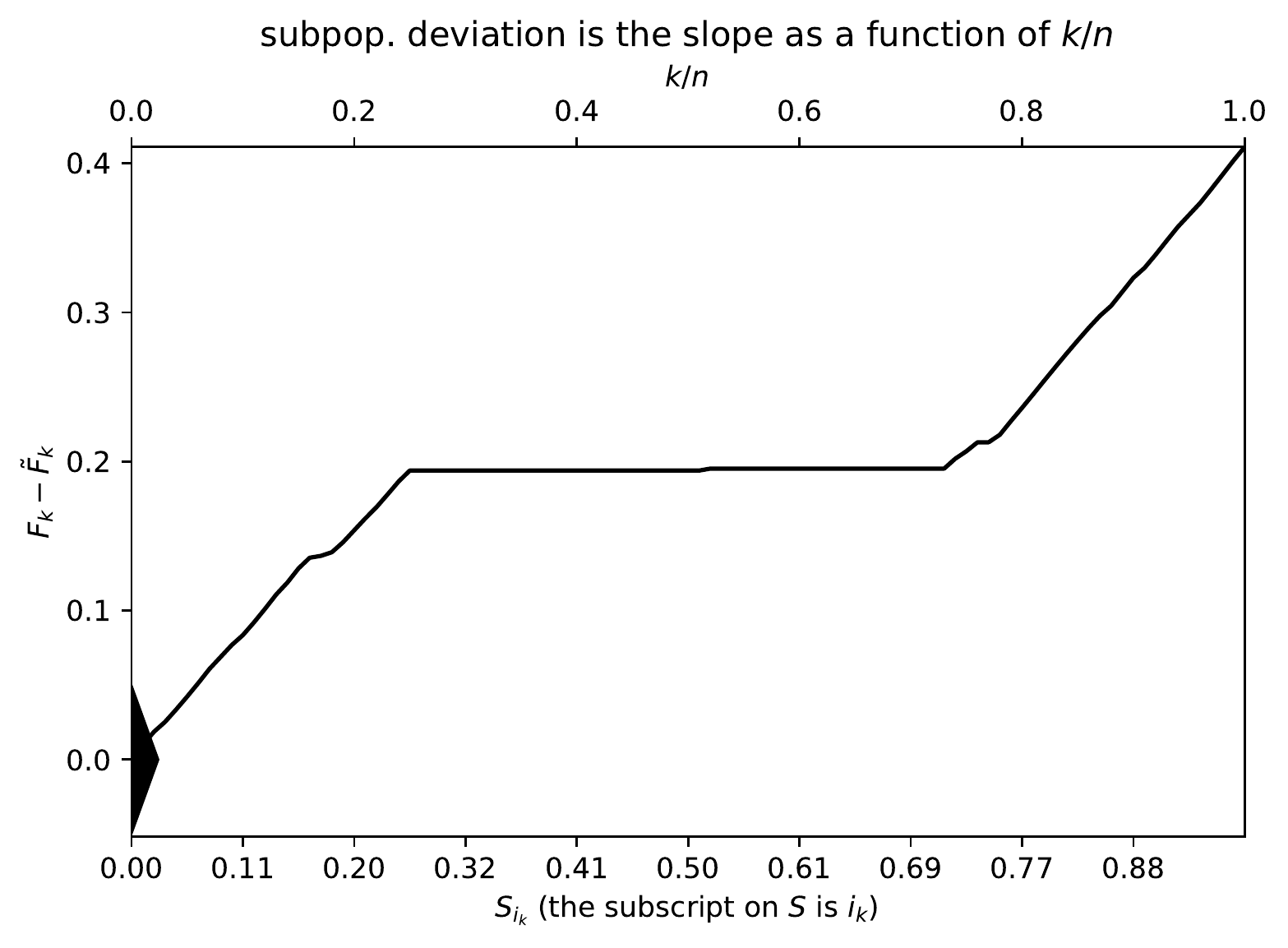}}
\quad\quad
(32) \parbox{\imsize}{\includegraphics[width=\imsize,
                                       trim={0pt 0pt 0pt 3pt}, clip]
{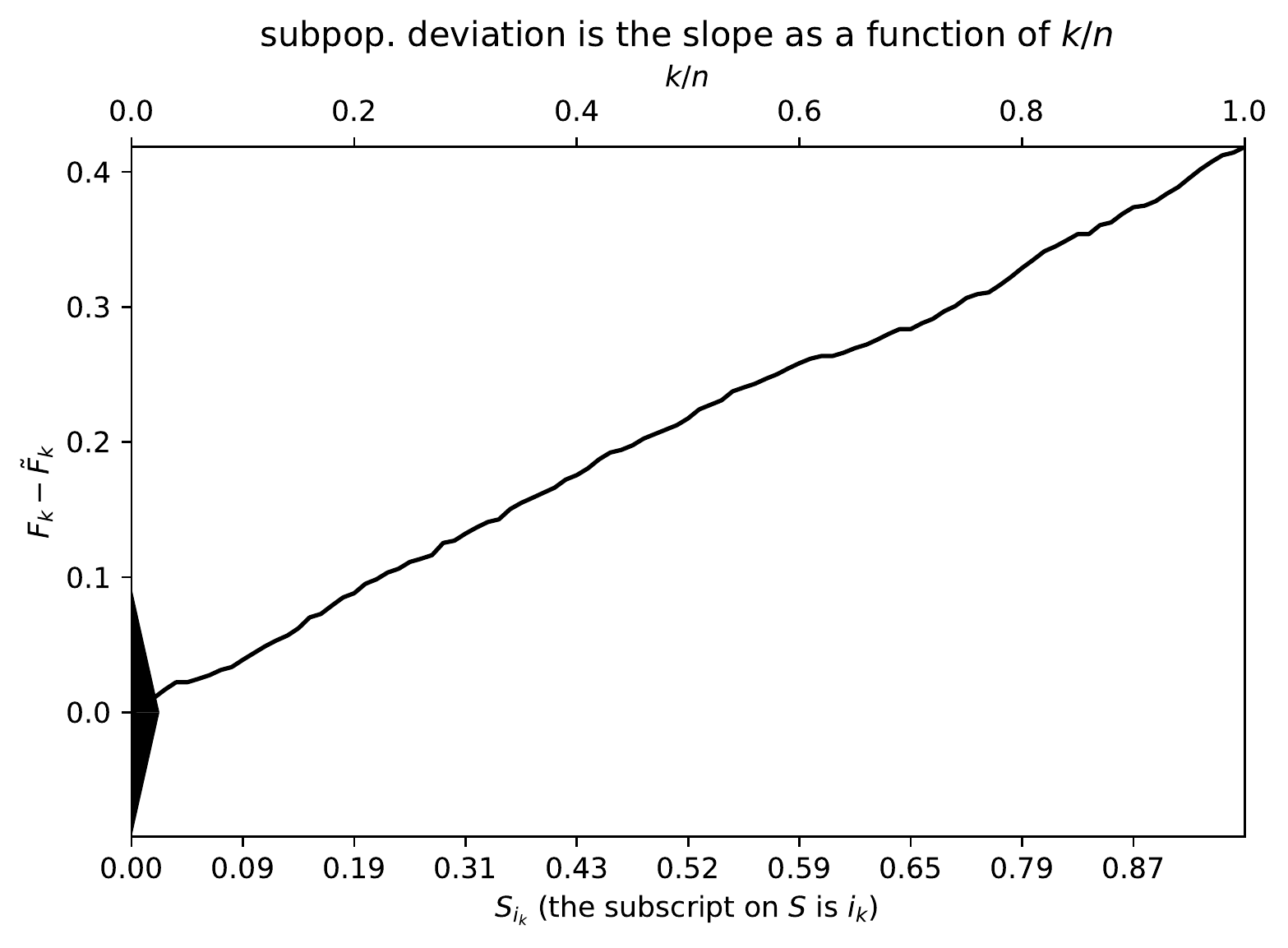}}

\parbox{\imsized}{\hfil \footnotesize $G$ = 0.4110; $H$ = 0.4110;
$G/\sigma$ = 15.90; $H/\sigma$ = 15.90}
\parbox{\imsized}{\hfil \footnotesize $G$ = 0.4185; $H$ = 0.4185;
$G/\sigma$ = 9.104; $H/\sigma$ = 9.104}

\vspace{\vertsep}

(4) \parbox{\imsize}{\includegraphics[width=\imsize,
                                      trim={0pt 0pt 0pt 2pt}, clip]
{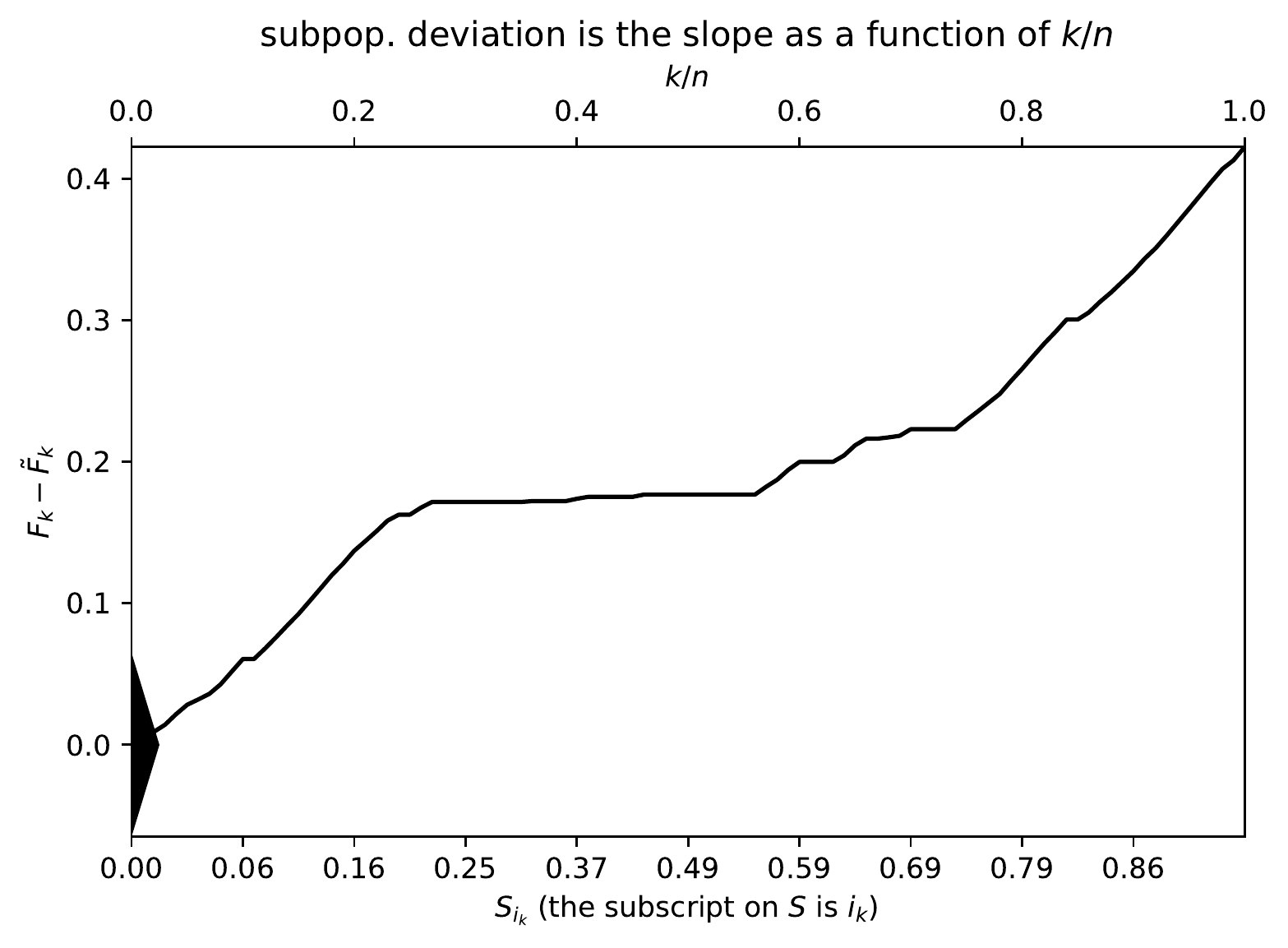}}
\quad\quad
(64) \parbox{\imsize}{\includegraphics[width=\imsize,
                                       trim={0pt 0pt 0pt 2pt}, clip]
{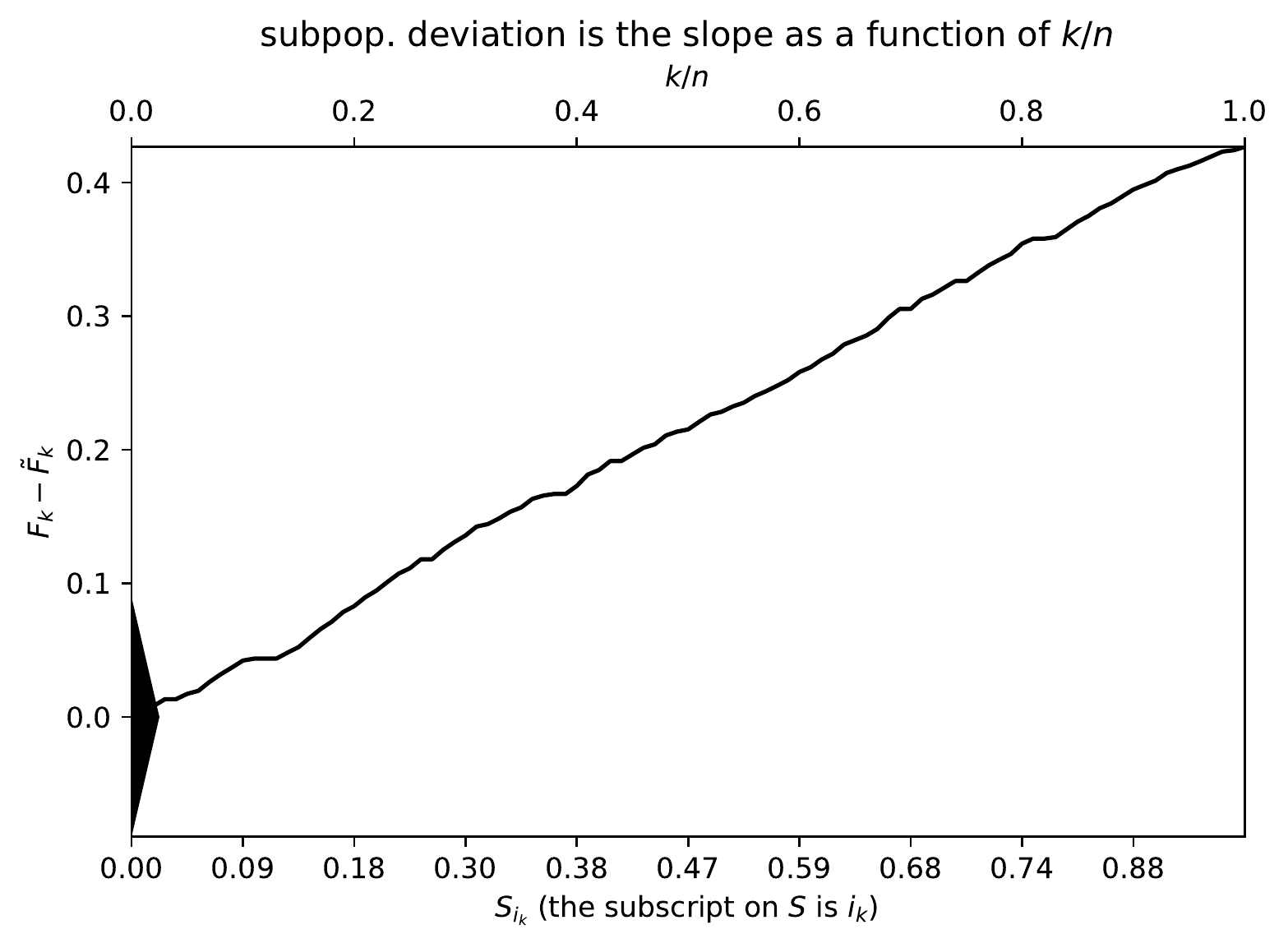}}

\parbox{\imsized}{\hfil \footnotesize $G$ = 0.4225; $H$ = 0.4225;
$G/\sigma$ = 13.03; $H/\sigma$ = 13.03}
\parbox{\imsized}{\hfil \footnotesize $G$ = 0.4267; $H$ = 0.4267;
$G/\sigma$ = 9.530; $H/\sigma$ = 9.530}

\vspace{\vertsep}

(8) \parbox{\imsize}{\includegraphics[width=\imsize,
                                      trim={0pt 0pt 0pt 2pt}, clip]
{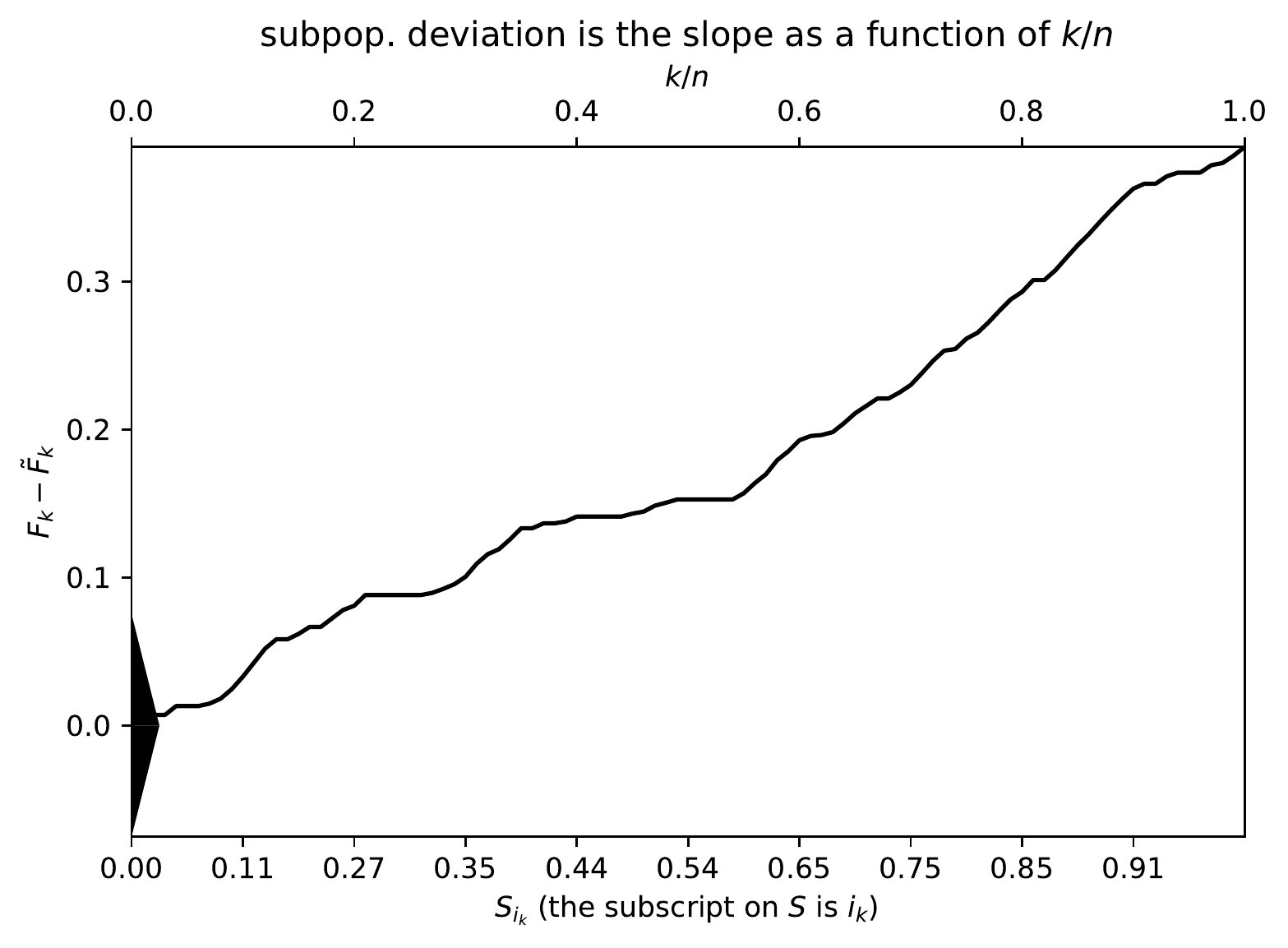}}
\quad\quad
(128) \parbox{\imsize}{\includegraphics[width=\imsize,
                                        trim={0pt 0pt 0pt 2pt}, clip]
{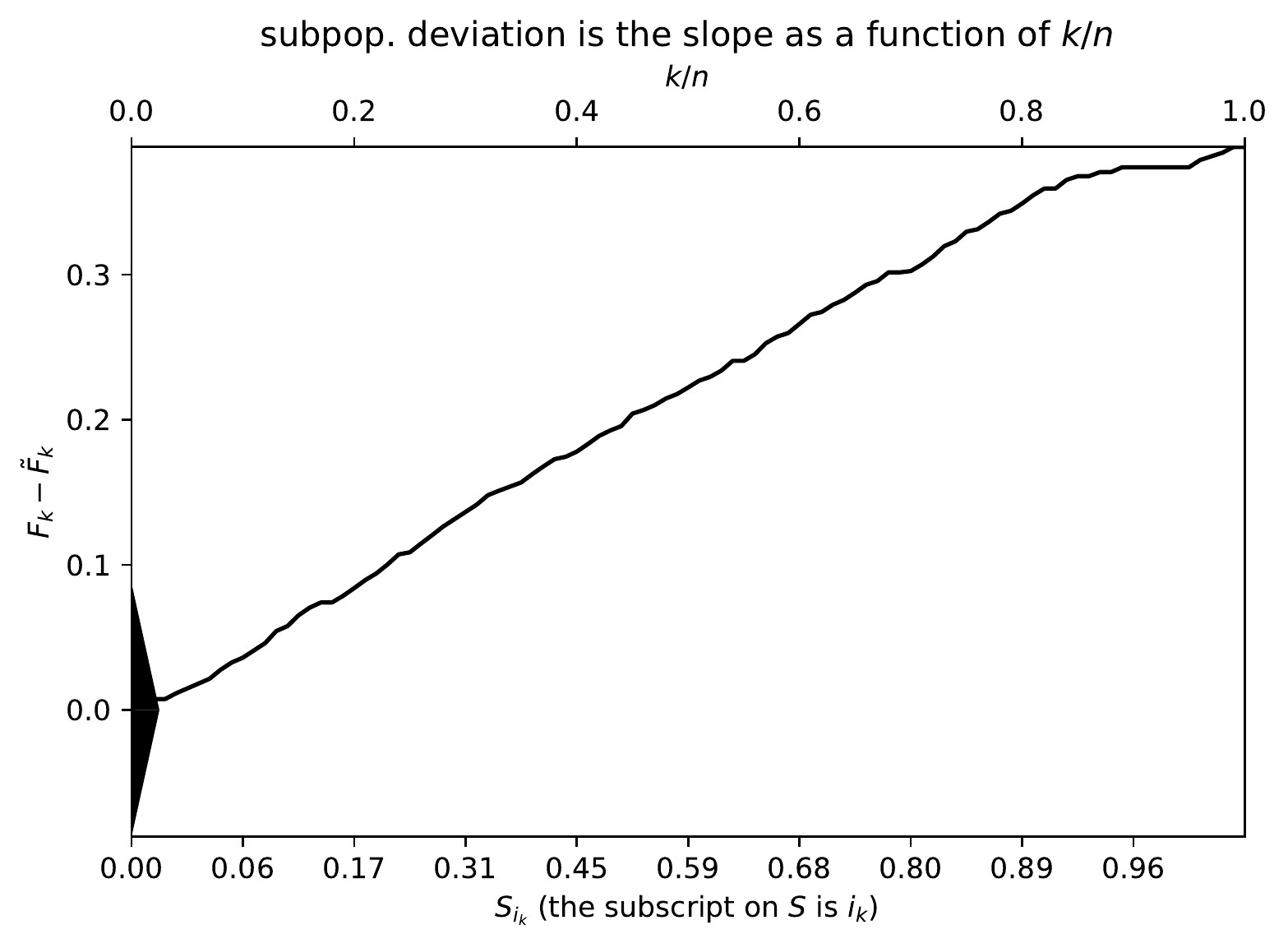}}

\parbox{\imsized}{\hfil \footnotesize $G$ = 0.3912; $H$ = 0.3912;
$G/\sigma$ = 10.41; $H/\sigma$ = 10.41}
\parbox{\imsized}{\hfil \footnotesize $G$ = 0.3880; $H$ = 0.3880;
$G/\sigma$ = 8.894; $H/\sigma$ = 8.894}

\vspace{\vertsep}

(16) \parbox{\imsize}{\includegraphics[width=\imsize,
                                       trim={0pt 0pt 0pt 2pt}, clip]
{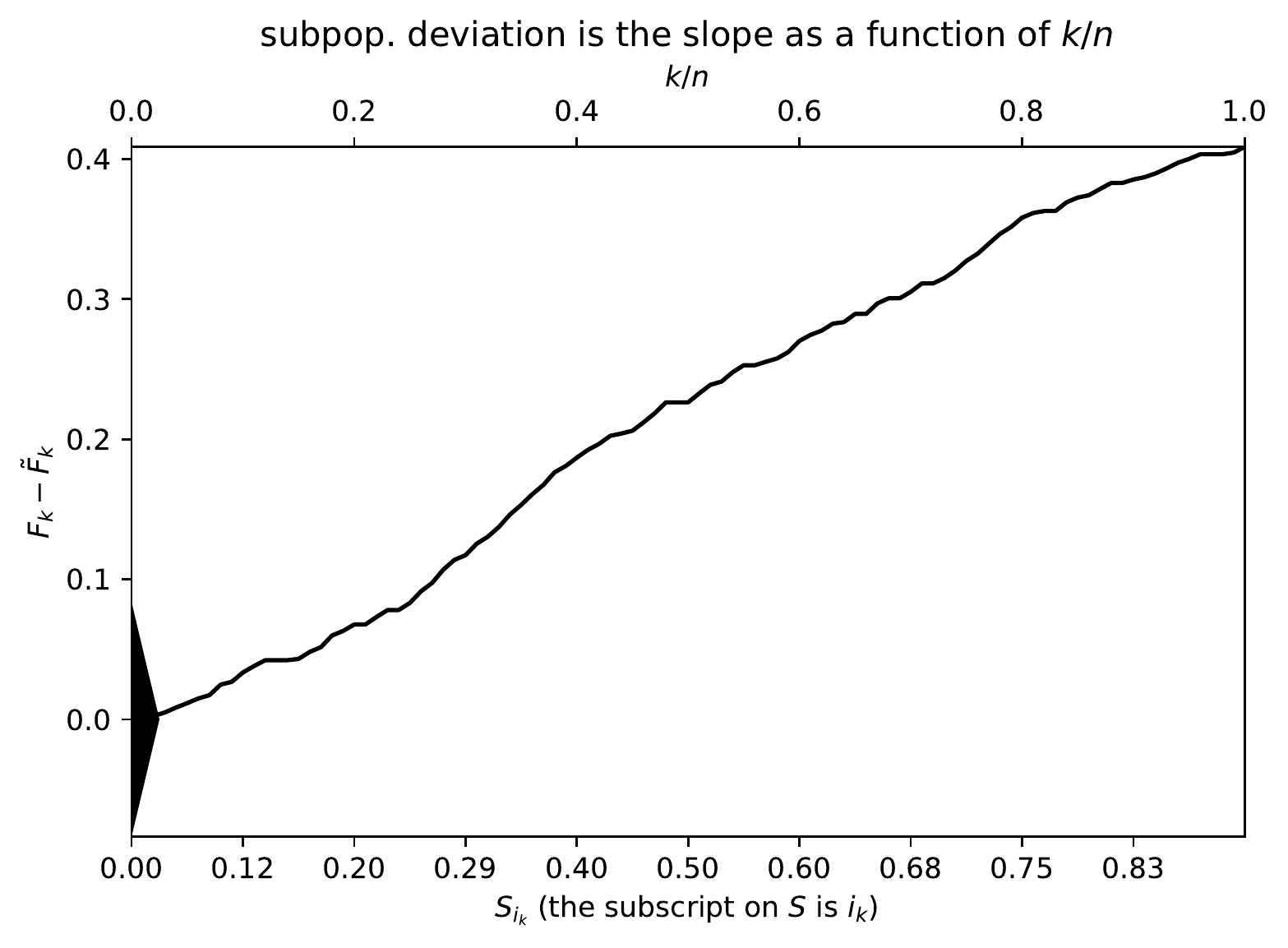}}
\quad\quad
(256) \parbox{\imsize}{\includegraphics[width=\imsize,
                                        trim={0pt 0pt 0pt 2pt}, clip]
{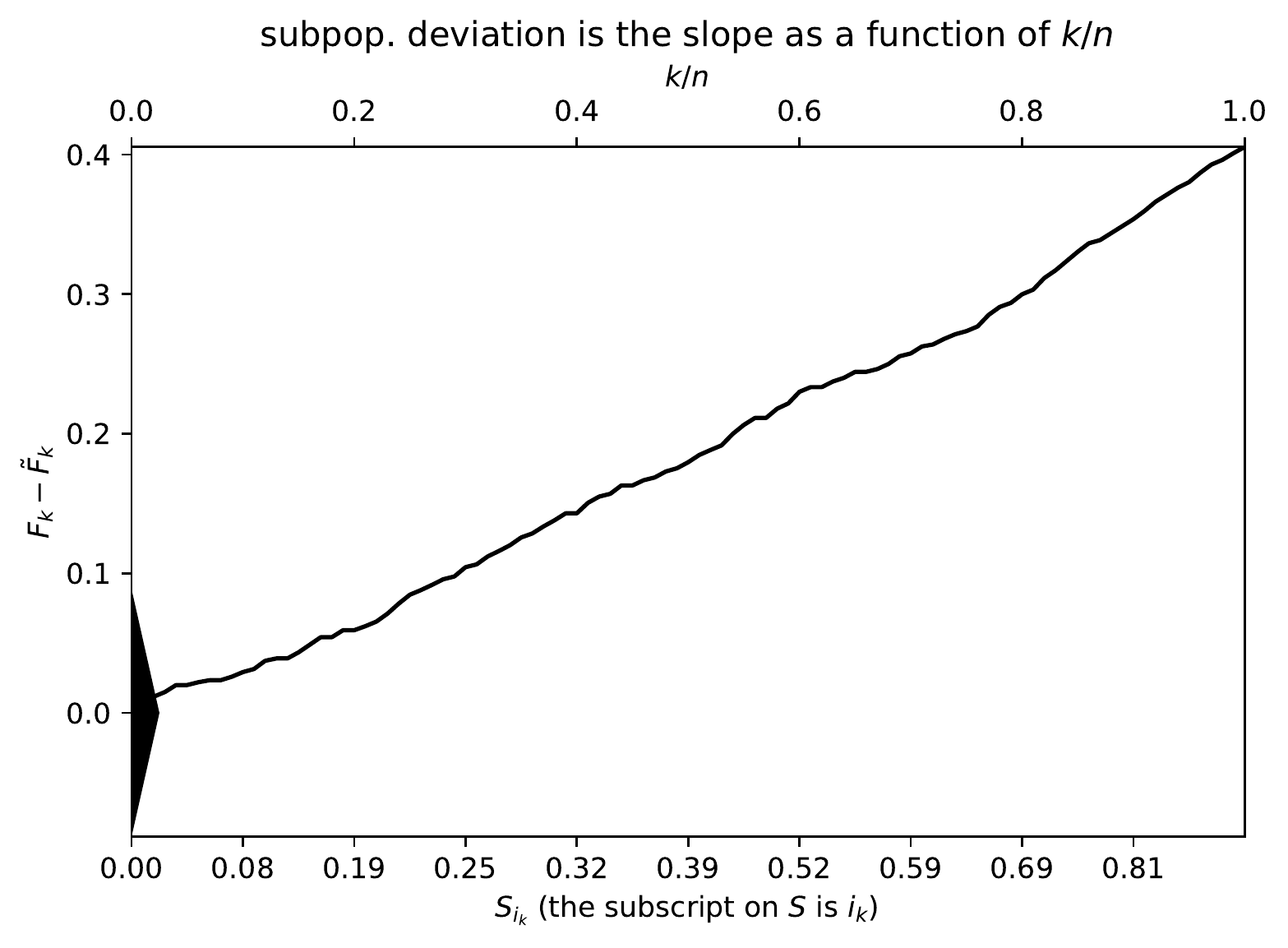}}

\parbox{\imsized}{\hfil \footnotesize $G$ = 0.4087; $H$ = 0.4087;
$G/\sigma$ = 9.768; $H/\sigma$ = 9.768}
\parbox{\imsized}{\hfil \footnotesize $G$ = 0.4055; $H$ = 0.4055;
$G/\sigma$ = 9.153; $H/\sigma$ = 9.153}

\end{centering}
\caption{Synthetic examples with a varying number of covariates,
reversing the order of the covariates}
\label{reverse}
\end{figure}

\begin{figure}
\begin{centering}
(512) \parbox{\imsize}{\includegraphics[width=\imsize]
{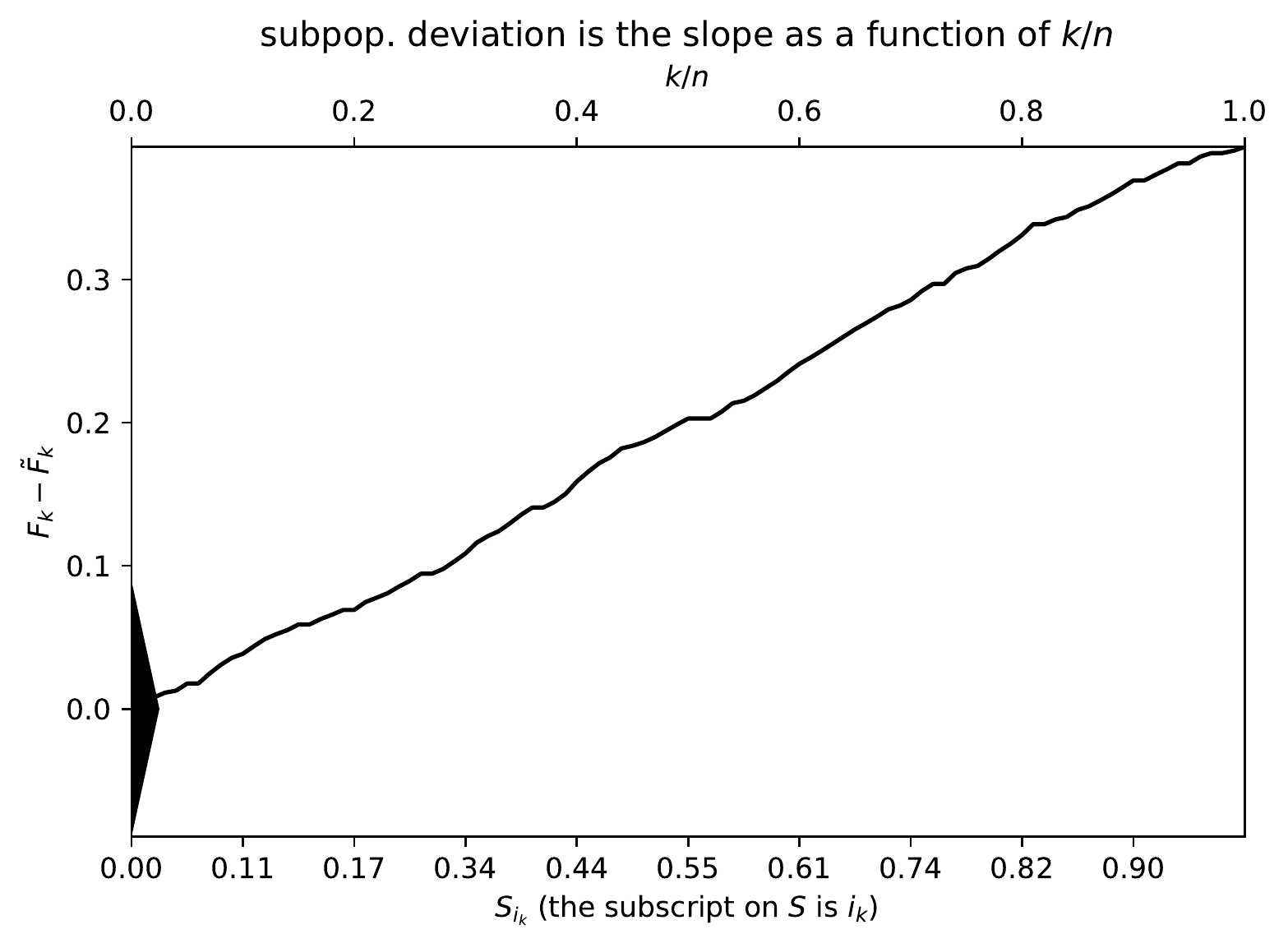}}
\quad
(2048) \parbox{\imsize}{\includegraphics[width=\imsize]
{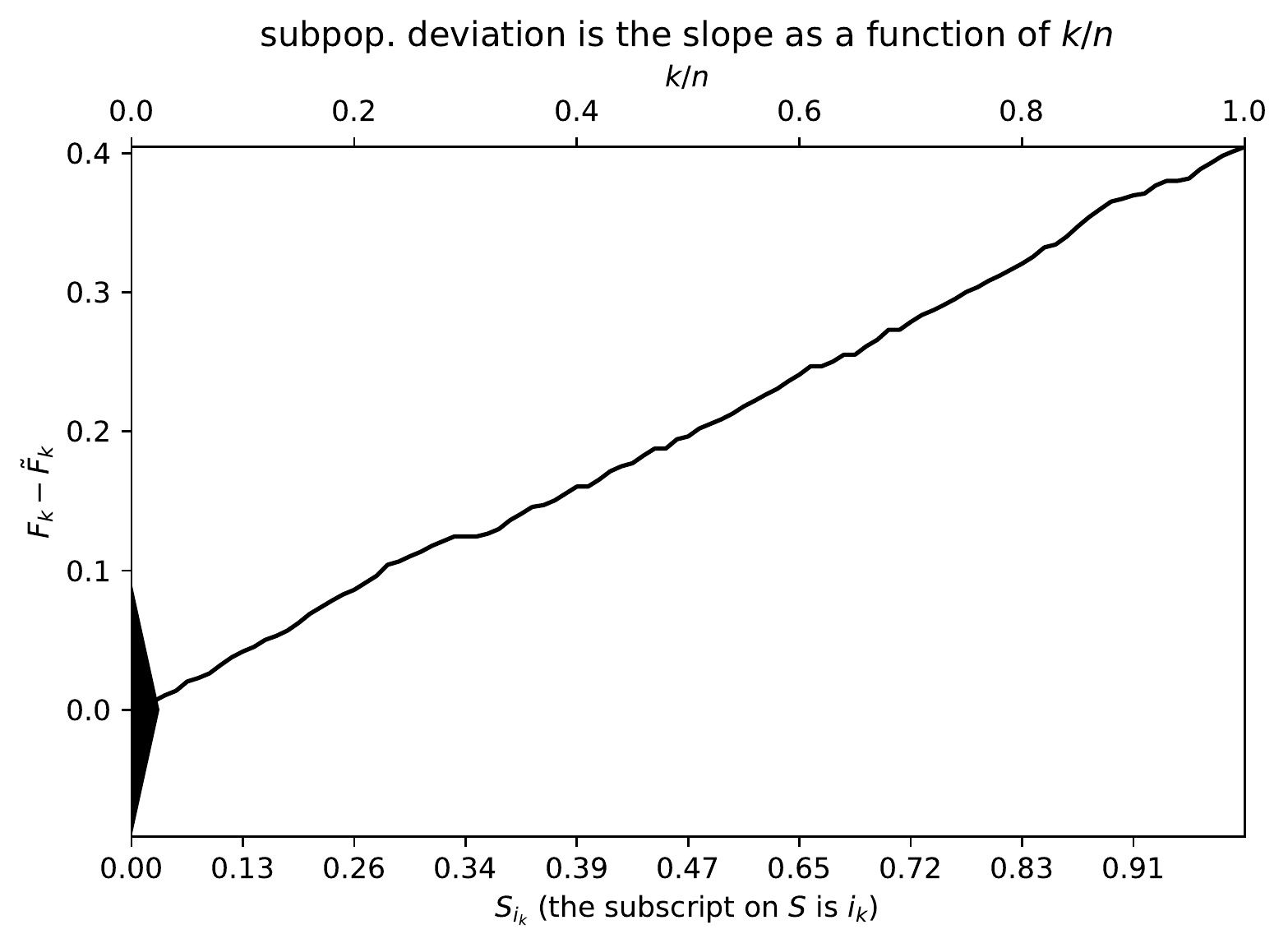}}

\parbox{\imsized}{\hfil \footnotesize $G$ = 0.3927; $H$ = 0.3927;
$G/\sigma$ = 8.798; $H/\sigma$ = 8.798}
\parbox{\imsized}{\hfil \footnotesize $G$ = 0.4044; $H$ = 0.4044;
$G/\sigma$ = 8.872; $H/\sigma$ = 8.872}

\vspace{\vertsep}

(1024) \parbox{\imsize}{\includegraphics[width=\imsize]
{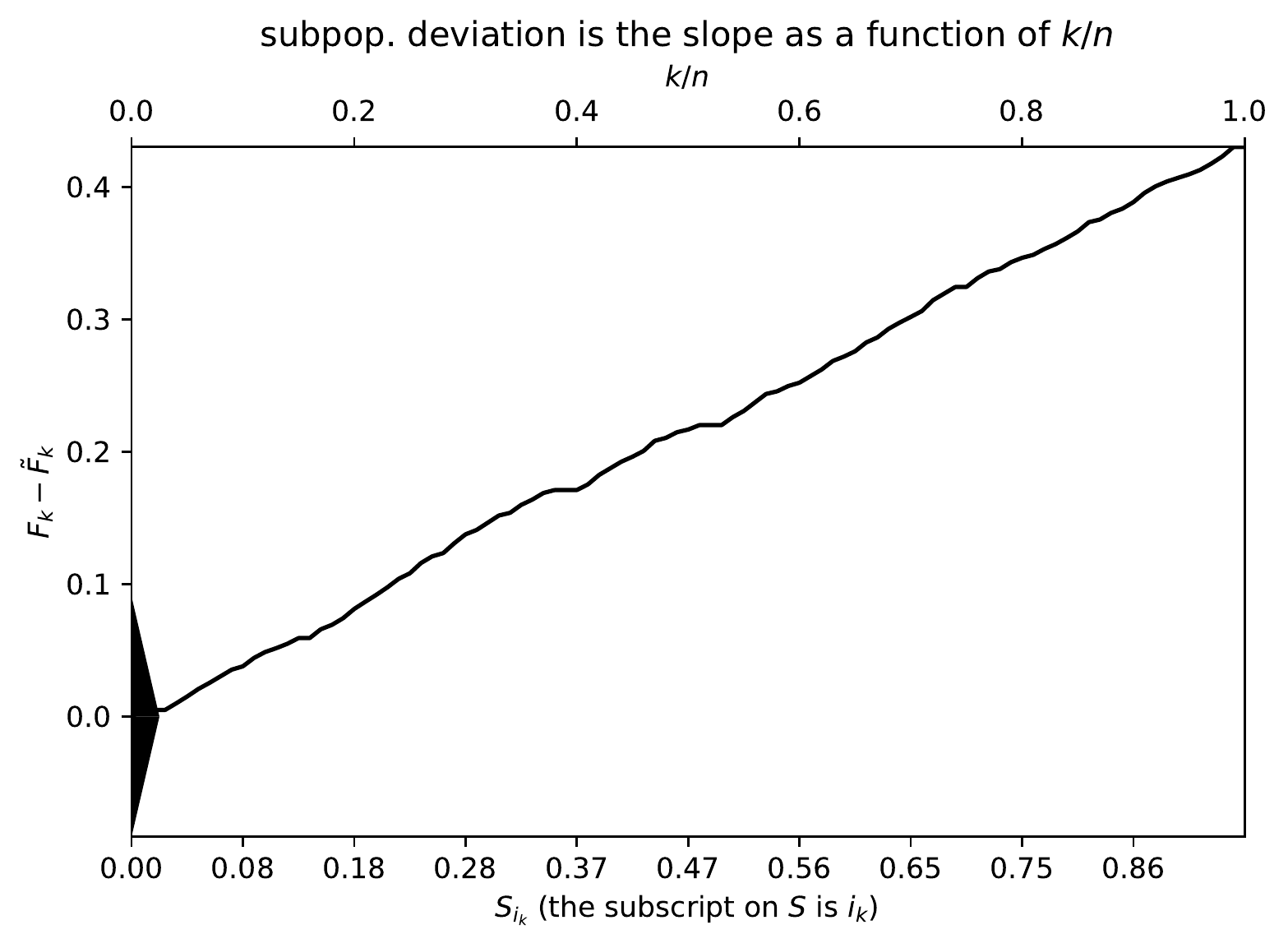}}
\quad
(4096) \parbox{\imsize}{\includegraphics[width=\imsize]
{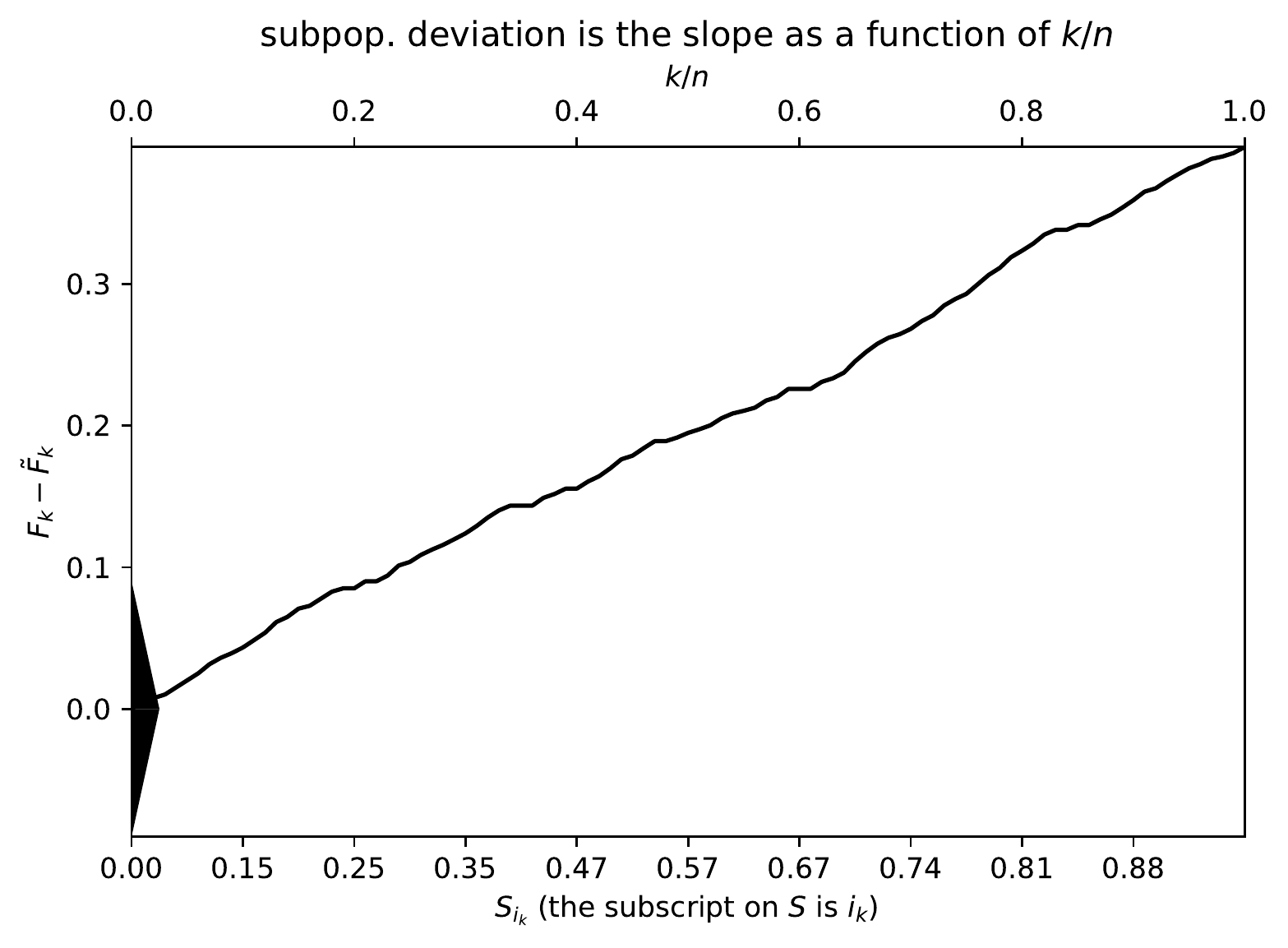}}

\parbox{\imsized}{\hfil \footnotesize $G$ = 0.4305; $H$ = 0.4305;
$G/\sigma$ = 9.496; $H/\sigma$ = 9.496}
\parbox{\imsized}{\hfil \footnotesize $G$ = 0.3966; $H$ = 0.3966;
$G/\sigma$ = 8.806; $H/\sigma$ = 8.806}
\end{centering}
\caption{Synthetic examples with even more covariates,
reversing the order of the covariates}
\label{reverses}
\end{figure}

\begin{figure}
\begin{centering}

(2) \parbox{\imsize}{\includegraphics[width=\imsize,
                                      trim={0pt 0pt 0pt 3pt}, clip]
{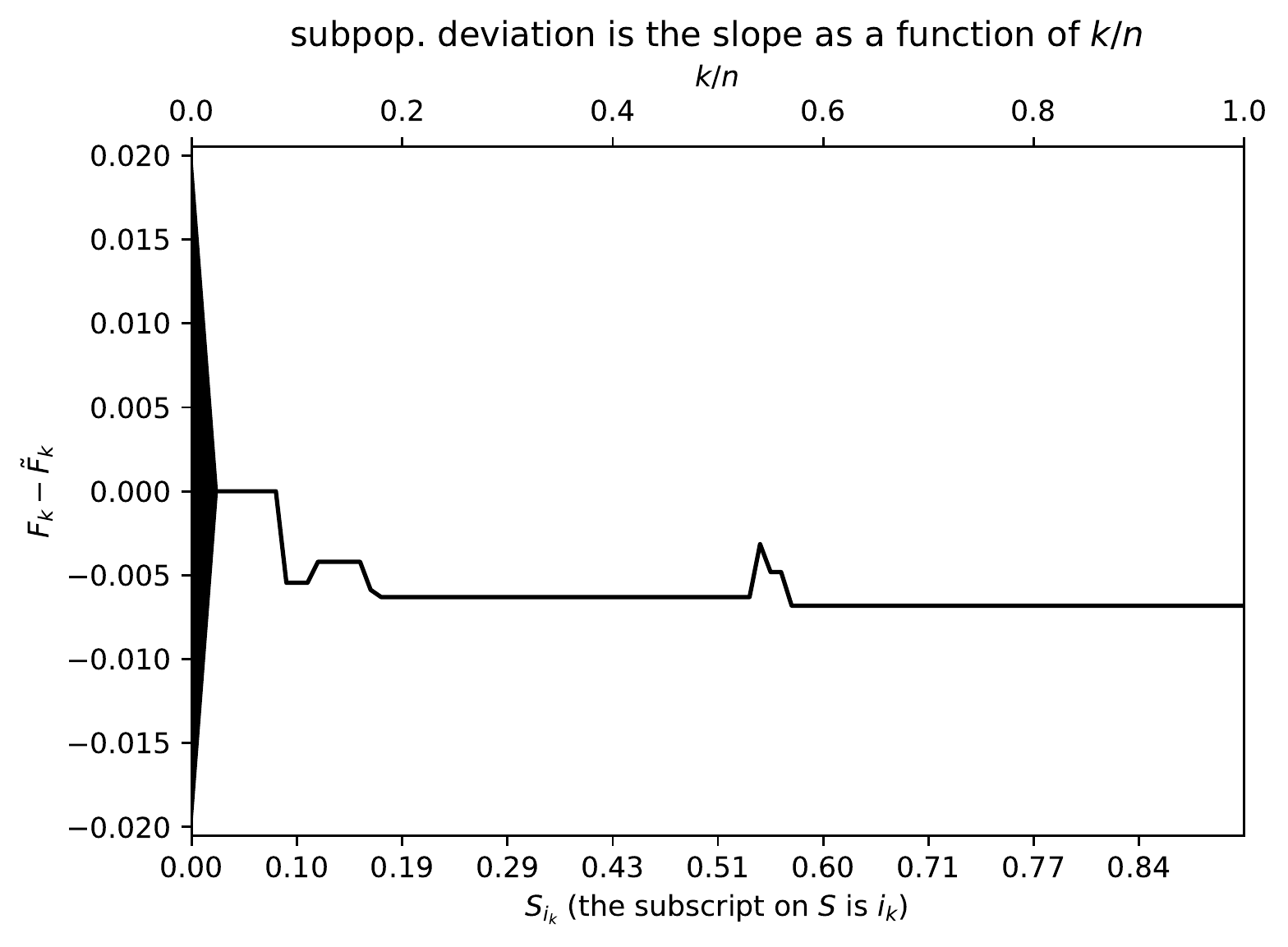}}
\quad\quad
(32) \parbox{\imsize}{\includegraphics[width=\imsize,
                                       trim={0pt 0pt 0pt 3pt}, clip]
{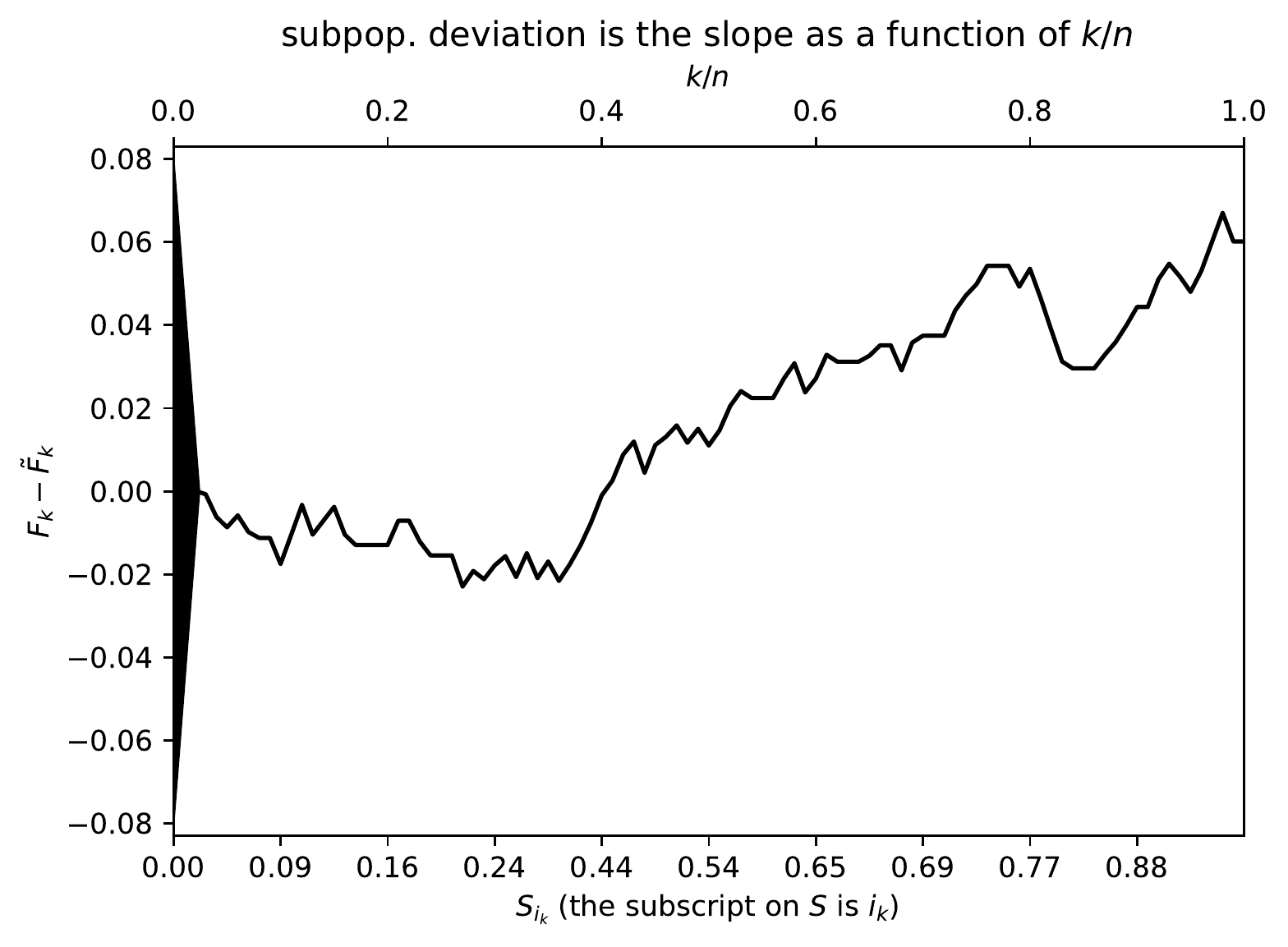}}

\parbox{\imsized}{\hfil \footnotesize $G$ = 0.006815; $H$ = 0.006815;
$G/\sigma$ = 0.6642; $H/\sigma$ = 0.6642}
\parbox{\imsized}{\hfil \footnotesize $G$ = 0.06699; $H$ = 0.08992;
$G/\sigma$ = 1.616; $H/\sigma$ = 2.169}

\vspace{\vertsep}

(4) \parbox{\imsize}{\includegraphics[width=\imsize,
                                      trim={0pt 0pt 0pt 2pt}, clip]
{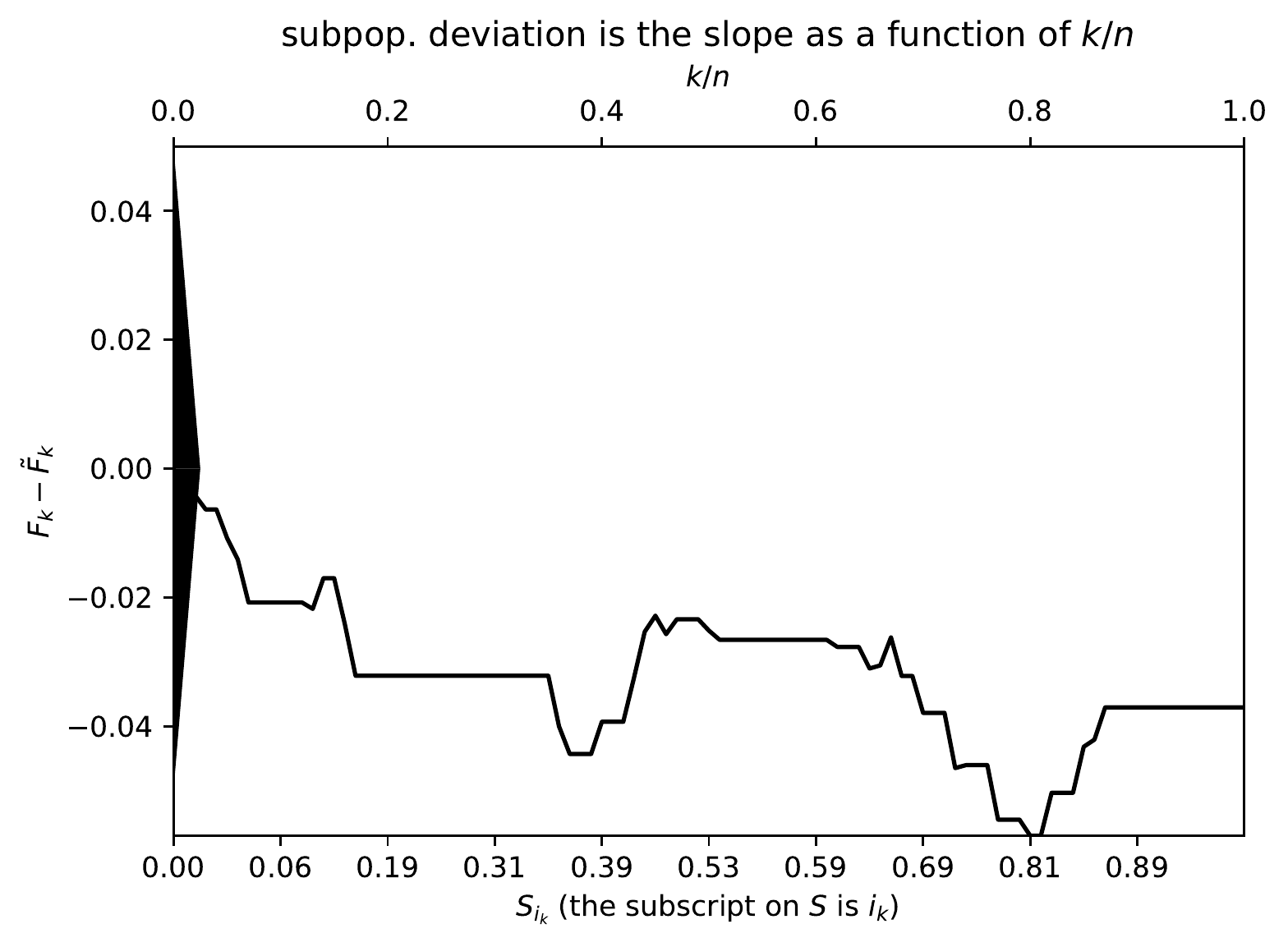}}
\quad\quad
(64) \parbox{\imsize}{\includegraphics[width=\imsize,
                                       trim={0pt 0pt 0pt 2pt}, clip]
{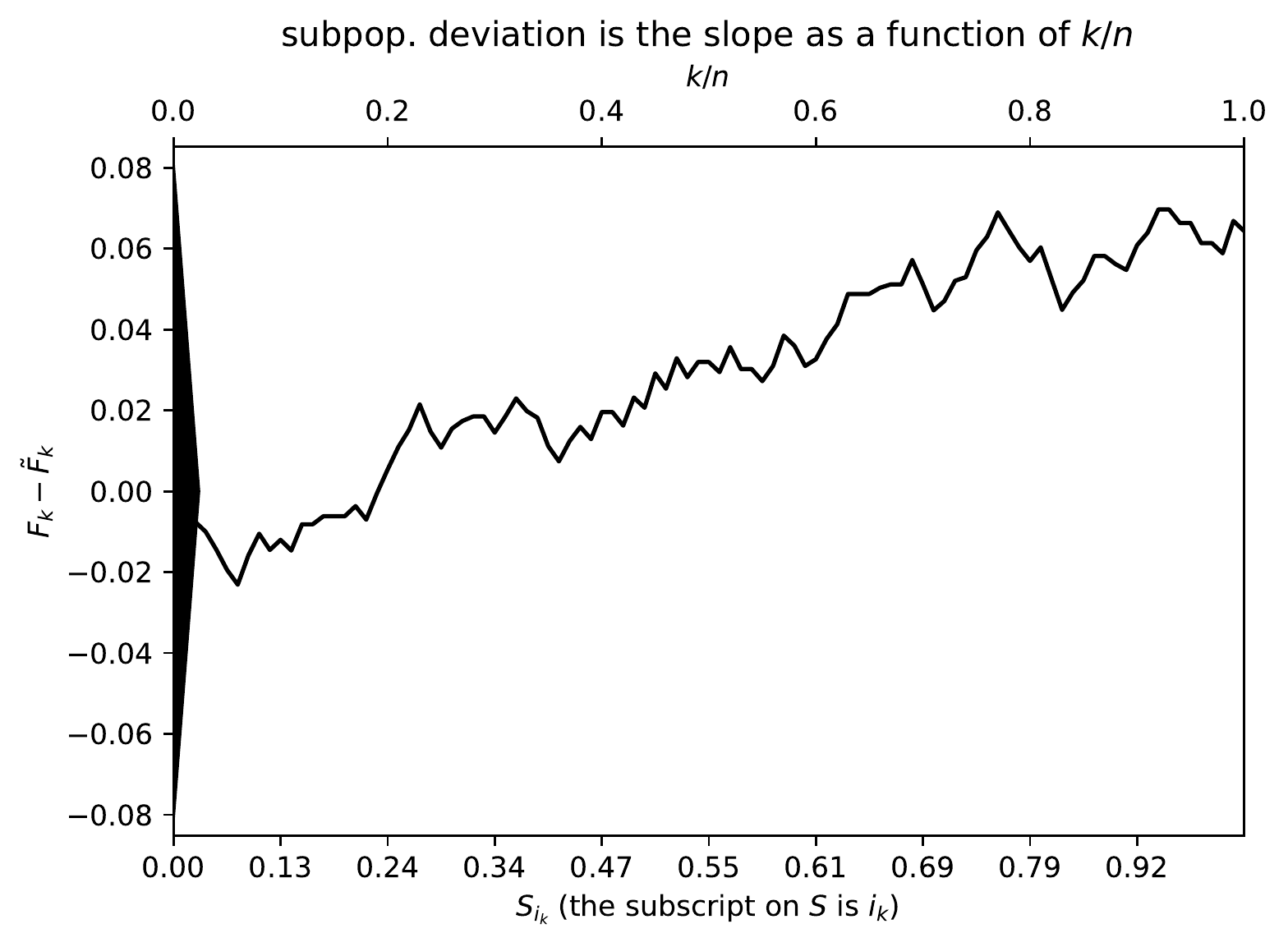}}

\parbox{\imsized}{\hfil \footnotesize $G$ = 0.0570; $H$ = 0.0570;
$G/\sigma$ = 2.281; $H/\sigma$ = 2.281}
\parbox{\imsized}{\hfil \footnotesize $G$ = 0.06966; $H$ = 0.09274;
$G/\sigma$ = 1.636; $H/\sigma$ = 2.178}

\vspace{\vertsep}

(8) \parbox{\imsize}{\includegraphics[width=\imsize,
                                      trim={0pt 0pt 0pt 2pt}, clip]
{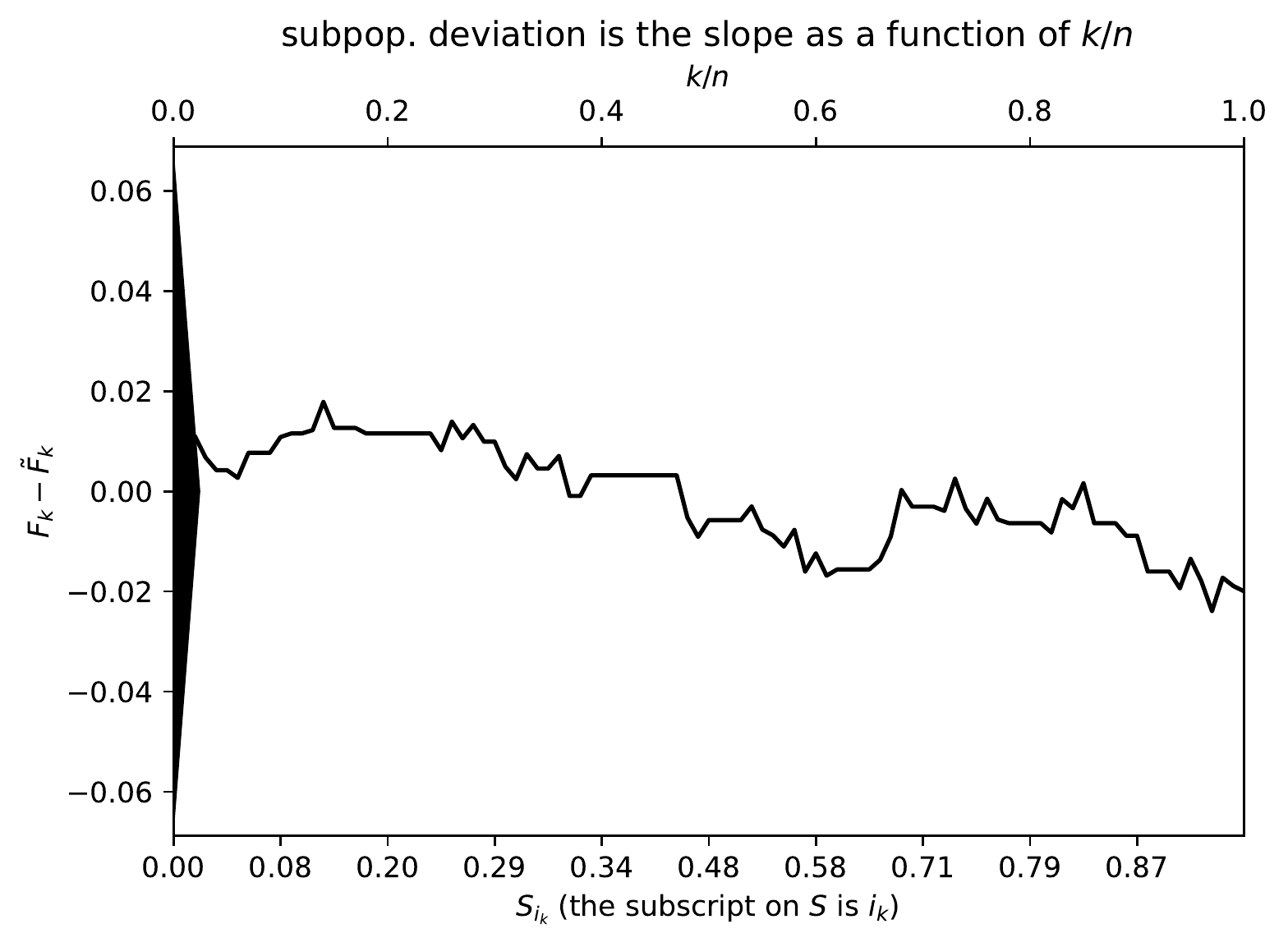}}
\quad\quad
(128) \parbox{\imsize}{\includegraphics[width=\imsize,
                                        trim={0pt 0pt 0pt 2pt}, clip]
{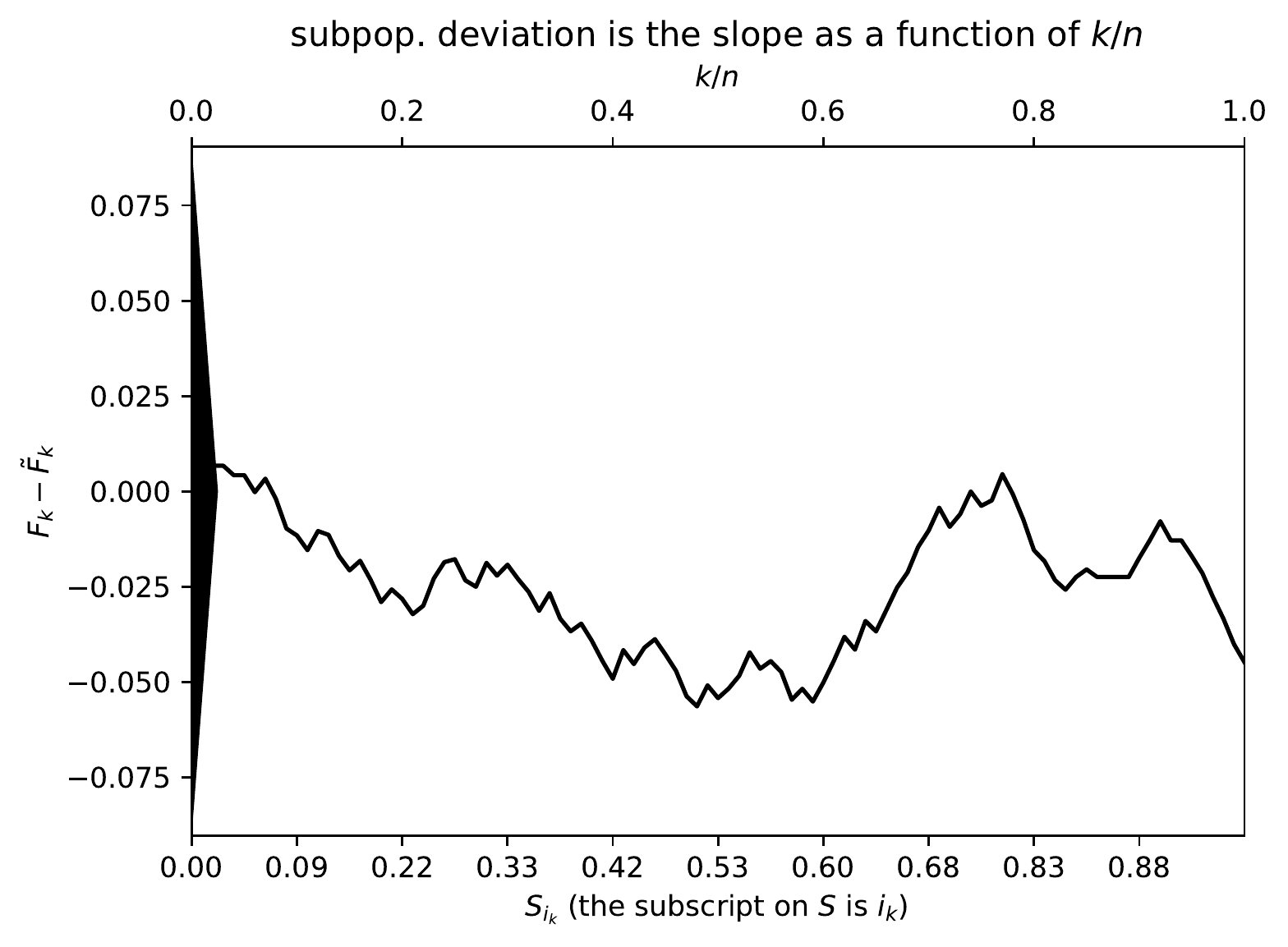}}

\parbox{\imsized}{\hfil \footnotesize $G$ = 0.02393; $H$ = 0.04177;
$G/\sigma$ = 0.6957; $H/\sigma$ = 1.214}
\parbox{\imsized}{\hfil \footnotesize $G$ = 0.05644; $H$ = 0.06315;
$G/\sigma$ = 1.249; $H/\sigma$ = 1.397}

\vspace{\vertsep}

(16) \parbox{\imsize}{\includegraphics[width=\imsize,
                                       trim={0pt 0pt 0pt 2pt}, clip]
{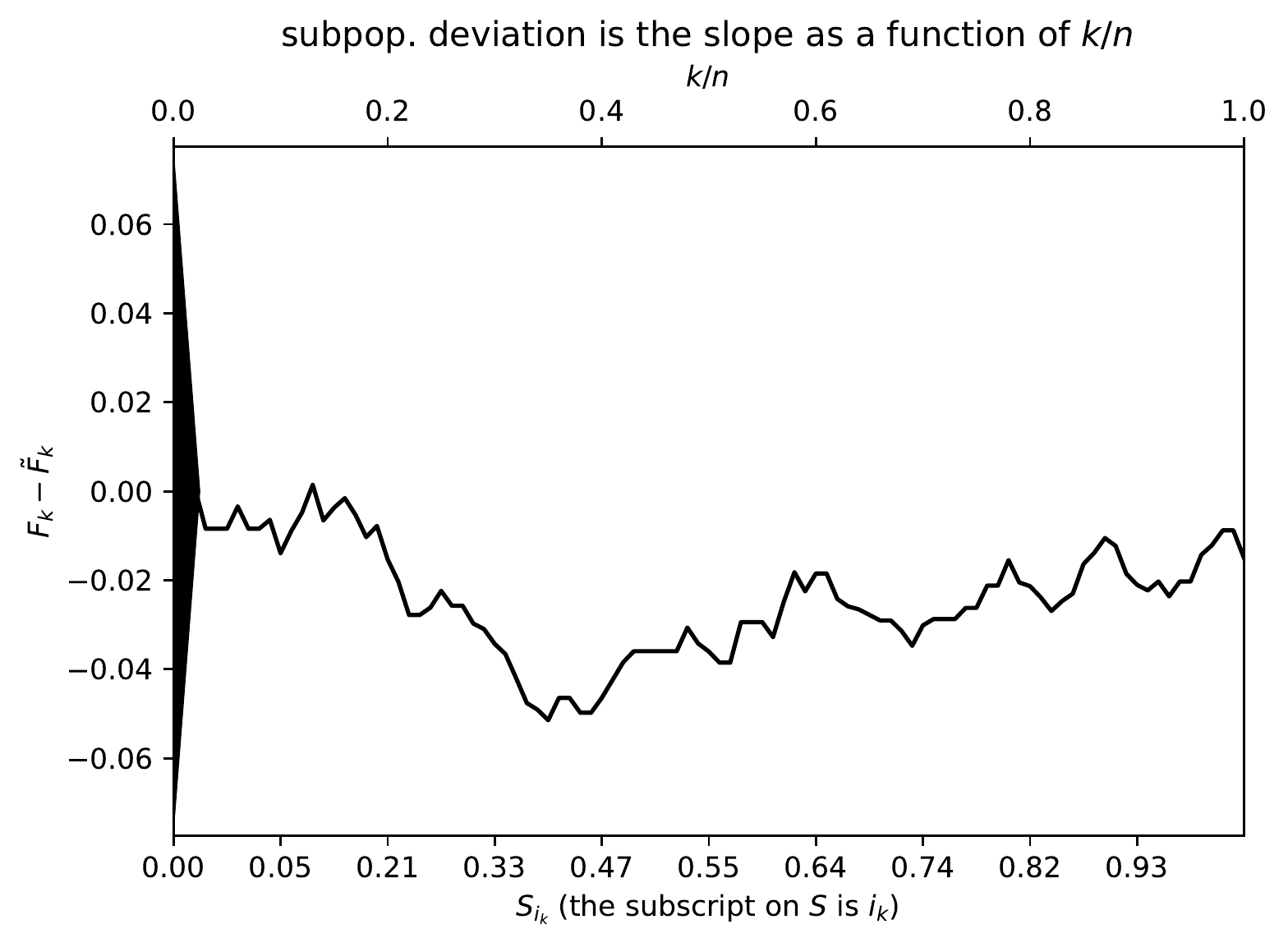}}
\quad\quad
(256) \parbox{\imsize}{\includegraphics[width=\imsize,
                                        trim={0pt 0pt 0pt 2pt}, clip]
{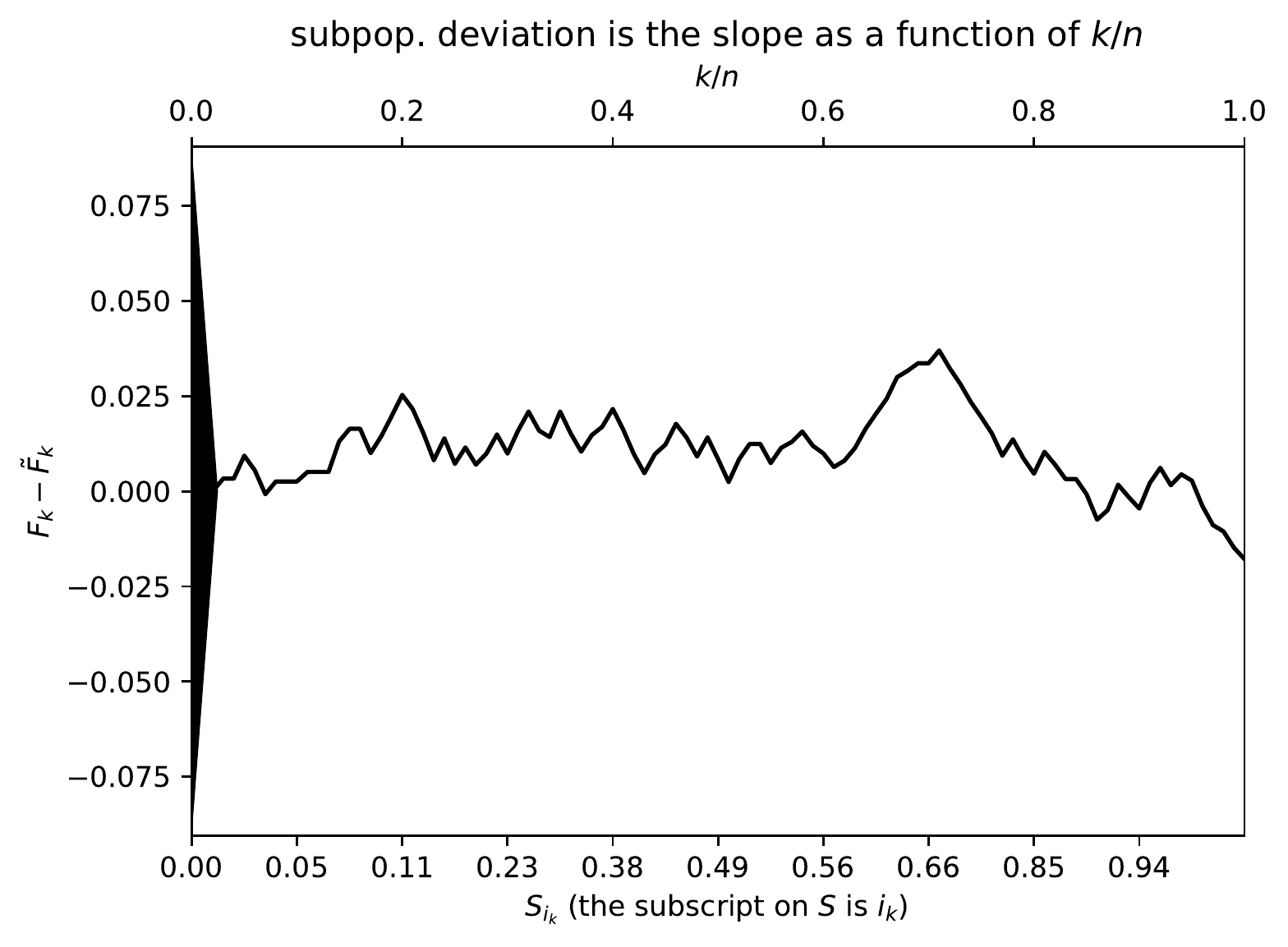}}

\parbox{\imsized}{\hfil \footnotesize $G$ = 0.05145; $H$ = 0.05289;
$G/\sigma$ = 1.329; $H/\sigma$ = 1.366}
\parbox{\imsized}{\hfil \footnotesize $G$ = 0.03697; $H$ = 0.05481;
$G/\sigma$ = 0.8171; $H/\sigma$ = 1.211}

\end{centering}
\caption{Synthetic examples with a varying number of covariates
but no significant deviation}
\label{randwalk}
\end{figure}

\begin{figure}
\begin{centering}
(512) \parbox{\imsize}{\includegraphics[width=\imsize]
{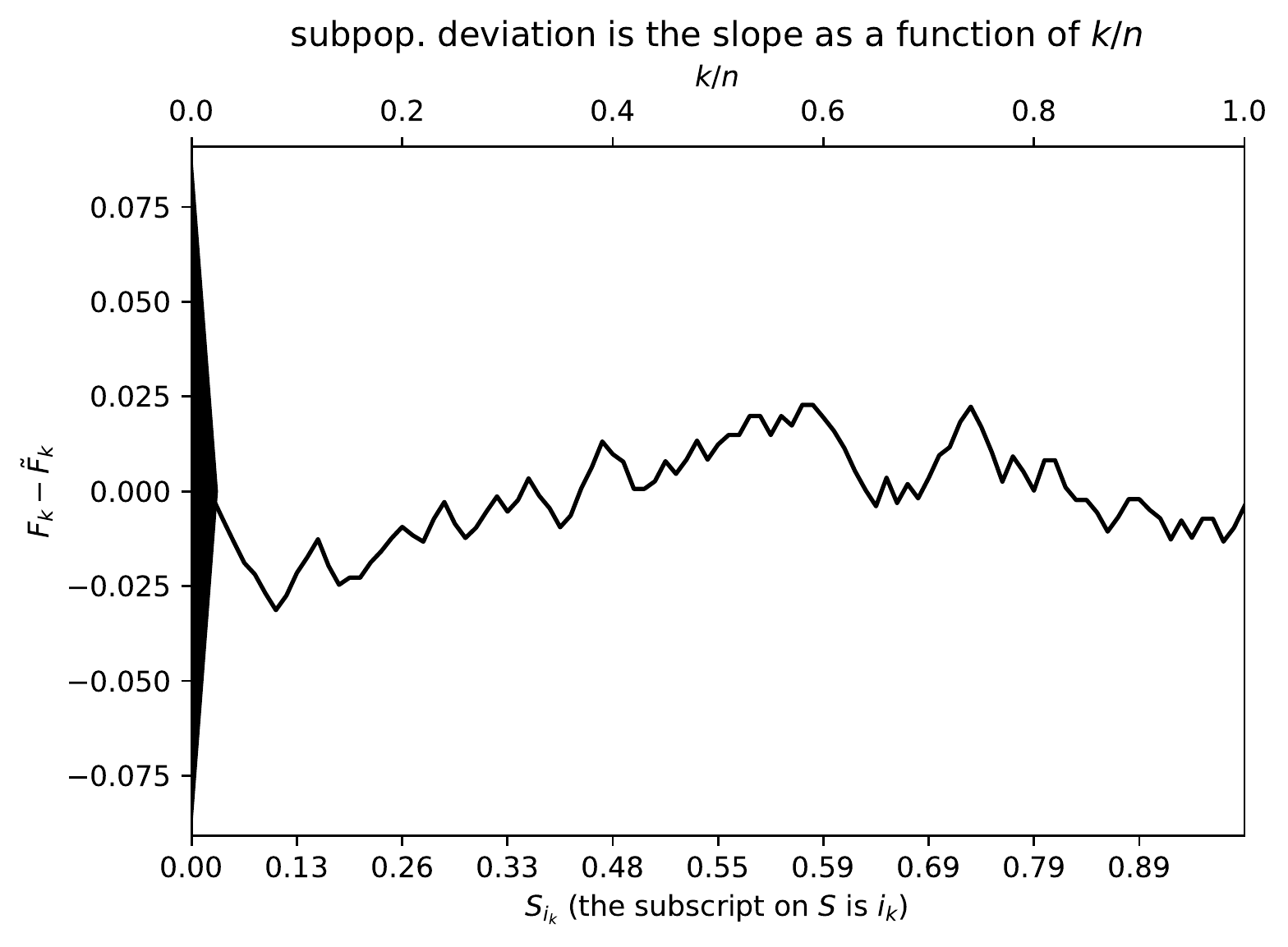}}
\quad
(2048) \parbox{\imsize}{\includegraphics[width=\imsize]
{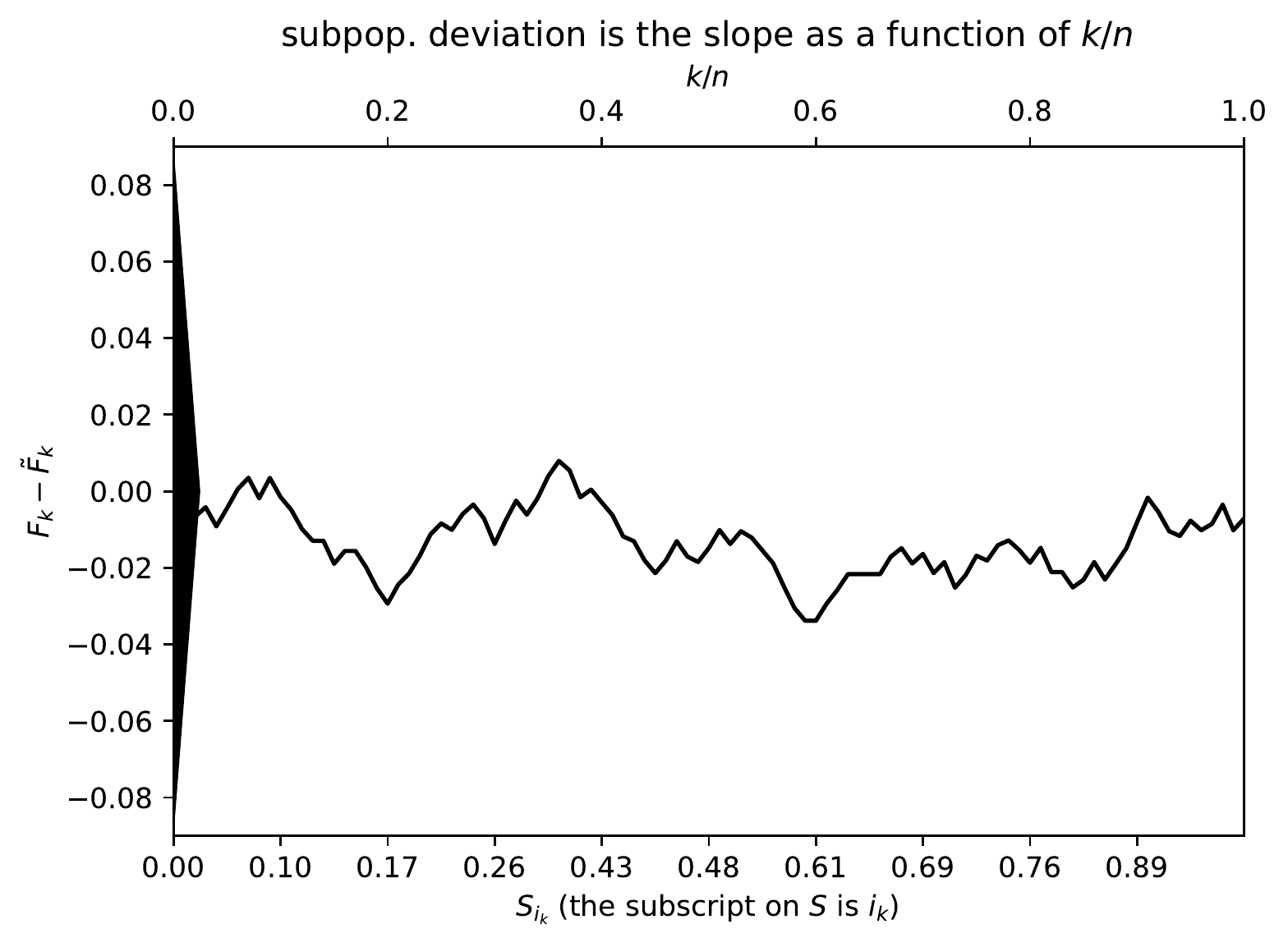}}

\parbox{\imsized}{\hfil \footnotesize $G$ = 0.03135; $H$ = 0.05414;
$G/\sigma$ = 0.6899; $H/\sigma$ = 1.192}
\parbox{\imsized}{\hfil \footnotesize $G$ = 0.03382; $H$ = 0.04174;
$G/\sigma$ = 0.7518; $H/\sigma$ = 0.9279}

\vspace{\vertsep}

(1024) \parbox{\imsize}{\includegraphics[width=\imsize]
{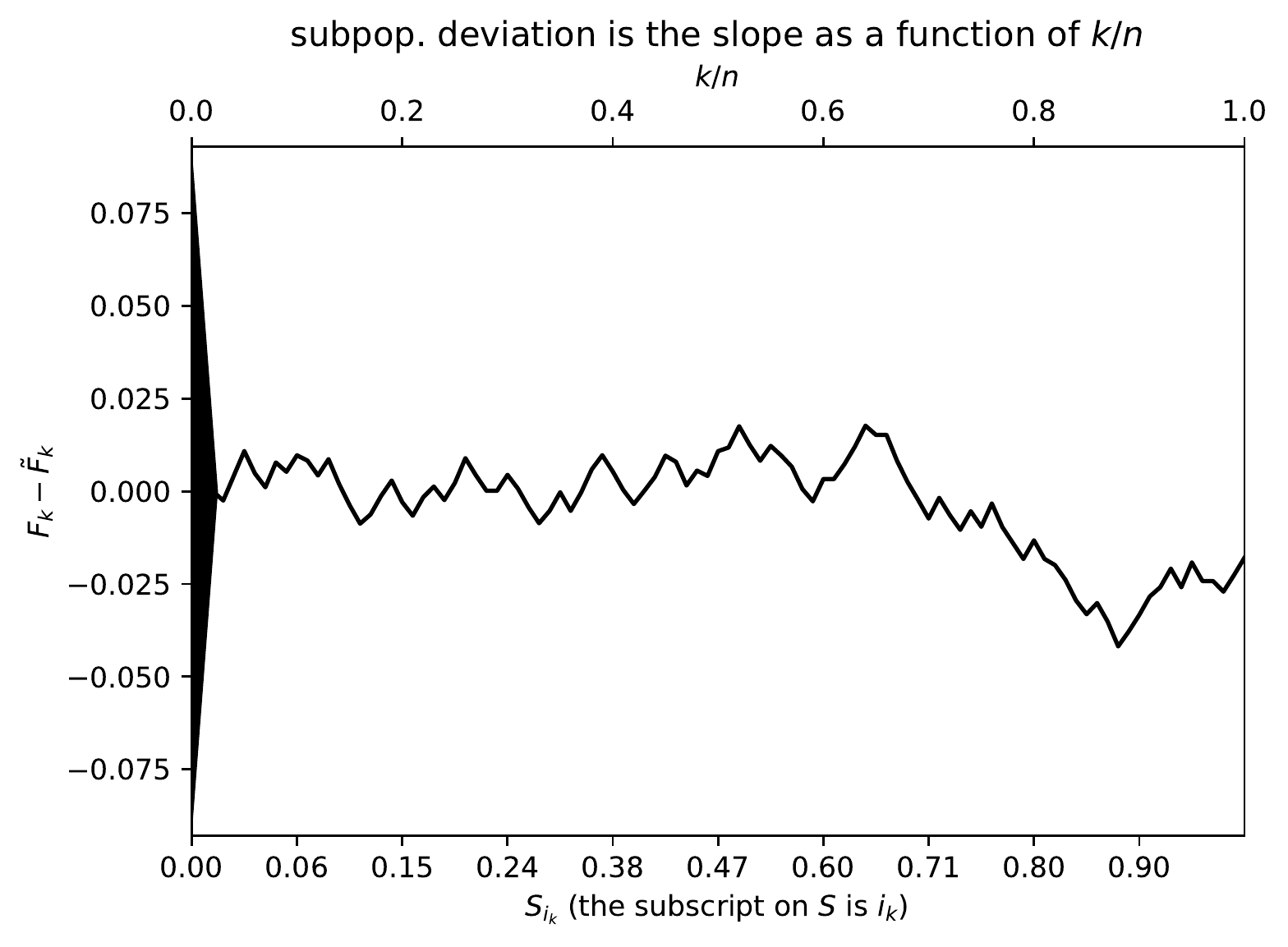}}
\quad
(4096) \parbox{\imsize}{\includegraphics[width=\imsize]
{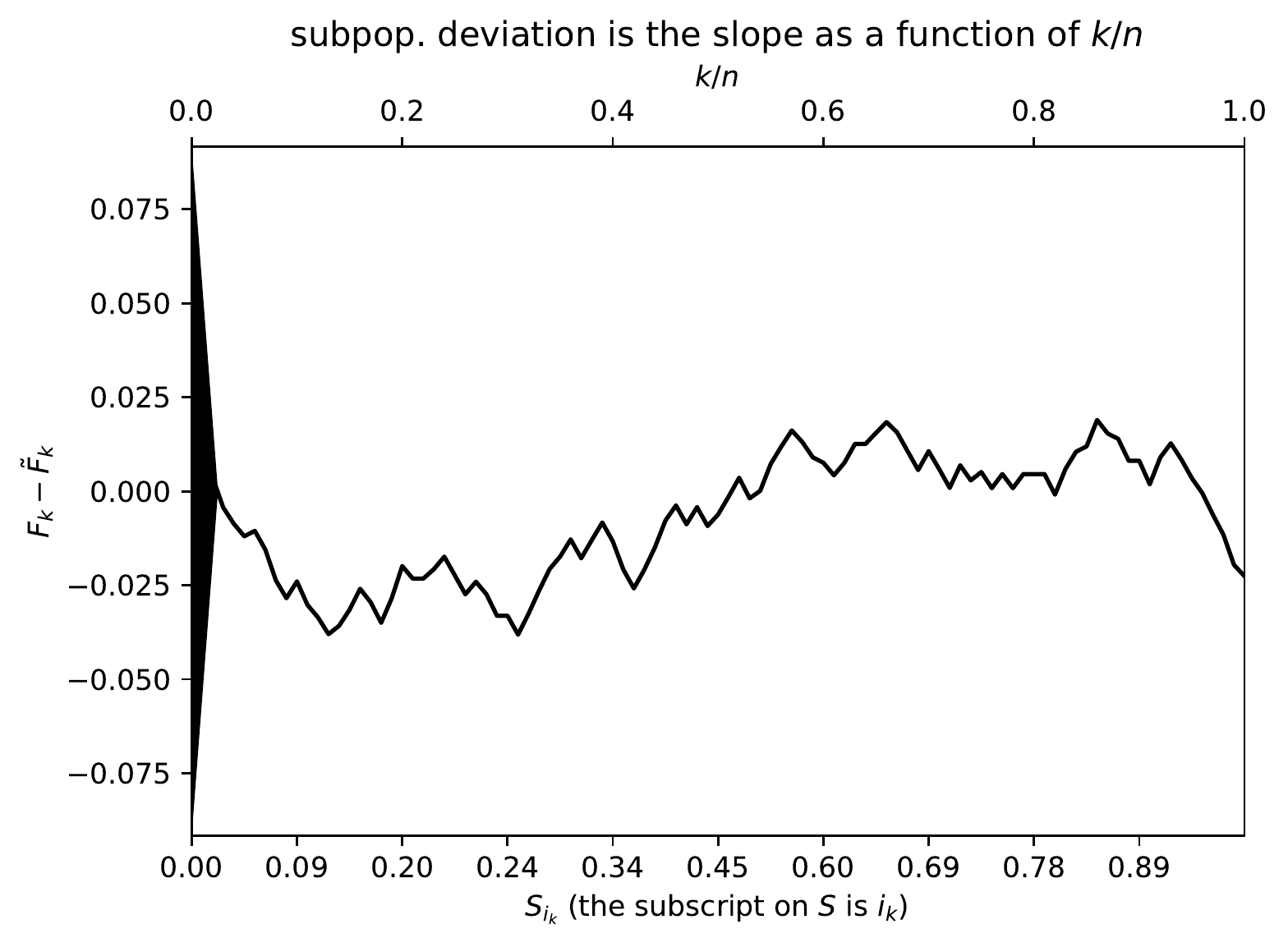}}

\parbox{\imsized}{\hfil \footnotesize $G$ = 0.04182; $H$ = 0.05950;
$G/\sigma$ = 0.9003; $H/\sigma$ = 1.281}
\parbox{\imsized}{\hfil \footnotesize $G$ = 0.03812; $H$ = 0.05705;
$G/\sigma$ = 0.8321; $H/\sigma$ = 1.245}
\end{centering}
\caption{Synthetic examples with even more covariates
but no significant deviation}
\label{randwalks}
\end{figure}

This subsection presents Figures~\ref{synth}--\ref{randwalks}.
These figures consider simple, easily understandable constructions,
detailed as follows:

Figures~\ref{synth}--\ref{randwalks} pertain to synthetic examples,
illustrating the methods of Section~\ref{methods}
in a controlled setting for which the geometry of the data is easy to grasp.
We consider a full population with $m =$ 1,000 members,
$n =$ 100 observations from a subpopulation,
and several different numbers of covariates,
namely $p =$ 2, 4, 8, 16, \dots, 4096.
We select the subpopulation uniformly at random
from the full population defined as follows.
We construct an $m \times p$ matrix $A$ whose entries
are independent and identically distributed (i.i.d.)\ draws
from the uniform distribution over the interval $(-1, 1)$.
We generate a $p \times 1$ vector $v$ whose entries are i.i.d.\ draws
from the standard normal distribution, so that $v$ points
in a uniformly random direction.
For the responses, we start with the Heaviside function applied
to every entry of the product of $A$ and $v$
(the Heaviside function is also known as the unit step function,
and takes the value 0 for negative arguments and the value 1
for positive arguments); for Figures~\ref{synth}--\ref{reverses}
(but not for Figures~\ref{randwalk} and~\ref{randwalks}),
we then modify the responses for the subpopulation to be 1 always, never 0.
Thus, the responses for the subpopulation are always 1
for Figures~\ref{synth}--\ref{reverses},
whereas the responses for the remaining 90\% of the full population
are 1 only about half the time.
For the covariates, we normalize the entries of $A$ to range from 0 to 1,
that is, the covariates take on the values given by the entries
of $(A+1)/2$, where $1$ denotes the $m \times p$ matrix
whose entries are all equal to 1.
We condition on (that is, control for) all the covariates.

Figures~\ref{synth} and~\ref{synths} plot the cumulative graphs
for the cases $p =$ 2, 4, 8, 16, \dots, 4096 indicated to the left
of each subfigure ($p$ is the number of covariates);
Figures~\ref{reverse} and~\ref{reverses} also plot the cumulative graphs
for the cases $p =$ 2, 4, 8, 16, \dots, 4096 indicated to the left
of each subfigure, but with the covariates controlled for in reverse order
(so that the Hilbert curve cycles through the covariates in the reverse order).
The normalized scalar summary statistics ($G/\sigma$ and $H/\sigma$)
display a significant decrease as $p$ increases through its smallest values.
Comparing Figure~\ref{synth} with Figure~\ref{reverse}
(as well as Figure~\ref{synths} with Figure~\ref{reverses})
shows that the normalized scalar summary statistics
are very similar when conditioning on the same sets of covariates, that is,
they change little when the order of the covariates is reversed.

Figures~\ref{randwalk} and~\ref{randwalks} plot the cumulative graphs
for the cases $p =$ 2, 4, 8, 16, \dots, 4096 indicated to the left
of each subfigure ($p$ is the number of covariates),
using data in which the responses for both the subpopulation
and the full population are 1 about half the time and 0 about half the time.
Since there is no statistically significant deviation
between the responses for the subpopulation
and the responses for the full population,
the cumulative graphs should and do look like driftless,
perfectly random walks.
The normalized scalar summary statistics ($G/\sigma$ and $H/\sigma$)
accordingly never indicate any statistically significant deviation,
with $G/\sigma$ never exceeding even twice its expected value of about 1.25
(the expected value is about $\sqrt{\pi/2} \approx 1.25$
under the null hypothesis of no deviation
between the probability distributions underlying the responses
for the subpopulation and the responses for the full population,
as detailed in Remark~7 of~\cite{tygert_full}).

\subsection{KDD Cup 1998}
\label{kddcup98}

\begin{figure}
\begin{centering}

(a) \parbox{\imsize}{\includegraphics[width=\imsize]
{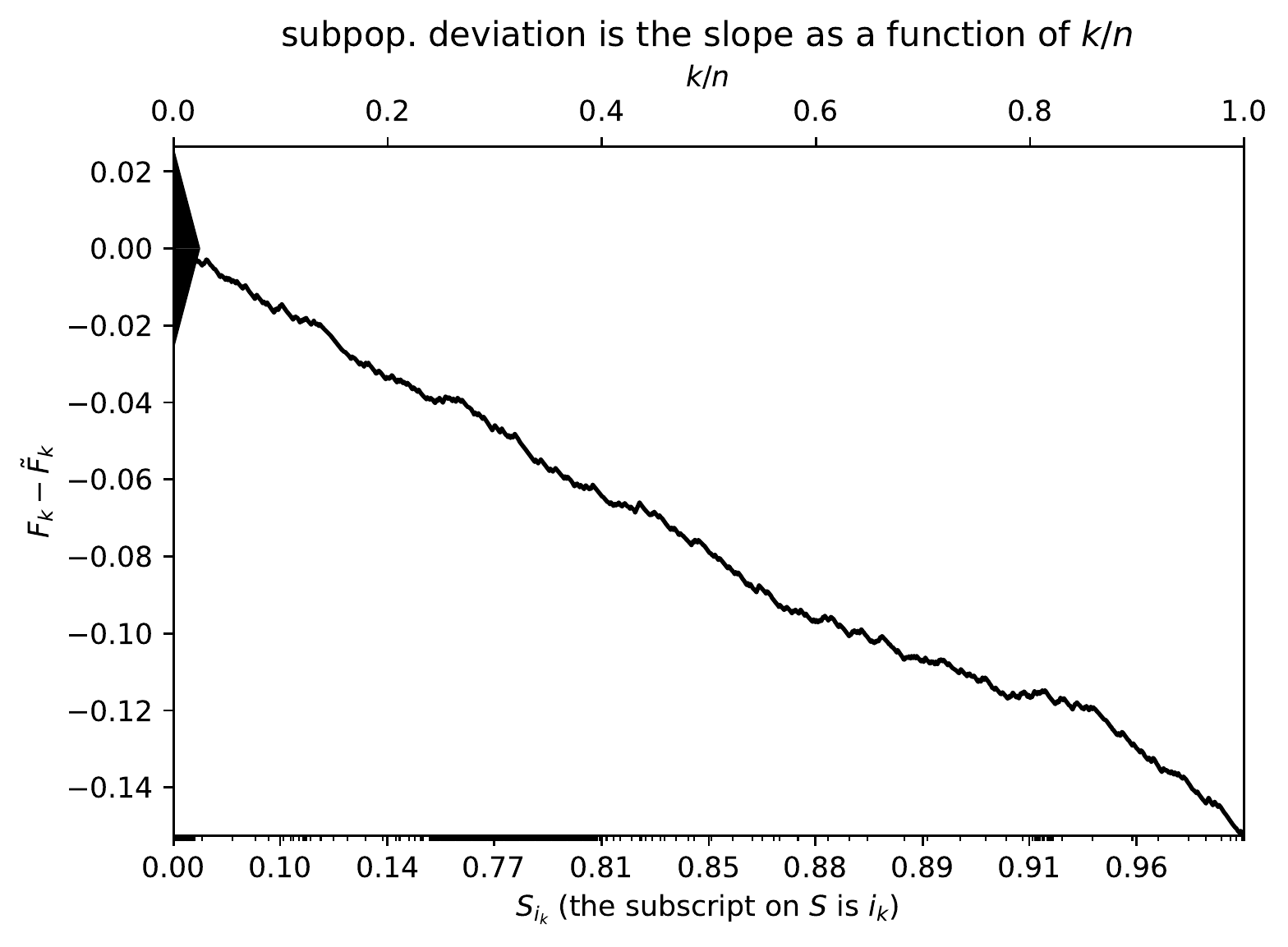}}
\quad\quad
(b) \parbox{\imsize}{\includegraphics[width=\imsize]
{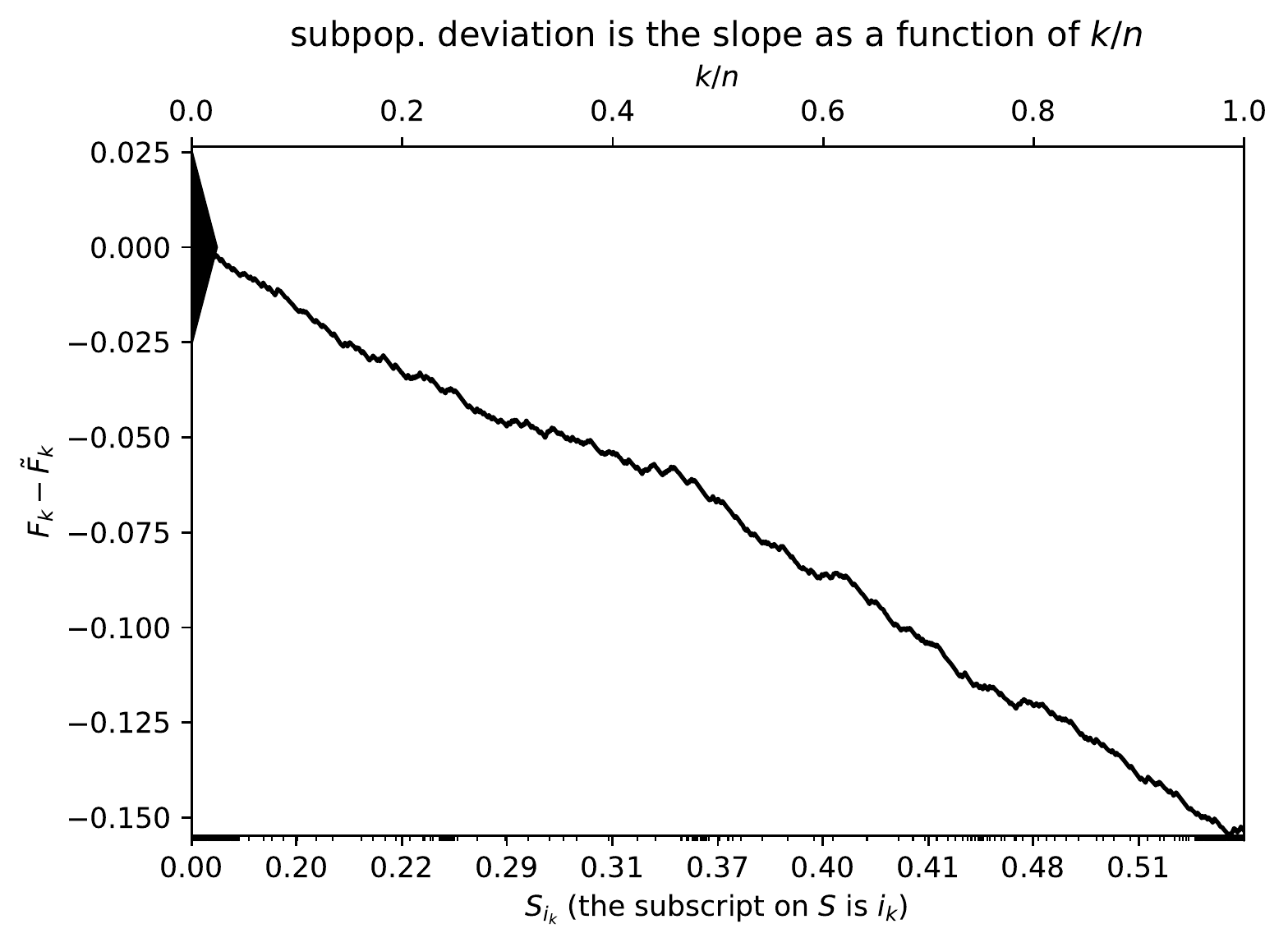}}

\parbox{\imsized}{\hfil \footnotesize $G$ = 0.1526; $H$ = 0.1526;
$G/\sigma$ = 11.54; $H/\sigma$ = 11.54}
\parbox{\imsized}{\hfil \footnotesize $G$ = 0.1547; $H$ = 0.1554;
$G/\sigma$ = 11.71; $H/\sigma$ = 11.76}

\vspace{\vertsep}

(c) \parbox{\imsize}{\includegraphics[width=\imsize]
{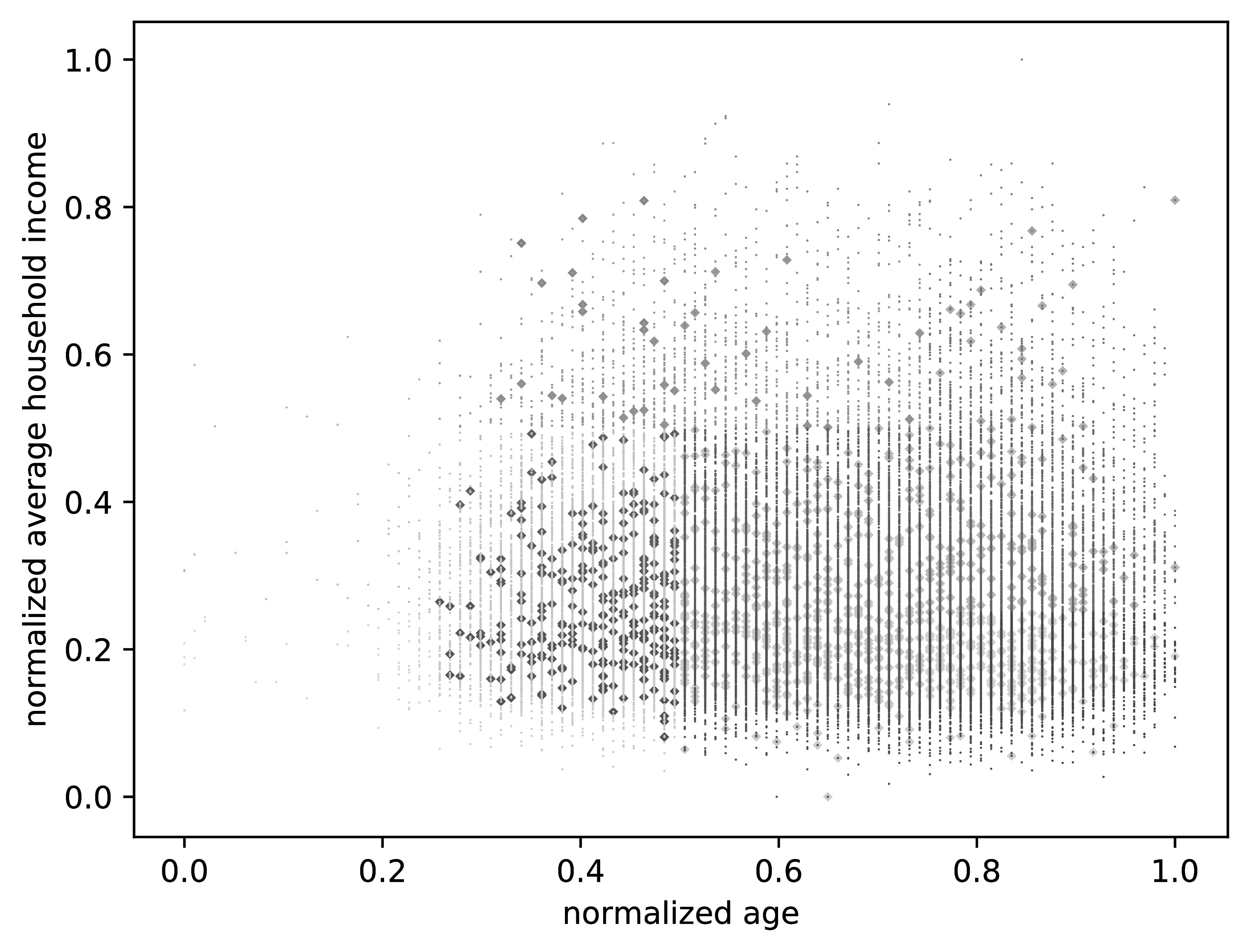}}
\quad\quad
(d) \parbox{\imsize}{\includegraphics[width=\imsize]
{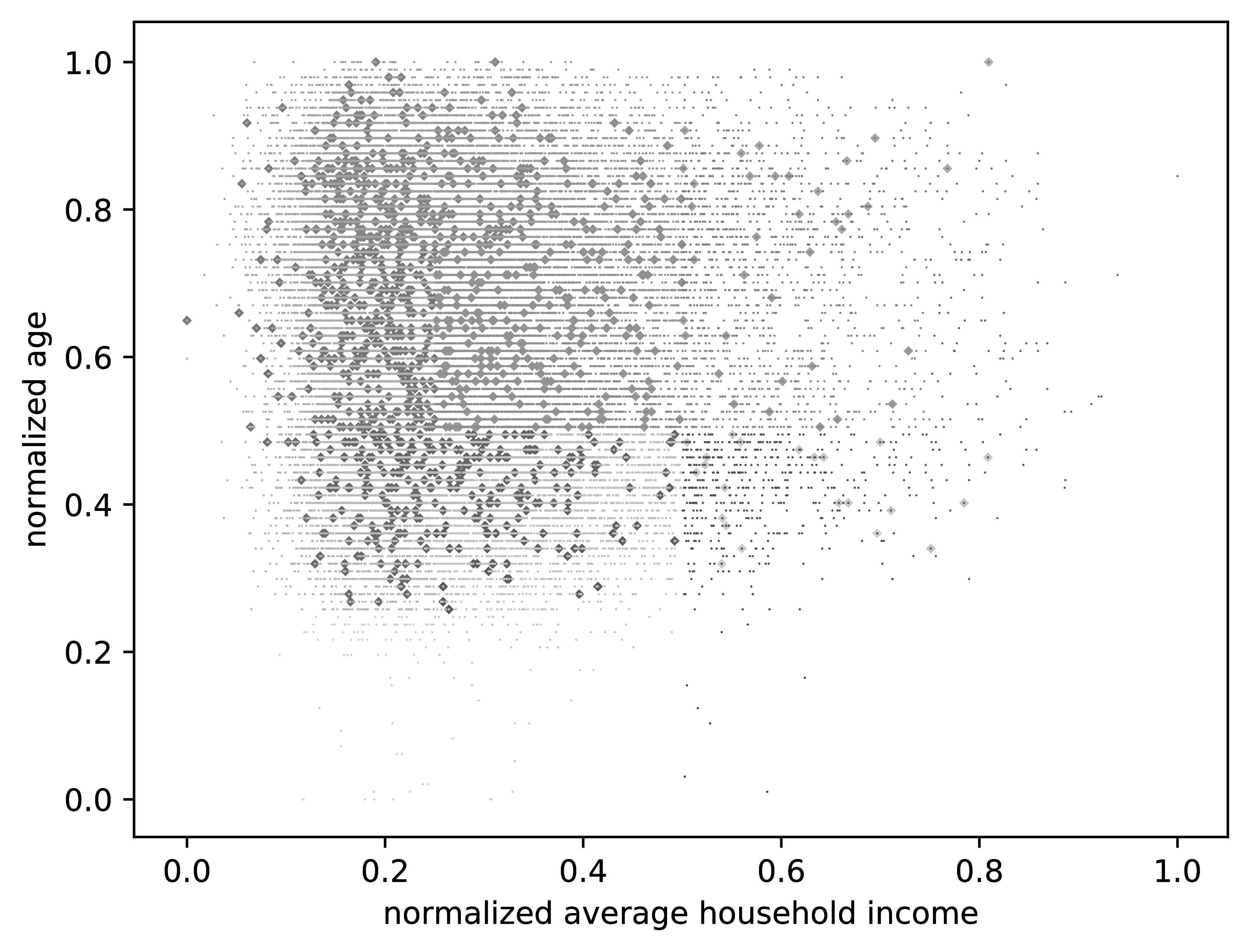}}

\vspace{\vertsep}

(e) \parbox{\imsize}{\includegraphics[width=\imsize]
{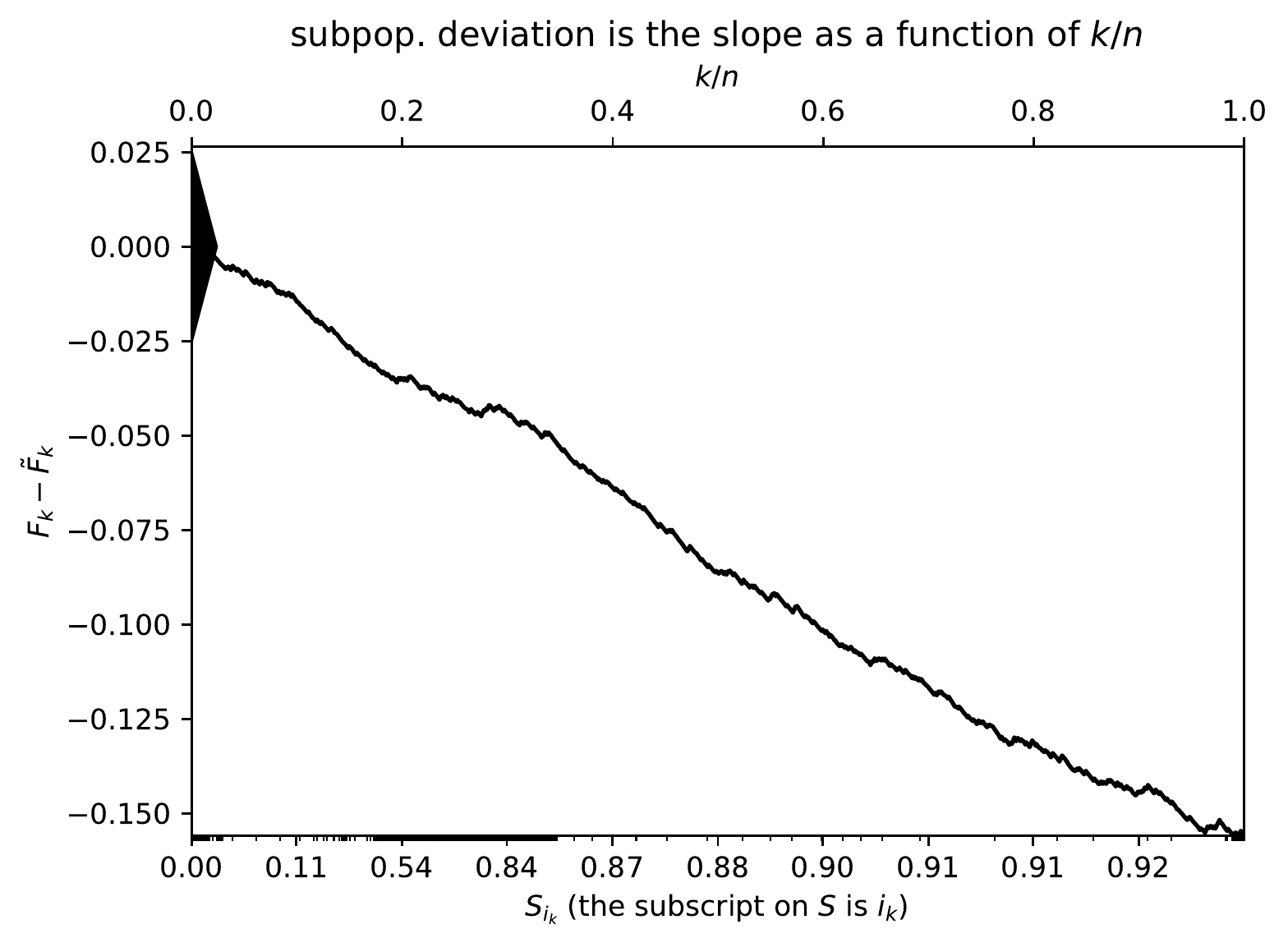}}
\quad\quad
(f) \parbox{\imsize}{\includegraphics[width=\imsize]
{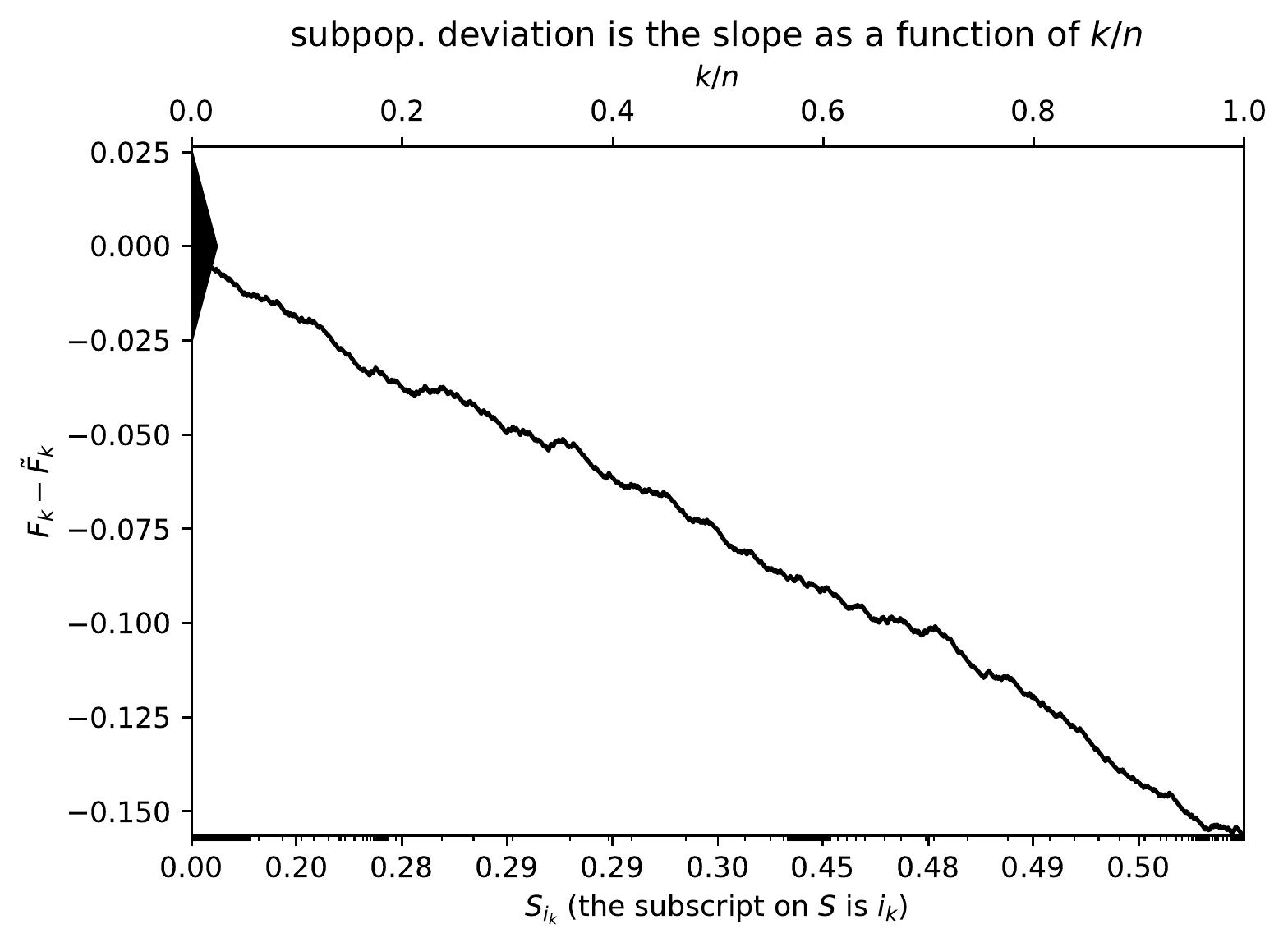}}

\parbox{\imsized}{\hfil \footnotesize $G$ = 0.1559; $H$ = 0.1559;
$G/\sigma$ = 11.80; $H/\sigma$ = 11.80}
\parbox{\imsized}{\hfil \footnotesize $G$ = 0.1565; $H$ = 0.1565;
$G/\sigma$ = 11.85; $H/\sigma$ = 11.85}

\vspace{\vertsep}

(g) \parbox{\imsize}{\includegraphics[width=\imsize]
{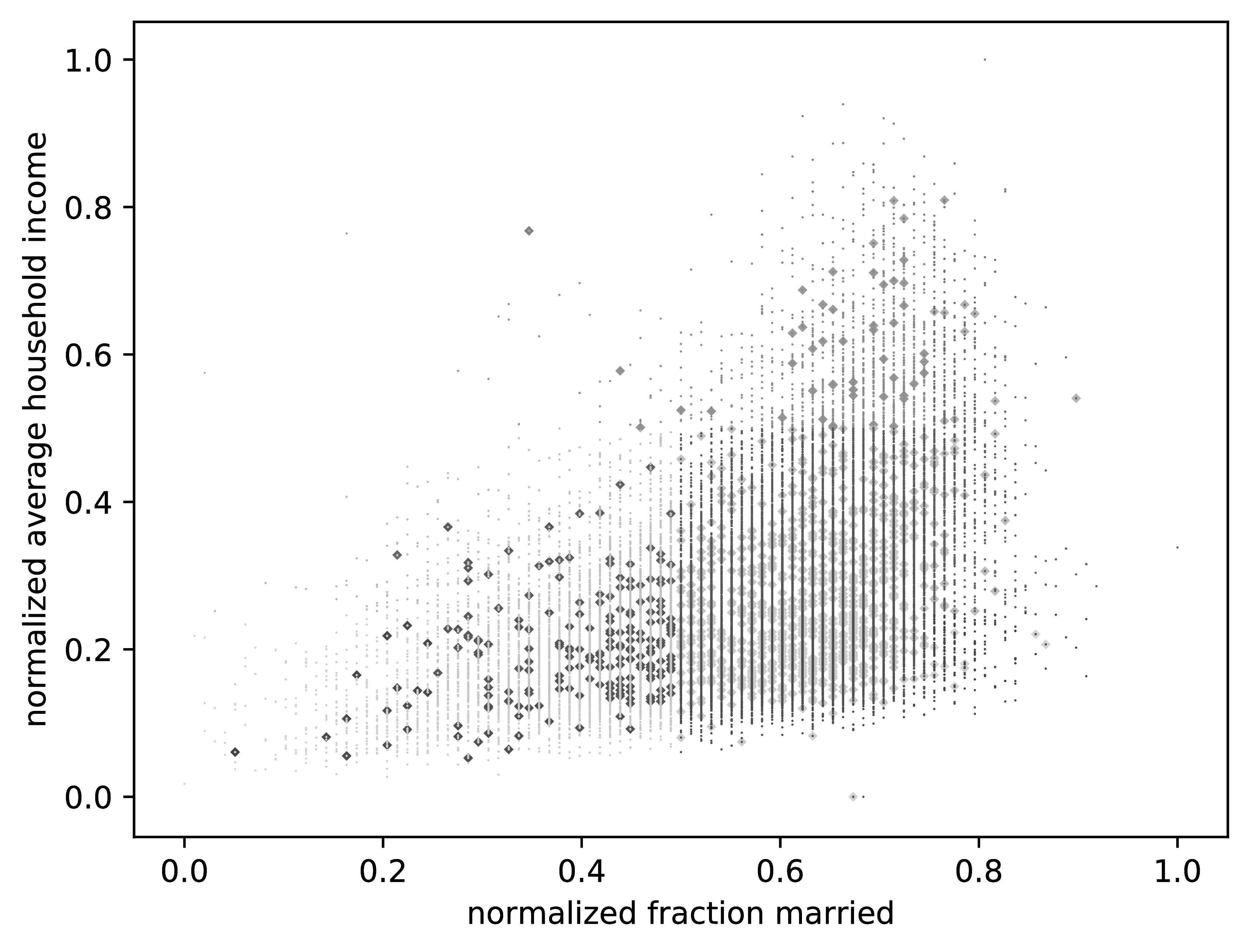}}
\quad\quad
(h) \parbox{\imsize}{\includegraphics[width=\imsize]
{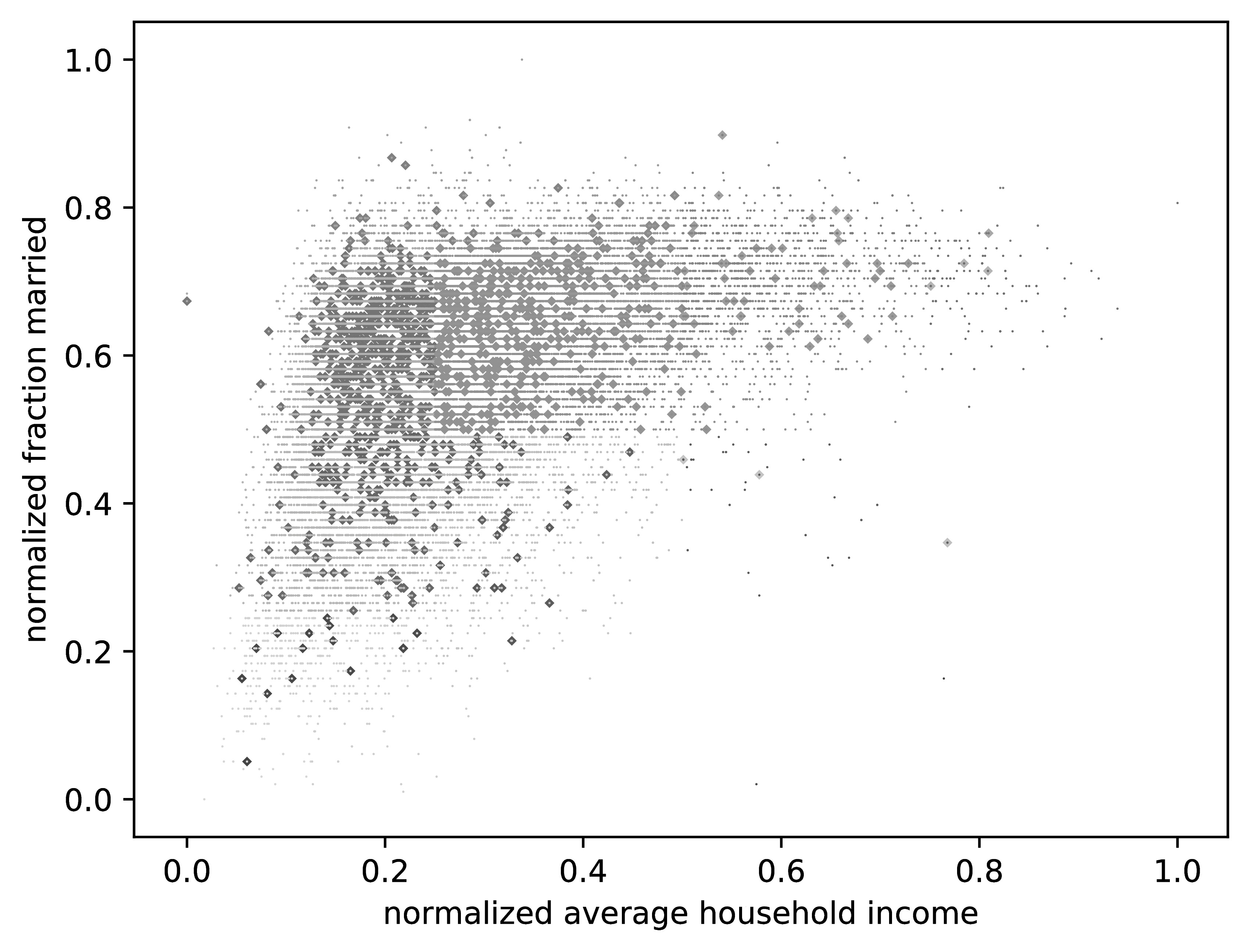}}

\end{centering}
\caption{Responses from folding cards versus the full population
($m =$ 47,117; $n =$ 1,236)}
\label{folding}
\end{figure}

\begin{figure}
\begin{centering}

(a) \parbox{\imsize}{\includegraphics[width=\imsize]
{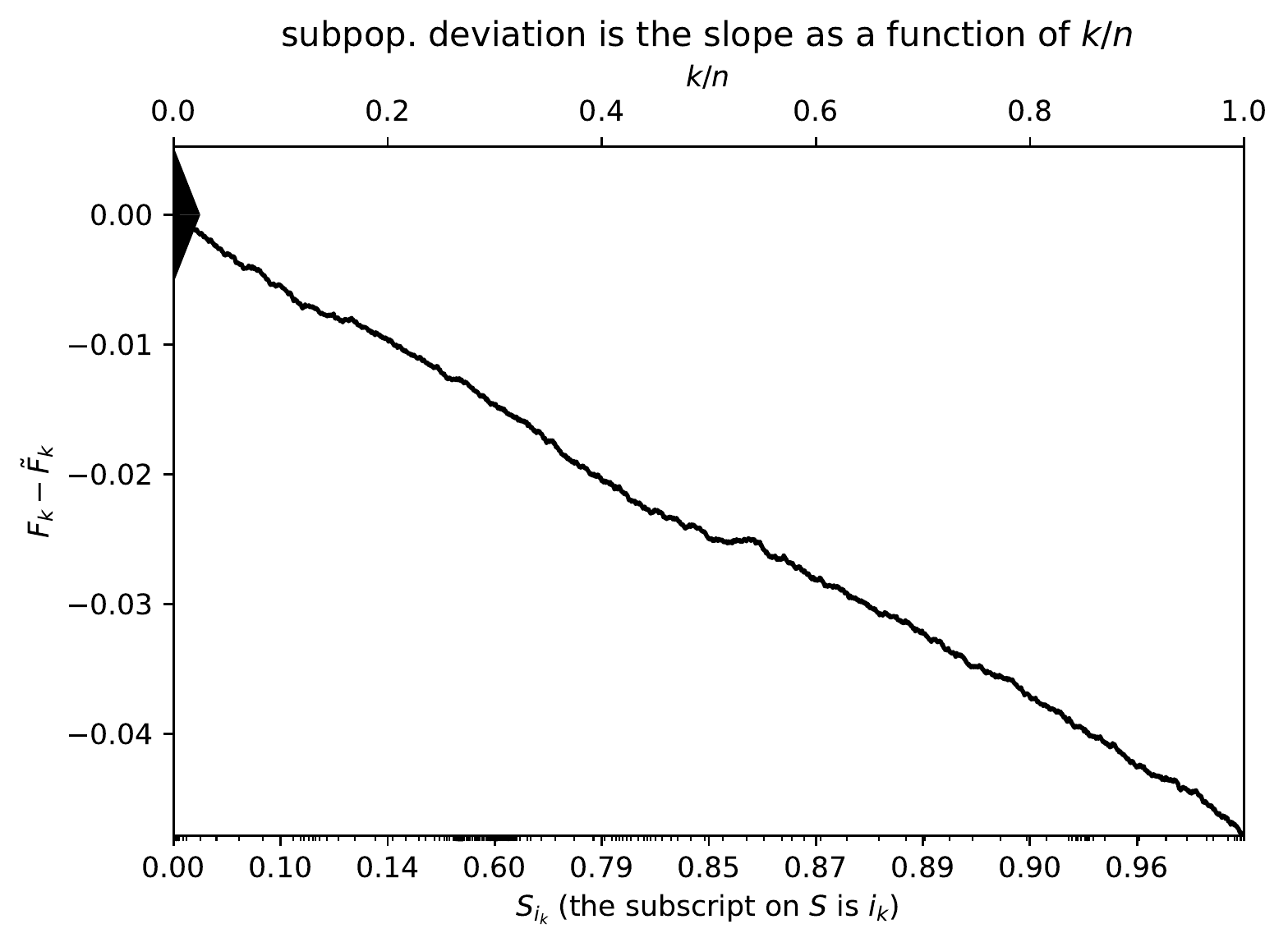}}
\quad\quad
(b) \parbox{\imsize}{\includegraphics[width=\imsize]
{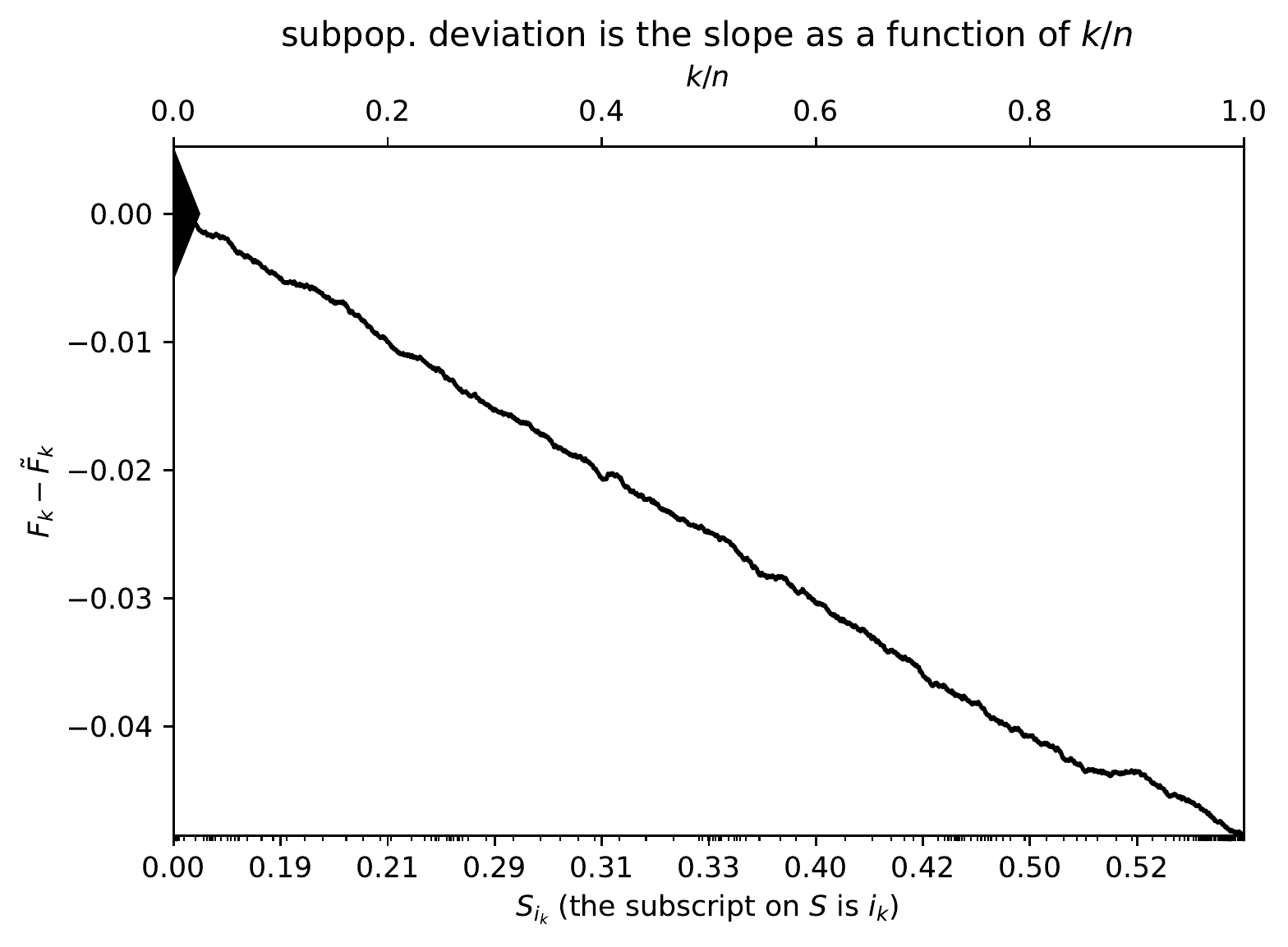}}

\parbox{\imsized}{\hfil \footnotesize $G$ = 0.04787; $H$ = 0.04787;
$G/\sigma$ = 18.27; $H/\sigma$ = 18.27}
\parbox{\imsized}{\hfil \footnotesize $G$ = 0.04850; $H$ = 0.04852;
$G/\sigma$ = 18.59; $H/\sigma$ = 18.60}

\vspace{\vertsep}

(c) \parbox{\imsize}{\includegraphics[width=\imsize]
{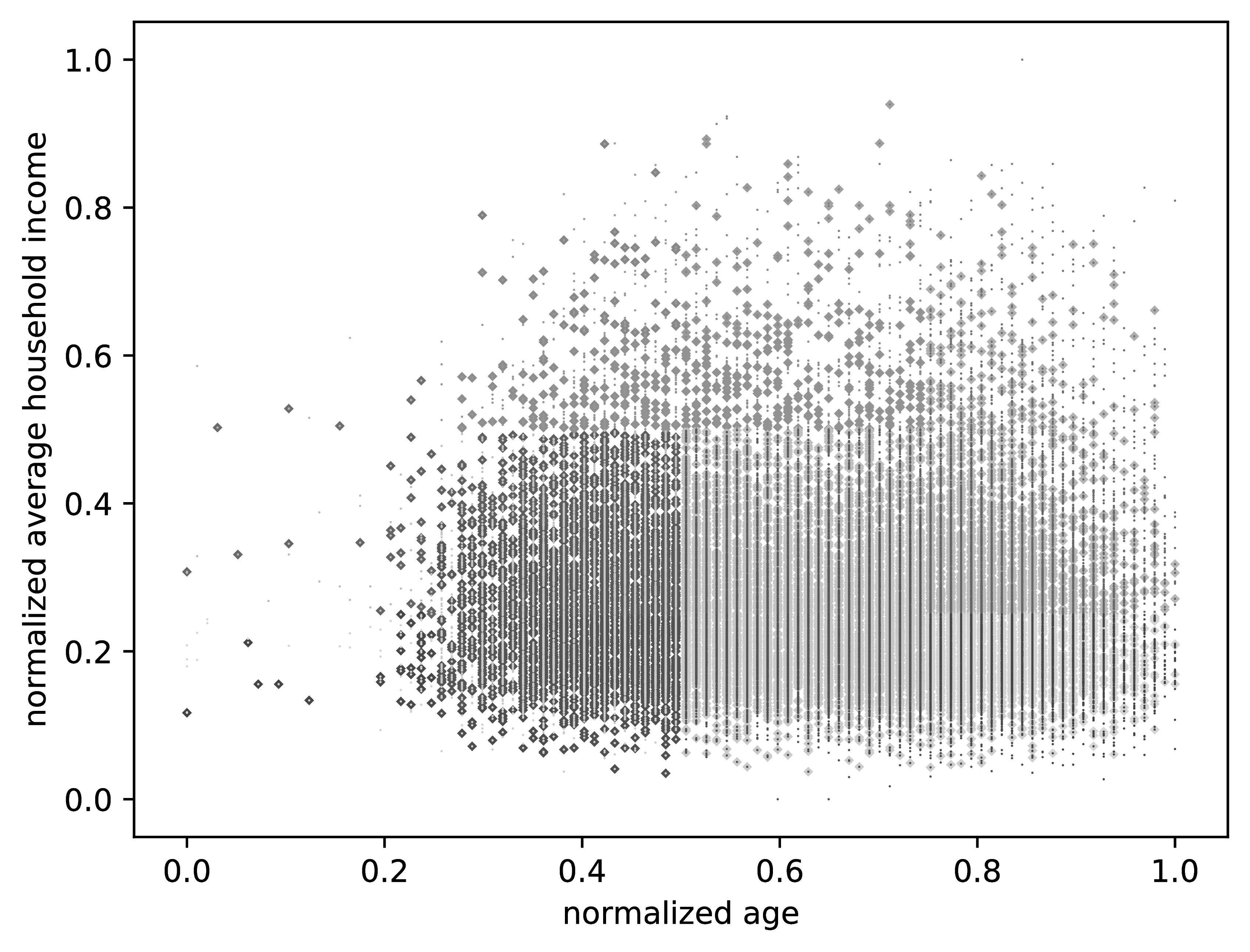}}
\quad\quad
(d) \parbox{\imsize}{\includegraphics[width=\imsize]
{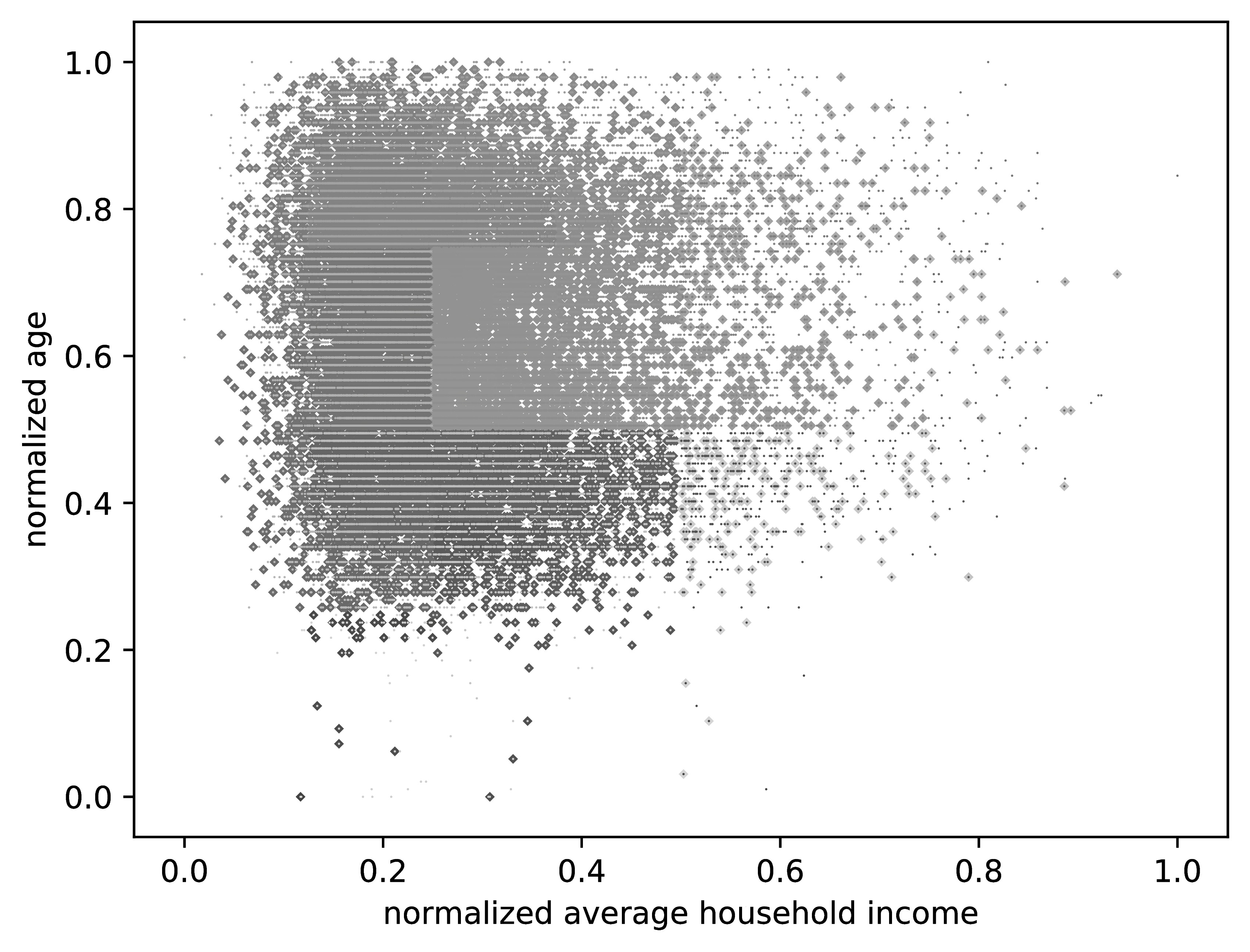}}

\vspace{\vertsep}

(e) \parbox{\imsize}{\includegraphics[width=\imsize]
{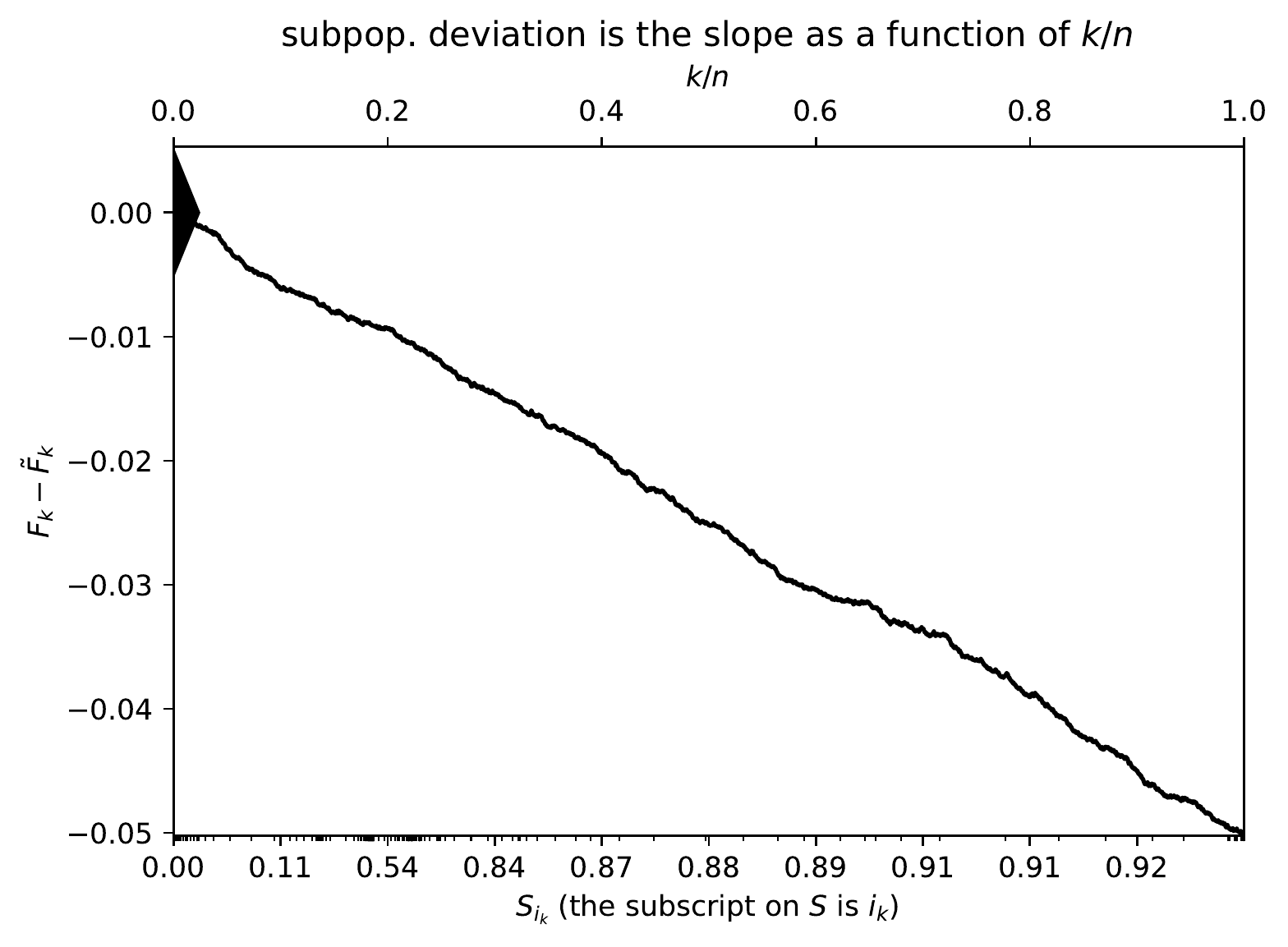}}
\quad\quad
(f) \parbox{\imsize}{\includegraphics[width=\imsize]
{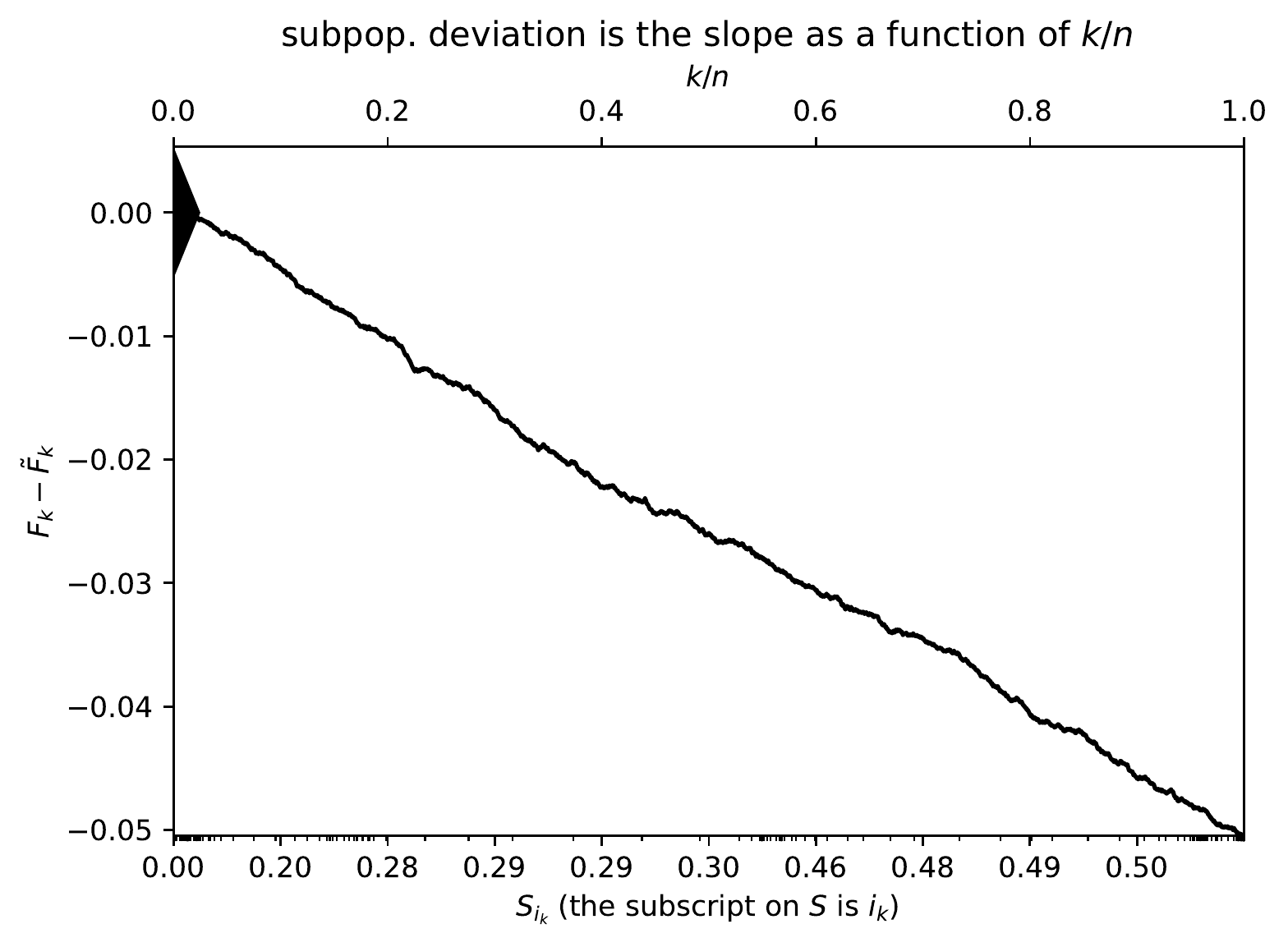}}

\parbox{\imsized}{\hfil \footnotesize $G$ = 0.05023; $H$ = 0.05023;
$G/\sigma$ = 18.92; $H/\sigma$ = 18.92}
\parbox{\imsized}{\hfil \footnotesize $G$ = 0.05046; $H$ = 0.05052;
$G/\sigma$ = 18.93; $H/\sigma$ = 18.95}

\vspace{\vertsep}

(g) \parbox{\imsize}{\includegraphics[width=\imsize]
{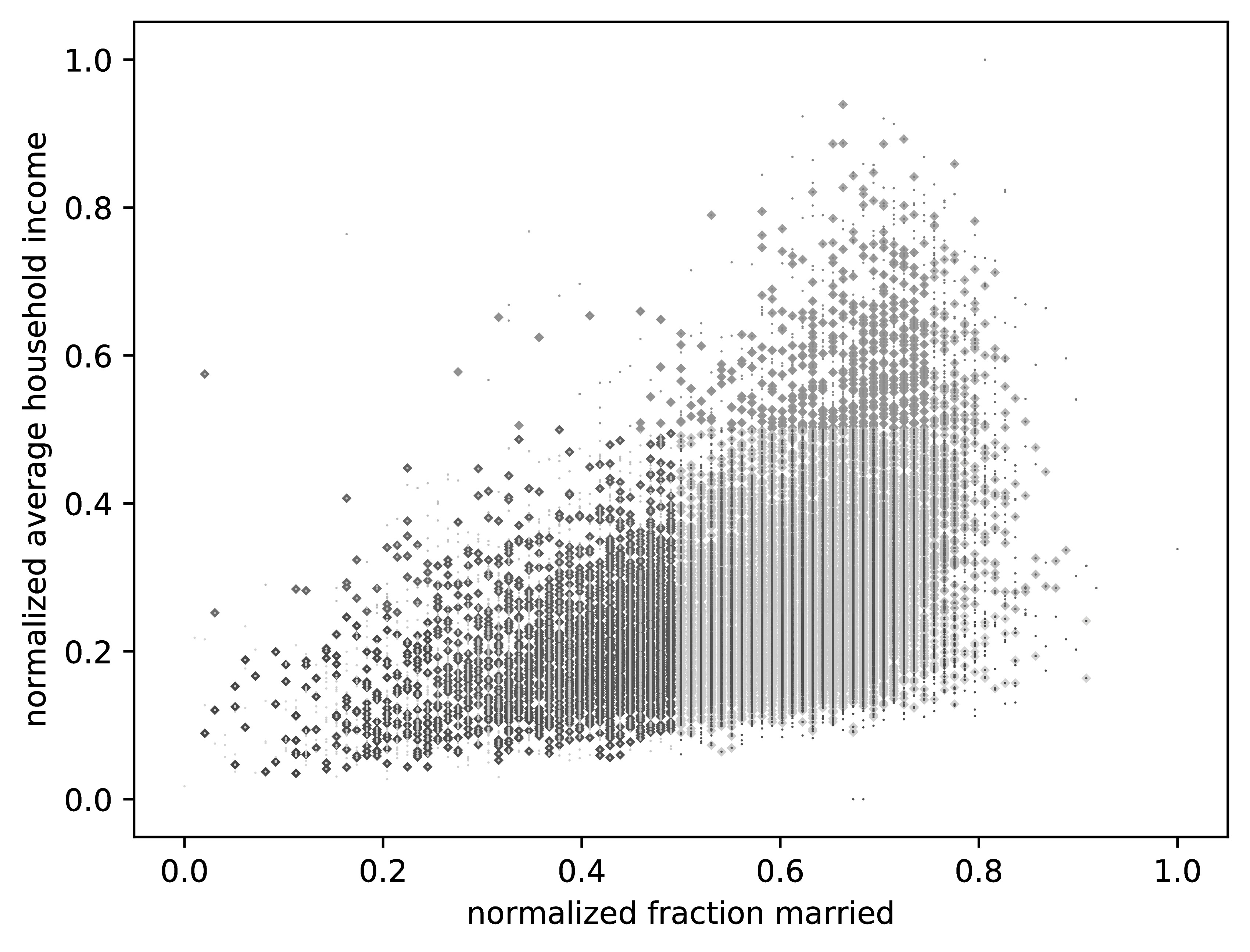}}
\quad\quad
(h) \parbox{\imsize}{\includegraphics[width=\imsize]
{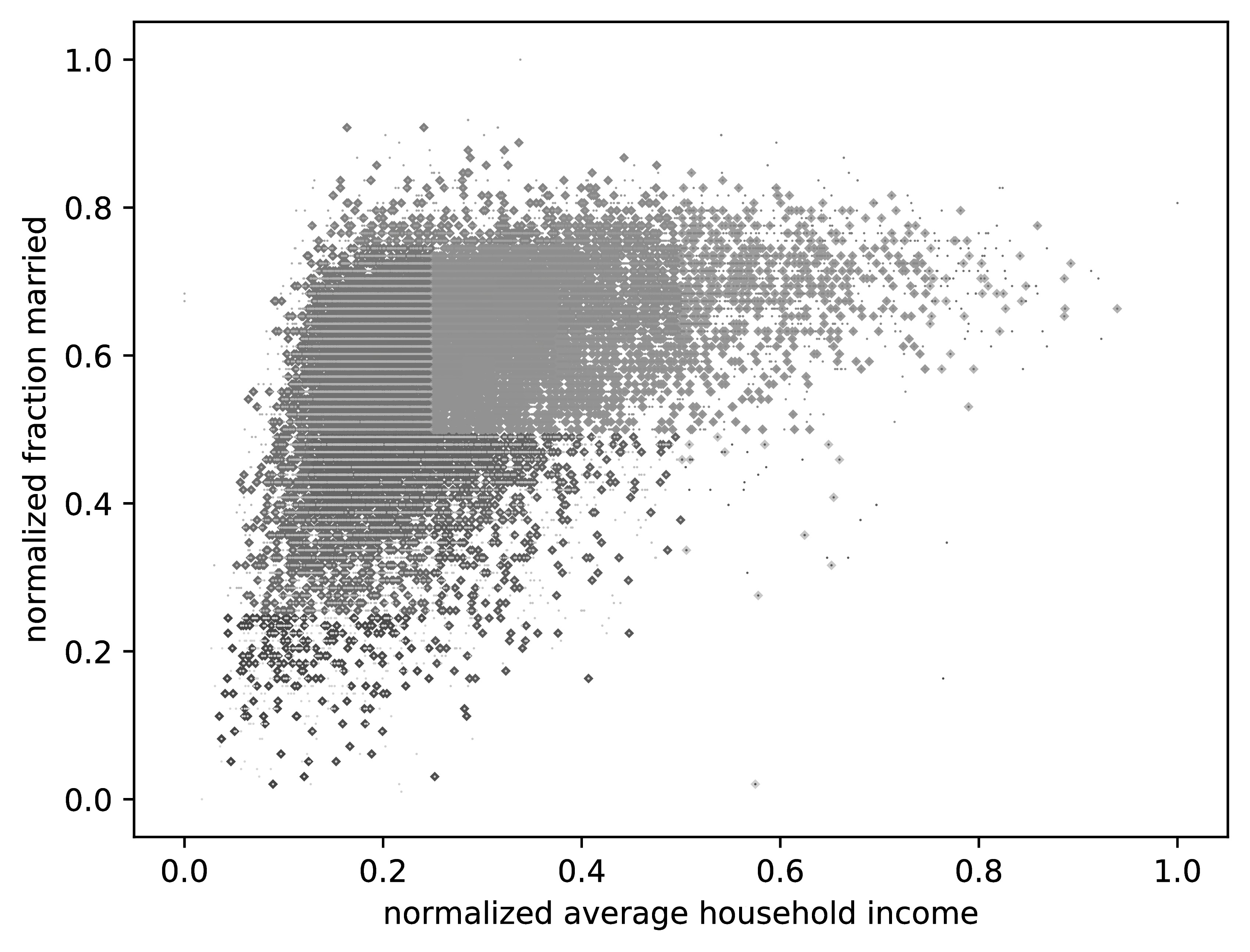}}

\end{centering}
\caption{Responses from normal cards versus the full population
($m =$ 47,117; $n =$ 15,866)}
\label{normal}
\end{figure}

\begin{figure}
\begin{centering}

(a) \parbox{\imsize}{\includegraphics[width=\imsize]
{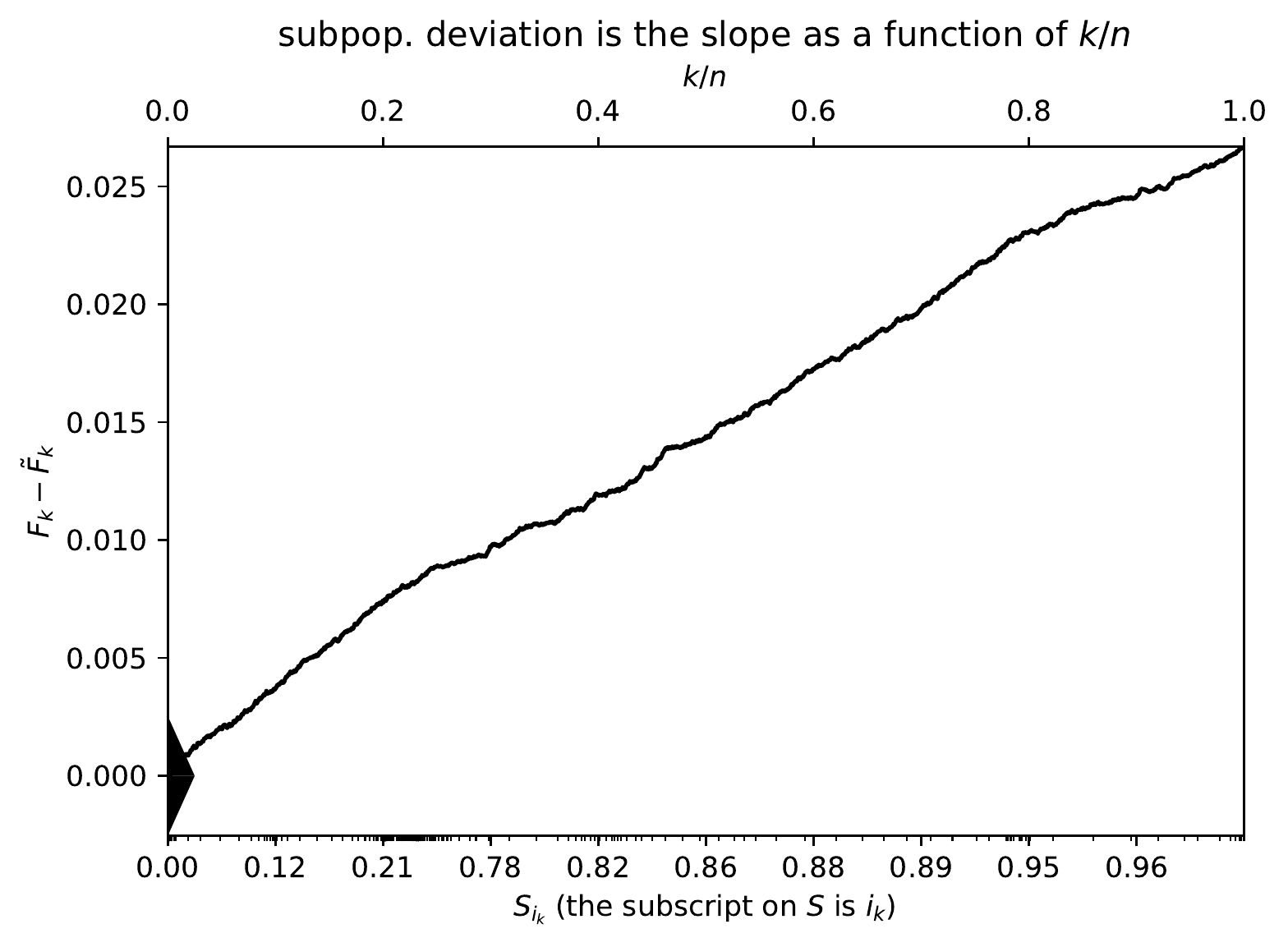}}
\quad\quad
(b) \parbox{\imsize}{\includegraphics[width=\imsize]
{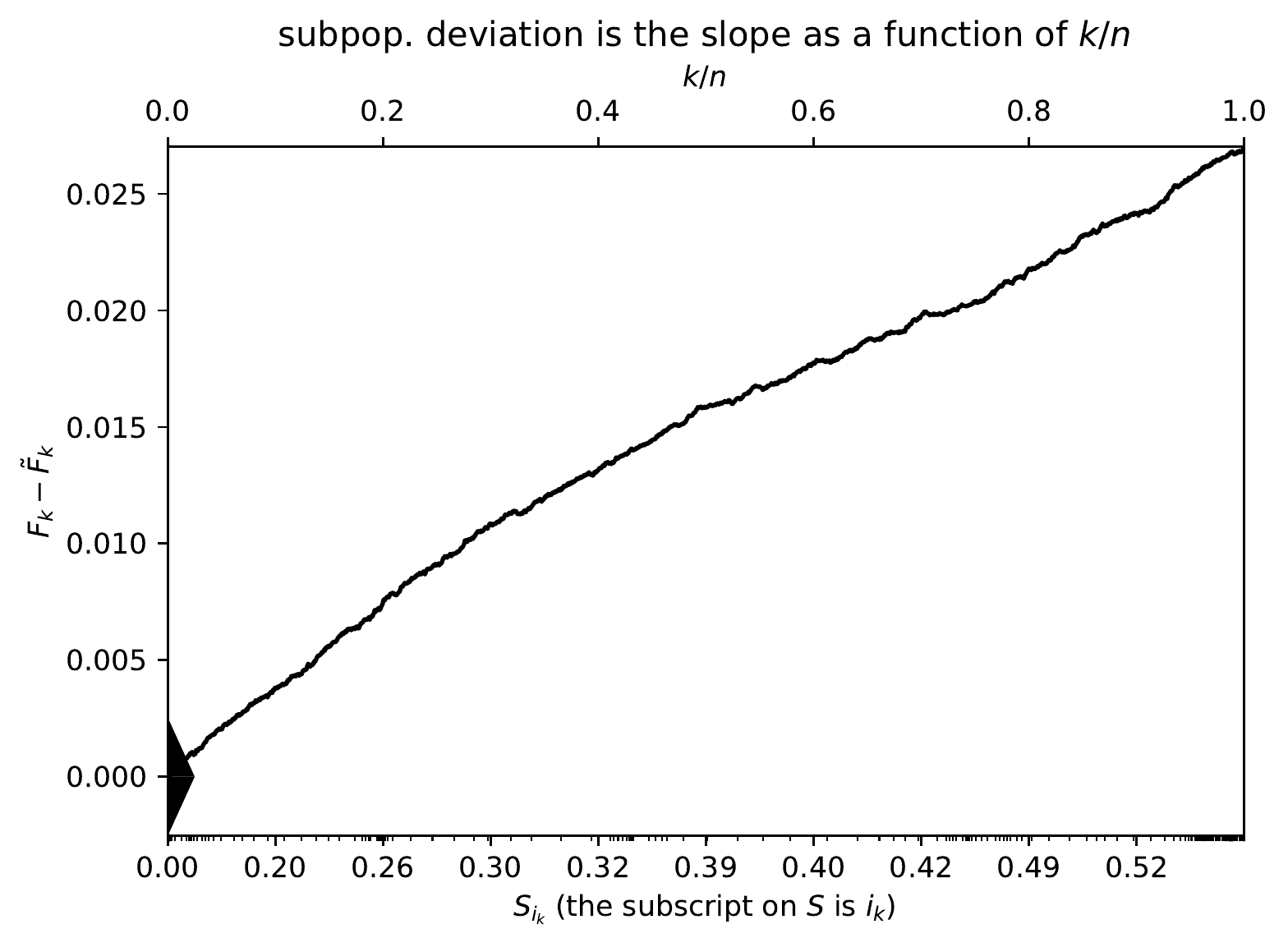}}

\parbox{\imsized}{\hfil \footnotesize $G$ = 0.02668; $H$ = 0.02668;
$G/\sigma$ = 21.01; $H/\sigma$ = 21.01}
\parbox{\imsized}{\hfil \footnotesize $G$ = 0.02702; $H$ = 0.02703;
$G/\sigma$ = 21.32; $H/\sigma$ = 21.33}

\vspace{\vertsep}

(c) \parbox{\imsize}{\includegraphics[width=\imsize]
{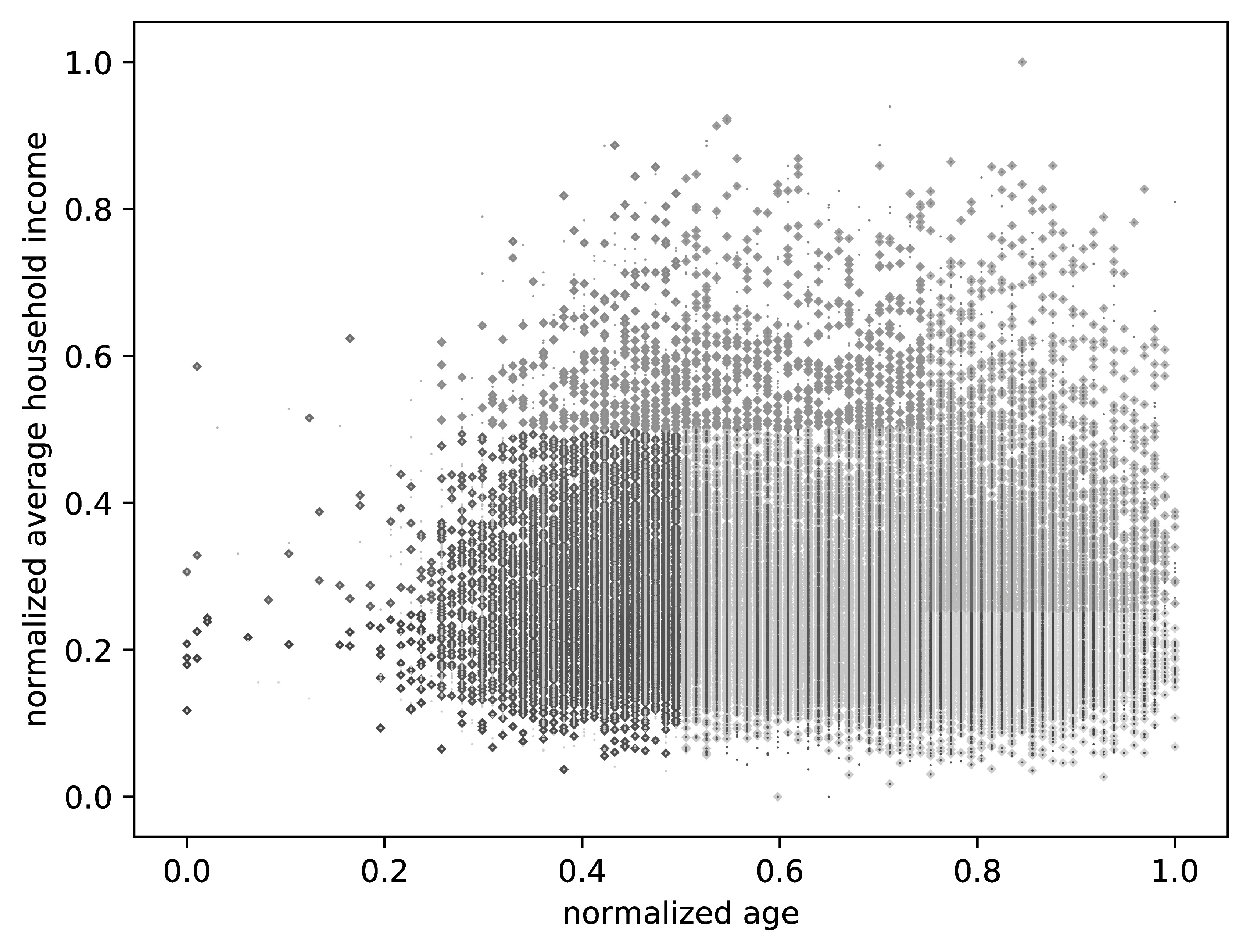}}
\quad\quad
(d) \parbox{\imsize}{\includegraphics[width=\imsize]
{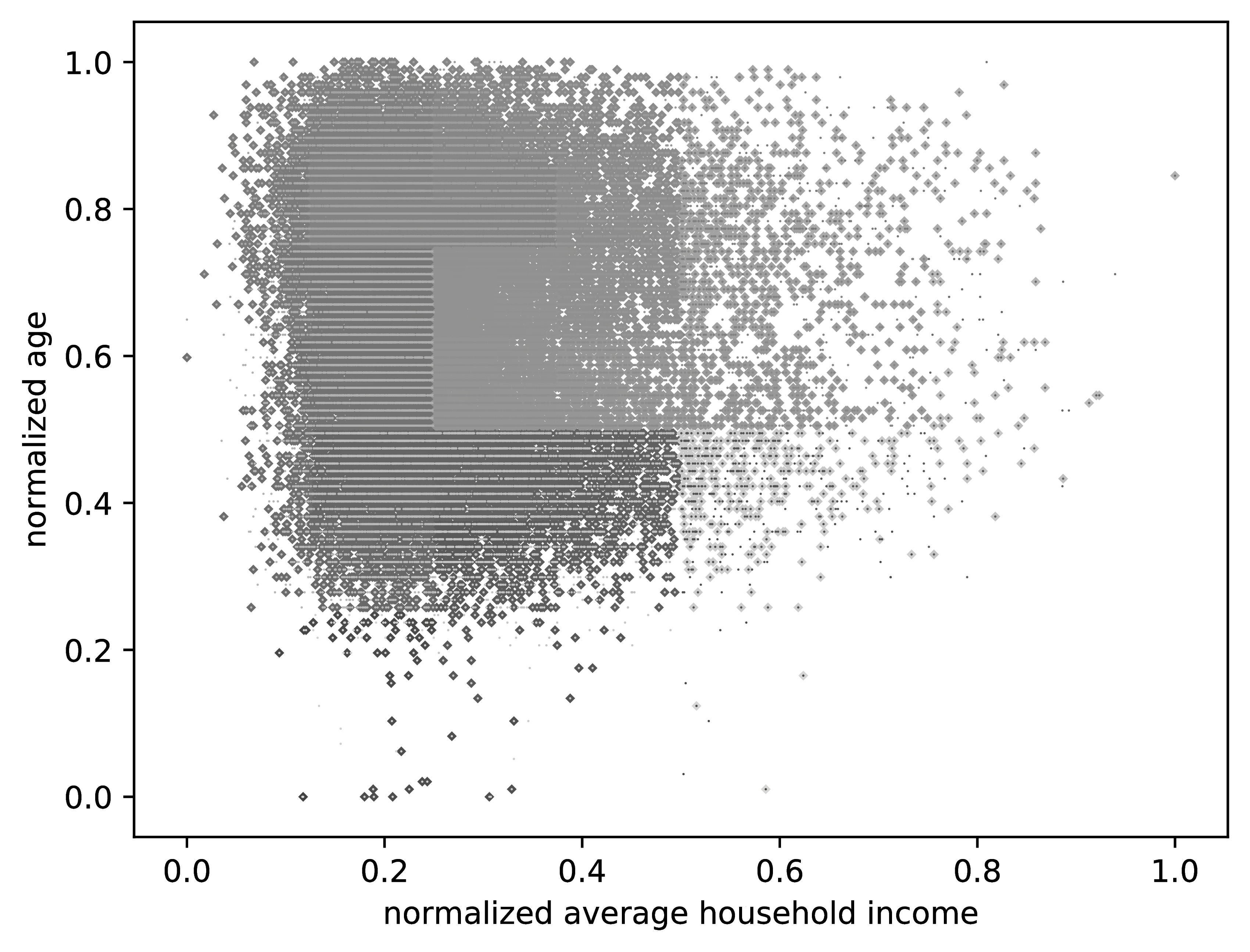}}

\vspace{\vertsep}

(e) \parbox{\imsize}{\includegraphics[width=\imsize]
{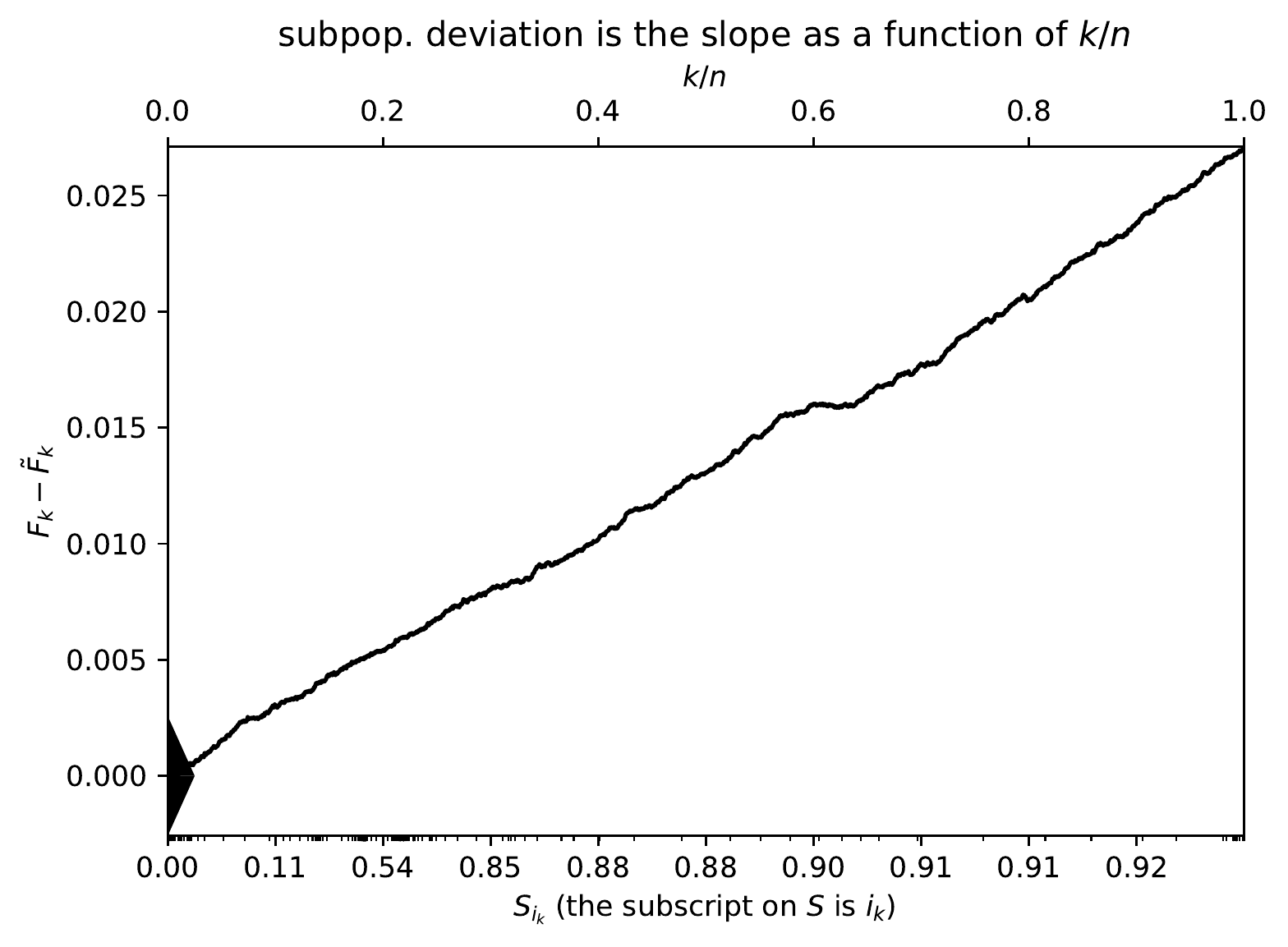}}
\quad\quad
(f) \parbox{\imsize}{\includegraphics[width=\imsize]
{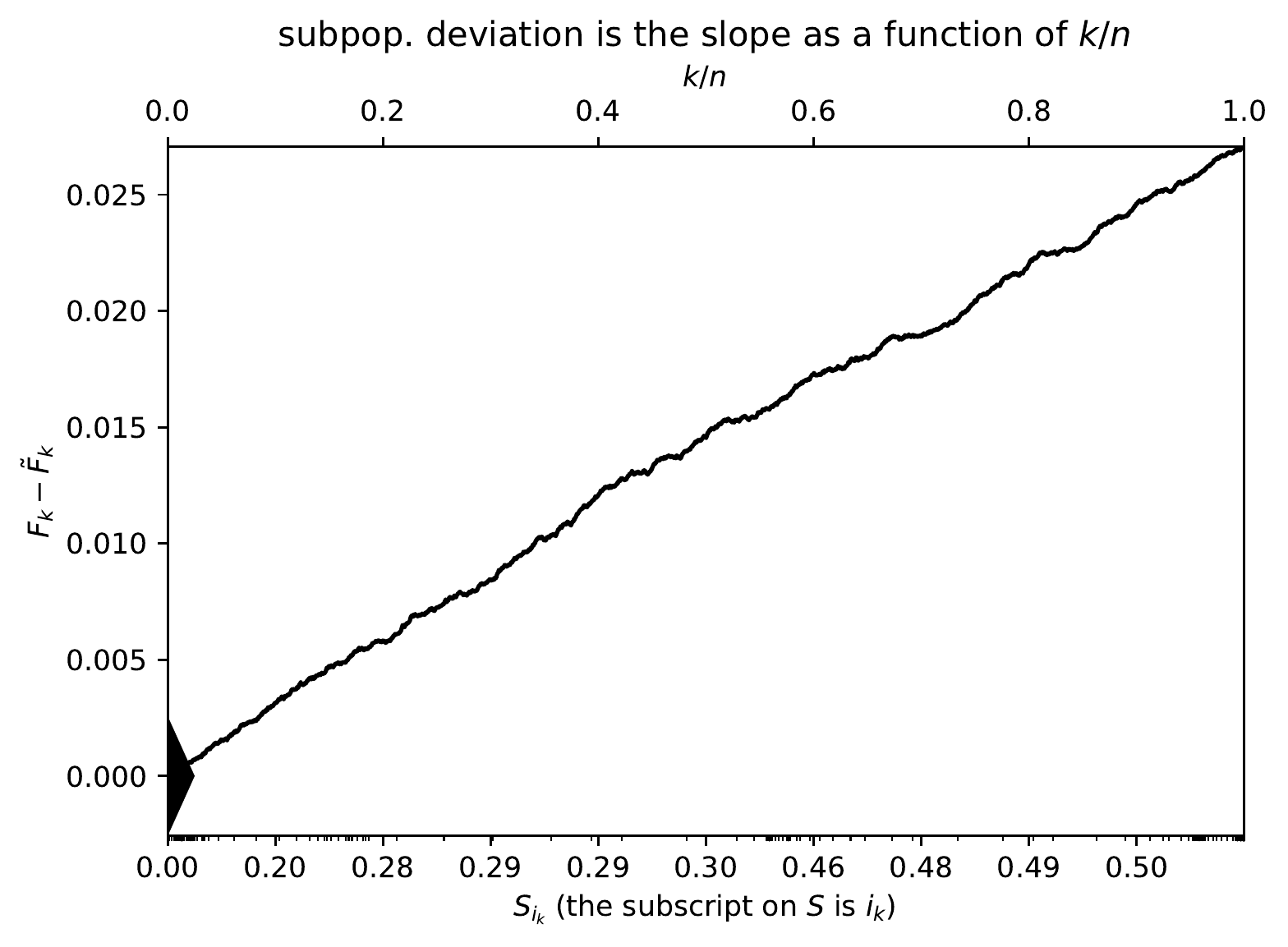}}

\parbox{\imsized}{\hfil \footnotesize $G$ = 0.02710; $H$ = 0.02714;
$G/\sigma$ = 21.02; $H/\sigma$ = 21.05}
\parbox{\imsized}{\hfil \footnotesize $G$ = 0.02706; $H$ = 0.02706;
$G/\sigma$ = 21.07; $H/\sigma$ = 21.07}

\vspace{\vertsep}

(g) \parbox{\imsize}{\includegraphics[width=\imsize]
{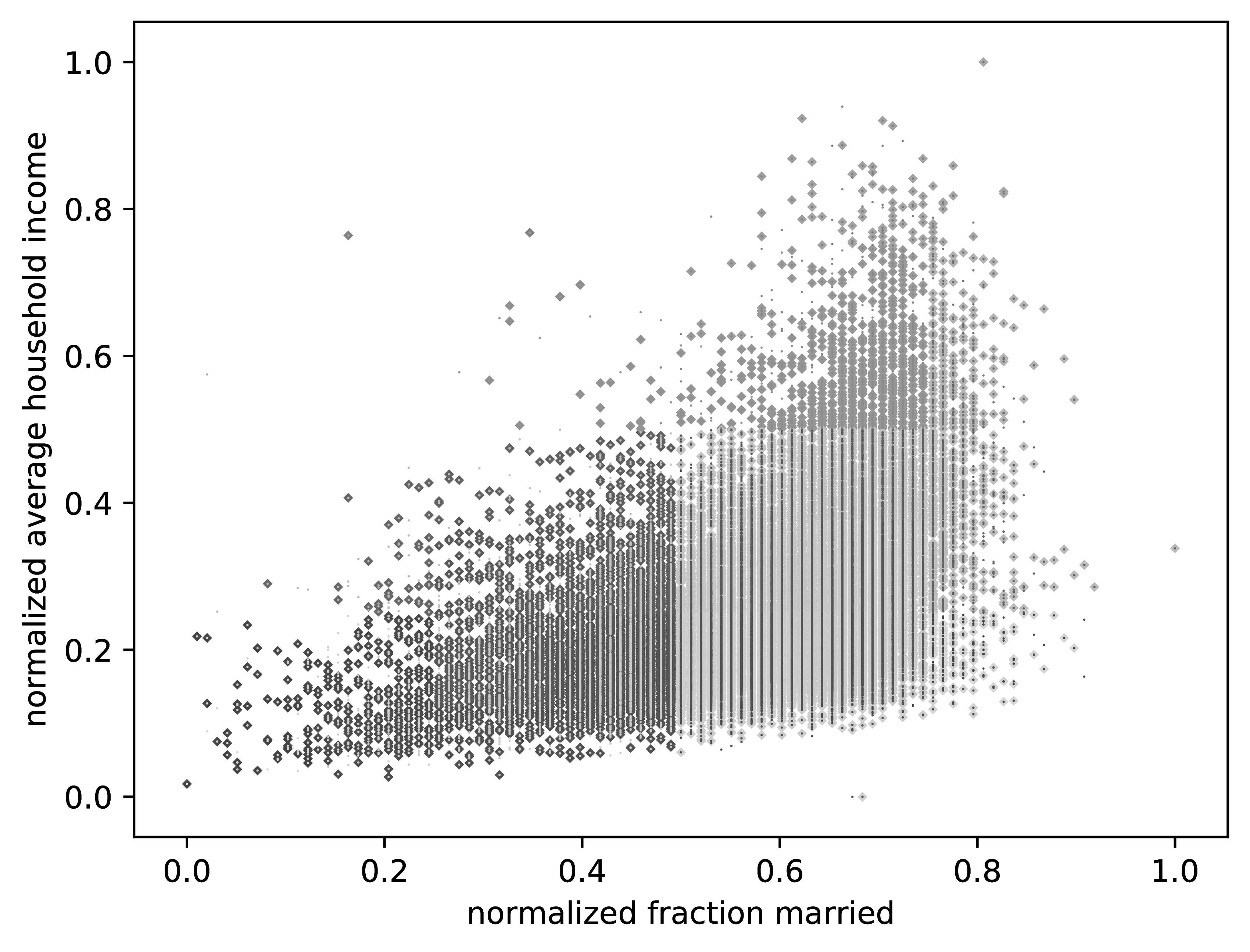}}
\quad\quad
(h) \parbox{\imsize}{\includegraphics[width=\imsize]
{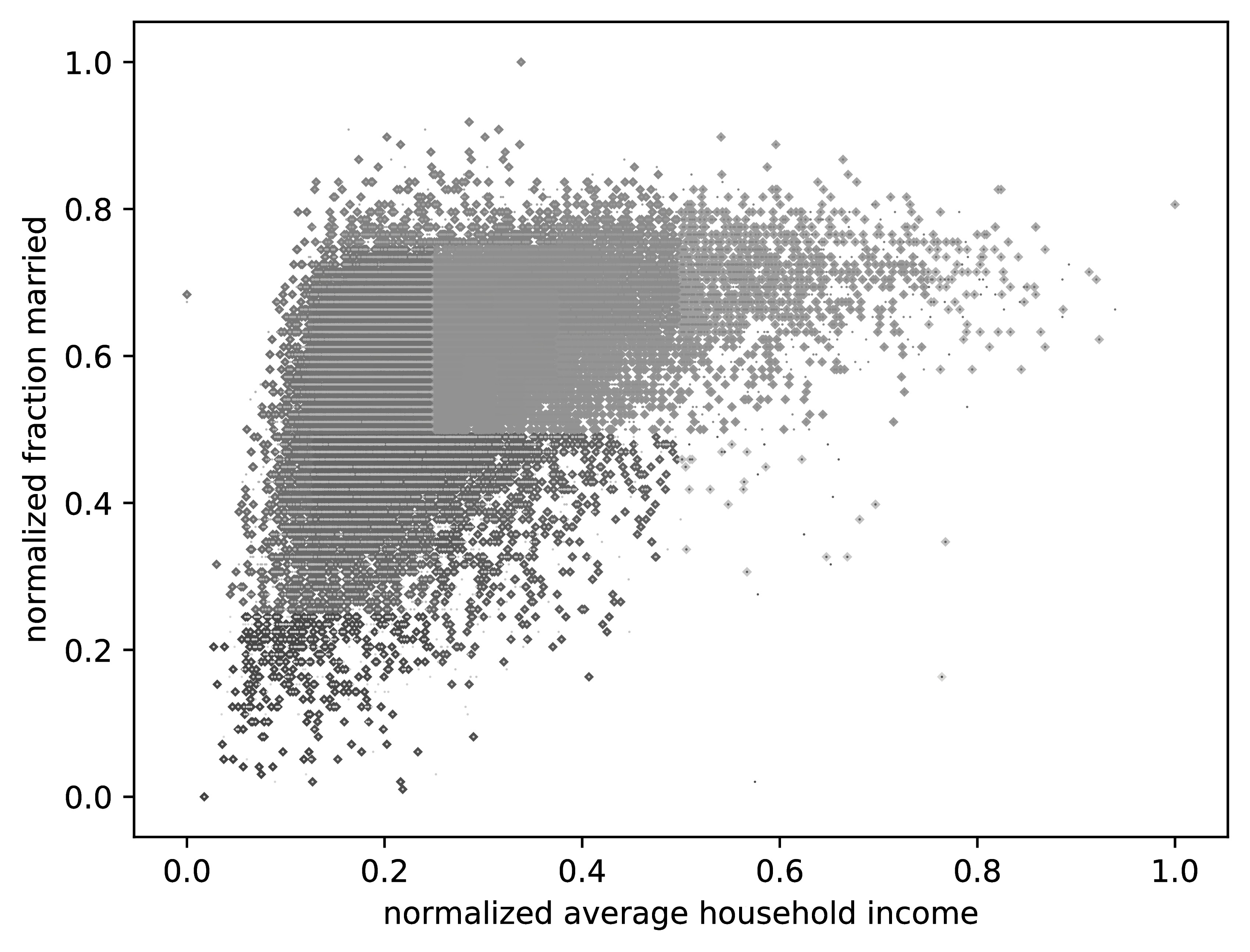}}

\end{centering}
\caption{Responses from those sent both types versus the full population
($m =$ 47,117; $n =$ 30,015)}
\label{both}
\end{figure}

\begin{figure}
\begin{centering}

(a) \parbox{\imsize}{\includegraphics[width=\imsize]
{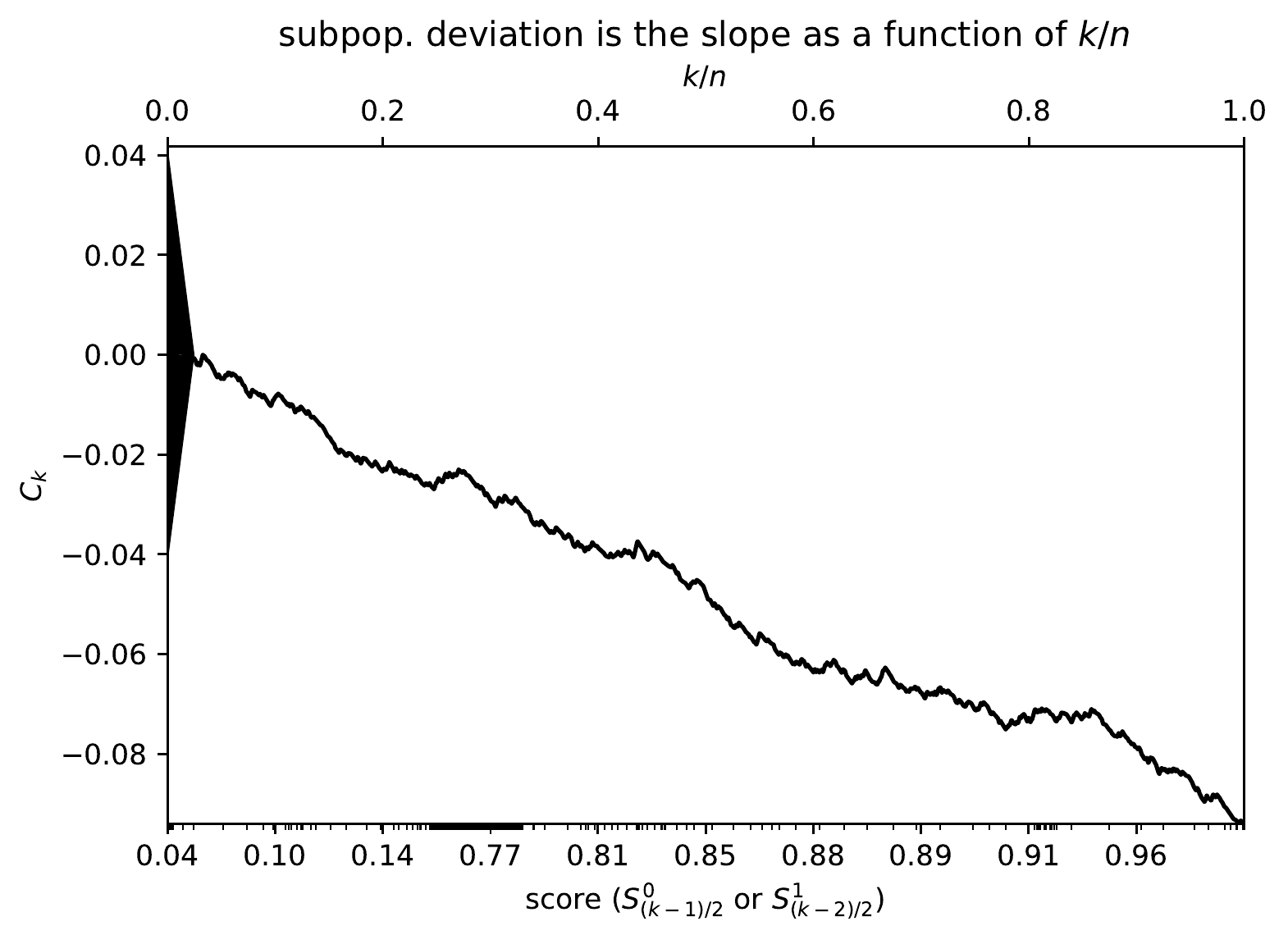}}
\quad\quad
(b) \parbox{\imsize}{\includegraphics[width=\imsize]
{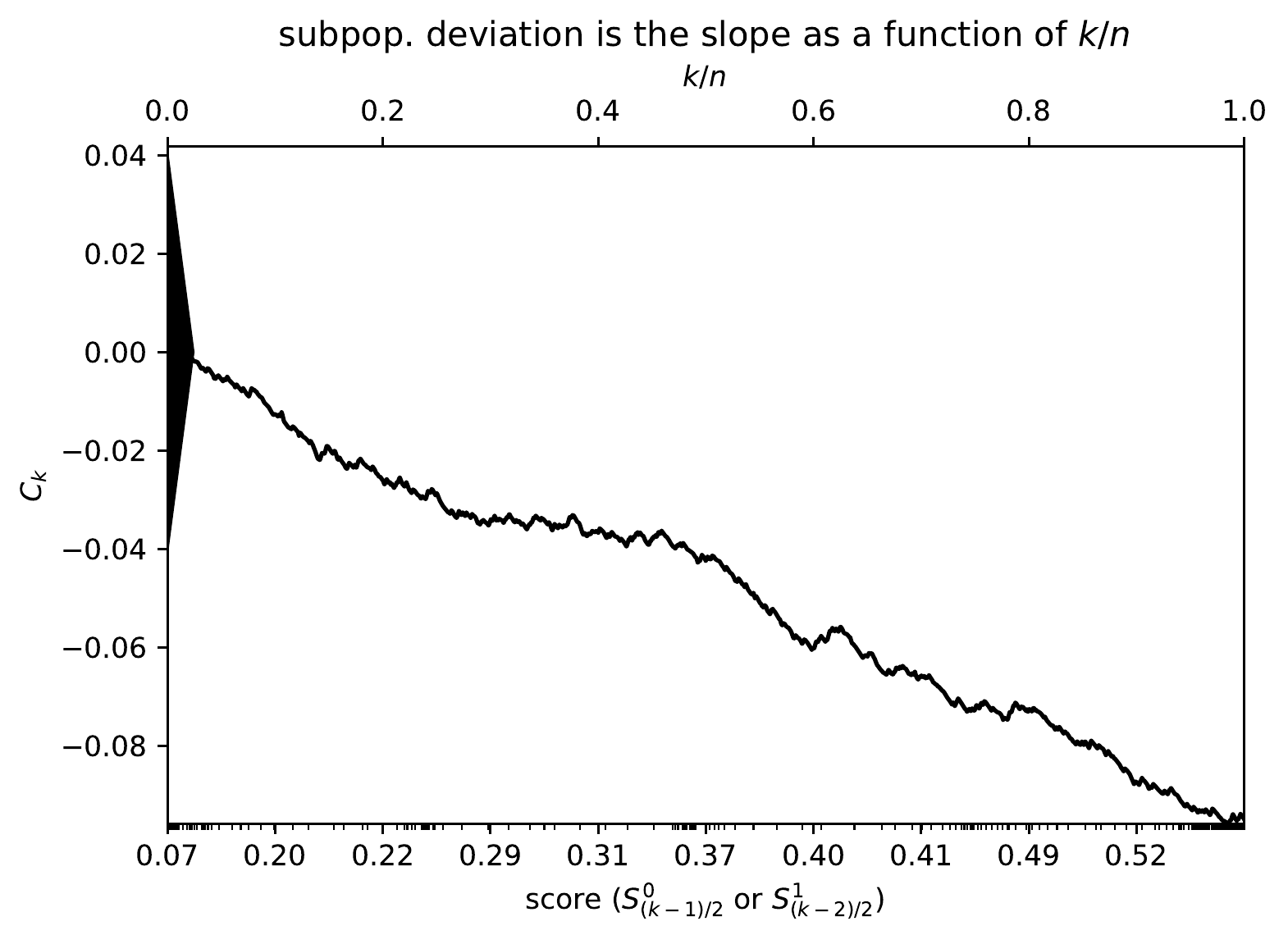}}

\parbox{\imsized}{\hfil \footnotesize $G$ = 0.09407; $H$ = 0.09442;
$G/\sigma$ = 4.509; $H/\sigma$ = 4.525}
\parbox{\imsized}{\hfil \footnotesize $G$ = 0.09591; $H$ = 0.09694;
$G/\sigma$ = 4.595; $H/\sigma$ = 4.644}

\vspace{\vertsep}

(c) \parbox{\imsize}{\includegraphics[width=\imsize]
{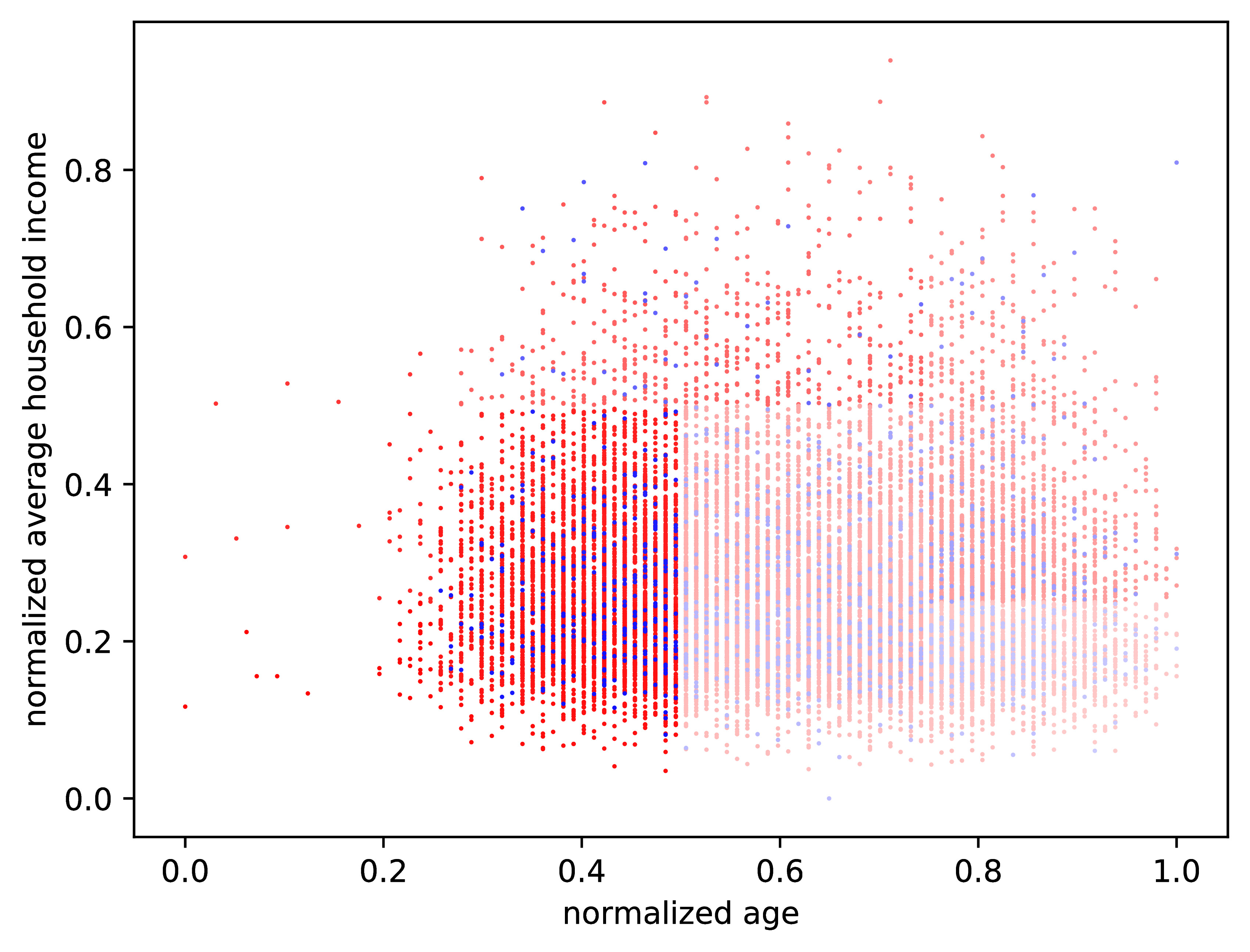}}
\quad\quad
(d) \parbox{\imsize}{\includegraphics[width=\imsize]
{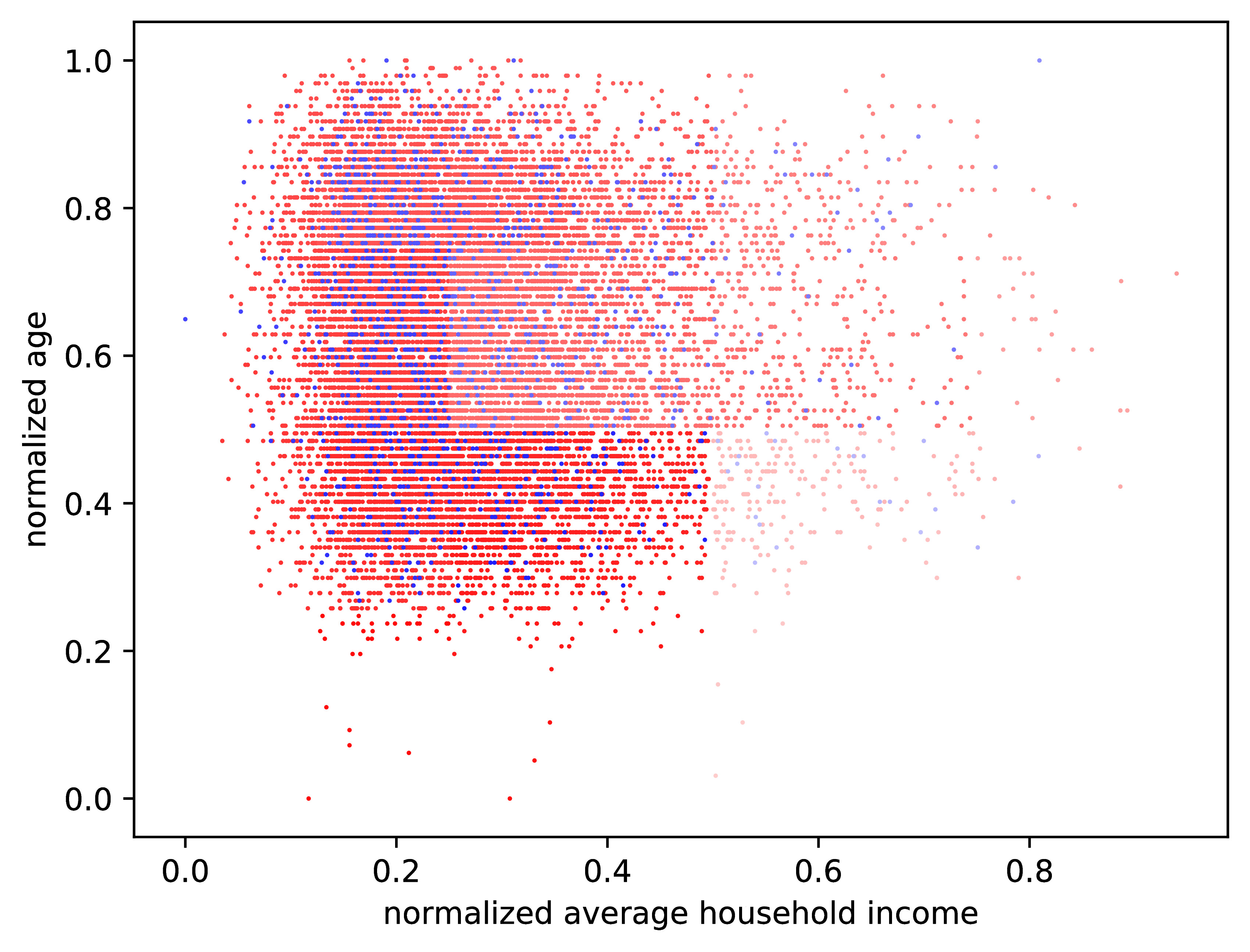}}

\vspace{\vertsep}

(e) \parbox{\imsize}{\includegraphics[width=\imsize]
{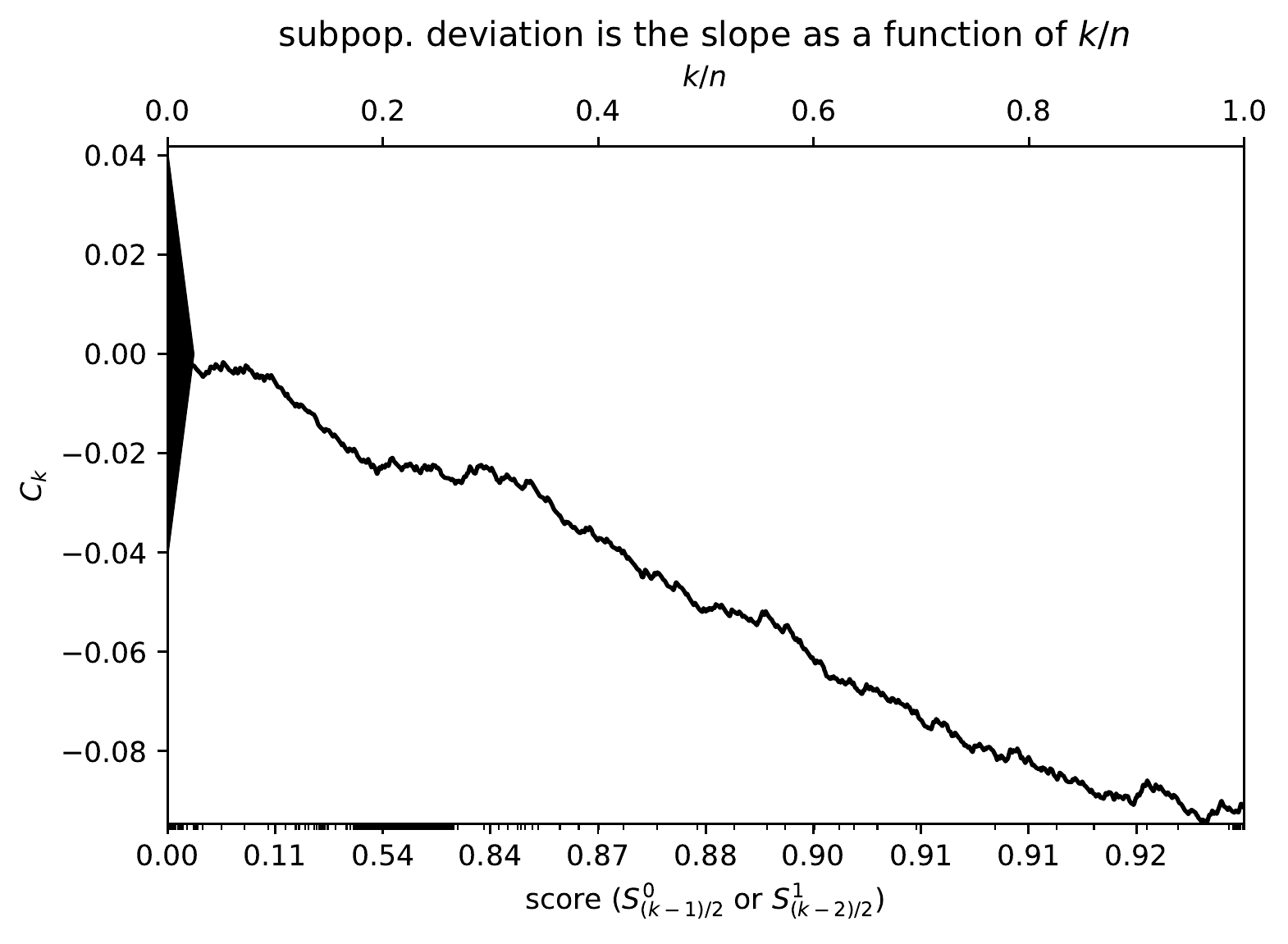}}
\quad\quad
(f) \parbox{\imsize}{\includegraphics[width=\imsize]
{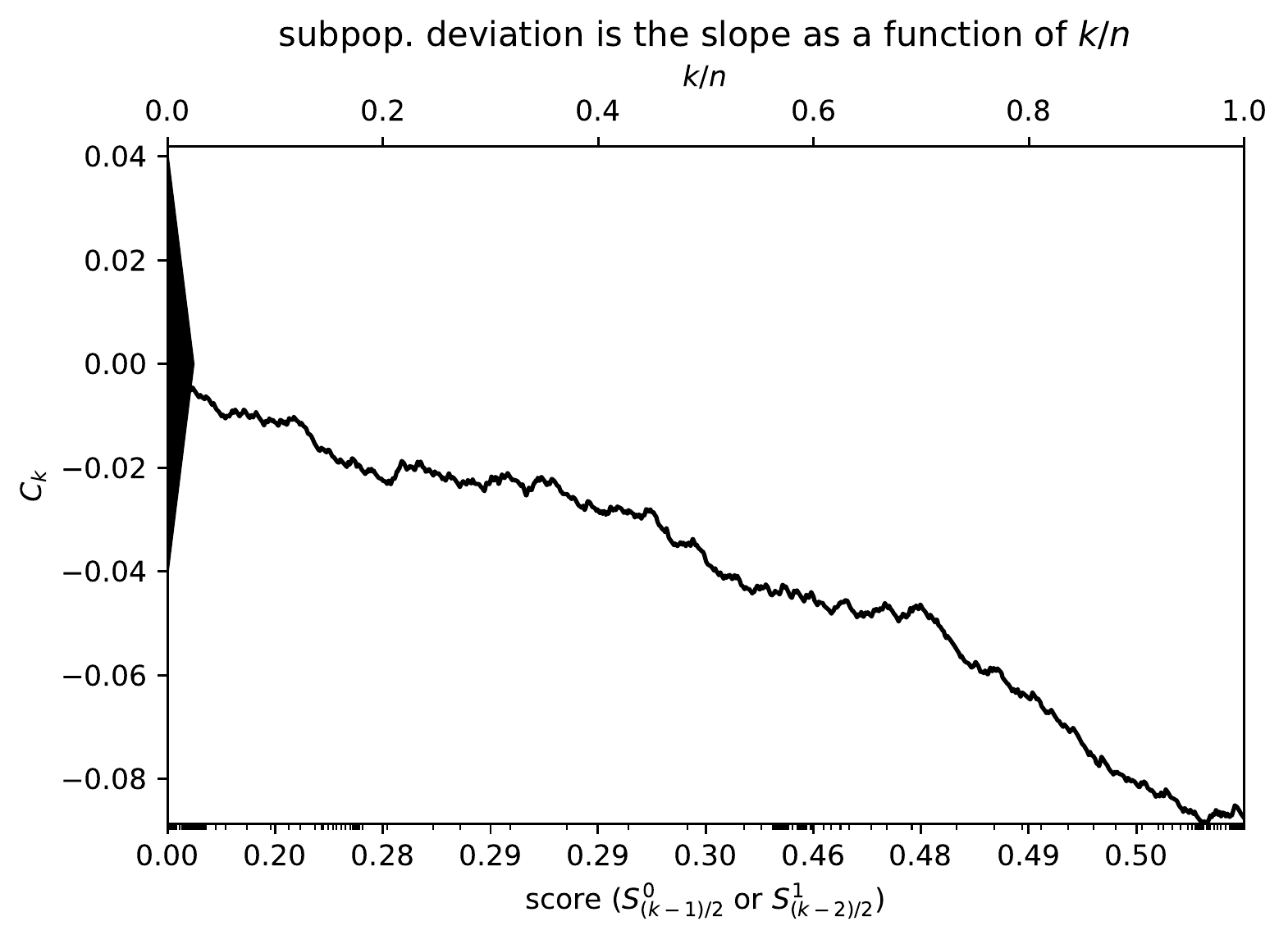}}

\parbox{\imsized}{\hfil \footnotesize $G$ = 0.09469; $H$ = 0.09469;
$G/\sigma$ = 4.527; $H/\sigma$ = 4.527}
\parbox{\imsized}{\hfil \footnotesize $G$ = 0.08869; $H$ = 0.08869;
$G/\sigma$ = 4.232; $H/\sigma$ = 4.232}

\vspace{\vertsep}

(g) \parbox{\imsize}{\includegraphics[width=\imsize]
{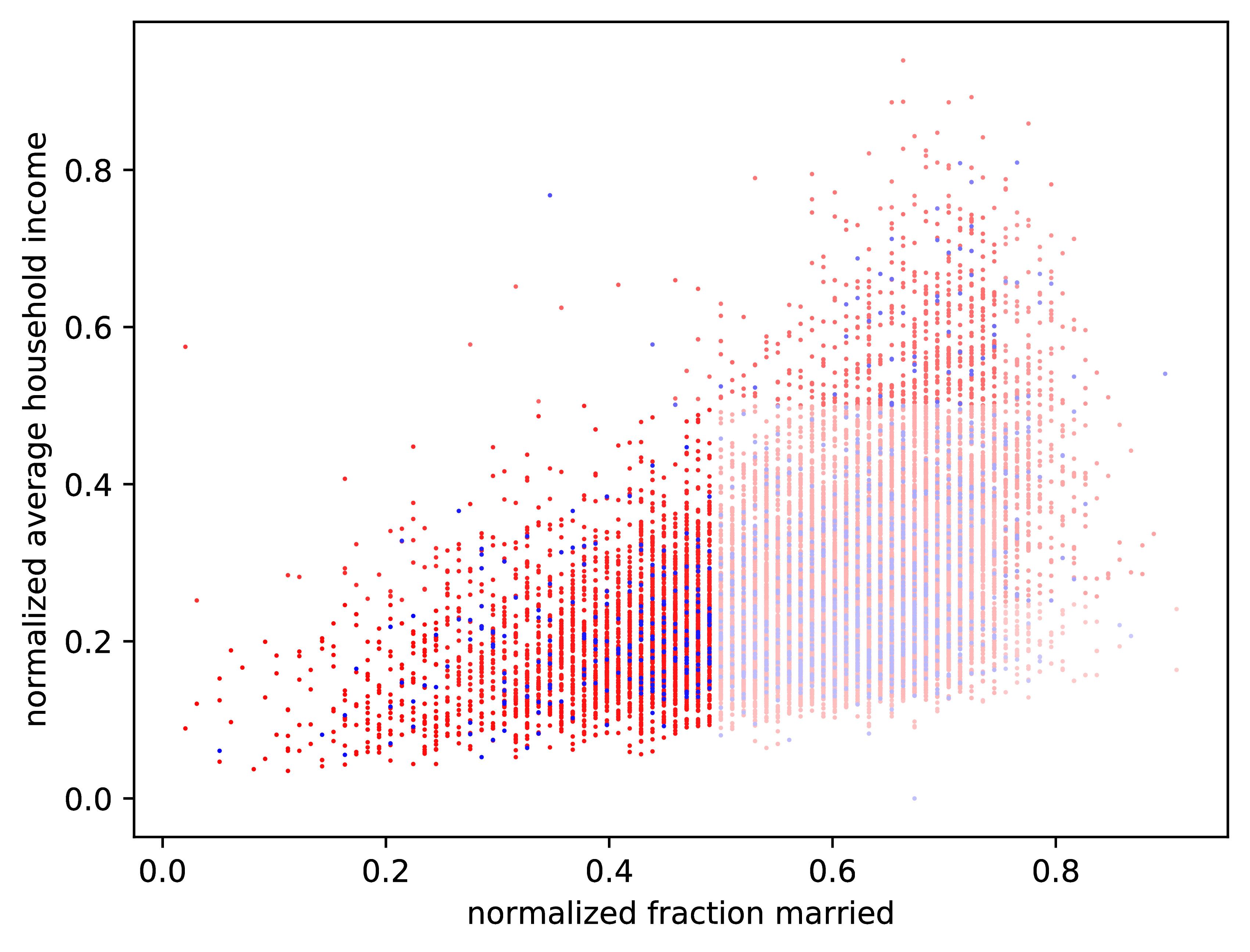}}
\quad\quad
(h) \parbox{\imsize}{\includegraphics[width=\imsize]
{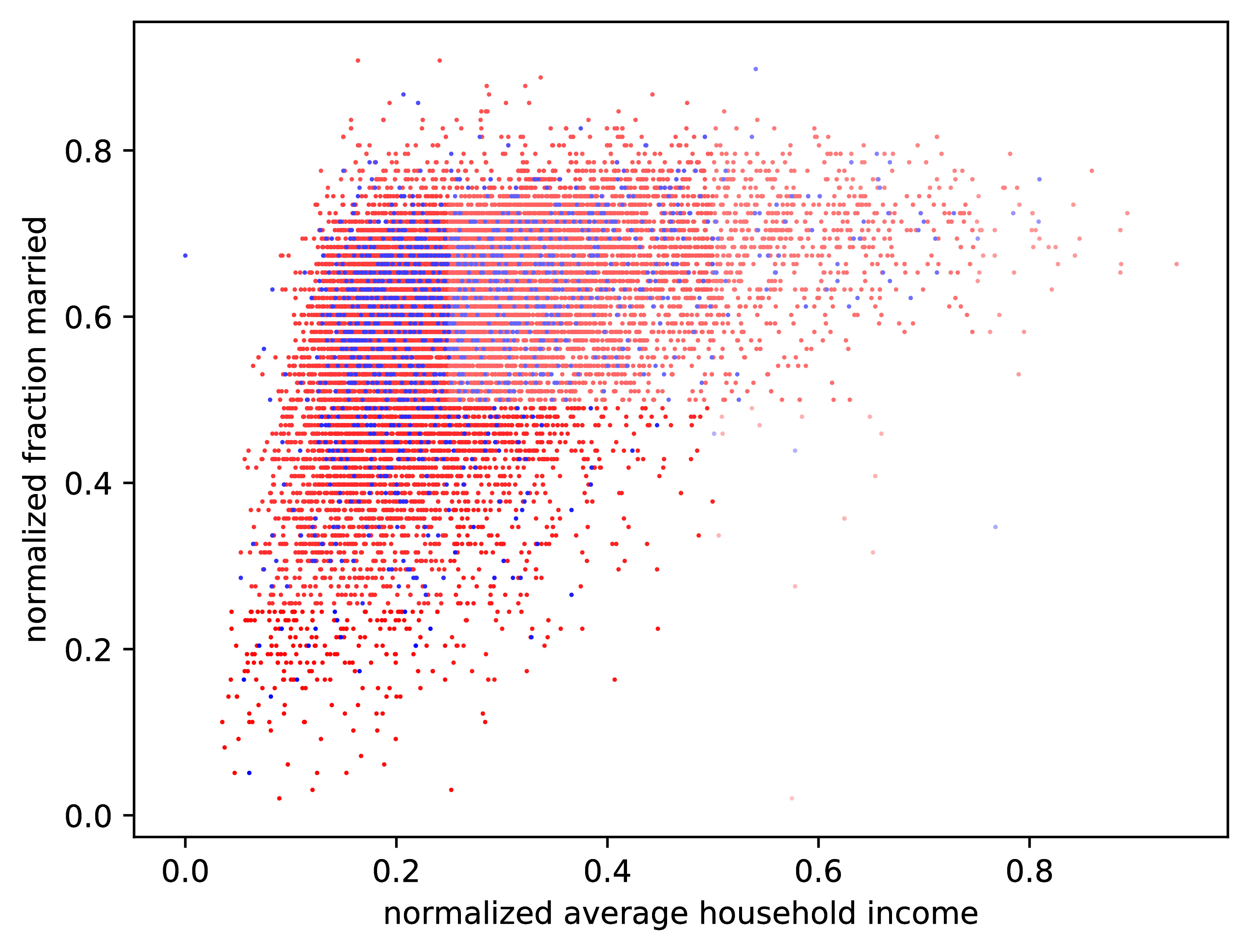}}

\end{centering}
\caption{Responses from folding cards versus normal cards
($n =$ 2,279 --- $n_0 =$ 1,236; $n_1 =$ 15,866)}
\label{folding_normal}
\end{figure}

\begin{figure}
\begin{centering}

(a) \parbox{\imsize}{\includegraphics[width=\imsize]
{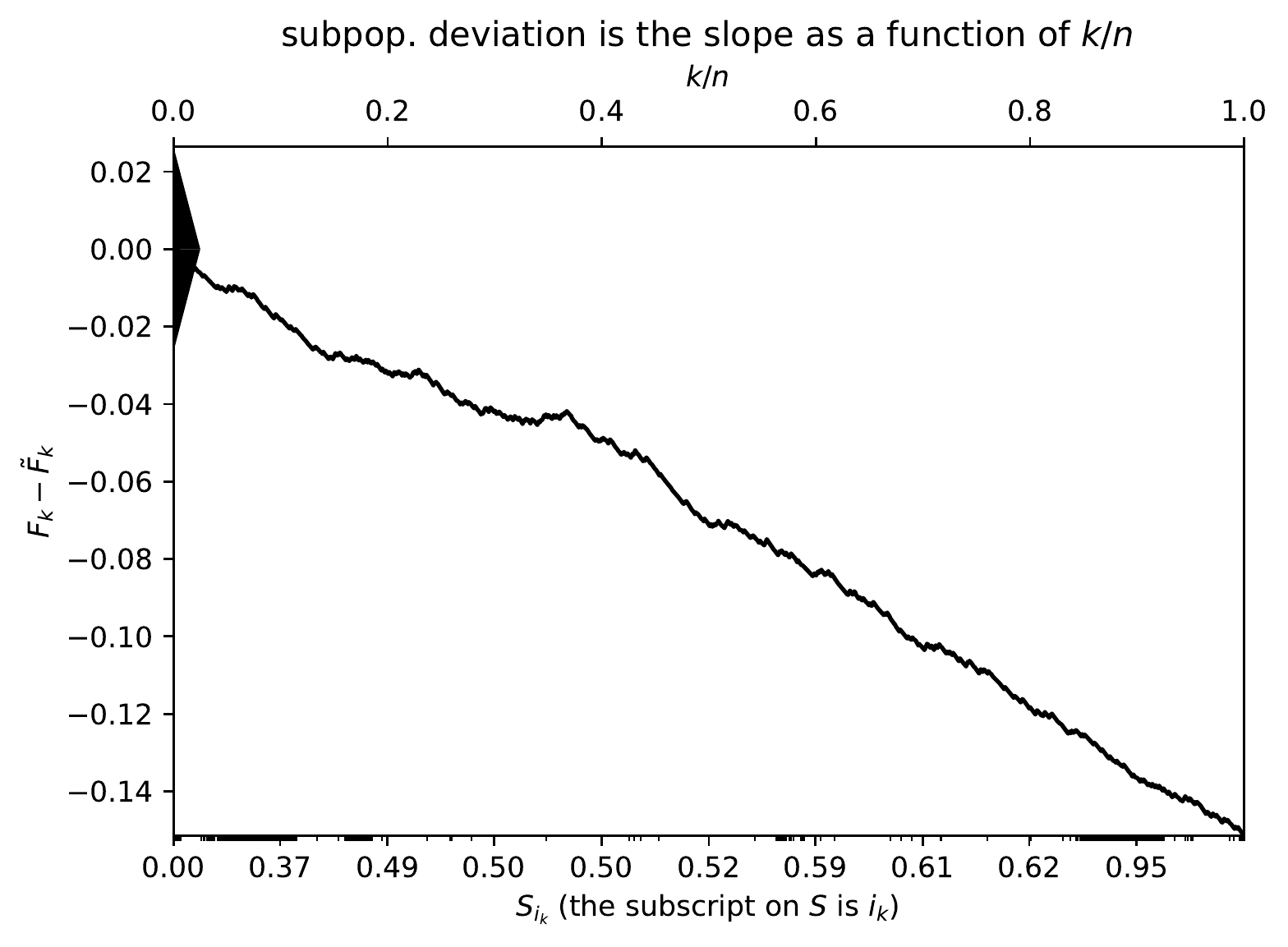}}
\quad\quad
(b) \parbox{\imsize}{\includegraphics[width=\imsize]
{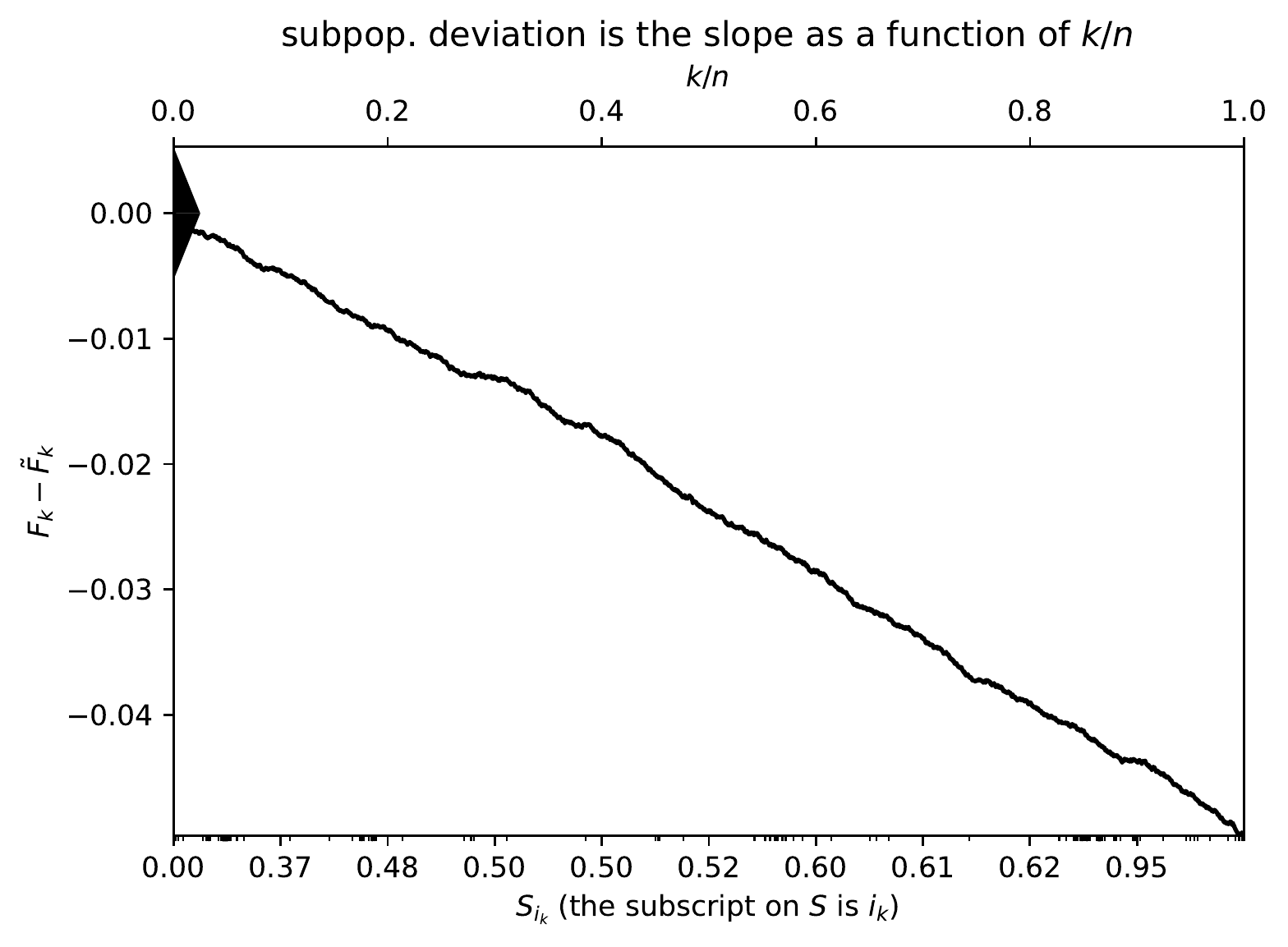}}

\parbox{\imsized}{\hfil \footnotesize $G$ = 0.1515; $H$ = 0.1515;
$G/\sigma$ = 11.47; $H/\sigma$ = 11.47}
\parbox{\imsized}{\hfil \footnotesize $G$ = 0.04964; $H$ = 0.04964;
$G/\sigma$ = 18.70; $H/\sigma$ = 18.70}

\vspace{\vertsep}

(c) \parbox{\imsize}{\includegraphics[width=\imsize]
{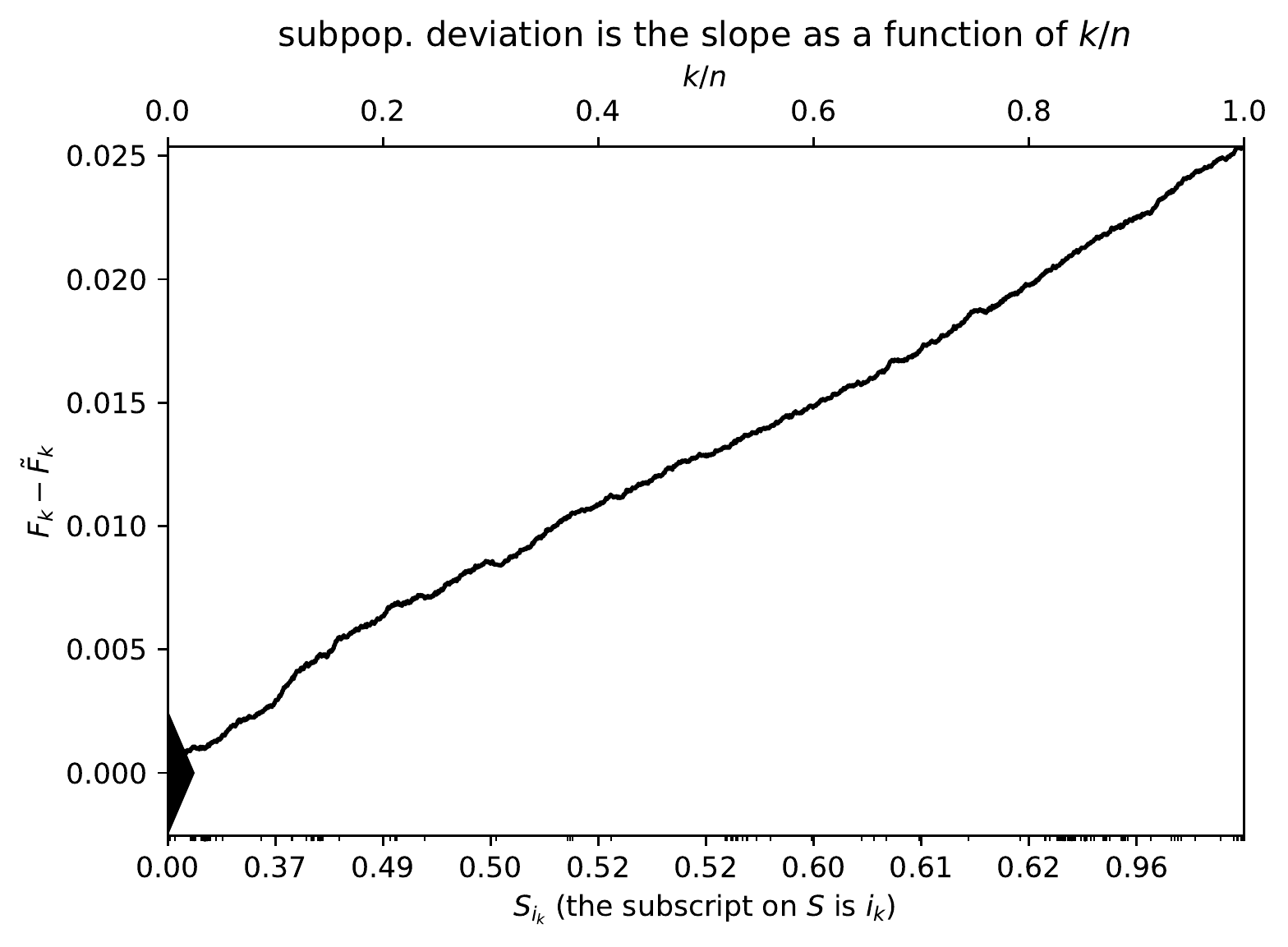}}
\quad\quad
(d) \parbox{\imsize}{\includegraphics[width=\imsize]
{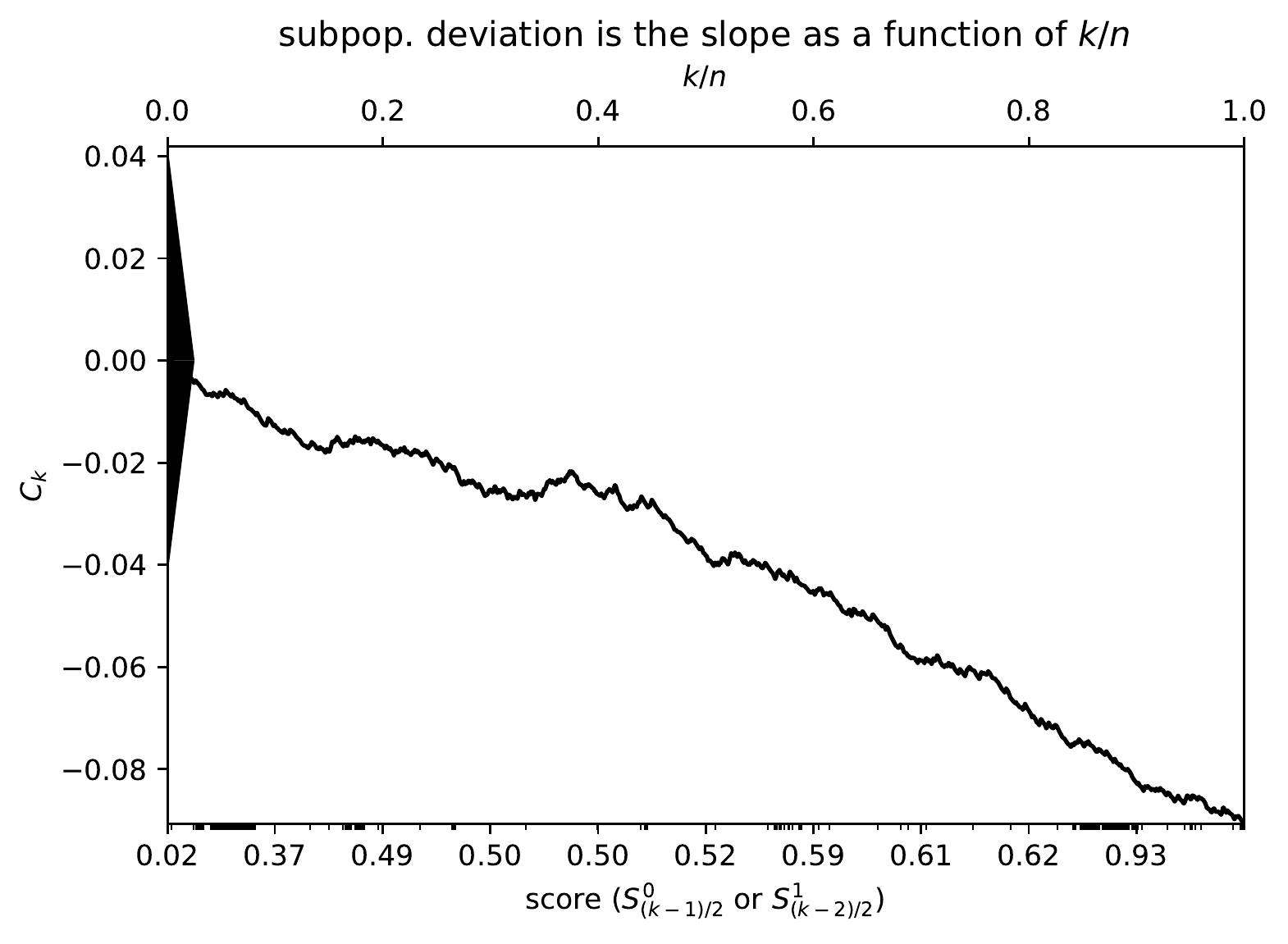}}

\parbox{\imsized}{\hfil \footnotesize $G$ = 0.02537; $H$ = 0.02537;
$G/\sigma$ = 19.95; $H/\sigma$ = 19.95}
\parbox{\imsized}{\hfil \footnotesize $G$ = 0.09074; $H$ = 0.09136;
$G/\sigma$ = 4.332; $H/\sigma$ = 4.362}

\end{centering}
\caption{Responses while controlling for the individual's age as well as
for the percent married and the average household income
in the individual's Census block ($m =$ 47,117);
the indicated subfigures compare
(a) folding cards only versus the full population ($n =$ 1,236),
(b) normal cards only versus the full population ($n =$ 15,866),
(c) those sent both folding and normal cards versus the full population
($n =$ 30,015), and (d)~folding cards only versus normal cards only
($n =$ 2,279 --- $n_0 =$ 1,236; $n_1 =$ 15,866)}
\label{allcovariates}
\end{figure}

This subsection presents Figures~\ref{folding}--\ref{allcovariates};
Table~\ref{labels} gives details about each subfigure
of Figures~\ref{folding}--\ref{folding_normal}.
These figures consider data from an experiment in direct-mail promotions
from back when snail-mail was a leading means of fundraising,
detailed as follows:

In 1994, a national veterans organization in the U.S. mailed many solicitations
for donations, trying different types of mailings to test
which would be best, and later contributed this data
to the 1998 Knowledge Discovery and Data-mining (KDD) Cup
competition.\footnote{The data from the 1998 KDD Cup is available
at \url{https://kdd.ics.uci.edu/databases/kddcup98/kddcup98.html}}
1,236 prospective donors received a mailing of folding cards,
15,866 received a mailing of normal cards,
and 30,015 got both mailings --- both folding cards and normal cards.
These three subsets form natural subpopulations of the full population
of 47,117.
All these numbers exclude those with missing ages (``AGE'' in the data set),
those with missing average household incomes in the associated Census block
(``IC3'' in the data), and those missing what fractions of householders
in the associated Census block are married (``MARR1'').
The figures report results of conditioning on various tuples formed
from combinations of these covariates
(ages, fractions married, and average household incomes
randomly perturbed by about 1 part in $10^8$ to guarantee uniqueness),
with each covariate normalized to range from 0 to 1 prior
to ordering via the Hilbert curve.
Prior to normalization, age is an integer
and the fraction married is an integer percentage.
Each response $R_i$ is simply whether the corresponding prospective donor
responded to the mailed solicitation (the mailing is considered effective
if it resulted in a response and ineffective otherwise,
with $R_i = 1$ in the former case and $R_i = 0$ in the latter case).
Figures~\ref{folding}--\ref{folding_normal} present four examples
conditioning on ordered pairs of covariates.
The first three, corresponding to Figures~\ref{folding}--\ref{both},
compare to the full population the single subpopulation indicated
in the caption to the corresponding figure;
these examples follow Subsection~\ref{subpop}.
The fourth example, corresponding to Figure~\ref{folding_normal},
compares directly the subpopulation sent folding cards only
to the subpopulation sent normal cards only;
this example follows Subsection~\ref{subpops}.
Table~\ref{labels} describes each subfigure
of Figures~\ref{folding}--\ref{folding_normal}.
Figure~\ref{allcovariates} presents analogues
of Figures~\ref{folding}--\ref{folding_normal},
conditioning on all three covariates rather than just the pairs
considered in Figures~\ref{folding}--\ref{folding_normal}.

Figures~\ref{folding}, \ref{normal}, \ref{both}, \ref{allcovariates}(a),
\ref{allcovariates}(b), and \ref{allcovariates}(c)
follow Subsection~\ref{subpop},
with $m$ being the number of members of the full population
and $n$ being the number in the subpopulation.
Figures~\ref{folding_normal} and \ref{allcovariates}(d)
follow Subsection~\ref{subpops},
with $n$ being the number of blocks for either one of the subpopulations,
$n_0$ being the number of members of subpopulation~0,
and $n_1$ being the number of members of subpopulation~1
(note that each block can contain multiple members).

In each of Figures~\ref{folding}--\ref{folding_normal},
the normalized scalar summary statistic $H/\sigma$
is always close across subfigures~(e) and~(f)
while also being either greater than in both subfigures~(e) and~(f)
or less than in both subfigures~(e) and~(f) when comparing
with subfigures~(a) and~(b) (greater in Figures~\ref{folding} and~\ref{normal},
less in Figures~\ref{both} and~\ref{folding_normal}).
Moreover, essentially the same remark holds separately
for the normalized scalar summary statistic $G/\sigma$.
This shows empirically that which covariates are conditioned on
tends to matter more than the ordering within that set of covariates.

\subsection{American Community Survey of the U.S.\ Census Bureau}
\label{acs}

\begin{figure}
\begin{centering}

(a) \parbox{\imsize}{\includegraphics[width=\imsize]
{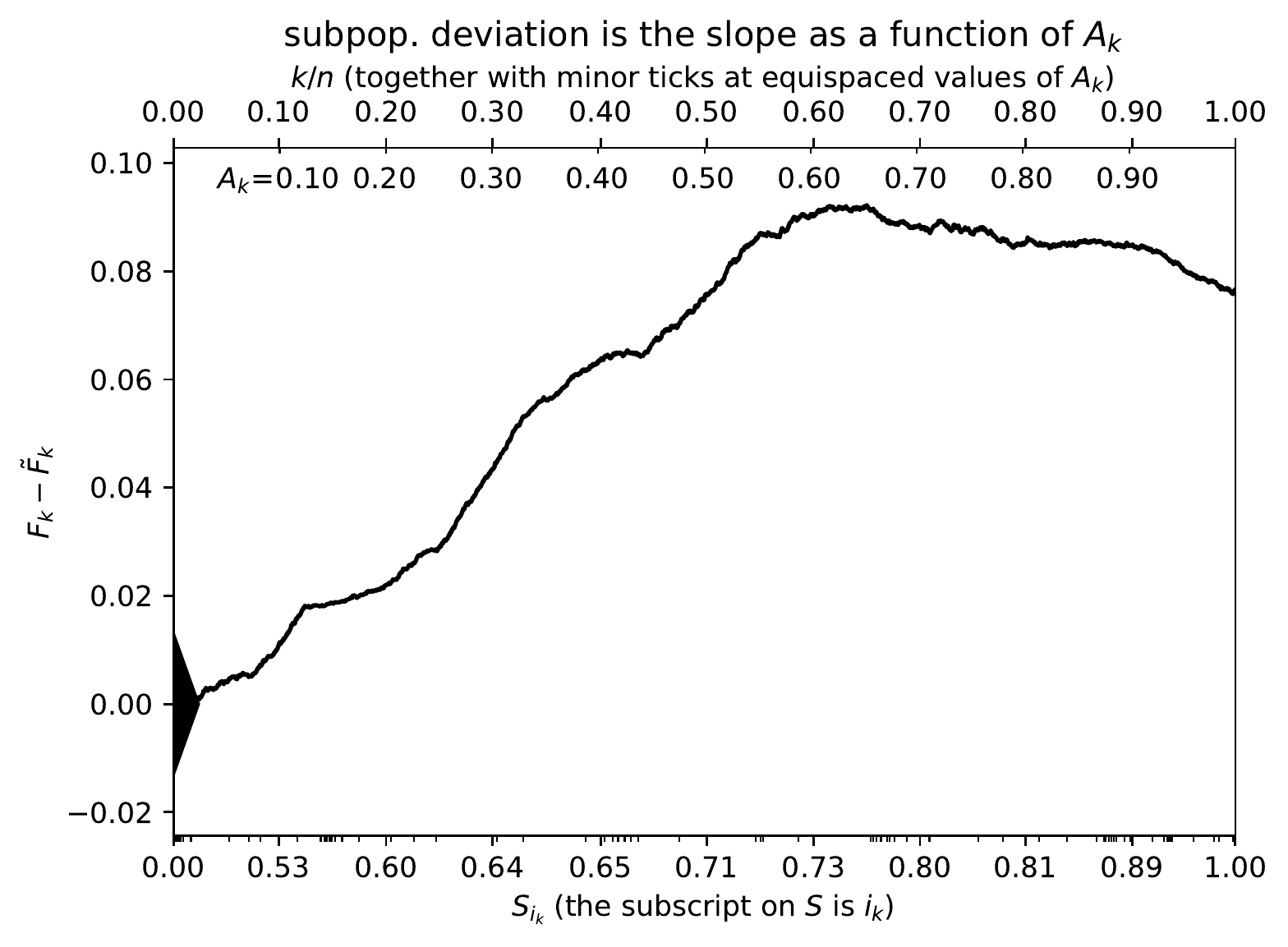}}
\quad\quad
(b) \parbox{\imsize}{\includegraphics[width=\imsize]
{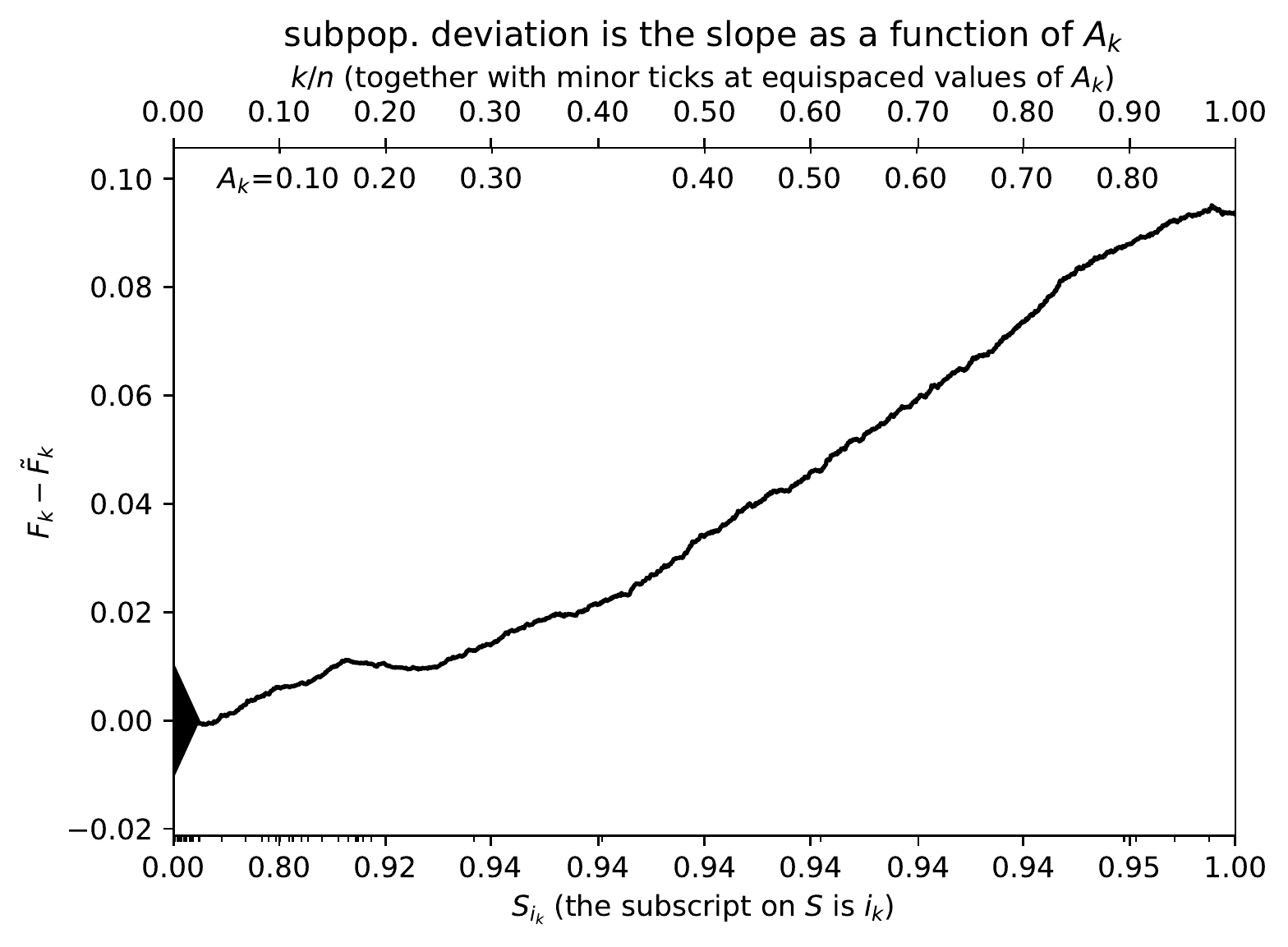}}

\parbox{\imsized}{\hfil \footnotesize $G$ = 0.09221; $H$ = 0.09267;
$G/\sigma$ = 13.38; $H/\sigma$ = 13.45}
\parbox{\imsized}{\hfil \footnotesize $G$ = 0.09514; $H$ = 0.09595;
$G/\sigma$ = 17.73; $H/\sigma$ = 17.88}

\vspace{\vertsep}

(c) \parbox{\imsize}{\includegraphics[width=\imsize]
{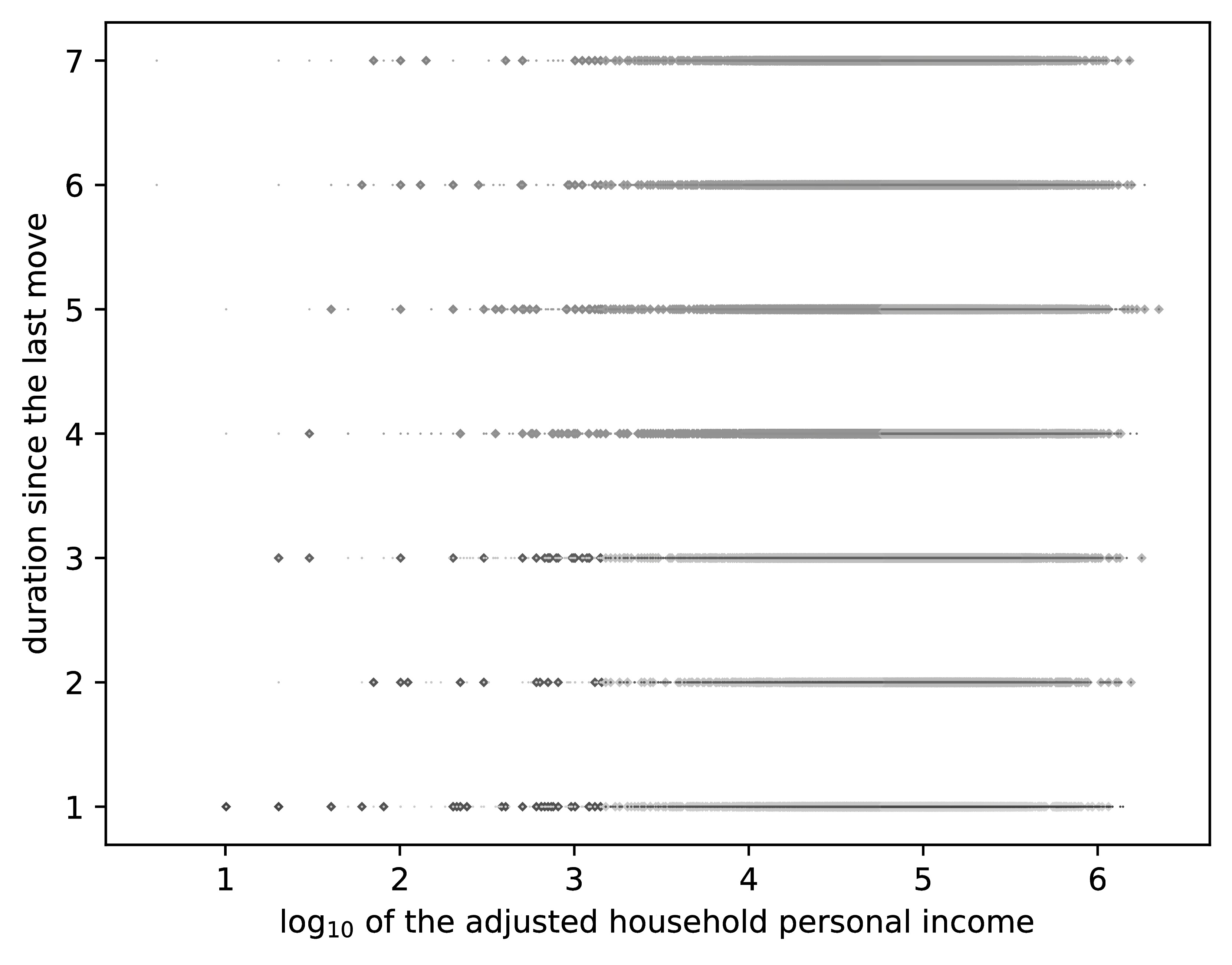}}
\quad\quad
(d) \parbox{\imsize}{\includegraphics[width=\imsize]
{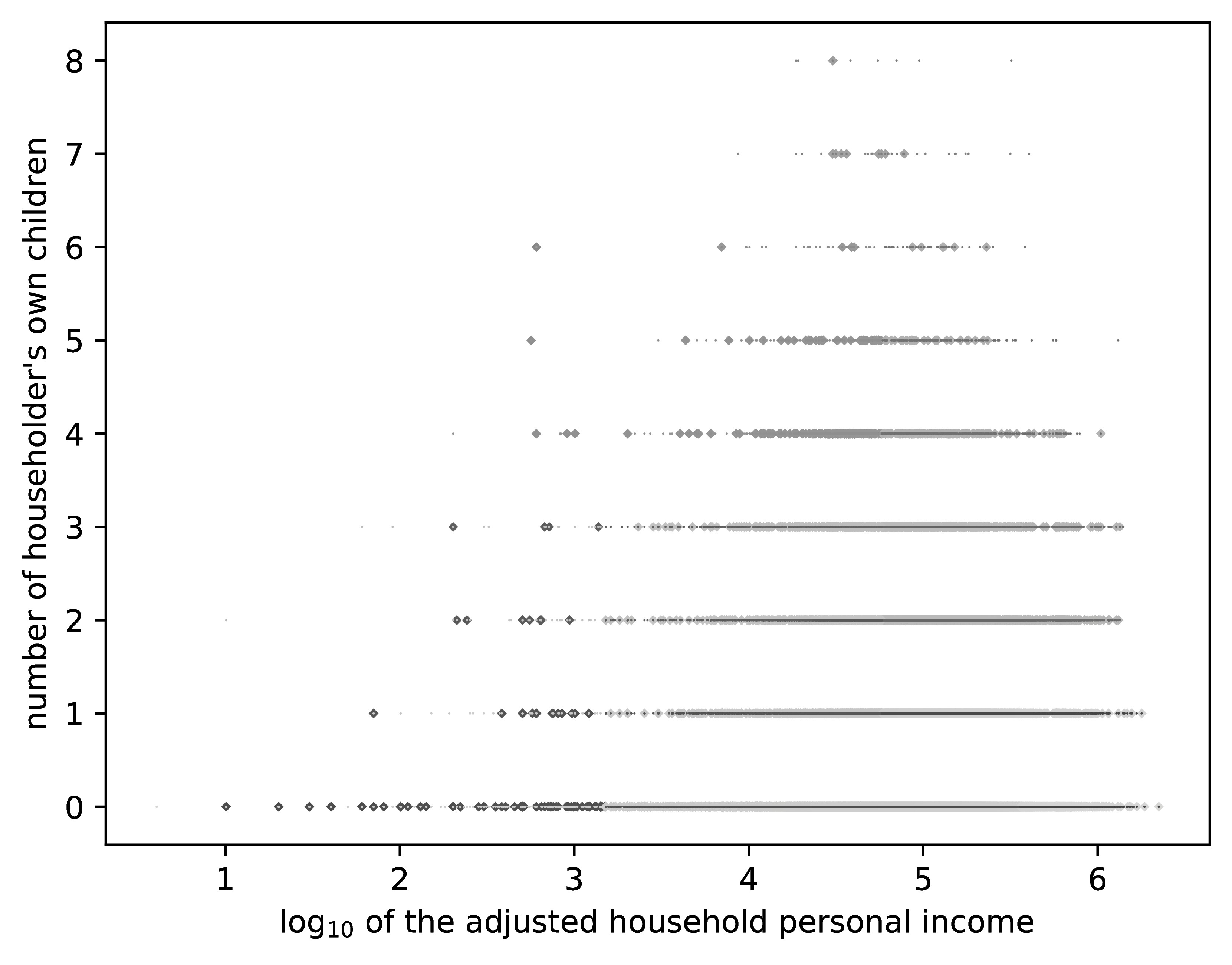}}

\vspace{\vertsep}

(e) \parbox{\imsize}{\includegraphics[width=\imsize]
{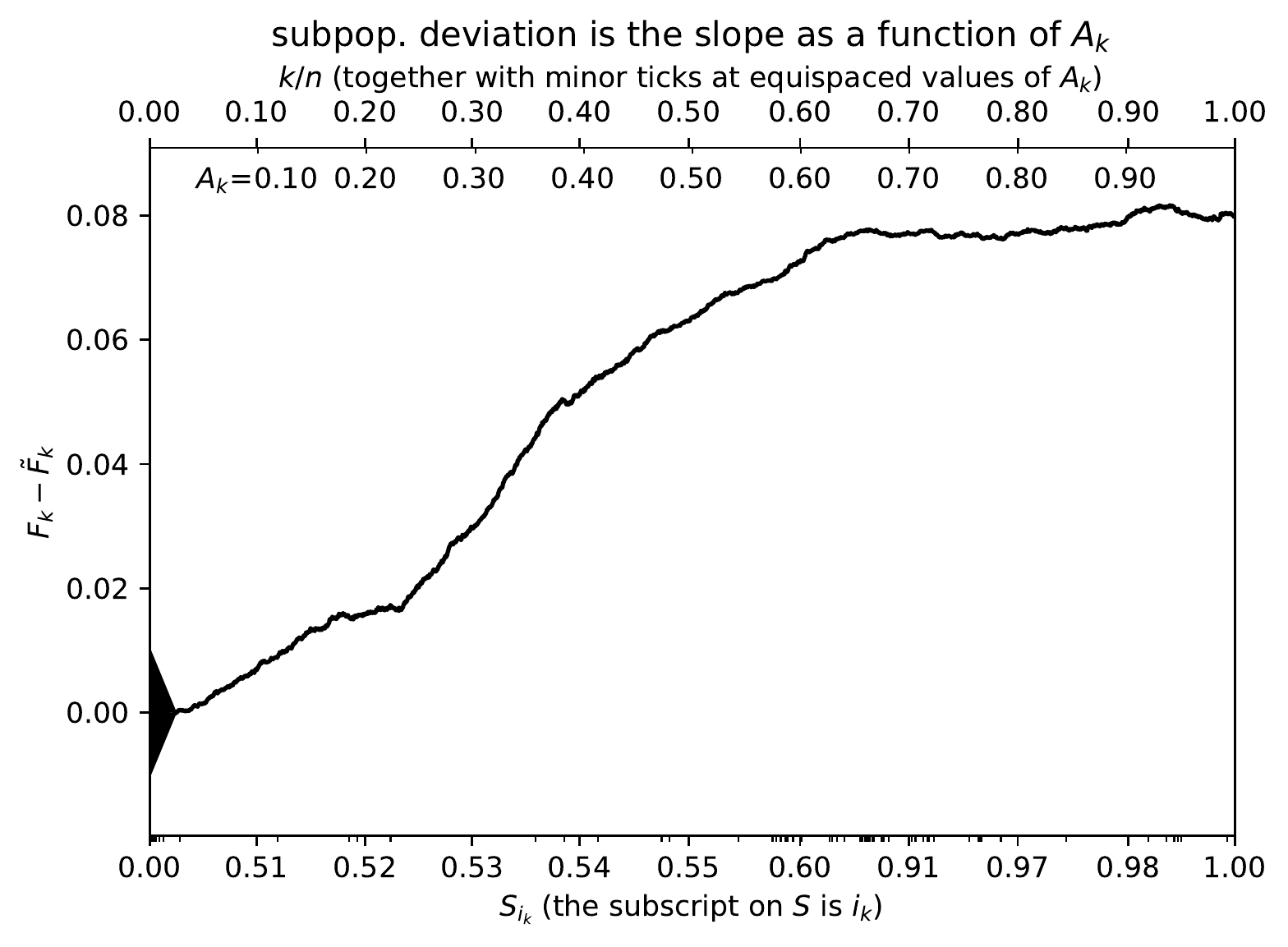}}
\quad\quad
(f) \parbox{\imsize}{\includegraphics[width=\imsize]
{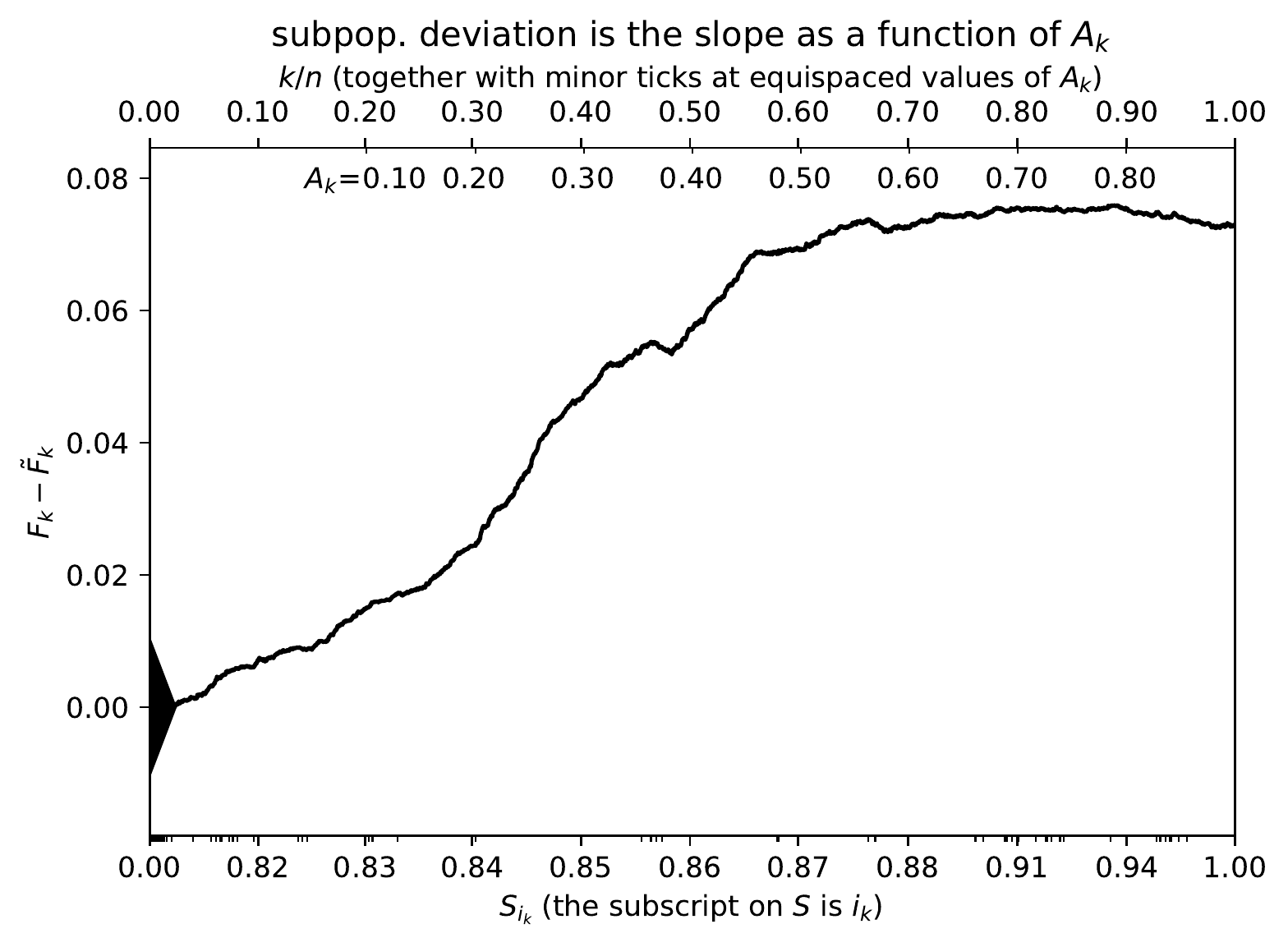}}

\parbox{\imsized}{\hfil \footnotesize $G$ = 0.08162; $H$ = 0.08209;
$G/\sigma$ = 15.40; $H/\sigma$ = 15.48}
\parbox{\imsized}{\hfil \footnotesize $G$ = 0.07591; $H$ = 0.07674;
$G/\sigma$ = 14.06; $H/\sigma$ = 14.22}

\end{centering}
\caption{Los Angeles County ($m =$ 134,094; $n =$ 35,364),
number of persons in the household;
notice that conditioning on ``MV'' and ``NOC'' in that order (e)
yields results more similar to conditioning on ``NOC'' and ``MV'' (f)
than to conditioning only on ``MV'' (a) or only on ``NOC'' (b) ---
the results of controlling for covariates appear to depend more on the choice
of the set of covariates than on the ordering of the conditioning
within a particular set. All cases condition also on the logarithm
of the adjusted household personal income, in addition to the combinations
of covariates ``MV'' and ``NOC'' mentioned here --- see Table~\ref{labelsw},
which details subfigures~(c) and~(d), too.}
\label{los_angeles}
\end{figure}

\begin{figure}
\begin{centering}

(a) \parbox{\imsize}{\includegraphics[width=\imsize]
{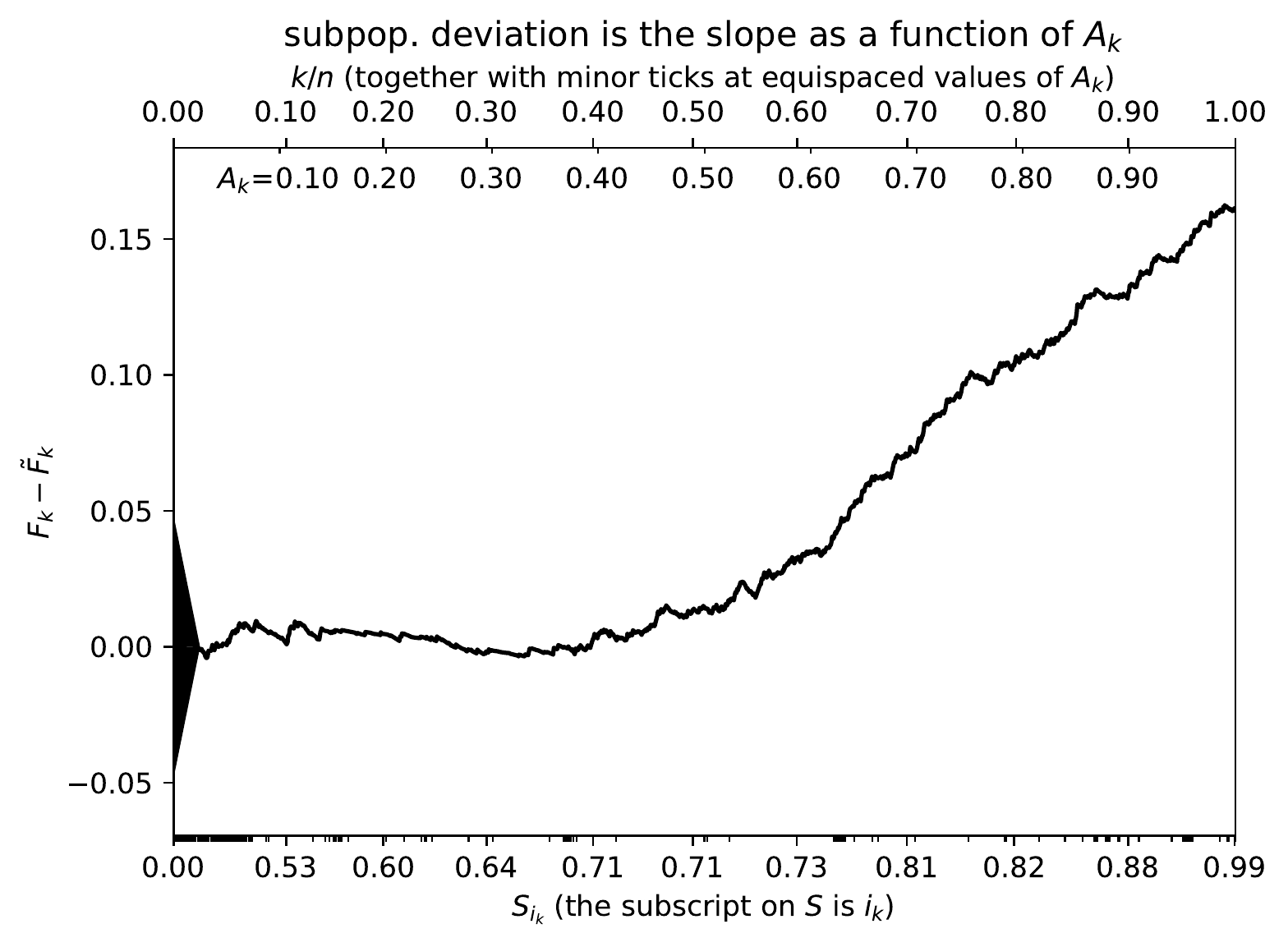}}
\quad\quad
(b) \parbox{\imsize}{\includegraphics[width=\imsize]
{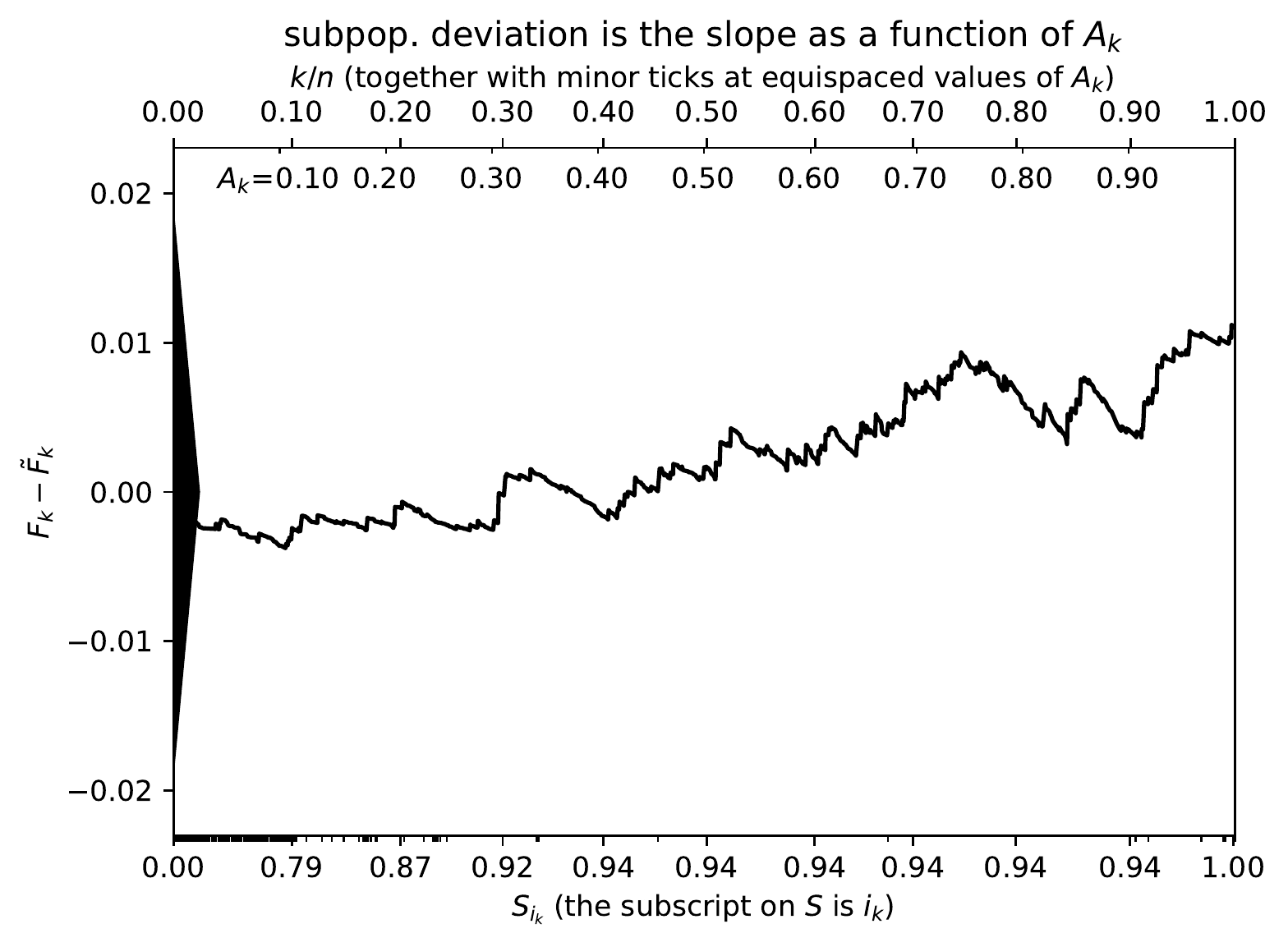}}

\parbox{\imsized}{\hfil \footnotesize $G$ = 0.1624; $H$ = 0.1665;
$G/\sigma$ = 6.706; $H/\sigma$ = 6.875}
\parbox{\imsized}{\hfil \footnotesize $G$ = 0.01118; $H$ = 0.01495;
$G/\sigma$ = 1.165; $H/\sigma$ = 1.557}

\vspace{\vertsep}

(c) \parbox{\imsize}{\includegraphics[width=\imsize]
{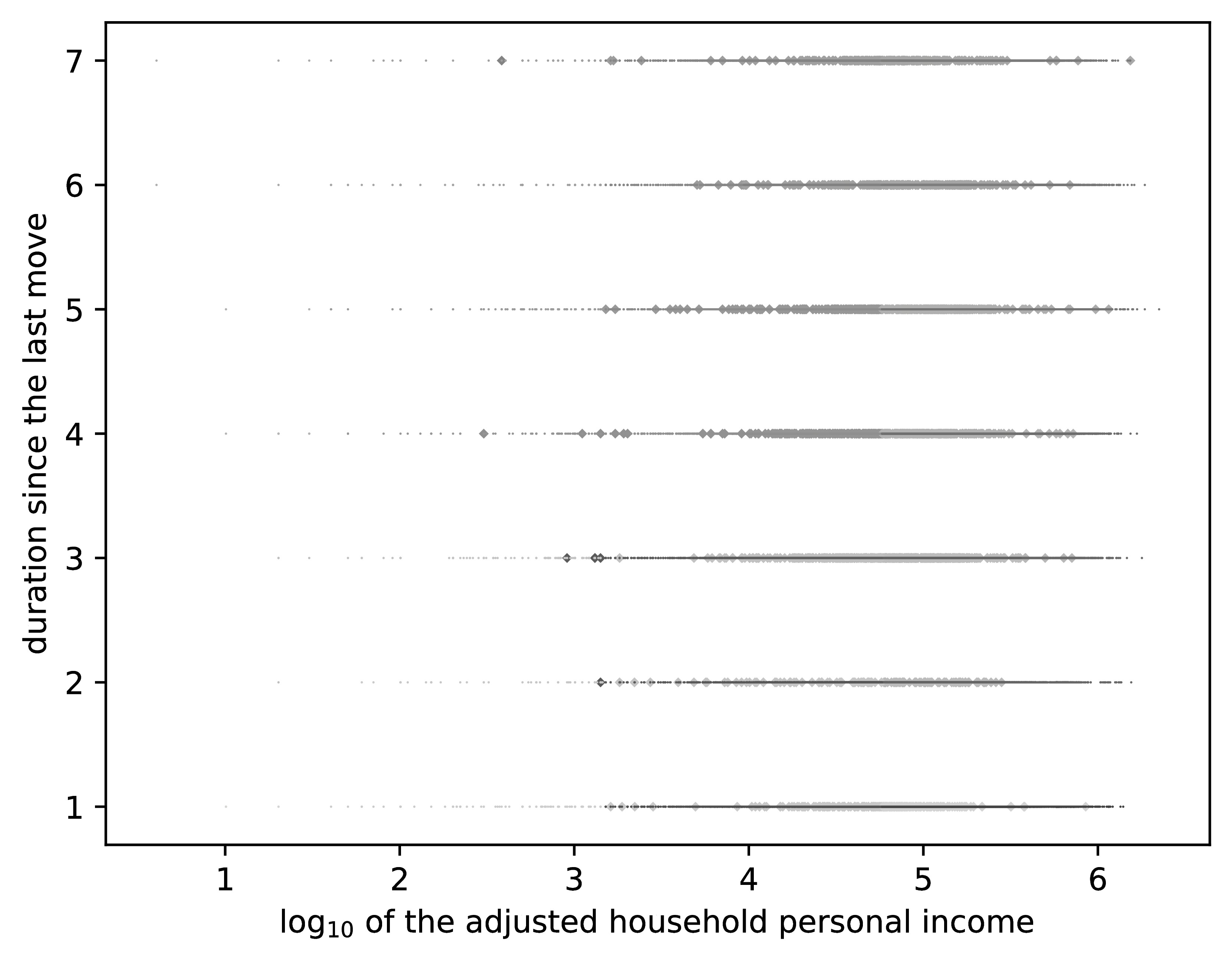}}
\quad\quad
(d) \parbox{\imsize}{\includegraphics[width=\imsize]
{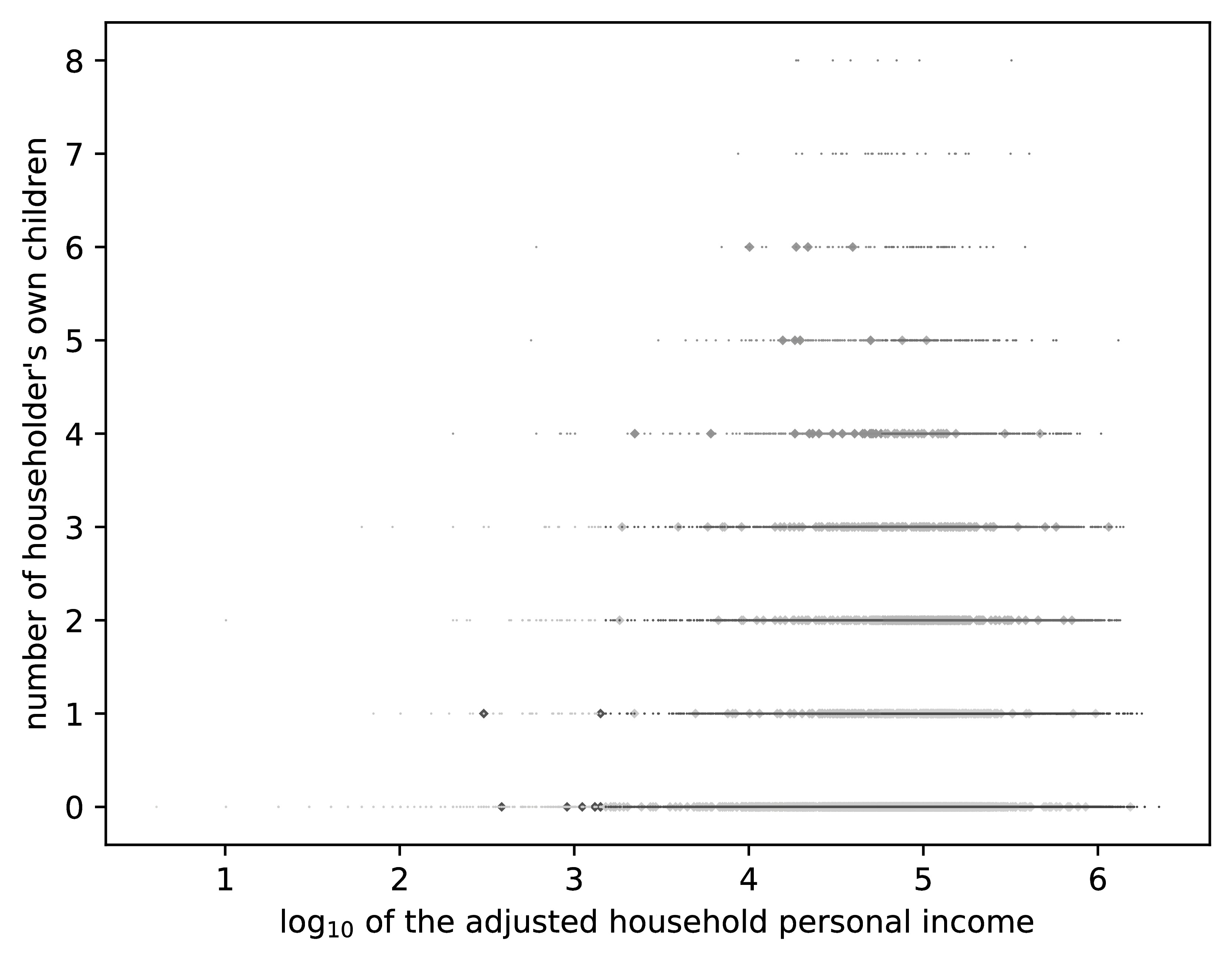}}

\vspace{\vertsep}

(e) \parbox{\imsize}{\includegraphics[width=\imsize]
{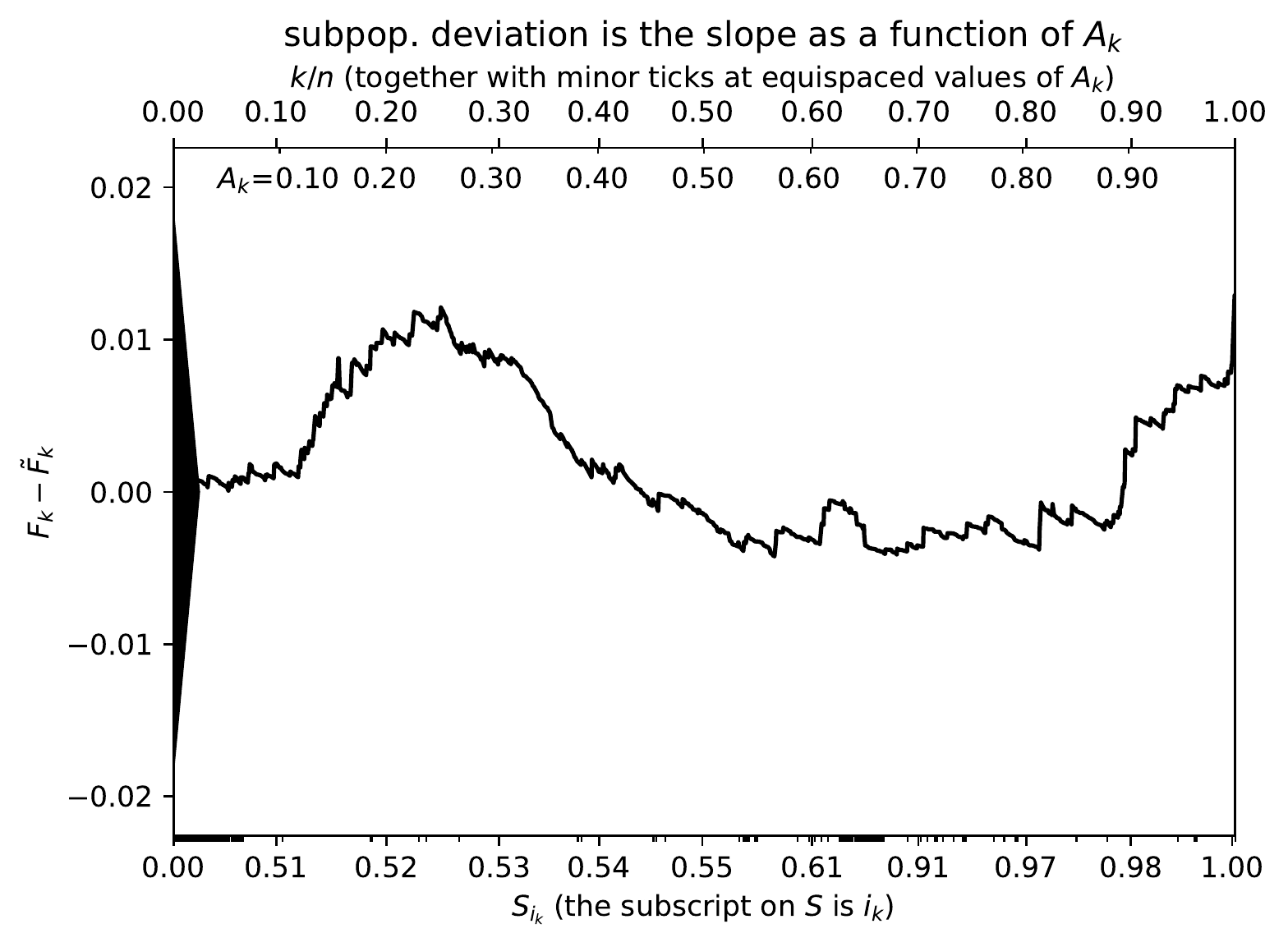}}
\quad\quad
(f) \parbox{\imsize}{\includegraphics[width=\imsize]
{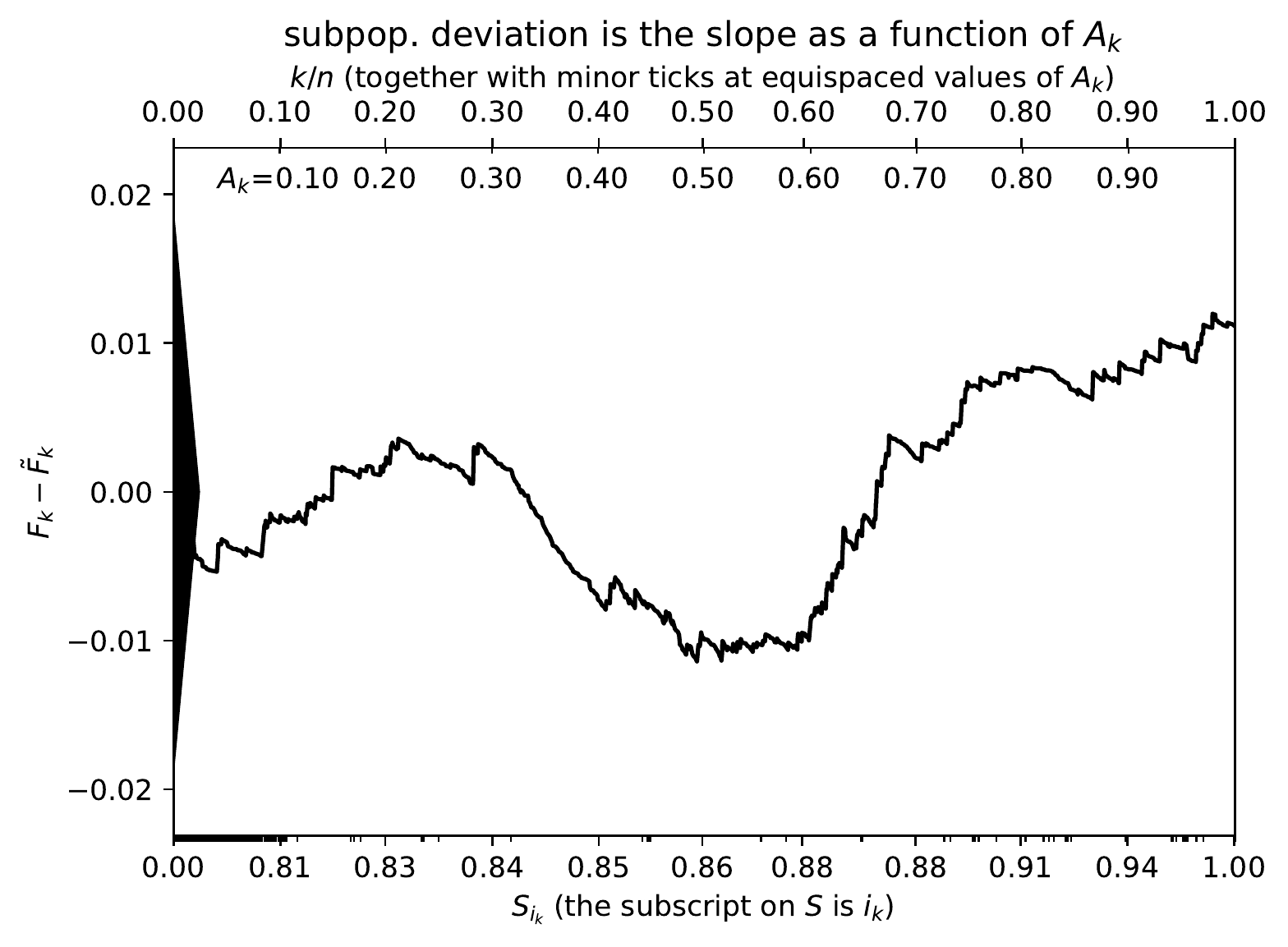}}

\parbox{\imsized}{\hfil \footnotesize $G$ = 0.01291; $H$ = 0.01714;
$G/\sigma$ = 1.371; $H/\sigma$ = 1.821}
\parbox{\imsized}{\hfil \footnotesize $G$ = 0.01198; $H$ = 0.02339;
$G/\sigma$ = 1.243; $H/\sigma$ = 2.426}

\end{centering}
\caption{San Joaquin County ($m =$ 134,094; $n =$ 2,282),
number of related children in the household;
notice that conditioning on ``MV'' and ``NOC'' in that order (e)
yields a graph more similar to conditioning on ``NOC'' and ``MV'' (f)
than to conditioning only on ``MV'' (a) or only on ``NOC'' (b) ---
again the graphs seem to depend more on the choice
of which set of covariates constitutes the controls
than on the ordering of the conditioning within a particular set.
All cases condition also on the logarithm
of the adjusted household personal income, in addition to the combinations
of covariates ``MV'' and ``NOC'' mentioned here --- see Table~\ref{labelsw},
which details subfigures~(c) and~(d), too.}
\label{san_joaquin}
\end{figure}

\begin{figure}
\begin{centering}

(a) \parbox{\imsize}{\includegraphics[width=\imsize]
{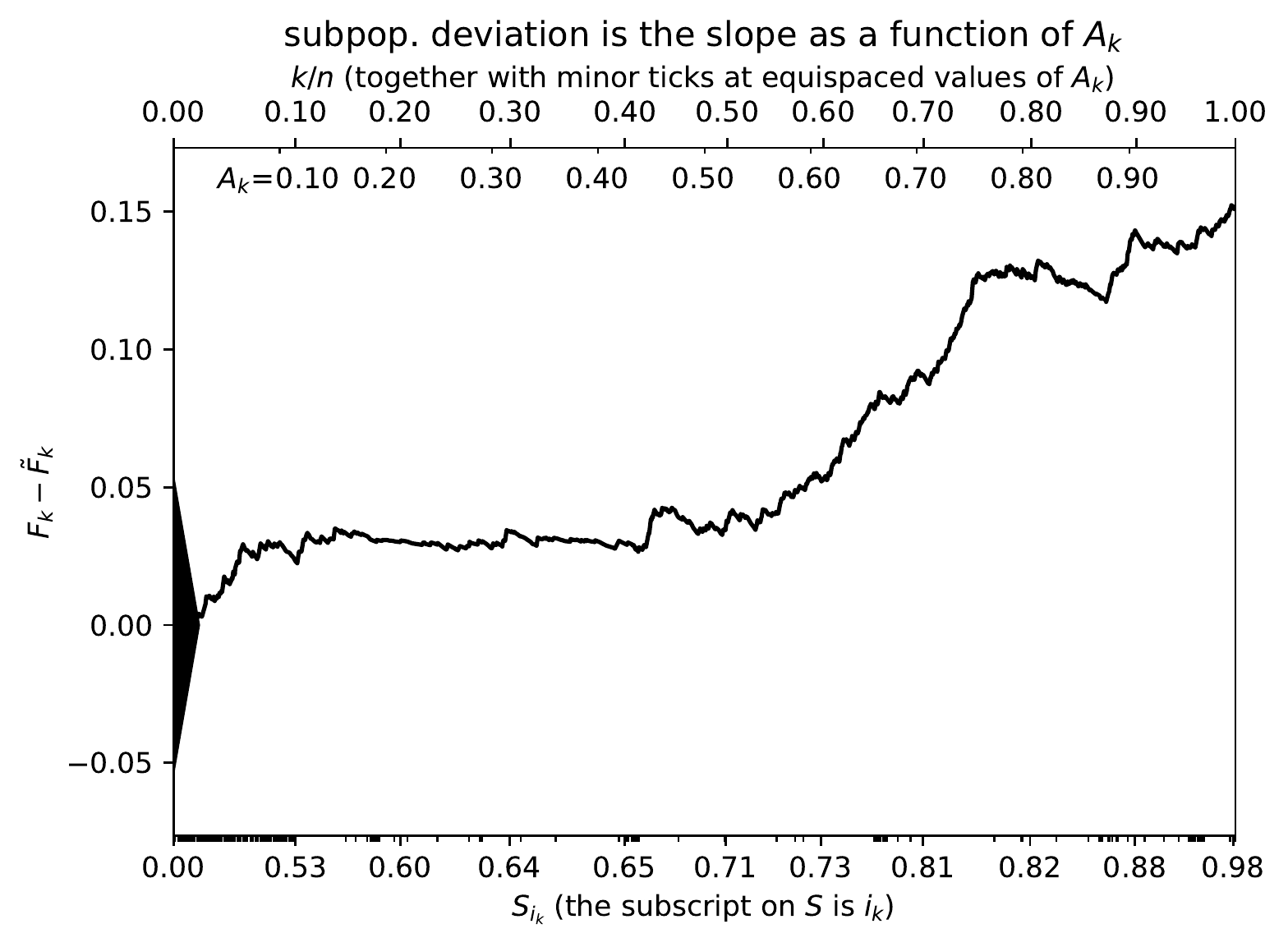}}
\quad\quad
(b) \parbox{\imsize}{\includegraphics[width=\imsize]
{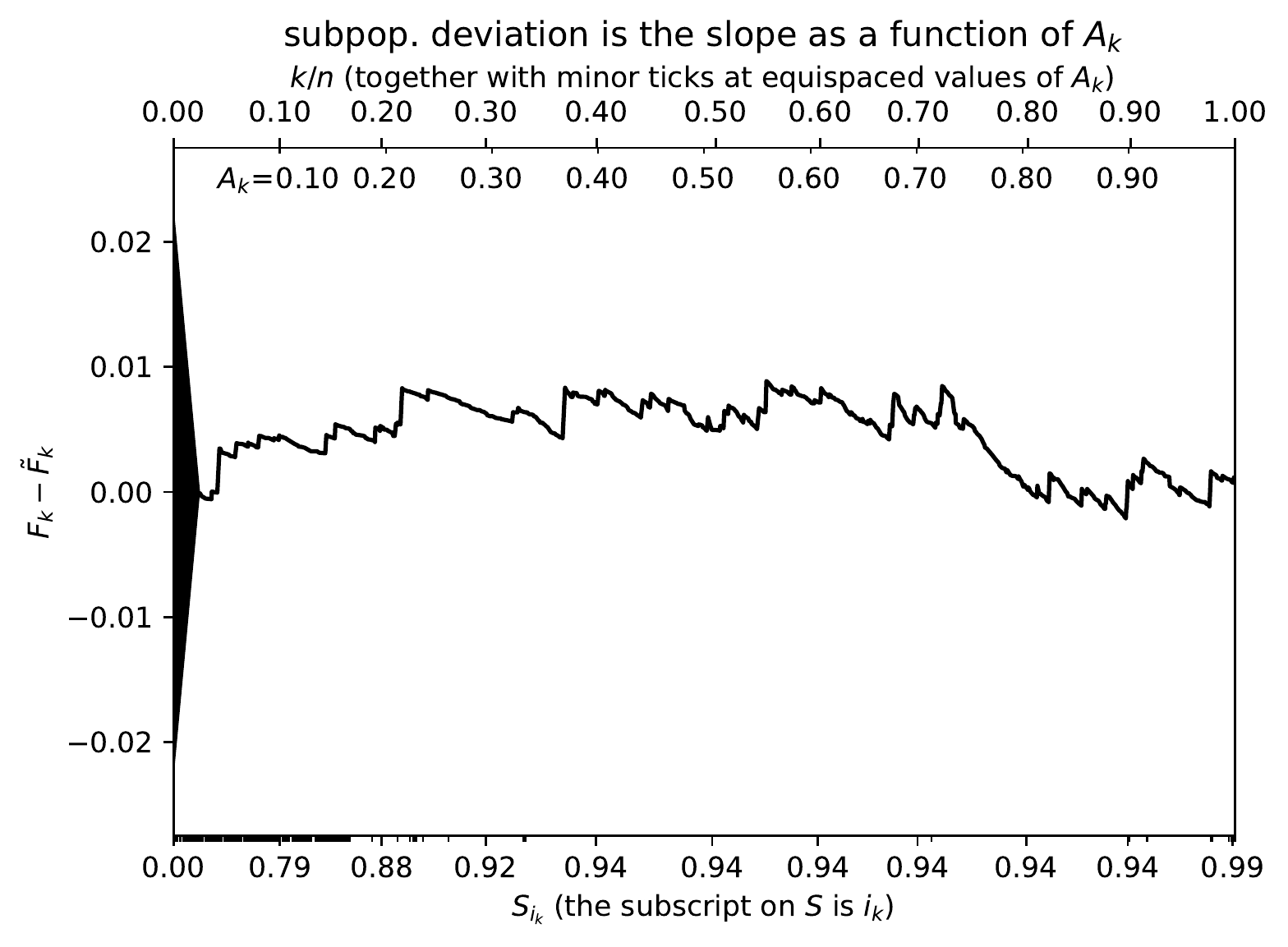}}

\parbox{\imsized}{\hfil \footnotesize $G$ = 0.1523; $H$ = 0.1546;
$G/\sigma$ = 5.470; $H/\sigma$ = 5.551}
\parbox{\imsized}{\hfil \footnotesize $G$ = 0.008848; $H$ = 0.01096;
$G/\sigma$ = 0.7724; $H/\sigma$ = 0.9568}

\vspace{\vertsep}

(c) \parbox{\imsize}{\includegraphics[width=\imsize]
{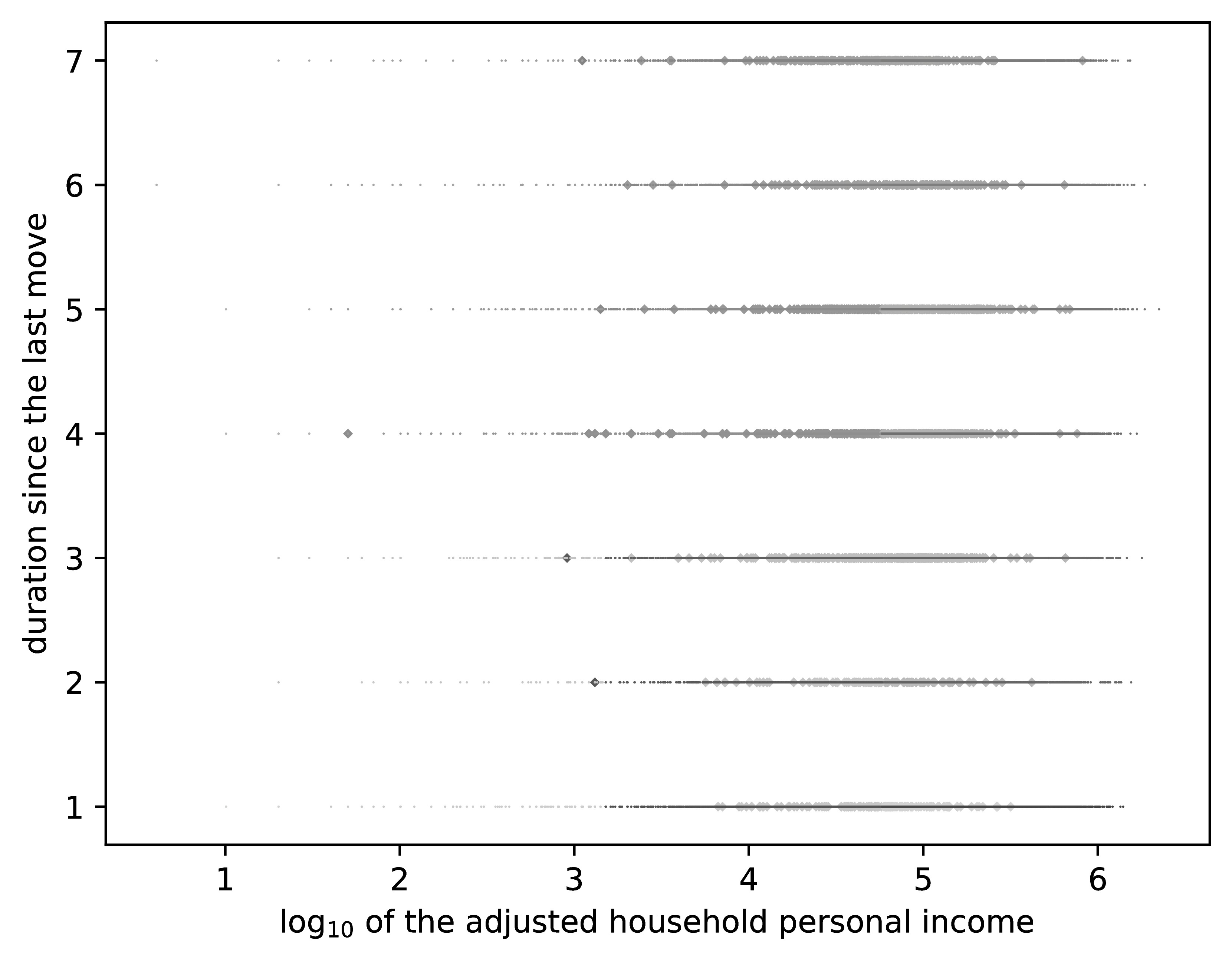}}
\quad\quad
(d) \parbox{\imsize}{\includegraphics[width=\imsize]
{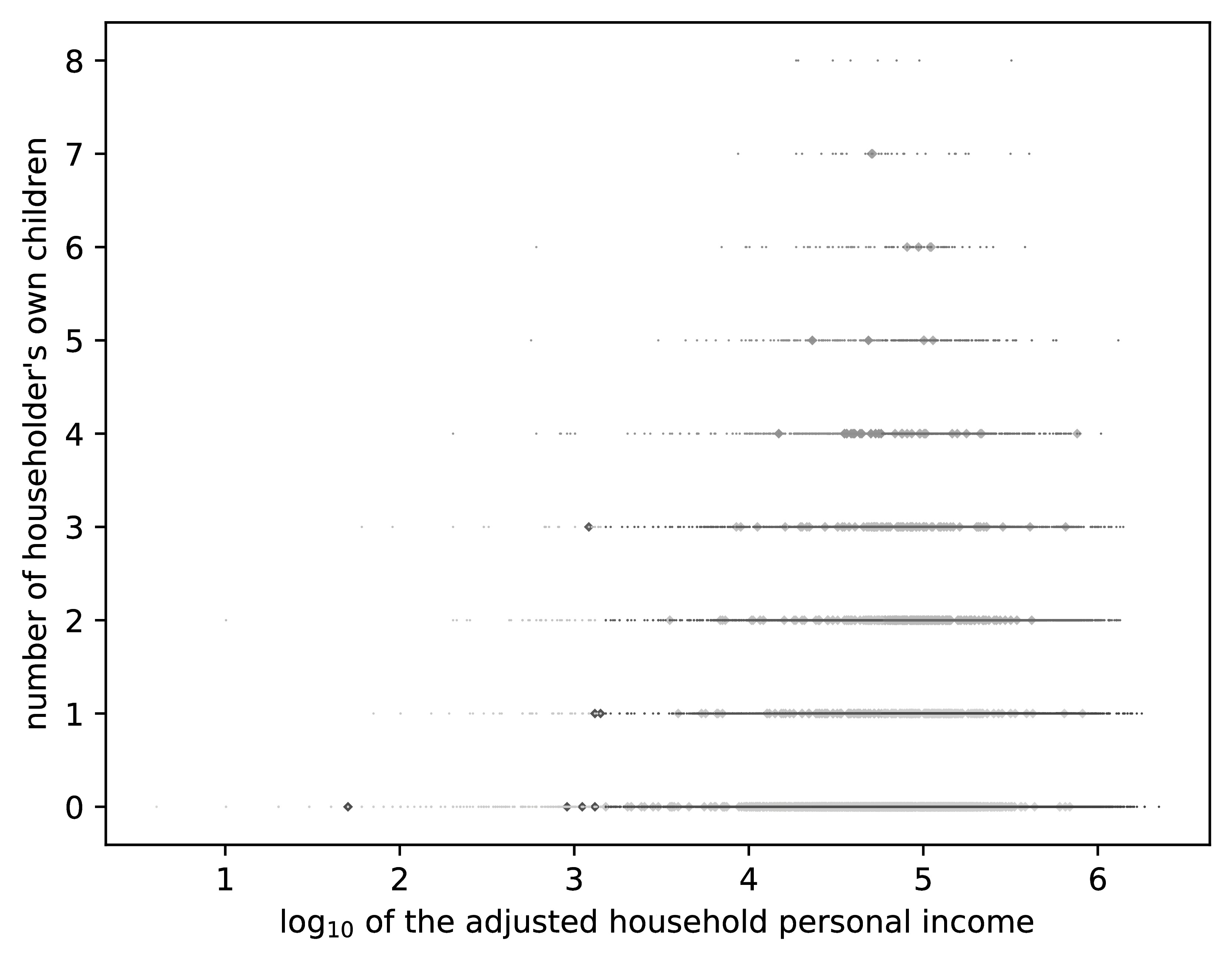}}

\vspace{\vertsep}

(e) \parbox{\imsize}{\includegraphics[width=\imsize]
{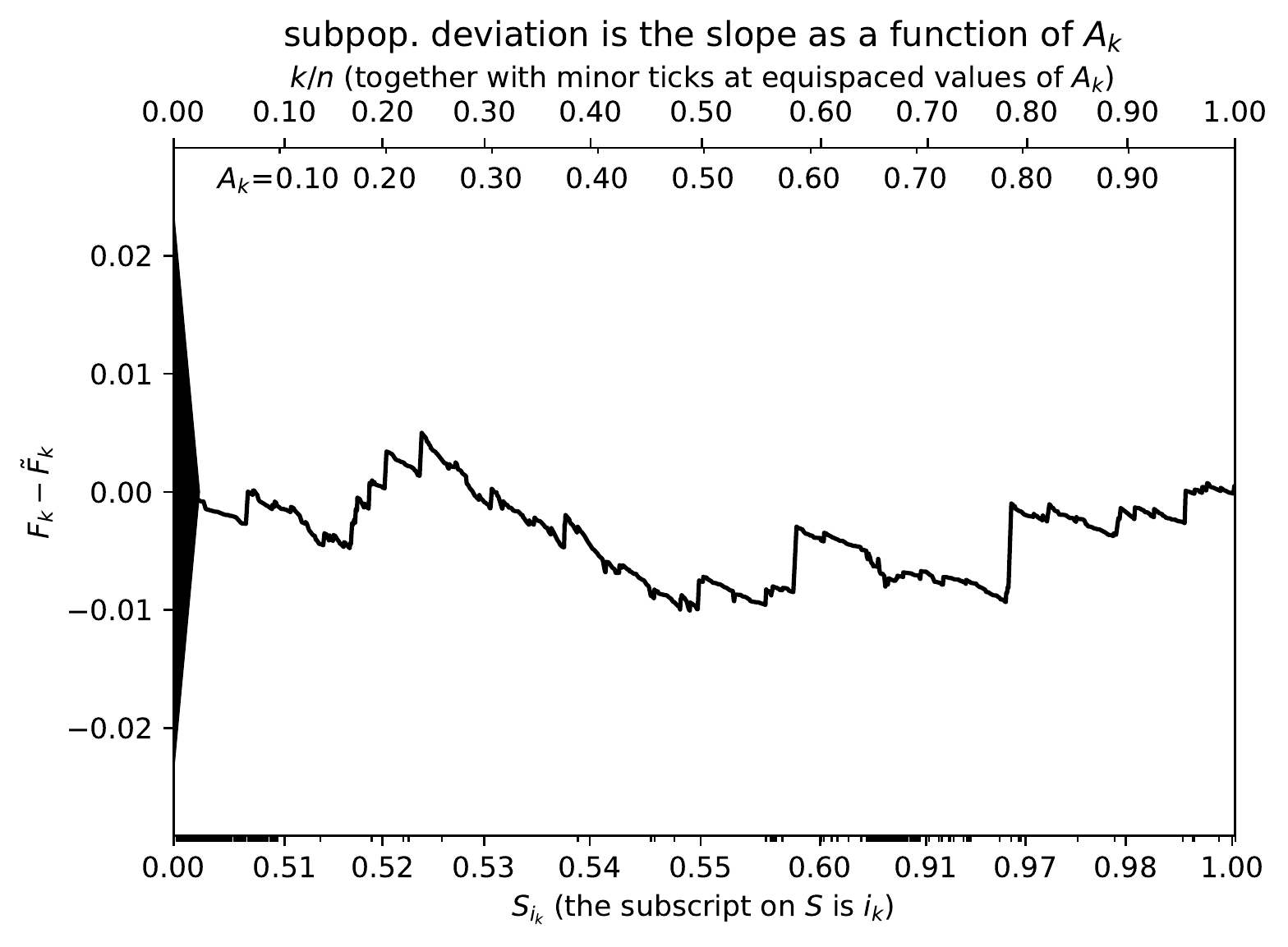}}
\quad\quad
(f) \parbox{\imsize}{\includegraphics[width=\imsize]
{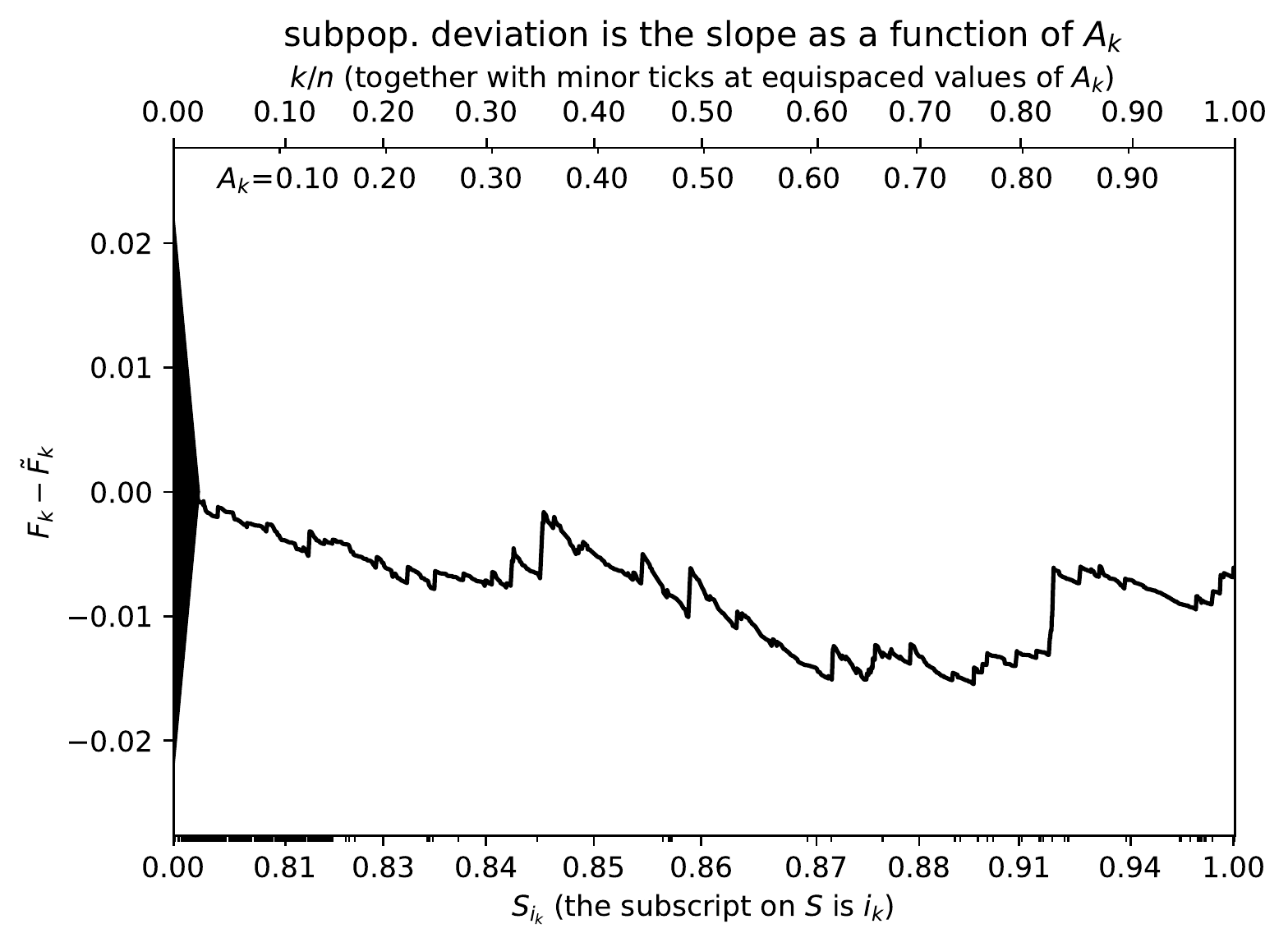}}

\parbox{\imsized}{\hfil \footnotesize $G$ = 0.01005; $H$ = 0.01505;
$G/\sigma$ = 0.8283; $H/\sigma$ = 1.241}
\parbox{\imsized}{\hfil \footnotesize $G$ = 0.01545; $H$ = 0.01602;
$G/\sigma$ = 1.341; $H/\sigma$ = 1.391}

\end{centering}
\caption{Stanislaus County ($m =$ 134,094; $n =$ 1,624),
number of related children in the household;
notice that conditioning on ``MV'' and ``NOC'' in that order (e) yields
metrics more similar to those when conditioning on ``NOC'' and ``MV'' (f)
than to those when conditioning only on ``MV'' (a) or only on ``NOC'' (b) ---
the Kolmogorov-Smirnov and Kuiper metrics when controlling
for covariates apparently depend more on the choice
of the set of covariates than on the ordering of the conditioning
within a particular set. All cases condition also on the logarithm
of the adjusted household personal income, in addition to the combinations
of covariates ``MV'' and ``NOC'' mentioned here --- see Table~\ref{labelsw},
which details subfigures~(c) and~(d), too.}
\label{stanislaus}
\end{figure}

\begin{figure}
\begin{centering}

(a) \parbox{\imsize}{\includegraphics[width=\imsize]
{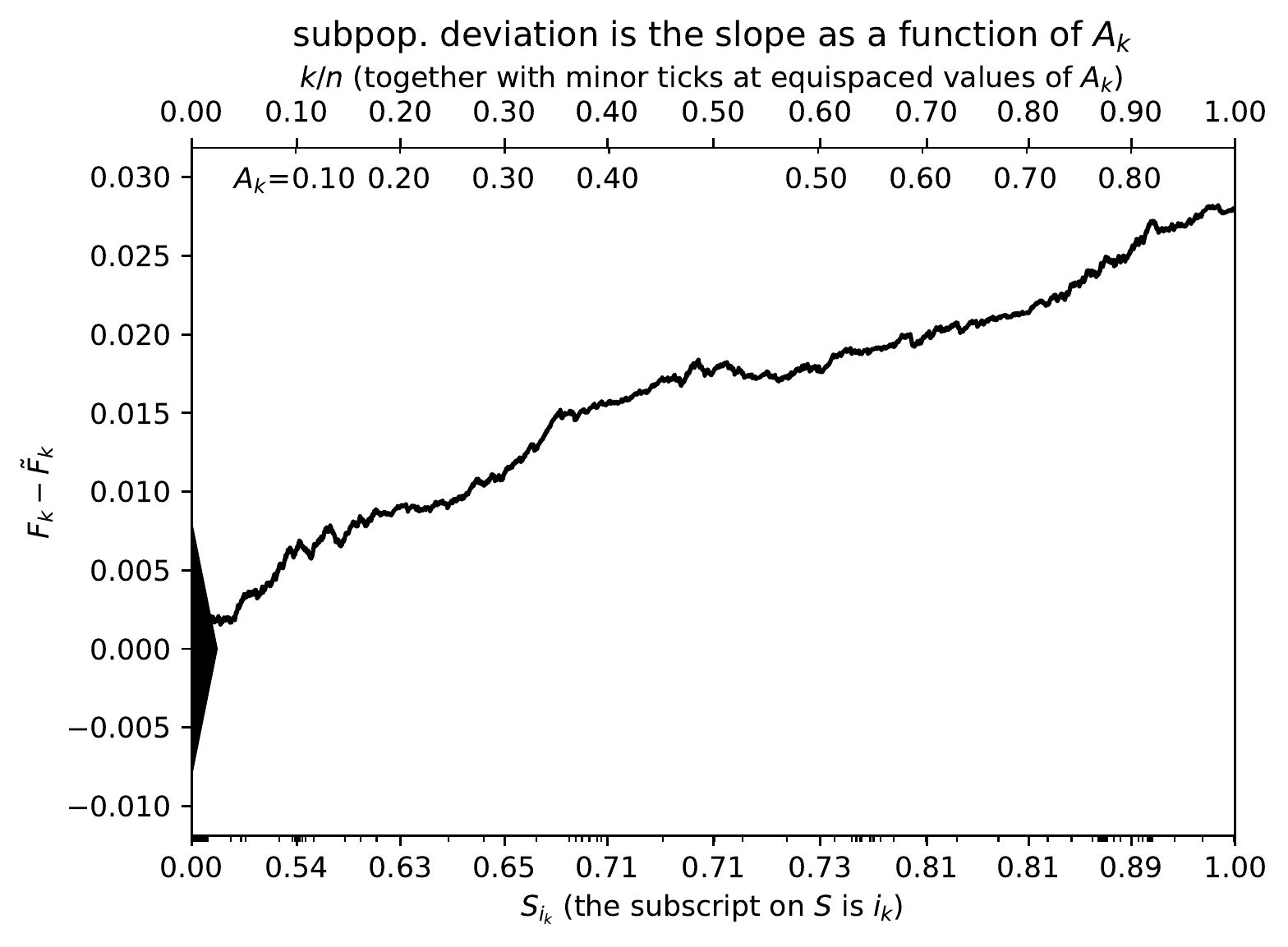}}
\quad\quad
(b) \parbox{\imsize}{\includegraphics[width=\imsize]
{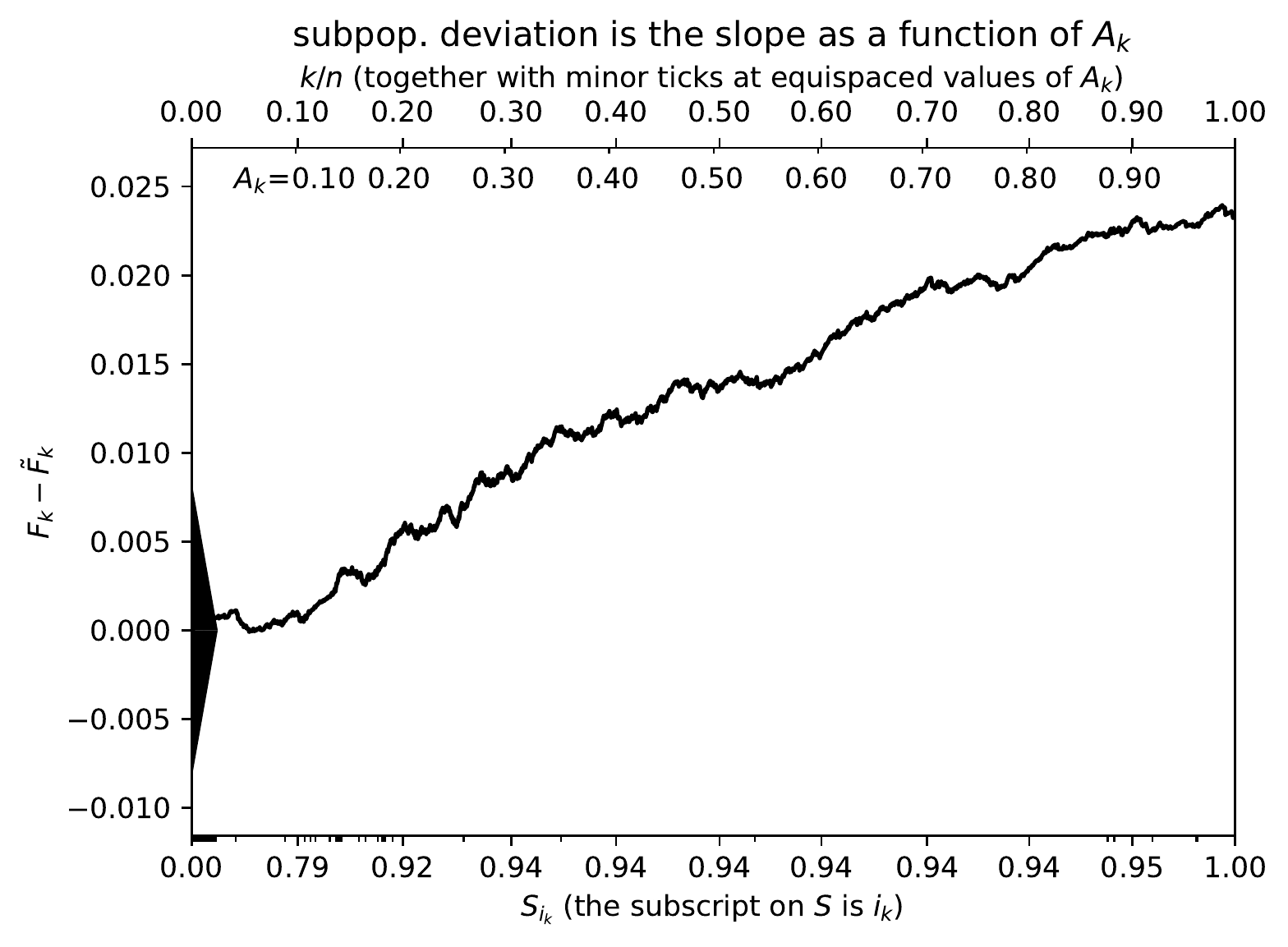}}

\parbox{\imsized}{\hfil \footnotesize $G$ = 0.02821; $H$ = 0.02825;
$G/\sigma$ = 6.840; $H/\sigma$ = 6.849}
\parbox{\imsized}{\hfil \footnotesize $G$ = 0.02394; $H$ = 0.02407;
$G/\sigma$ = 5.738; $H/\sigma$ = 5.769}

\vspace{\vertsep}

(c) \parbox{\imsize}{\includegraphics[width=\imsize]
{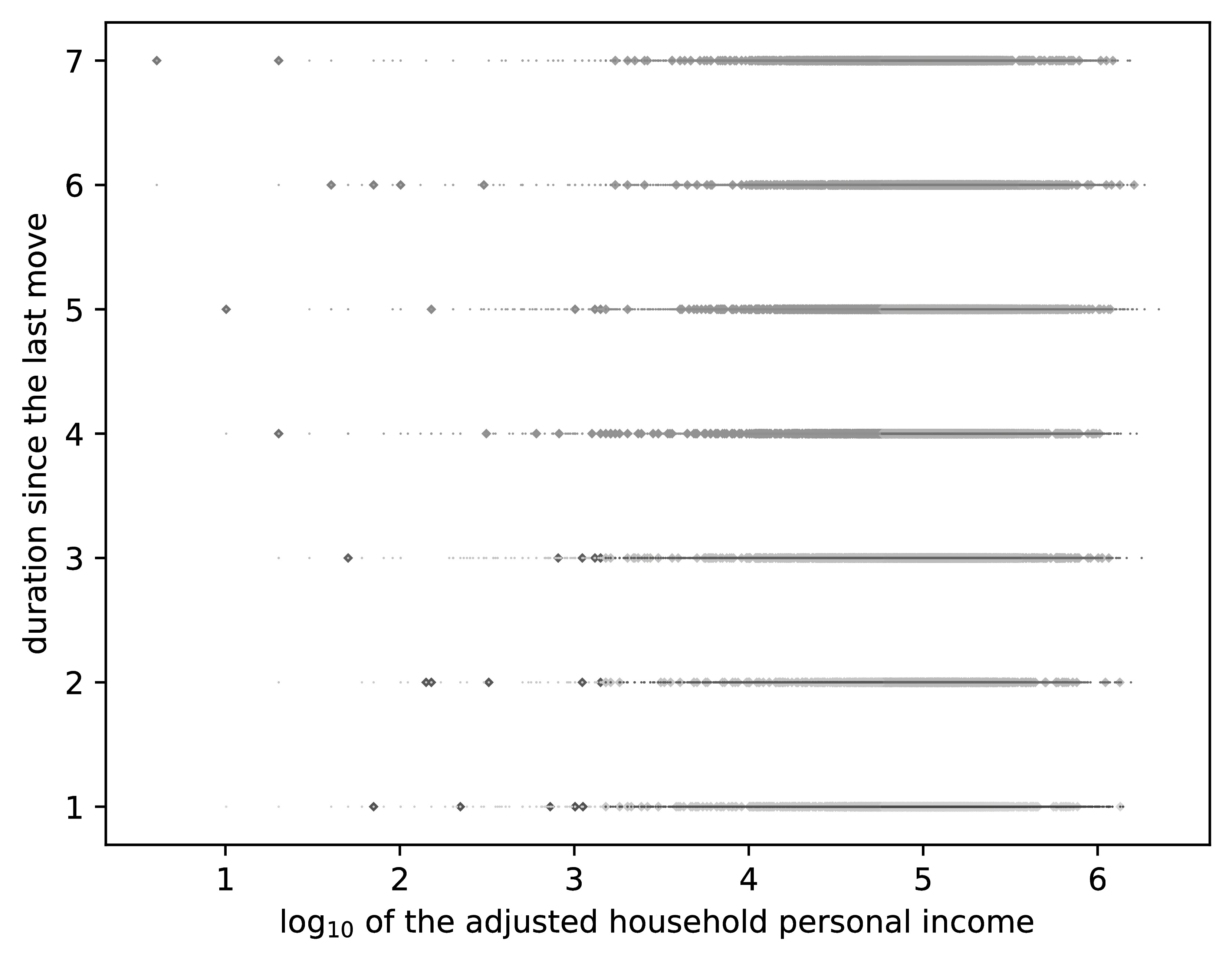}}
\quad\quad
(d) \parbox{\imsize}{\includegraphics[width=\imsize]
{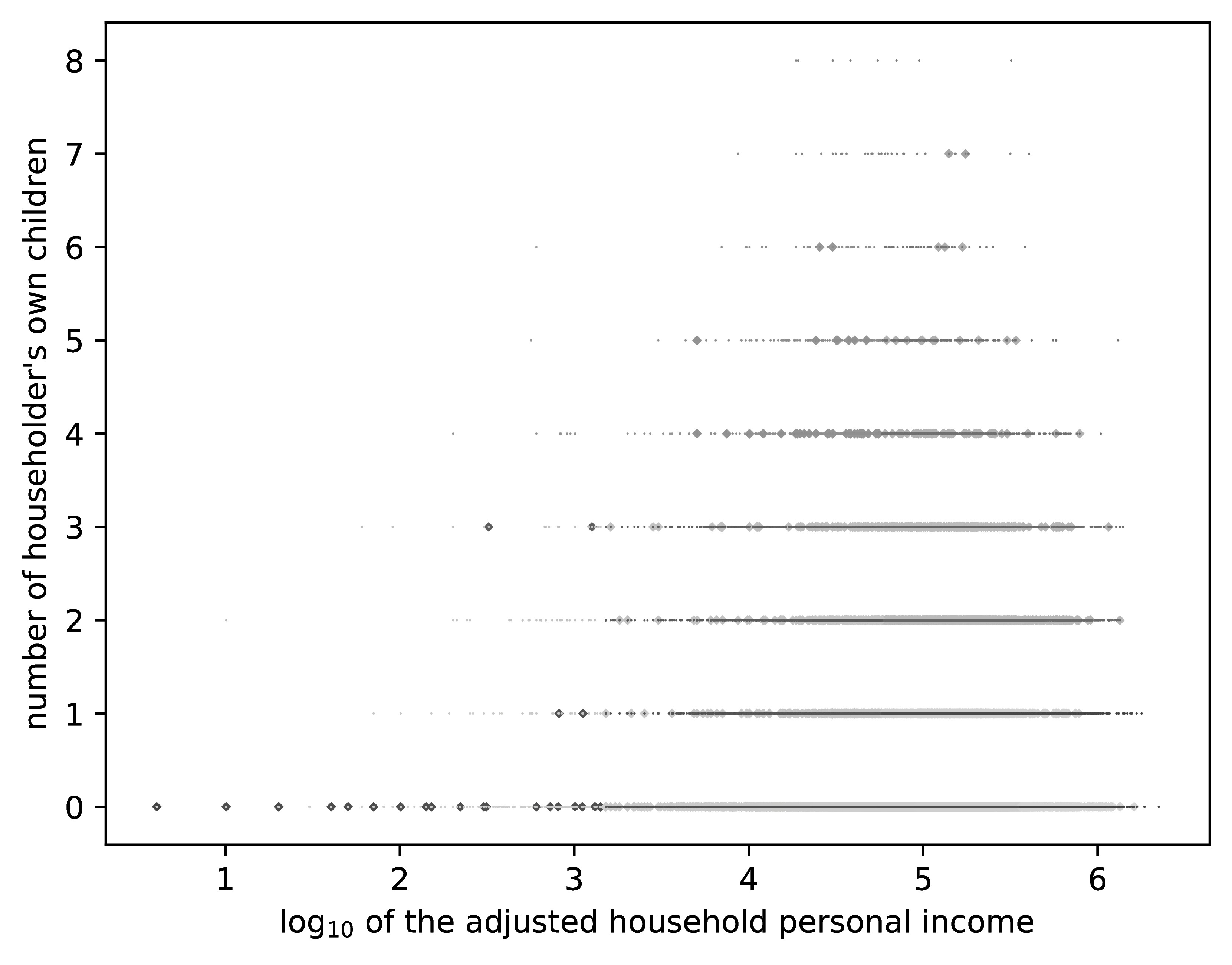}}

\vspace{\vertsep}

(e) \parbox{\imsize}{\includegraphics[width=\imsize]
{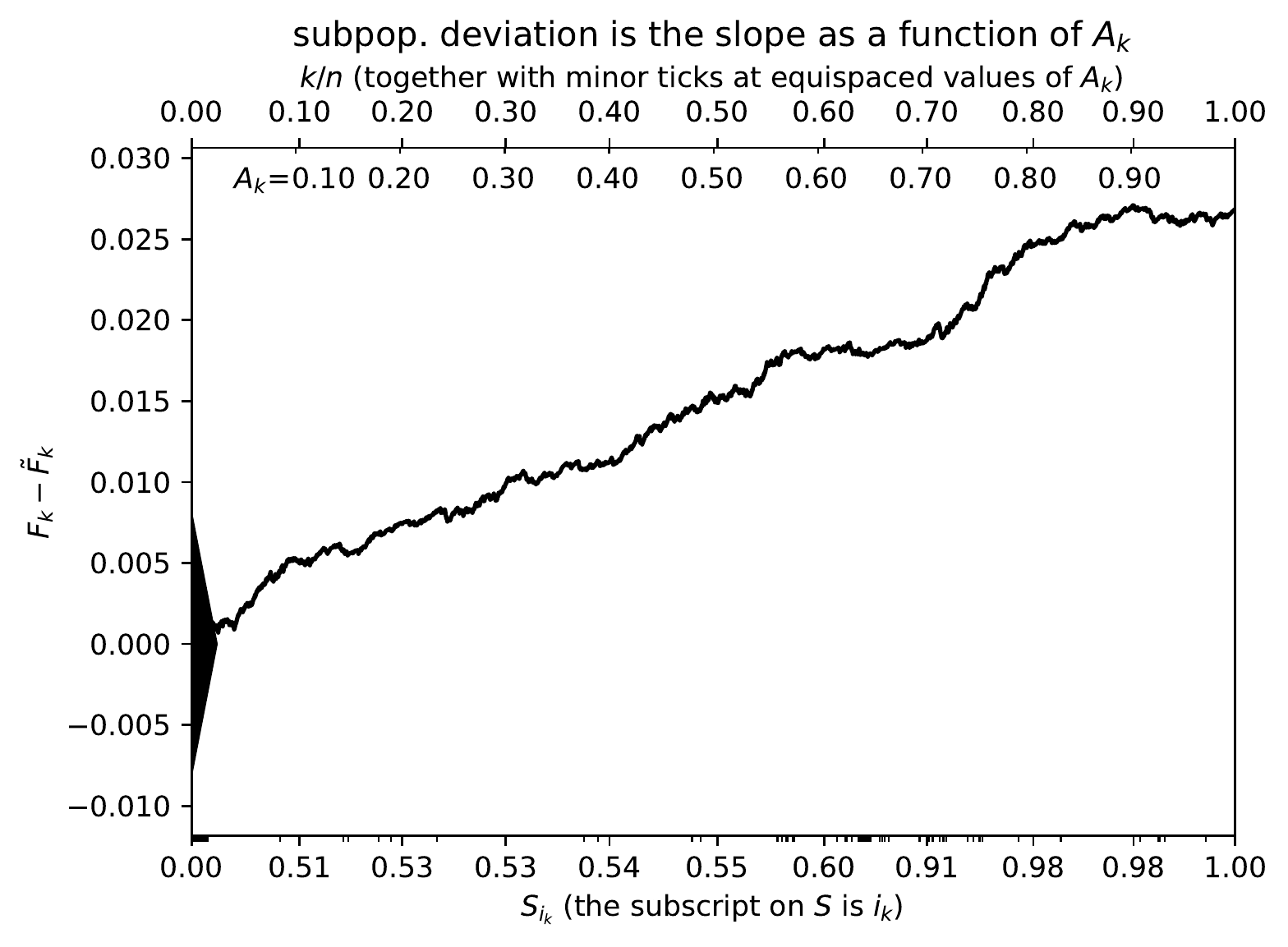}}
\quad\quad
(f) \parbox{\imsize}{\includegraphics[width=\imsize]
{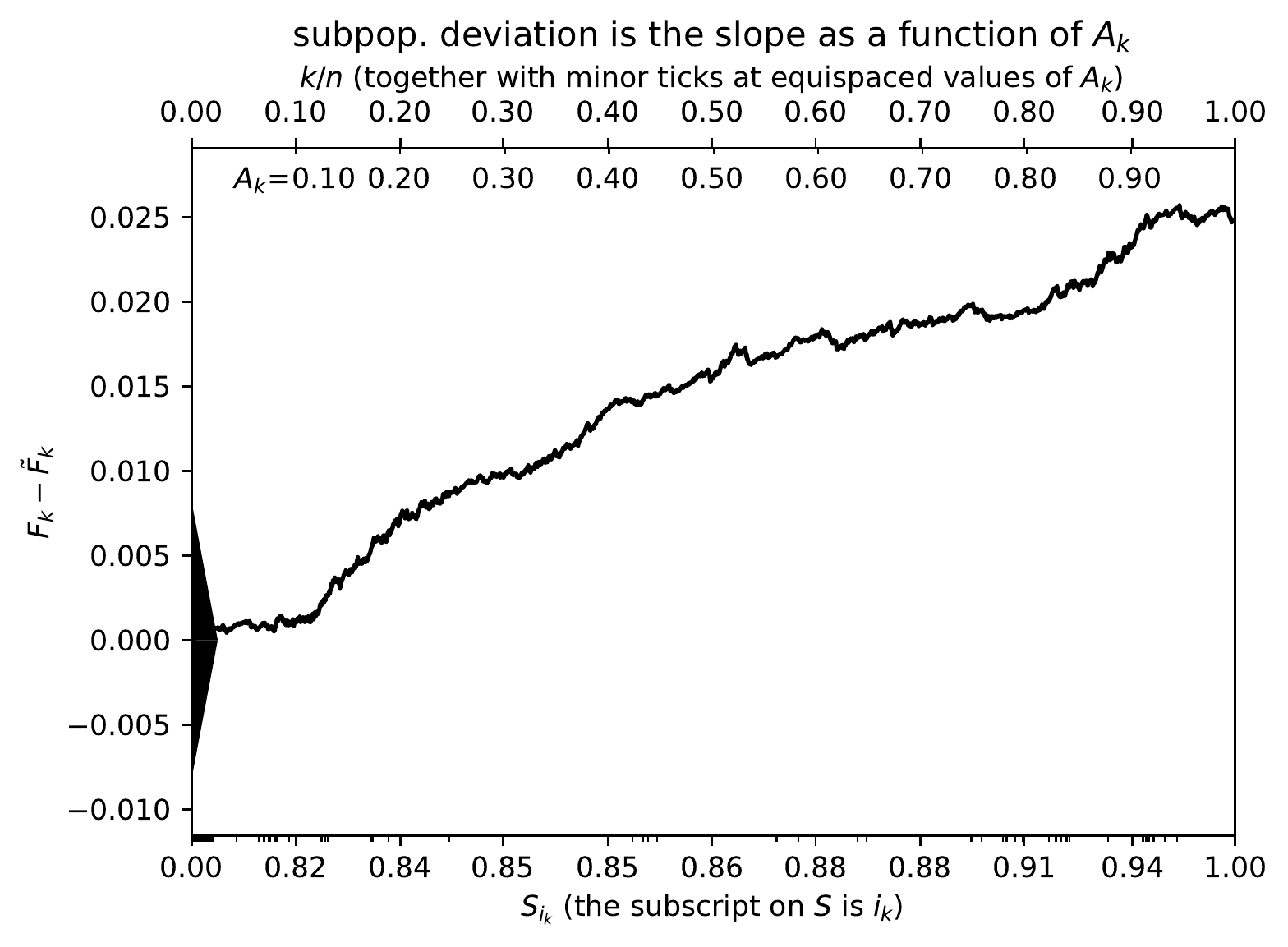}}

\parbox{\imsized}{\hfil \footnotesize $G$ = 0.02709; $H$ = 0.02713;
$G/\sigma$ = 6.521; $H/\sigma$ = 6.529}
\parbox{\imsized}{\hfil \footnotesize $G$ = 0.02570; $H$ = 0.02573;
$G/\sigma$ = 6.287; $H/\sigma$ = 6.295}

\end{centering}
\caption{Orange County ($m =$ 134,094; $n =$ 10,680),
household has high-speed (broadband) access to the Internet;
all four cumulative graphs are fairly similar.
See Table~\ref{labelsw} for detailed descriptions of the subfigures.}
\label{orange}
\end{figure}

\begin{figure}
\begin{centering}

(a) \parbox{\imsize}{\includegraphics[width=\imsize]
{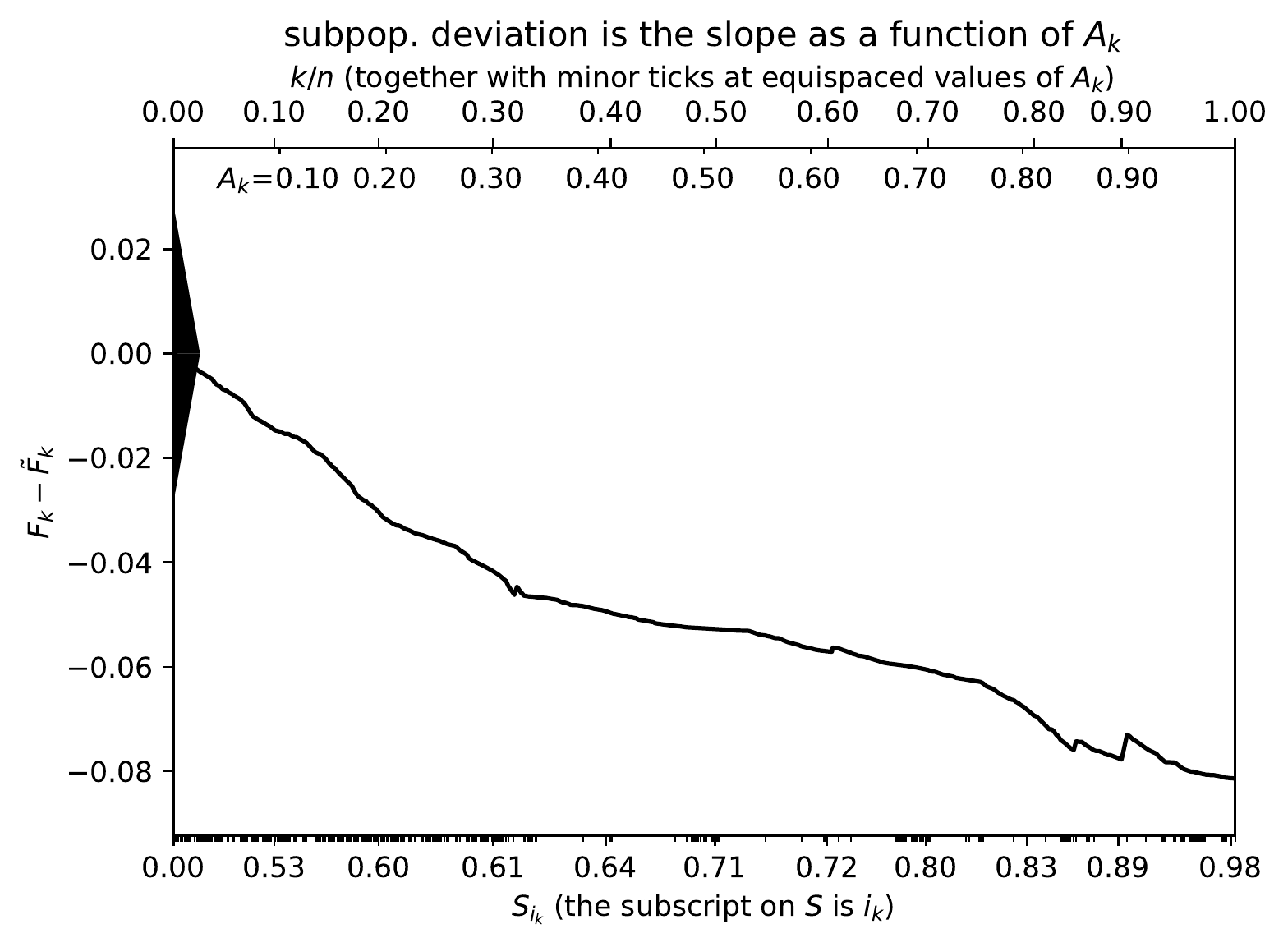}}
\quad\quad
(b) \parbox{\imsize}{\includegraphics[width=\imsize]
{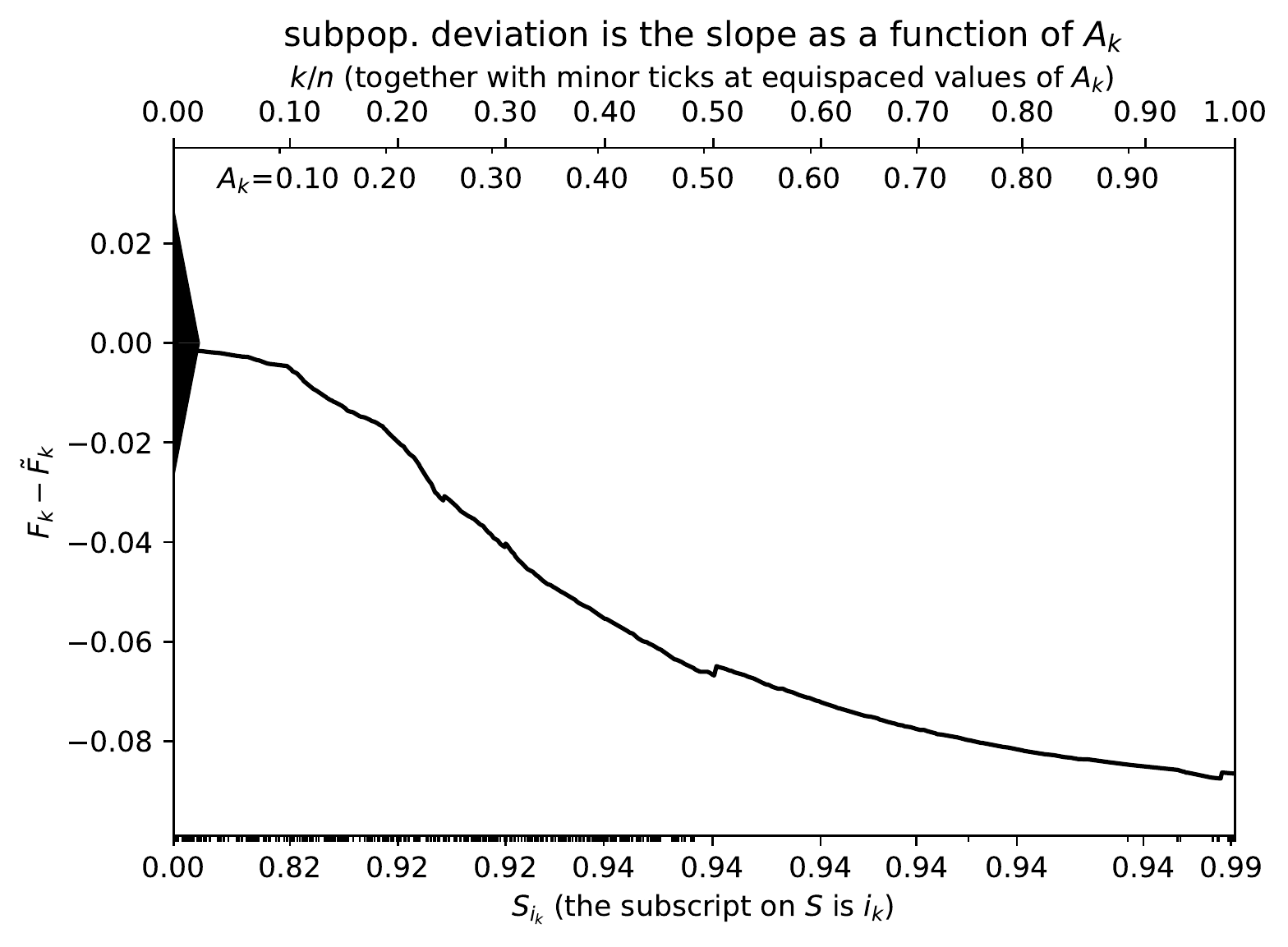}}

\parbox{\imsized}{\hfil \footnotesize $G$ = 0.08140; $H$ = 0.08140;
$G/\sigma$ = 5.721; $H/\sigma$ = 5.721}
\parbox{\imsized}{\hfil \footnotesize $G$ = 0.08751; $H$ = 0.08751;
$G/\sigma$ = 6.321; $H/\sigma$ = 6.321}

\vspace{\vertsep}

(c) \parbox{\imsize}{\includegraphics[width=\imsize]
{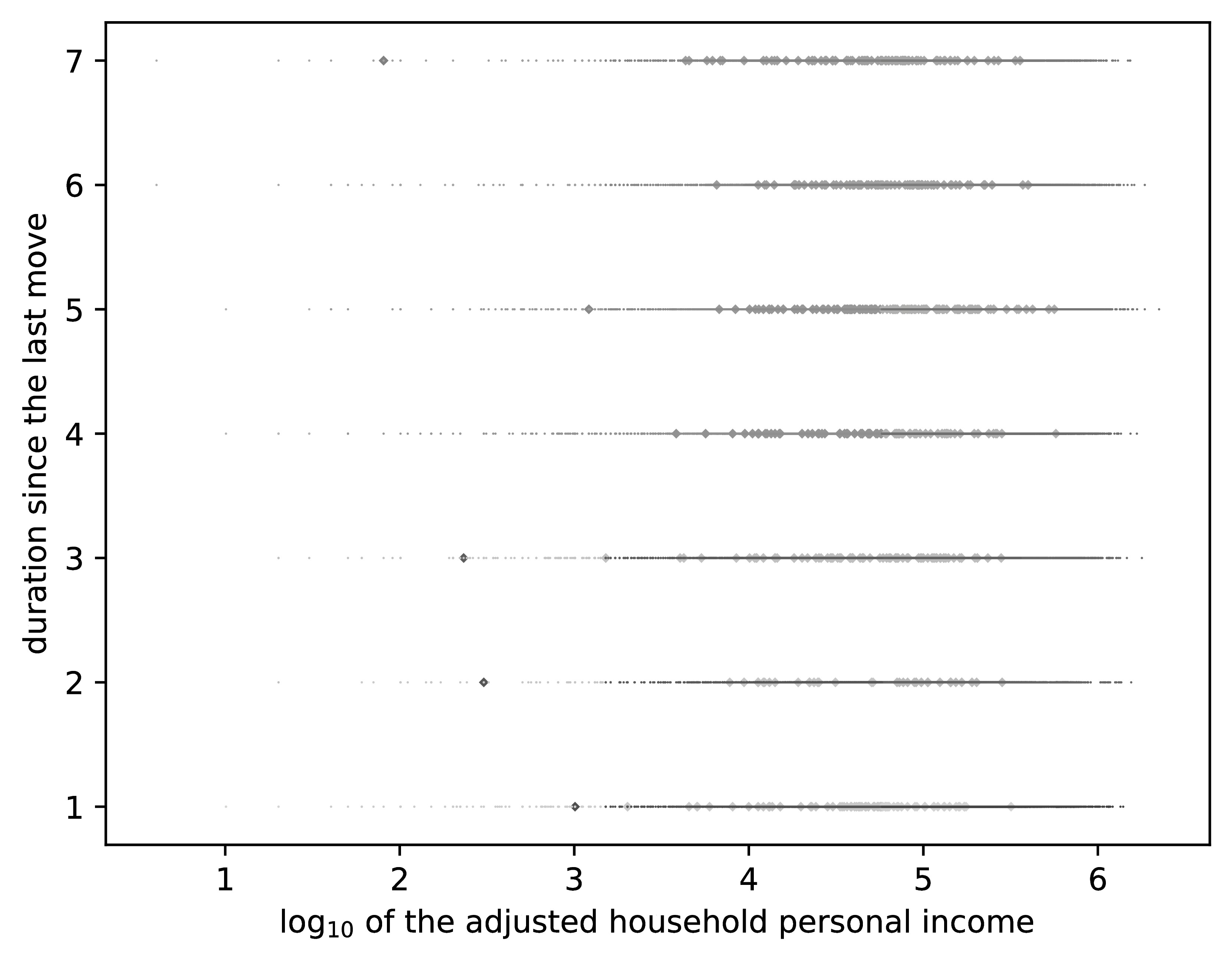}}
\quad\quad
(d) \parbox{\imsize}{\includegraphics[width=\imsize]
{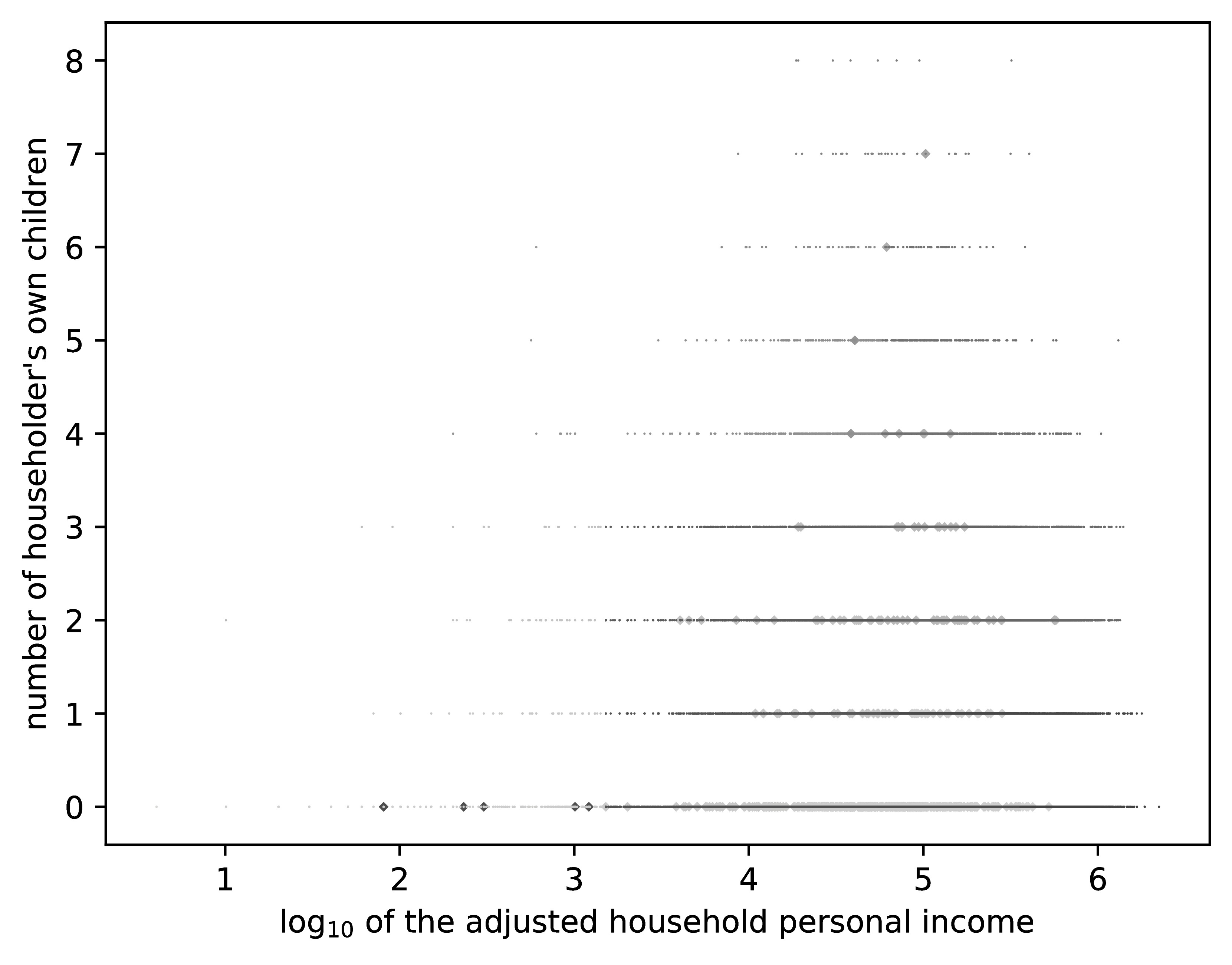}}

\vspace{\vertsep}

(e) \parbox{\imsize}{\includegraphics[width=\imsize]
{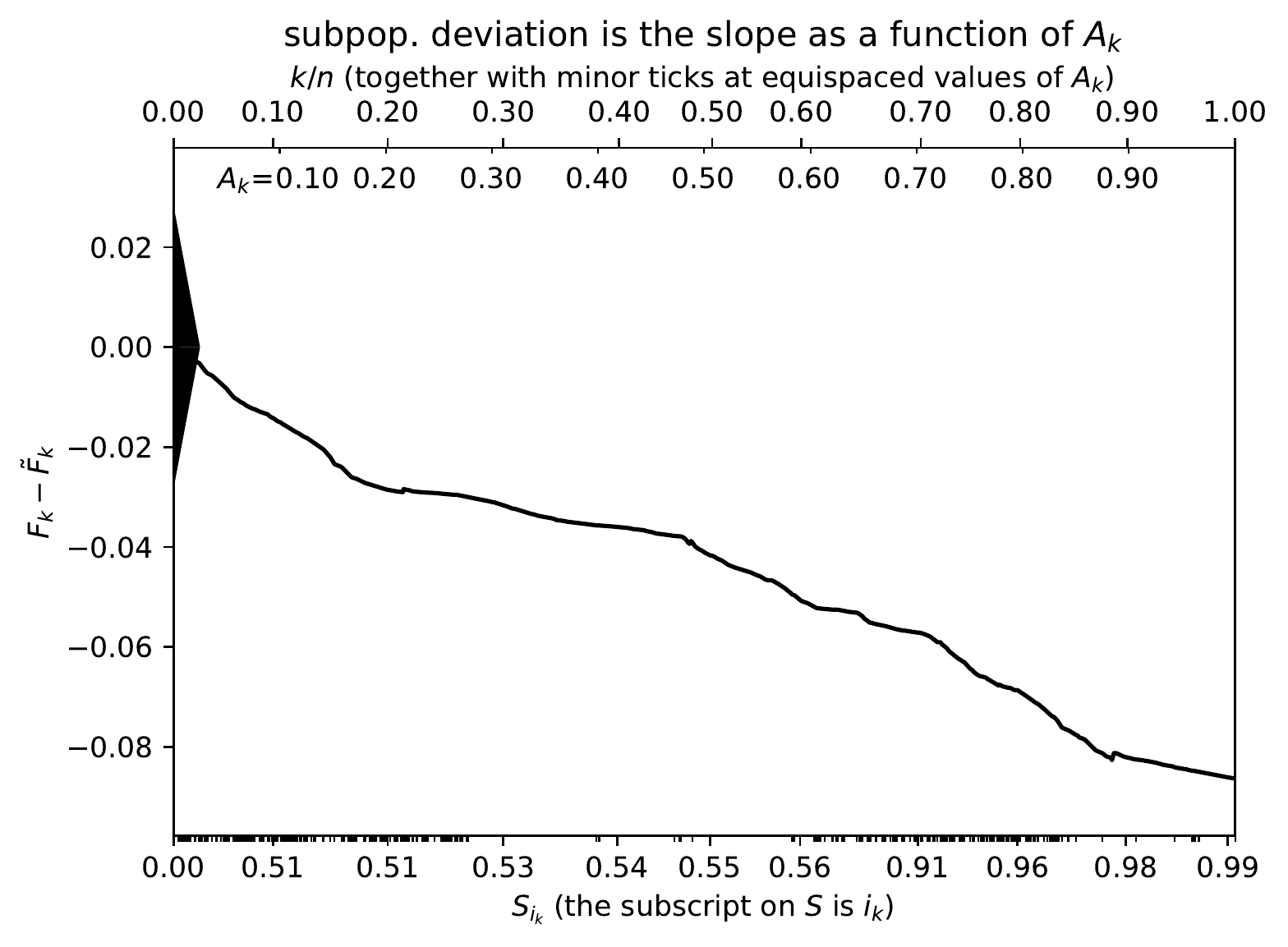}}
\quad\quad
(f) \parbox{\imsize}{\includegraphics[width=\imsize]
{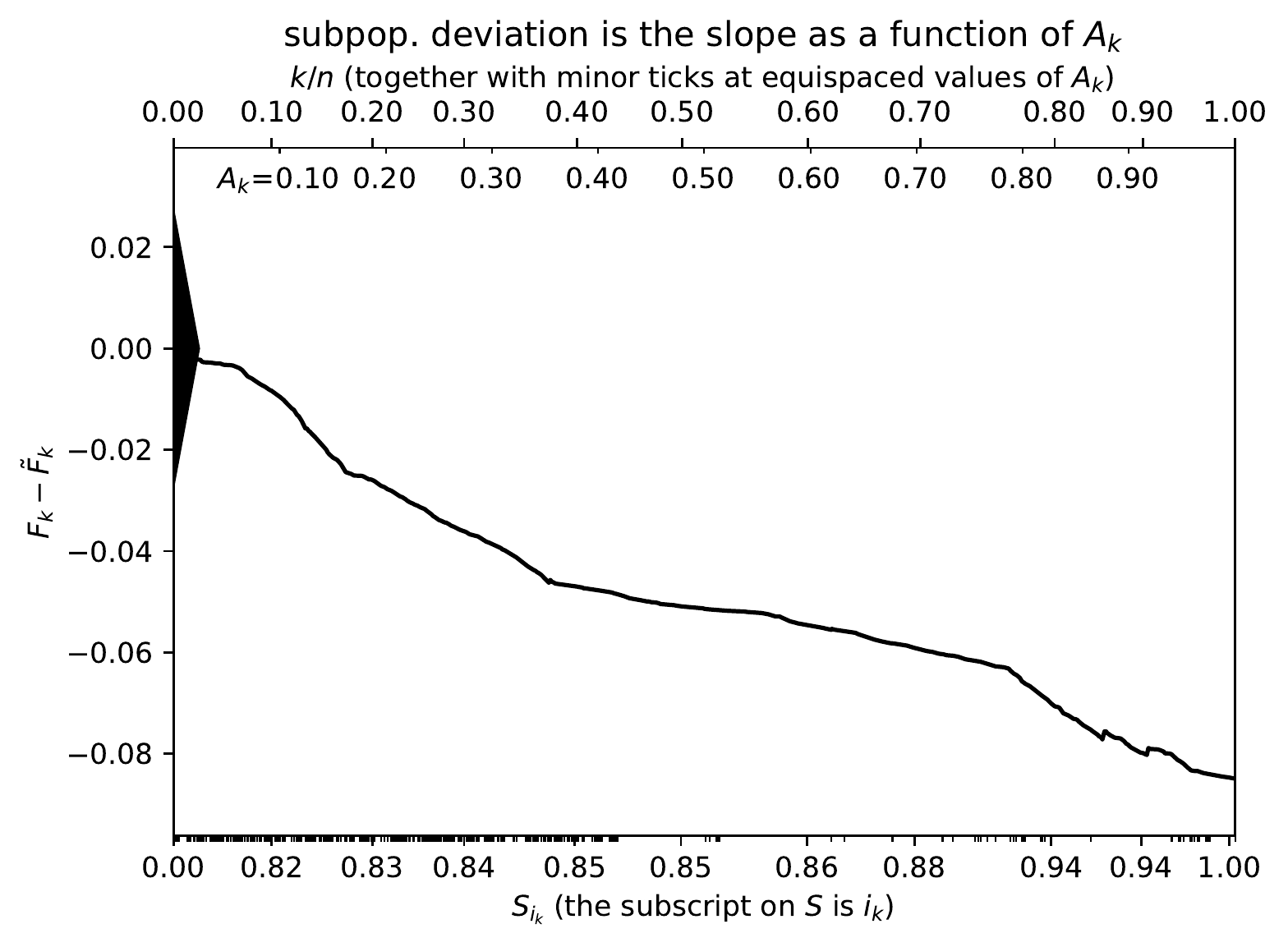}}

\parbox{\imsized}{\hfil \footnotesize $G$ = 0.08636; $H$ = 0.08636;
$G/\sigma$ = 6.083; $H/\sigma$ = 6.083}
\parbox{\imsized}{\hfil \footnotesize $G$ = 0.08488; $H$ = 0.08488;
$G/\sigma$ = 6.008; $H/\sigma$ = 6.008}

\end{centering}
\caption{Humboldt County ($m =$ 134,094; $n =$ 583),
household includes someone older than 13 who speaks English very well.
Once again, conditioning on ``MV'' and ``NOC'' in that order (e) yields
metrics more similar to those when conditioning on ``NOC'' and ``MV'' (f)
than to those when conditioning only on ``MV'' (a) or only on ``NOC'' (b) ---
the Kolmogorov-Smirnov and Kuiper metrics when controlling
for covariates seem to depend more on the choice
of the set of covariates than on the ordering of the conditioning
within a particular set. All cases condition also on the logarithm
of the adjusted household personal income, in addition to the combinations
of covariates ``MV'' and ``NOC'' mentioned here --- see Table~\ref{labelsw},
which details subfigures~(c) and~(d), too.}
\label{humboldt}
\end{figure}

\begin{figure}
\begin{centering}

(a) \parbox{\imsize}{\includegraphics[width=\imsize]
{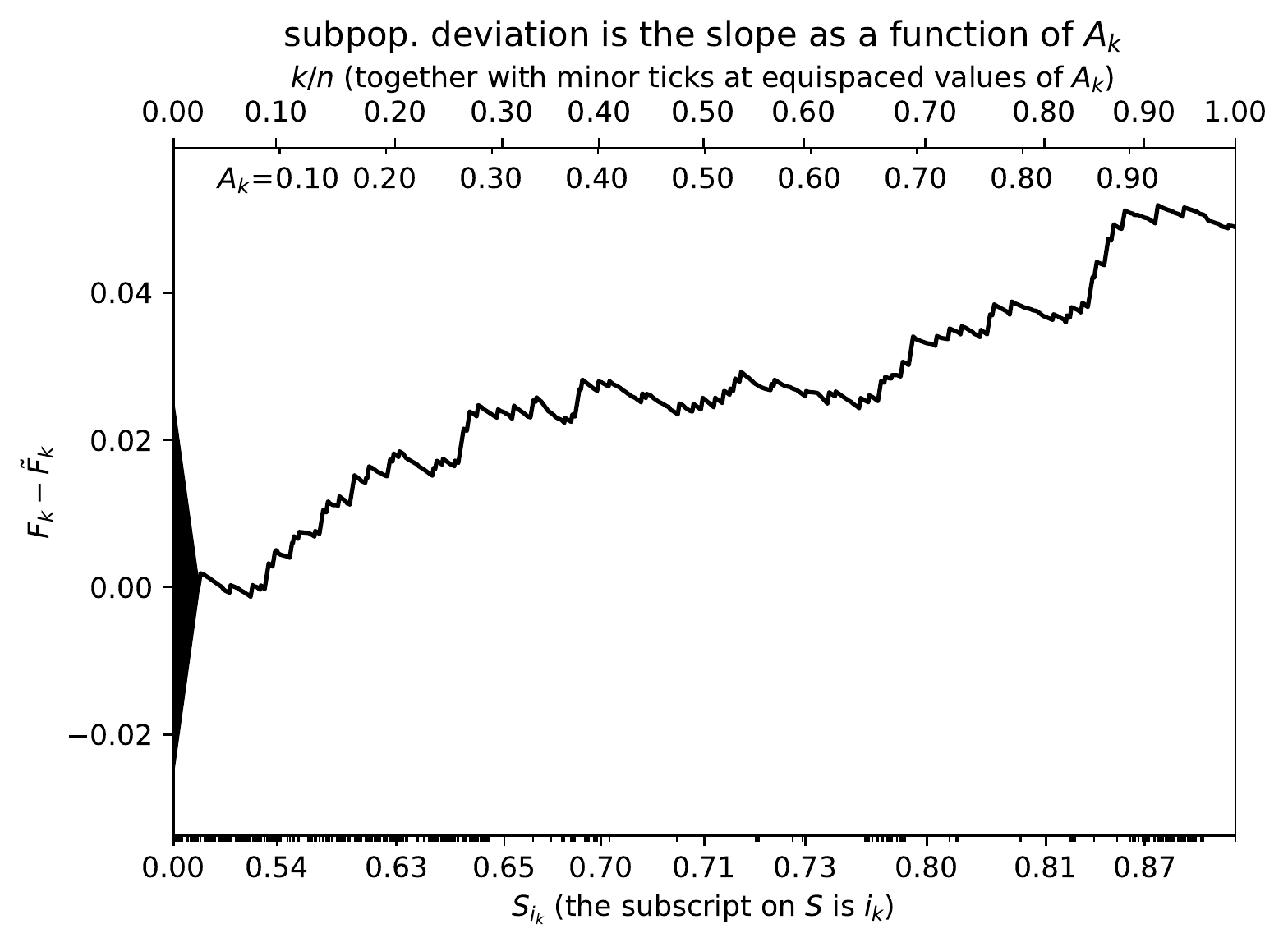}}
\quad\quad
(b) \parbox{\imsize}{\includegraphics[width=\imsize]
{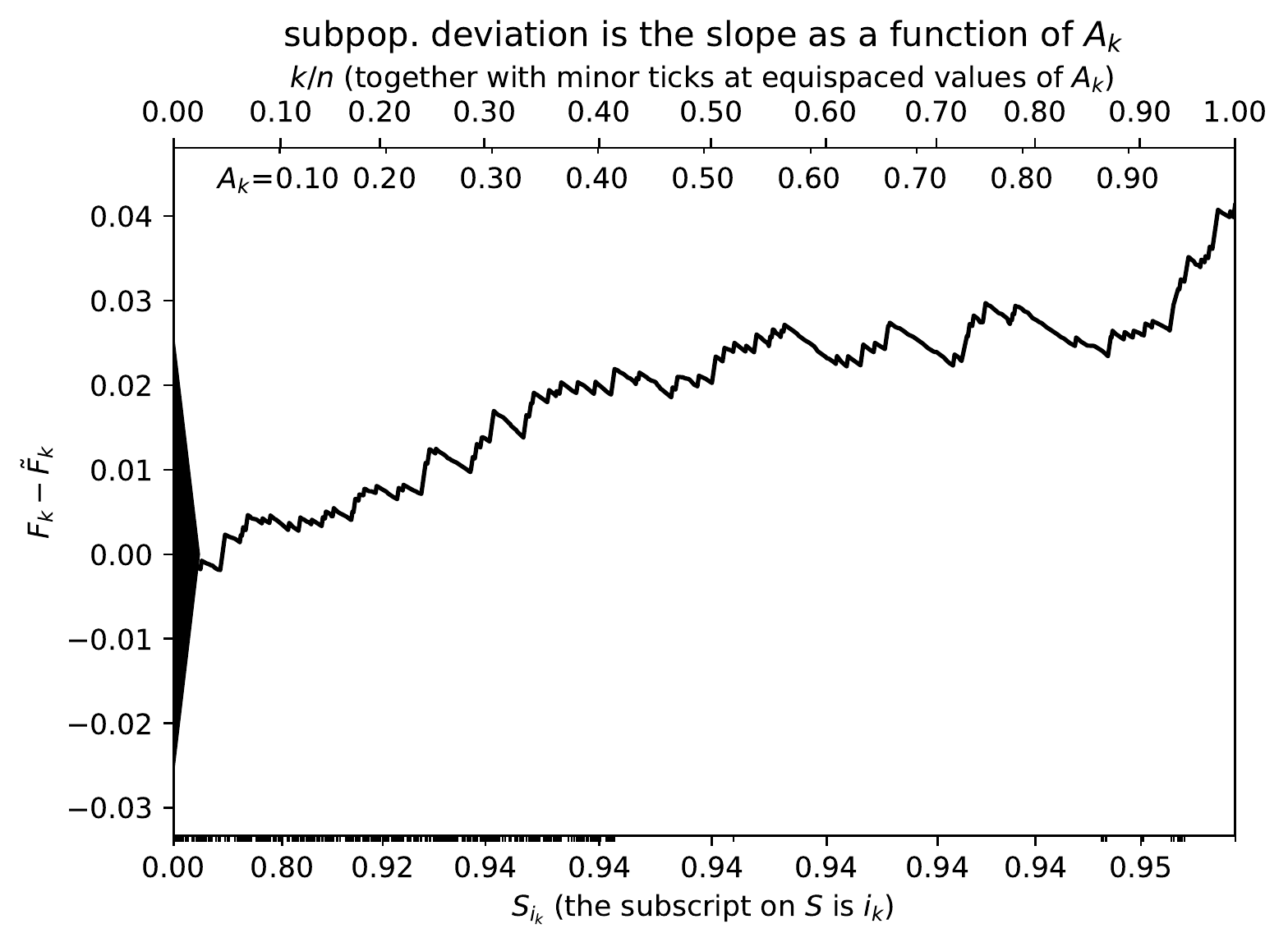}}

\parbox{\imsized}{\hfil \footnotesize $G$ = 0.05191; $H$ = 0.05346;
$G/\sigma$ = 3.995; $H/\sigma$ = 4.114}
\parbox{\imsized}{\hfil \footnotesize $G$ = 0.04128; $H$ = 0.04314;
$G/\sigma$ = 3.114; $H/\sigma$ = 3.254}

\vspace{\vertsep}

(c) \parbox{\imsize}{\includegraphics[width=\imsize]
{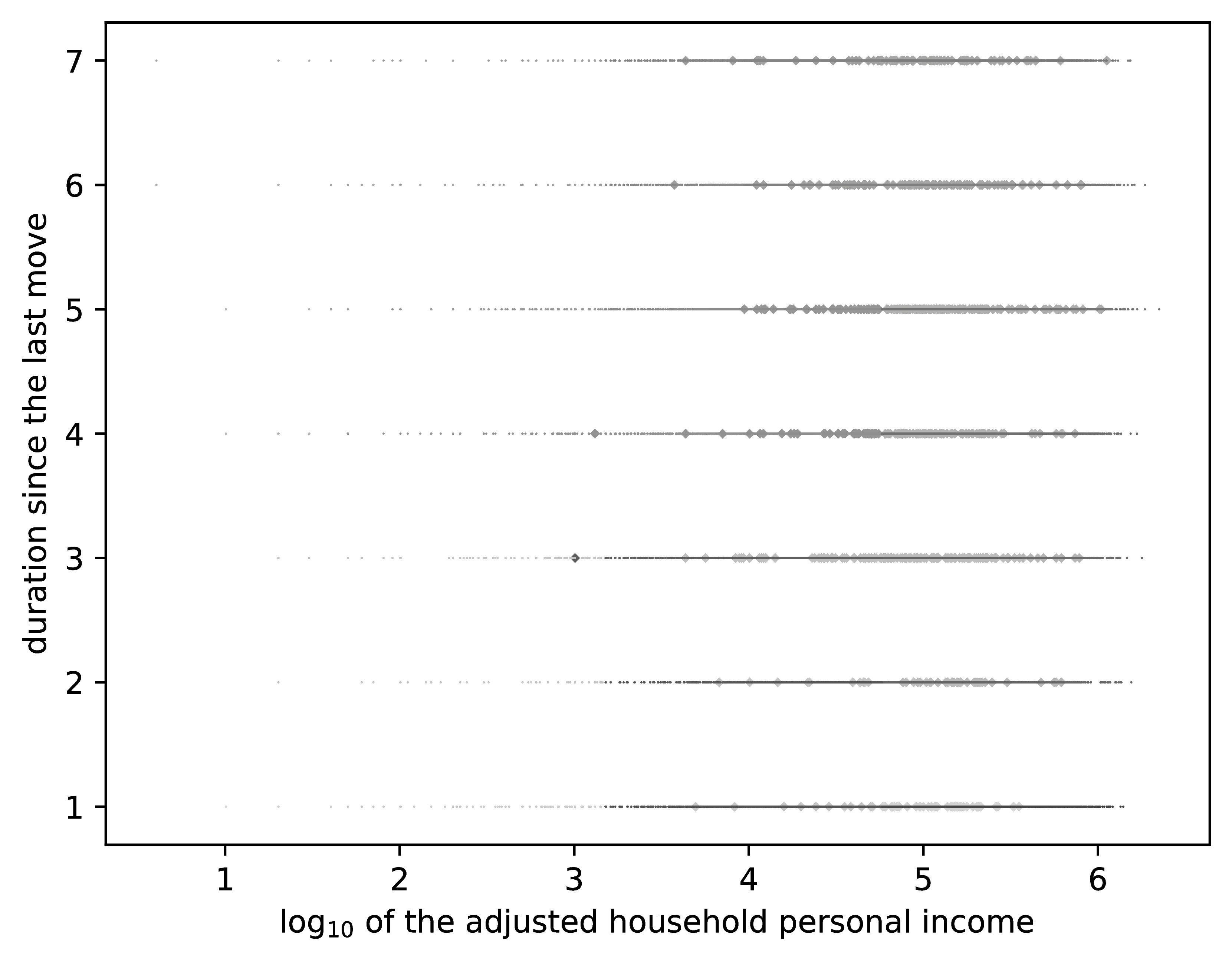}}
\quad\quad
(d) \parbox{\imsize}{\includegraphics[width=\imsize]
{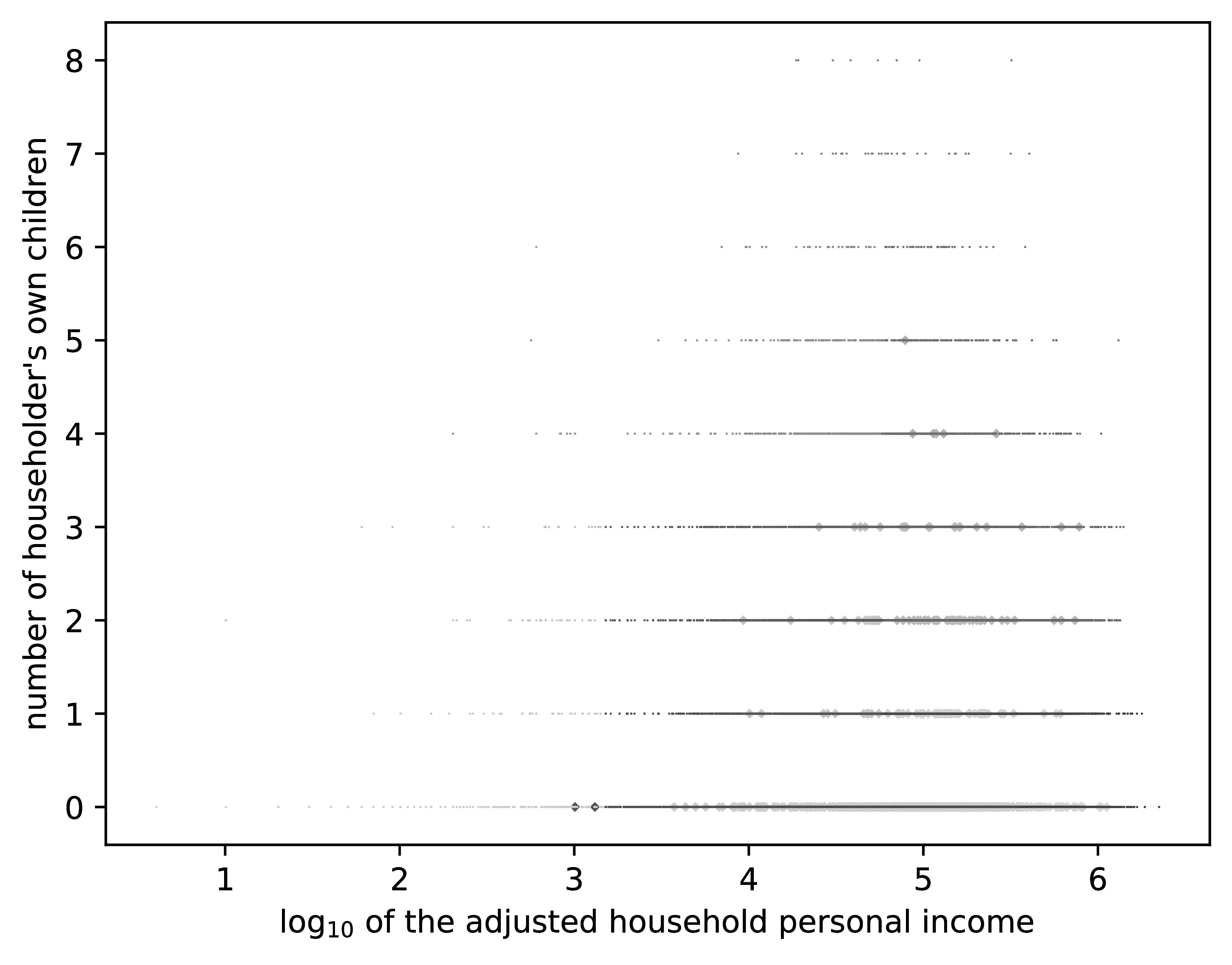}}

\vspace{\vertsep}

(e) \parbox{\imsize}{\includegraphics[width=\imsize]
{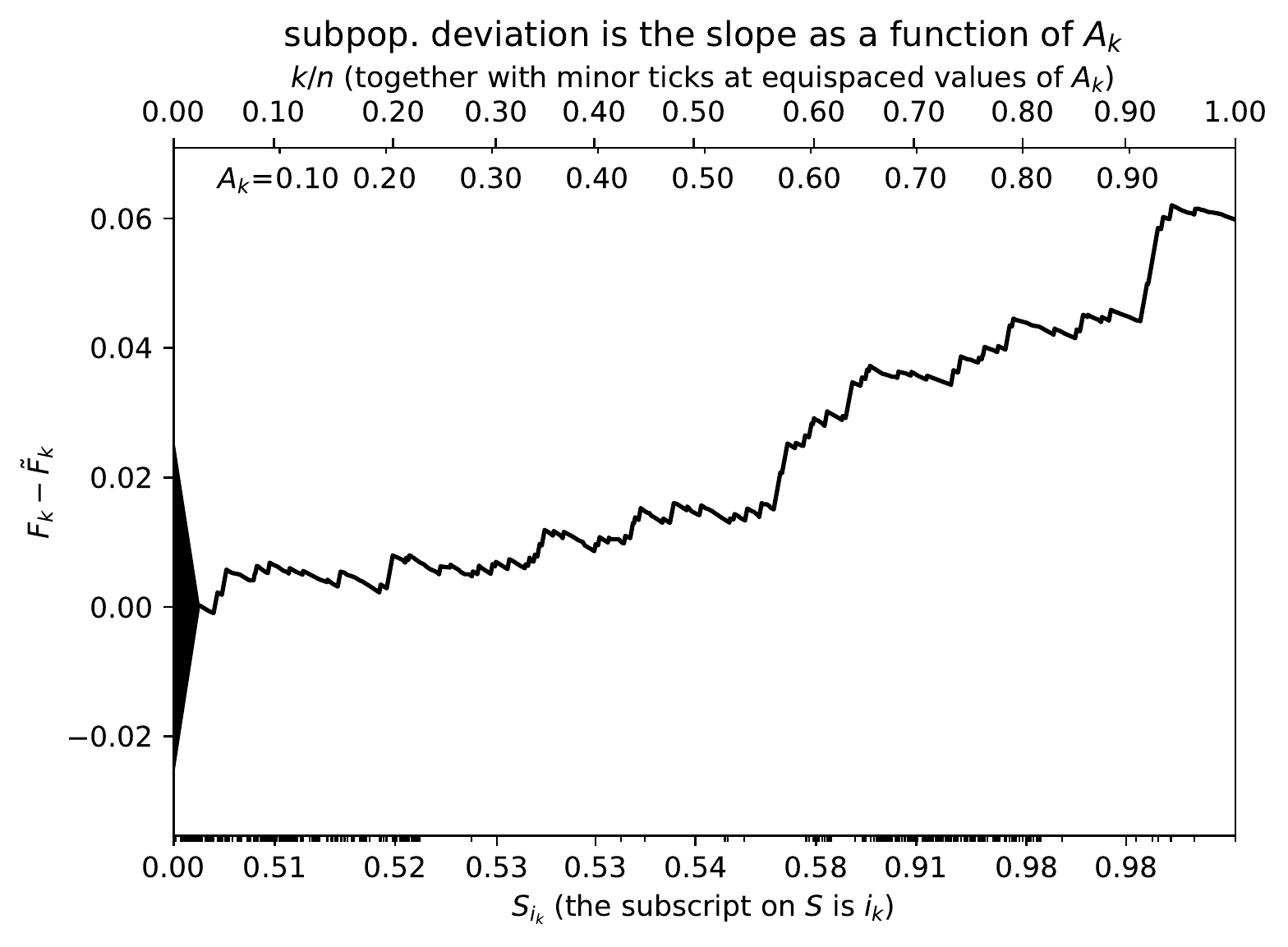}}
\quad\quad
(f) \parbox{\imsize}{\includegraphics[width=\imsize]
{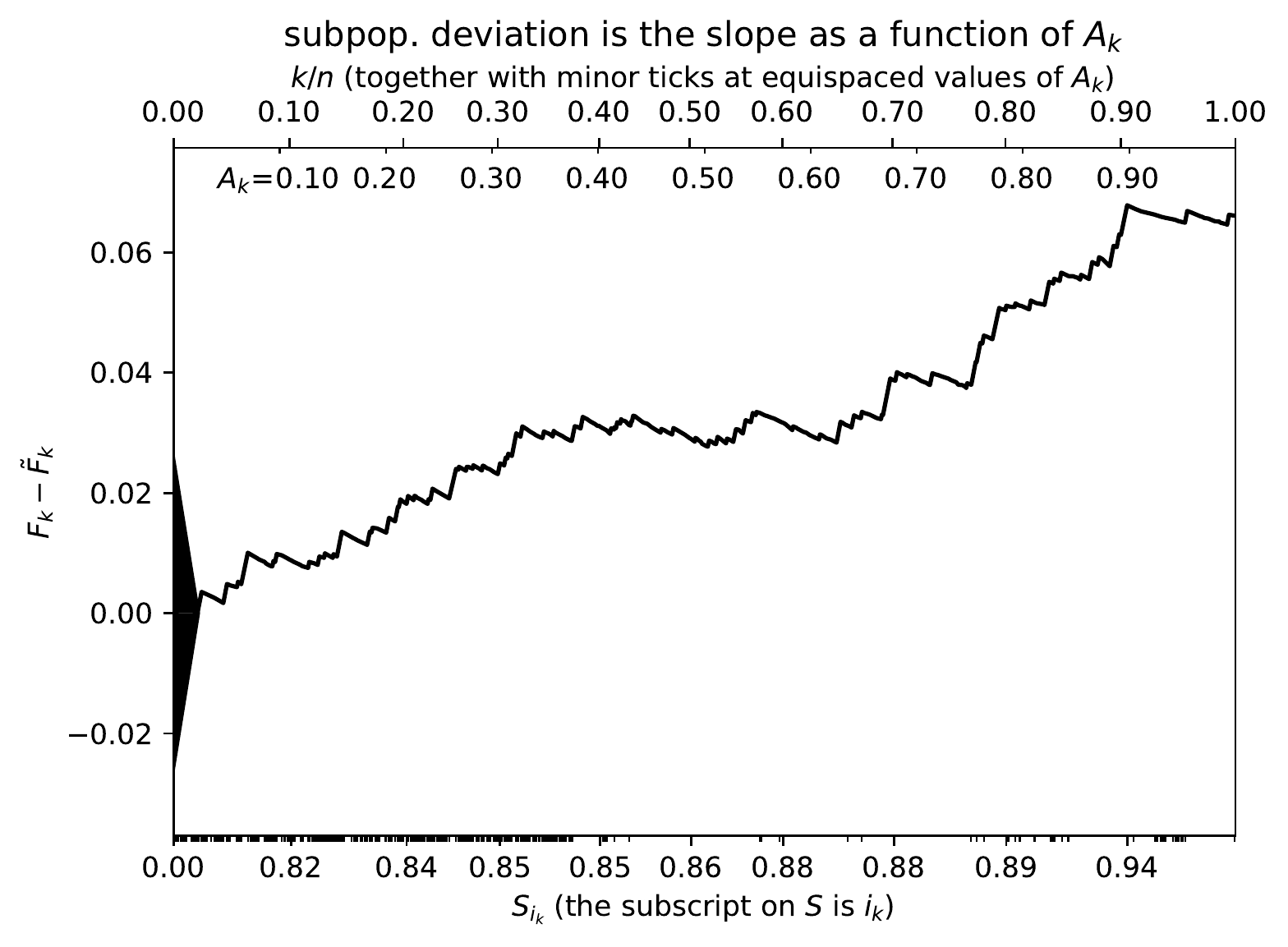}}

\parbox{\imsized}{\hfil \footnotesize $G$ = 0.06203; $H$ = 0.06320;
$G/\sigma$ = 4.685; $H/\sigma$ = 4.773}
\parbox{\imsized}{\hfil \footnotesize $G$ = 0.06783; $H$ = 0.06957;
$G/\sigma$ = 4.935; $H/\sigma$ = 5.061}

\end{centering}
\caption{Napa County ($m =$ 134,094; $n =$ 679),
household obtains access to the Internet via satellite.
Yet again, conditioning on ``MV'' and ``NOC'' in that order (e) yields
metrics closer to those when conditioning on ``NOC'' and ``MV'' (f)
than to those when conditioning only on ``MV'' (a) or only on ``NOC'' (b) ---
the Kolmogorov-Smirnov and Kuiper metrics when controlling
for covariates appear to depend more on the choice
of the set of covariates than on the ordering of the conditioning
within a particular set. All cases condition also on the logarithm
of the adjusted household personal income, in addition to the combinations
of covariates ``MV'' and ``NOC'' mentioned here --- see Table~\ref{labelsw},
which details subfigures~(c) and~(d), too.}
\label{napa}
\end{figure}

This subsection presents Figures~\ref{los_angeles}--\ref{napa};
Table~\ref{labelsw} gives details about each subfigure
of Figures~\ref{los_angeles}--\ref{napa}.
These figures consider weighted sampling of the U.S.\ population,
detailed as follows:

As with most years, in 2019 the U.S.\ Census Bureau collected
detailed information on an extensive sample of households,
in its American Community Survey.\footnote{The data
from the American Community Survey is available
at \url{https://www.census.gov/programs-surveys/acs/microdata.html}}
The sampling in this survey is weighted; to analyze this data,
we retain only those households whose weights (``WGTP'' in the microdata)
are strictly positive (thus also eliminating consideration of group quarters),
and discard any household whose personal income (``HINCP'') is zero
or for which the adjustment factor to income (``ADJINC'') is flagged
as missing (thus eliminating consideration of vacant addresses, too).
For the full population, we use the households from all counties
in the state of California put together; for the subpopulation,
we use the households from the individual county specified in the caption
to the corresponding figure.
We consider conditioning on various tuples of covariates, with each covariate
normalized to range from 0 to 1 prior to ordering via the Hilbert curve
(the normalization simply divides by the maximum of all values).
All tuples considered include the logarithm
of the adjusted household personal income
(the adjusted income is ``HINCP'' times ``ADJINC,'' divided by one million
when ``ADJINC'' omits its decimal point in the integer-valued microdata),
randomly perturbed by about one part in $10^8$ to ensure uniqueness.
The two other covariates considered as controls are
(1) the data set's ordinal, integer encoding of the duration
since the last move (``MV'' in the microdata) and
(2) the number of the householder's own children (``NOC'' in the microdata).
For the responses $R_1$,~$R_2$, \dots, $R_m$ (where $m =$ 134,094),
we use the variates specified at the beginnings of the captions of the figures.
As in Subsection~\ref{subpop}, $m$ is the number of members
of the full population, while $n$ is the number in the subpopulation
(specified in the caption to each figure).
Figures~\ref{los_angeles}--\ref{napa} present six examples.
Table~\ref{labelsw} describes each subfigure.

As in Subsections~\ref{synthetic} and~\ref{kddcup98},
Figures~\ref{los_angeles}--\ref{napa} indicate that 
the normalized scalar summary statistics ($G/\sigma$ and $H/\sigma$)
depend more on which covariates are conditioned on
than on the ordering of the conditioning within a particular set of covariates.
The captions of the figures discuss these results in greater detail.

\clearpage

\begin{table}
\caption{Description of subfigures
for Figures~\ref{folding}--\ref{folding_normal};
the averages and percentages mentioned in the table refer to the individual's
Census block, and every covariate is normalized to range from 0 to 1.
The conditioning (via the Hilbert curve) happens in the order specified
in the table.}
\label{labels}
\begin{center}
\begin{tabular}{rl}
subfigure & description \\\hline
(a) & conditioning on the individual's age and the average household income
too \\
(b) & conditioning on the average household income and the individual's age
too \\
(c) & the average household income versus the individual's age \\
(d) & the individual's age versus the average household income \\
(e) & conditioning on the percent married and the average household income
too \\
(f) & conditioning on the average household income and the percent married
too \\
(g) & the average household income versus the percent married \\
(h) & the percent married versus the average household income \\\hline
& The points' intensities indicate the total ordering given
by the Hilbert curve. \\
& The grayscale plots, which compare a subpopulation to the full population, \\
(c), (d), (g), (h) &
use large points for the subpopulation and small for the full population. \\
& The color plots, which compare two subpopulations (numbered 0 and 1), \\
& use blue points for subpopulation 0 on top of red points for subpopulation 1.
\end{tabular}
\end{center}
\end{table}

\begin{table}
\caption{Description of subfigures for Figures~\ref{los_angeles}--\ref{napa};
all variates pertain to the Census-defined individual household
under consideration.}
\label{labelsw}
\begin{center}
\begin{tabular}{rl}
subfigure & description \\\hline
(a) & conditioning on the log of the income
and on the duration since last moving \\
(b) & conditioning on the log of the income
and on the number of children \\
(c) & the duration since last moving versus the logarithm of the income \\
(d) & the number of children versus the logarithm of the income \\
(e) & conditioning on the duration since last moving,
on the number of children, \\
& and on the logarithm of the income, in that order \\
(f) & conditioning on the number of children,
on the duration since last moving, \\
& and on the logarithm of the income, in that order \\\hline
& The points' intensities indicate the total ordering given
by the Hilbert curve. \\
(c), (d) & The grayscale plots, which compare a subpopulation
to the full population, \\
& use large points for the subpopulation and small for the full population.
\end{tabular}
\end{center}
\end{table}

\clearpage

\clearpage

\subsection{Future outlook}
\label{outlook}

This subsection discusses potential alternatives to and generalizations of
the space-filling curves used in the present paper.

The Hilbert curve is ideal for adhering to the geometry of the data given
by the $L^1$ metric of the ambient space of covariates
(the $L^1$ distance is also known as the ``Manhattan,'' ``city block,''
or ``taxicab'' metric, as reviewed, for example,
by~\cite{cormen-leiserson-rivest-stein}).
The Hilbert curve avoids mixing together different coordinates
as the Euclidean metric does.
However, the Hilbert curve is agnostic to the intrinsic geometry
of the data set, except insofar as applying the inverse mapping $g$
from Subsection~\ref{Hilbert} encodes the density of data points
in various parts of the ambient space.
One way to account more extensively for the intrinsic geometry of the data
would be to find an approximate solution to the traveling salesman problem,
yielding a total order for the data points, as reviewed, for example,
by~\cite{cormen-leiserson-rivest-stein}.
In the examples above which condition on one discrete variable
together with one variable that takes on a fairly dense set of values
continuously throughout its range, the solution
to the traveling salesman problem yields the snake raster ordering.
The snake raster scan simply stratifies the data into separate slices,
one for each value of the discrete variable, then concatenates these slices
together in the order given by the values of the discrete variable.
The methods of Subsections~\ref{subpop} and~\ref{subpops} thus degenerate
when using the ordering from the solution to the traveling salesman problem,
essentially analyzing each stratum
(one for each value of the discrete covariate) separately.
Nevertheless, using an approximate solution to the traveling salesman problem
does account for the intrinsic geometry of the data set and may be helpful
when conditioning on many covariates or with other data sets that are
very sparse. Still, the degeneration of the solution
to the traveling salesman problem in even the simplest real examples
given above indicates that more meaningfully accounting
for the intrinsic geometry of the data set might require better ideas.
Perhaps future work will address this issue.

\section{Conclusion}
\label{conclusion}

Traditional methods of controlling for or conditioning on specified covariates
depend on fairly arbitrary and often debatable choices,
whether using binning, separation, segmentation, stratification, or smoothing,
or using regression analysis via parametric or semi-parametric modeling.
In contrast, the graphical method and scalar summary statistics
proposed in the present paper are fully automatic,
unique to the extent that the Hilbert curve is unique,
that is, unique modulo the ordering of the dimensions
corresponding to the covariates.
The Hilbert curve induces a total ordering on the values
taken by the vector of covariates, naturally yielding one-dimensional scores
for use in the methodology of~\cite{tygert_full} and~\cite{tygert_two}
(the methodology of~\cite{tygert_full} and~\cite{tygert_two}
is tailor-made for one-dimensional scores);
the continuity of the curve as a mapping from one dimension
to multiple dimensions ensures that the cumulative approach
of~\cite{tygert_full} and~\cite{tygert_two} works nicely
for conditioning on the multiple covariates, too.
The Hilbert curve does depend on the ordering of the dimensions
associated with the covariates, yet the empirical results
of Section~\ref{results} above indicate that this ordering
has relatively little impact.
Thus, the graphical method and summary statistics of the present paper
leave no parameters to tune or knobs to twist,
aside from setting the ordering of the covariates,
and the ordering of the covariates appears to have fairly little impact
in experiments, in any case.
The methodology of the present paper is fully non-parametric
and fully automated.

\section*{Acknowledgements}

We would like to thank Mike Rabbat and Arthur Szlam for helpful discussions.

\newpage

\bibliography{multidim}
\bibliographystyle{vancouver}

\end{document}